\pdfoutput=1
\documentclass[12pt,letterpaper]{article}
\usepackage[utf8]{inputenc}
\usepackage{amsmath,amssymb,amsfonts,amsthm,mathtools}
\usepackage{bm,latexsym,graphicx,url,float,xcolor}
\usepackage{setspace}
\usepackage{sectsty}
\sectionfont{\centering}
\subsectionfont{\centering}
\usepackage[top=1in,bottom=1in,left=1in,right=1in]{geometry}
\usepackage[colorlinks=true,linkcolor=blue,citecolor=blue,urlcolor=blue,hypertexnames=false]{hyperref}
\usepackage[authoryear]{natbib}

\makeatletter
\newif\if@suppcites
\newwrite\supp@aux
\def\supp@routecites#1{%
  \@for\supp@key:=#1\do{%
    \if@suppcites
      \@ifundefined{supp@main@\supp@key}%
        {\immediate\write\supp@aux{\string\citation{\supp@key}}}%
        {\immediate\write\@auxout{\string\citation{\supp@key}}}%
    \else
      \expandafter\gdef\csname supp@main@\supp@key\endcsname{}%
      \immediate\write\@auxout{\string\citation{\supp@key}}%
    \fi}}
\renewcommand\NAT@sort@cites[1]{%
  \let\NAT@cite@list\@empty
  \@for\@citeb:=#1\do{\expandafter\NAT@star@cite\@citeb\@@}%
  \if@filesw\expandafter\supp@routecites\expandafter{\NAT@cite@list}\fi
  \@ifnum{\NAT@sort>\z@}{%
    \expandafter\NAT@sort@cites@\expandafter{\NAT@cite@list}%
  }{}%
}
\newcommand\StartSupplementCitations{%
  \if@filesw
    \immediate\openout\supp@aux=\jobname-supp.aux
    \immediate\write\supp@aux{\string\relax}%
    \immediate\write\supp@aux{\string\bibstyle{chicago}}%
    \immediate\write\supp@aux{\string\bibdata{reference-gen}}%
  \fi
  \global\@suppcitestrue}
\newcommand\SupplementReferences{%
  \if@filesw\immediate\closeout\supp@aux\fi}
\makeatother

\numberwithin{equation}{section}
\theoremstyle{plain}
\newtheorem{theorem}{Theorem}[section]
\newtheorem{assumption}{Assumption}[section]
\newtheorem{proposition}{Proposition}[section]

\newtheorem{lemma}{Lemma}[section]
\newtheorem{fact}{Fact}[section]
\theoremstyle{definition}
\newtheorem{algorithm}{Algorithm}[section]

\newtheorem{remark}{Remark}[section]
\theoremstyle{plain}

\newcounter{bean}

\newcommand{\Xone}{\mathcal{X}_1}
\newcommand{\Xtwo}{\mathcal{X}_2}
\newcommand{\dom}{\Xone\times\Xtwo}
\newcommand{\E}{E}
\DeclareMathOperator{\Var}{Var}
\DeclareMathOperator{\Cov}{Cov}
\newcommand{\ind}{\mathbf{1}}
\newcommand{\dto}{\stackrel{d}{\rightarrow}}
\newcommand{\pto}{\stackrel{p}{\rightarrow}}
\newcommand{\tauh}{\widehat{\tau}}
\newcommand{\Dh}{\widehat{D}}
\newcommand{\Sh}{\widehat{S}}
\newcommand{\Gammah}{\widehat{\Gamma}}
\newcommand{\ah}{\widehat{m}}
\newcommand{\sigh}{\widehat{\sigma}}
\newcommand{\rhoh}{\widehat{\varrho}}
\newcommand{\Psiabs}{\Psi_{\mathrm{abs}}}
\newcommand{\Psimax}{\Psi_{\mathrm{max}}}
\newcommand{\Csym}[1]{\mathcal{C}_{#1}^{\mathrm{sym}}}
\newcommand{\Cord}[1]{\mathcal{C}_{#1}^{\geq}}
\newcommand{\bwc}{k}
\newcommand{\projfield}{\Pi}
\makeatletter
\newcommand{\eqlabelprev}[1]{{\protected@edef\@currentlabel{\theequation}\label{#1}}}
\makeatother

\providecommand{\keywords}[1]{\textit{Keywords:} #1}

\title{Asymptotic Theory  of Integrated Pairwise-Difference Functionals of Kernel Estimators, with Applications to  Treatment Effect Heterogeneity}
\author{Julius Owusu\thanks{Department of Economics, Concordia University. Email: julius.owusu@concordia.ca.}}
\date{November 2026}

\begin{document}
\onehalfspacing
\maketitle
\addtocontents{toc}{\protect\setcounter{tocdepth}{-2}}
\begin{abstract}
\thispagestyle{empty}
This paper develops asymptotic theory for integrated pairwise-difference statistics constructed from kernel estimators, with applications to treatment effect heterogeneity. The statistics integrate absolute or positive-part differences between kernel estimates evaluated at pairs of points. Their two-point structure falls outside existing one-point \(L_1\)-type functional theory. We establish asymptotic normality using a Hájek-type projection of a Poissonized statistic and develop feasible centering and variance estimators. We consider two applications. First, we test whether treatment effects vary with one covariate block while holding another fixed. Second, we establish root-\(N\) inference for the directional Gini mean difference of the conditional treatment effect function, which measures heterogeneity in outcome units. The framework also accommodates tests of monotone effect modification and estimated pseudo-outcomes. Simulations corroborate the asymptotic theory and show good finite-sample performance. An application to 401(k) eligibility and net financial assets illustrates the usefulness of the proposed tests.
\vspace{0.3cm}
\\
\end{abstract}

\setlength{\parindent}{3ex}
\newpage

\setcounter{page}{1}
\doublespacing
\section{Introduction}\label{sec:intro}

Many questions in econometrics concern how an unknown function varies across its arguments. Examples include whether treatment effects vary with individual characteristics \citep{crump2008nonparametric}, whether particular covariates contribute to a conditional mean function \citep{lavergne2000nonparametric}, and how household demand responds to expenditure and demographic characteristics \citep{blundell2003nonparametric}. Integrated pairwise difference functionals naturally quantify such variation, capturing both magnitude and direction. When the underlying function is estimated nonparametrically, however, the asymptotic properties of these functionals are not immediate. Unlike conventional integrated functionals of kernel estimators evaluated at a single point, integrated pairwise-difference statistics involve jointly smoothed estimates at two points and therefore fall outside existing one-point \(L_1\)-type functional asymptotic theory. This paper develops asymptotic theory for integrated pairwise-difference statistics constructed from kernel estimators. The results provide tools to test and measure variation in unknown functions, with treatment effect heterogeneity as a leading application.

In the analysis of treatment effect heterogeneity, an important question is not simply whether treatment effects vary, but which covariates account for that variation. Identifying these covariates is important for constructing welfare-maximizing treatment assignment rules and assessing the external validity of treatment effect estimates. If treatment effects do not vary with one block of covariates conditional on another, a treatment rule that already uses the second block gains nothing from incorporating the first. Similarly, transporting an average treatment effect to another population requires reweighting only on the second block, provided the conditional treatment effect is transportable \citep{dahabreh2020extending}. Let \(X=(X_1,X_2)\) denote a partition of covariates and \(\tau(x_1,x_2)\) the conditional average treatment effect (CATE). As an application of the asymptotic theory developed below, we test whether \(\tau(x_1,x_2)\) varies with \(x_1\) holding \(x_2\) fixed, and conversely, denoting these hypotheses by \(H_0^1\) and \(H_0^2\). This conditional approach addresses a fundamental limitation of tests based on marginal treatment effects, such as \(\tau^{\mathrm{marg}}_1(x_1)=\E[\tau(X_1, X_2)\mid X_1=x_1]\), which average over one covariate block before assessing variation in the other. Such marginalization can conceal genuine treatment effect heterogeneity or attribute variation to covariates that do not independently modify treatment effects. For example, suppose \(\tau(x_1,x_2)=1+\beta(x_1-1/2)(x_2-1/2)\), where $\beta\neq 0$, and \(X_1\) and \(X_2\) are independent and uniformly distributed. Although treatment effects vary with both covariates, both marginal treatment effect functions are constant at one, so tests based on these functions, hereafter referred to as marginal tests, fail to detect this heterogeneity. Conversely, if treatment effects depend only on \(X_2\) and the covariate blocks are correlated, \(\tau^{\mathrm{marg}}_1(x_1)\) can vary with \(x_1\) even though \(X_1\) does not independently modify treatment effects. In our simulations, this leads a marginal test to spuriously attribute heterogeneity to \(X_1\). The integrated pairwise difference statistics avoid these shortcomings by assessing treatment effect variation across values of one covariate block while holding the other fixed, rather than averaging over it.

The proposed statistics integrate absolute or positive-part differences between kernel estimates of \(\tau\) evaluated at different values of one covariate block while holding the other fixed. For absolute differences, the population counterpart is a directional Gini mean difference \citep{yitzhaki2003gini} of the functions \(\tau(\cdot,x_2)\), which equals zero if and only if \(H_0^1\) holds. For positive-part differences over ordered pairs, it equals zero if and only if these functions satisfy the corresponding monotonicity restriction. Beyond hypothesis testing, the absolute-difference functional measures treatment-effect heterogeneity in outcome units and corresponds to a directional Gini mean difference of the CATE function. 

The paper makes four main contributions. First, we establish asymptotic normality and consistency for the two-point \(L_1\)-type functionals. Because each integrand involves kernel estimates at two potentially distant points, the arguments used to establish existing one-point \(L_1\) asymptotic results do not apply directly. The difficulty is analogous to the distinction between
the asymptotic theory of sample averages and that of order-two U-statistics. Once the sample size is Poissonized, the integrands of a one-point $L_1$ statistic are independent whenever their kernel windows are disjoint (eg., see \citealp{gine2003bm}). Our statistics
instead integrate over pairs of evaluation points, so two integrands can remain dependent when just one window from each pair overlaps, regardless of the separation between the remaining points, as with two U-statistic terms sharing one index. These pairs determine the leading variance and rule out the one-point theory's blocking argument. We overcome this difficulty through a Hájek-type projection of the Poissonized statistic that reduces the pairwise structure to terms depending on individual evaluation points. A finite-sample centering correction for bias, combined with a variance estimator that uses realized kernel weights, improves the accuracy of the normal approximation. Second, we establish consistency and asymptotic normality for integrated pairwise-difference statistics based on pseudo-outcomes with estimated nuisance parameters, providing conditions under which first-stage estimation does not affect their limiting distributions under either $H_0^1$ or $H_0^2$. A drift lemma shows that, under these null hypotheses, first-stage estimation error affects the absolute-difference statistic only at quadratic order, whereas one-point positive-part statistics such as that of \citet[Theorem~4.3]{shi2019testing} are affected at first order. These results accommodate a root-$N$-consistent parametric first-stage estimator on the same sample, provided either a doubly robust working model or the propensity model for the inverse probability weighted pseudo-outcome is correctly specified, without requiring an asymptotically linear representation of the first-stage estimator. They also accommodate nuisance functions estimated on an independent sample under a product-rate condition distinct from the root-$N$ requirement in double machine learning \citep{chernozhukov2018double}. In both settings, the resulting tests retain their asymptotic level. Third, we develop a multiple testing procedure to determine which covariate blocks account for variation in treatment effects by jointly testing $H_0^1$ and $H_0^2$, while controlling the familywise error rate (FWER). Rejecting \(H_0^1\) provides evidence that \(X_1\) modifies treatment effects conditional on \(X_2\). Fourth, we develop bounds on the magnitude of treatment effect variation in outcome units and establish root-\(N\) inference for the directional Gini mean difference.

The hypotheses $H_0^1$ and $H_0^2$ coincide with those studied in nonparametric significance testing \citep{delgado2001significance,fanli1996consistent,lavergne2000nonparametric,lavergne2015significance} and conditional mean independence testing \citep{lundborg2024projected,scheidegger2022weighted}, when the treatment effect pseudo-outcome is used as the response variable, as in the tests of effect invariance across environments of \citet{saengkyongam2023effect}. Among recent methods, the projected covariance measure (PCM) of \citet{lundborg2024projected} and the weighted generalized covariance measure (WGCM) of \citet{scheidegger2022weighted} are state of the art in conditional mean independence testing. These methods can detect alternatives at faster rates in directions captured by their respective statistics. Both estimate a direction (a projection for PCM or a weight function for WGCM) on one subsample and evaluate the test statistic on another, using sample splitting to ensure the independence required for validity.
Our proposed statistics instead have nondegenerate null limiting distributions \textit{without} sample splitting, allowing every observation to contribute to both CATE estimation and testing, without making the result depend on a random partition of the sample. In simulations, our joint $L_1$ pairwise absolute-difference test has competitive power in settings favorable to PCM and WGCM, while remaining well calibrated and powerful in settings where they perform less well. The WGCM, whose target is a weighted covariance between $X_1$ and the CATE, has no power when the effect varies symmetrically in $X_1$, whereas our proposed joint test detects this heterogeneity and attributes it to $X_1$; and with the additive learners of its replication code, the PCM has no power against a pure interaction between the two blocks, which our test also detects. With both known and estimated pseudo-outcomes, the proposed test maintains rejection probabilities close to the nominal level under the null, whereas the PCM, even when aggregated over multiple random splits to reduce split dependence, is markedly conservative.

The $L_1$ pairwise absolute-difference framework also yields a direct measure of the magnitude of treatment effect heterogeneity, a one-sided statistic for testing monotone effect modification, and asymptotic level control when nuisance parameters are estimated on the same sample.  In an application to the effect of 401(k) eligibility on net financial assets \citep{chernozhukov2004effects}, treatment effects vary with income holding age fixed across all specifications, and with age holding income fixed at every recommended bandwidth. After accounting for the noise-induced bias, the estimated magnitude of income-related heterogeneity, approximately 2,700 USD, is distinguishable from zero, whereas that of age-related heterogeneity is not.

Our application to treatment effect heterogeneity attribution also relates to the broader literature on testing treatment effect heterogeneity. Most existing tests of treatment effect heterogeneity consider either an omnibus null, under which the CATE is constant across all covariates, or a marginal null, under which the CATE averaged over the remaining covariates is constant in a chosen subvector. The best linear predictor of \citet{chernozhukov2018generic}, whose proxy is learned on an auxiliary sample and whose inference is aggregated over many sample splits, and the rank-weighted average treatment effects of \citet{yadlowsky2025evaluating}, both built on a proxy learned from all covariates, are omnibus in that use. The integrated conditional moment tests of \citet{lu2026orthogonal} and \citet{lapenta2026machine} and the linear-functional tests of \citet{dukes2026nonparametric} test constancy given a chosen subvector, with estimated nuisances; they are omnibus when the subvector is $(X_1, X_2)$ and marginal when it is $X_1$. \citet{chang2015nonparametric} and \citet{hsu2017consistent} test zero, sign and dominance nulls. Leave-covariate-out importance measures (\citealp{hines2022variable} and \citealp{williamson2023general}) share the nulls $H_0^1$ and $H_0^2$ but have an influence function that degenerates at them, and the sign test of \citet{shi2019testing}, whose null is implied by $H_0^1$, has rejection probability tending to zero at every point of $H_0^1$. Level results with estimated nuisances exist for a one-point kernel statistic with a parametric first stage \citep{shi2019testing} and for a Kolmogorov--Smirnov statistic of inverse-probability-weighted moments indexed by hypercubes, with a series logit propensity score estimated on the same sample \citep{hsu2017consistent}. Estimators of the CATE on a subvector with an estimated first stage are given by \citet{abrevaya2015estimating}, \citet{fan2022estimation}, \citet{lee2017doubly} and \citet{semenova2021debiased}. Our general statistic also accommodates tests of monotone effect modification, relating it to tests of regression monotonicity \citep{chetverikov2019testing,ghosal2000testing}. Our asymptotic analysis builds on the theory of integrated $L_1$ statistics of kernel estimators evaluated at individual points \citep{gine2003bm,chang2015nonparametric,lee2013testing}, which we extend to pairwise differences. Our asymptotic proofs rely on the modern Poissonization technique employed in the one-point integrated $L_1$ theory of \citet{gine2003bm}, \citet{chang2015nonparametric} and \citet{lee2013testing} and H\'ajek-type projection \citep[Chap.~11]{vanderVaart2000asymptotic}.

The remainder of the paper is organized as follows. Section~\ref{sec:stat} sets out the setting, the hypotheses, and the test statistics. Section~\ref{sec:asym} develops the asymptotic theory: the centering, the limit variances, asymptotic normality, size and consistency, the extension to estimated nuisance parameters (Section~\ref{sec:asym-nuisance}), inference on the magnitudes (Section~\ref{sec:app-gini}) and the implementation (Section~\ref{sec:impl}). Section~\ref{sec:mc} reports Monte Carlo evidence on size, power, and attribution, and on the coverage of the magnitude intervals. Section~\ref{sec:app-emp} applies the tests to the effect of 401(k) eligibility on net financial assets, and Section~\ref{sec:concl} concludes. The proofs of the asymptotic results, the results for estimated nuisance parameters, and the magnitude results appear in Sections~\ref{osec:proofs}, \ref{osec:proofs-nui}, and \ref{osec:proofs-gini} of the Online Appendix. The Online Appendix contains auxiliary lemmas, further simulation designs and implementation details; its sections, equations, results, tables and figures are numbered with the prefix S.

\section{Setting, hypotheses and statistics}\label{sec:stat}

\subsection{Setup and estimators}\label{sec:stat-setup}
The data are $W_i=(Y_i, T_i, X_{1i}, X_{2i}, V_i)$, $i\le N<\infty$, with a binary treatment $T_i\in\{0,1\}$, an real-valued outcome $Y_i\in\mathcal{Y}\subset\mathbb R$, covariates $X_i=(X_{1i},X_{2i})\in\dom\subset\mathbb R^{d_1}\times\mathbb R^{d_2}$ along which heterogeneity is assessed, partitioned by the analyst into two continuously distributed blocks that need not be independent, and further covariates $V_i$ needed for identification, possibly none. Let $d:=d_1+d_2$, and suppose $\Xone,\Xtwo$ are compact with nonempty interior. With $Y(t)$ the potential outcome under treatment $t$, the object of interest is
\begin{equation}\label{eq:cate}
\tau(x_1,x_2):=\E\big[Y(1)-Y(0)\mid X_1=x_1,X_2=x_2\big],\qquad (x_1,x_2)\in\dom .
\end{equation}

\begin{assumption}[Sampling]\label{ass:sampling}
$\{W_i\}_{i=1}^N$ are independent and identically distributed draws from the law of $W$.
\end{assumption}

\begin{assumption}[SUTVA and unconfoundedness]\label{ass:unconf}
(a) $Y=TY(1)+(1-T)Y(0)$; (b) $(Y(1),Y(0))$ is independent of $T$ conditionally on $(X_1,X_2,V)$.
\end{assumption}

\begin{assumption}[Overlap]\label{ass:overlap}
$(X_1,X_2)$ has a Lebesgue density $f$, and for some $\underline p\in(0,1/2)$, $r_t(x,v):=\Pr(T=t\mid X=x,V=v)\ge\underline p$, $t\in\{0,1\}$, and $f(x)\ge\underline p$ for all $x$ in a neighborhood of $\dom$.
\end{assumption}
Imposing Assumption \ref{ass:overlap} on a neighborhood of the evaluation region avoids boundary bias and the ensuing corrections. With $\mu_t(x,v):=\E[Y\mid T=t, X=x, V=v]$, the \textit{inverse probability weighted (IPW)} and \textit{doubly robust (DR)} pseudo-outcomes are
\begin{equation}\label{eq:psi-ipw}
\zeta^{\mathrm{IPW}}:=\Big[\frac{\ind(T=1)}{r_1}-\frac{\ind(T=0)}{r_0}\Big]Y,\quad
\zeta^{\mathrm{DR}}:=\mu_1-\mu_0+\frac{\ind(T=1)(Y-\mu_1)}{r_1}-\frac{\ind(T=0)(Y-\mu_0)}{r_0},
\end{equation}
\eqlabelprev{eq:psi-dr}with $r_t,\mu_t$ evaluated at $(X,V)$ and  $\zeta$ denotes either.

\begin{proposition}[Identification]\label{prop:id}\label{prop:ipw-reduced}\label{lem:ipw-full}
Under Assumptions~\ref{ass:sampling}--\ref{ass:overlap}, $\tau(x)=\E[\zeta\mid X=x]$ for every $x\in\dom$, and  without $V$, $\tau(x)=\E[Y\mid T=1,X=x]-\E[Y\mid T=0,X=x]$.
\end{proposition}

The proof is in Section~\ref{osec:id} of the Online Appendix. Because the CATE conditional on \(X\) averages over \(V\), variation in this CATE implies that the CATE conditional on \((X, V)\) is not constant, but does not establish which covariates drive that variation. Conversely, a constant CATE conditional on \(X\) does not rule out heterogeneity conditional on \((X,V)\), since averaging over \(V\) may conceal it. 

Let $K_1,K_2$ be product kernels on $\mathbb R^{d_1}$ and $\mathbb R^{d_2}$ with bandwidths $h_1,h_2$, $\eta:=h_1^{d_1}h_2^{d_2}$\label{eq:eta} the joint smoothing volume, $\mathcal K(x;X):=K_1((x_1-X_1)/h_1)K_2((x_2-X_2)/h_2)$ and $\mathcal K_i(x):=\mathcal K(x;X_i)$. We estimate the CATE using Nadaraya--Watson kernel regression of the pseudo-outcome, smoothing jointly over both covariate blocks, i.e., 
\begin{equation}\label{eq:tauhat-psi}
\tauh(x):=\sum_{i=1}^N\ell_i(x)\zeta_i,\qquad \ell_i(x):=\mathcal K_i(x)\Big/\sum_{l=1}^N\mathcal K_l(x),
\end{equation}
or, without $V$, by
\begin{equation}\label{eq:tauhat}
\tauh(x):=\frac{1}{N\eta}\sum_{i=1}^{N}Y_i\Big[\frac{T_i}{\widehat P_1(x)}-\frac{1-T_i}{\widehat P_0(x)}\Big]\mathcal K_i(x),\qquad \widehat P_t(x):=\frac1{N\eta}\sum_{i=1}^N\ind(T_i=t)\mathcal K_i(x),
\end{equation}
where $\widehat P_t$ estimates $P_t(x):=\Pr(T=t\mid X=x)f(x)$\eqlabelprev{eq:Pt}. The pointwise scaled variance of \eqref{eq:tauhat} is
\begin{equation}\label{eq:rho}
\varrho(x):=\lim_{N\to\infty}N\eta\Var\big(\tauh(x)\big)=\Big[\sum_{t\in\{0,1\}}\frac{\varsigma_t^2(x)}{P_t(x)}\Big]\int K_1^2\int K_2^2 ,
\end{equation}
where $\varsigma_t^2(x):=\Var(Y\mid X=x,T=t)$ and $\int K_j^2:=\int_{\mathbb R^{d_j}}K_j(u)^2\,du$, $j=1,2$, is the integrated squared kernel. For \eqref{eq:tauhat-psi} it is $\varrho_\zeta:=\varsigma_\zeta^2(x)f(x)^{-1}\int K_1^2\int K_2^2$, where $\varsigma_\zeta^2(x):=\Var(\zeta\mid X=x)$.

\subsection{Hypotheses}\label{sec:app-hom}
For treatment effect heterogeneity attribution, we consider directional nulls 
\begin{equation}\label{eq:H01}
H_0^1:\ \tau(x_1,x_2)=\tau(x_1',x_2)\ \ \forall\,x_1,x_1',x_2,\qquad
H_0^2:\ \tau(x_1,x_2)=\tau(x_1,x_2')\ \ \forall\,x_1,x_2,x_2' ,
\end{equation}
with $x_j,x_j'\in\mathcal X_j$. Under $H_0^1$, $X_1$ does not modify the treatment effect at fixed $X_2$, though the effect may vary with $X_2$,  with the mirror $H_0^{2}$. The monotone null $H_0^{1,\ge}$ is that $\tau(\cdot,x_2)$ is non-increasing in $X_1$, componentwise, for every $X_2$, with the mirror $H_0^{2,\ge}$.

When \(X_1\) and \(X_2\) are \textit{independent}, \(H_0^1\) is stronger than constancy of the marginalized CATE function \(\tau_1^{\mathrm{marg}}(x_1):=\E[\tau(x_1,X_2)\mid X_1=x_1]\), since averaging over \(X_2\) can conceal variation in the CATE across values of \(X_1\). In particular, variation in different directions across values of \(X_2\) can cancel upon integration. For instance, this occurs when
 \begin{equation}\label{eq:interaction-example}
 \tau(x_1,x_2)=1+\beta\,(x_1-\tfrac12)(x_2-\tfrac12),\qquad X_1\perp X_2,\qquad X_j\sim U[0,1],\qquad \beta\neq0 .
 \end{equation}
Both marginalized CATE functions are constant at one even though \(H_0^1\) and \(H_0^2\) are false. Consequently, tests of marginal constancy using either covariate block cannot detect the heterogeneity, whereas tests using both blocks can detect it without identifying which block contributes to the variation. 

Thus, to identify which covariate blocks contribute to the variation, we recommend testing $H_0^1$ and $H_0^2$ jointly using the stepdown multiple testing procedure (MTP) of \citet{holm1979simple}, which controls the FWER regardless of dependence between the tests. Let $p_{(1)}\le p_{(2)}$ denote the ordered $p$-values. Reject the hypothesis associated with $p_{(1)}$ if $\tilde p_{(1)}:=2p_{(1)}\le\alpha$, and reject the other if, in addition, $\tilde p_{(2)}:=\max\{2p_{(1)},p_{(2)}\}\le\alpha$. A rejection of $H_0^1$ establishes, at the familywise level, that $X_1$ modifies the effect at fixed $X_2$, so that a rule or a reweighting for external validation that uses $X_2$ should also use $X_1$; rejecting both nulls puts both blocks in the rule. %

\subsection{The statistics and the magnitudes}\label{sec:stat-stats}
Let $\Gamma_1(x_1,x_1',x_2):=\tau(x_1,x_2)-\tau(x_1',x_2)$ and $\Gamma_2(x_1,x_2,x_2'):=\tau(x_1,x_2)-\tau(x_1,x_2')$\label{eq:Gamma} be the pairwise-difference functions and $\Gammah_1,\Gammah_2$ their estimates. A configuration $c$ is a triple at which $\Gamma_j(\cdot)$ is evaluated, with evaluation points $c_1,c_2\in\dom$, so that $\Gamma_j(c)=\tau(c_1)-\tau(c_2)$. A configuration set is a measurable  $\mathcal C_1\subseteq\Xone^2\times\Xtwo$ or $\mathcal C_2\subseteq\Xone\times\Xtwo^2$ with nonempty interior. We consider the symmetric sets $\Csym1:=\Xone^2\times\Xtwo$ and $\Csym2:=\Xone\times\Xtwo^2$, and the ordered sets $\Cord1:=\{(x_1,x_1',x_2)\in\Csym1:x_1\ge x_1'\}$ and $\Cord2:=\{(x_1,x_2,x_2')\in\Csym2:x_2\ge x_2'\}$, where inequalities are interpreted componentwise. We define the \textit{band of $\mathcal C_j$ }as the subset $\{c\in\mathcal C_j:\|c_1-c_2\|_\infty\le h_j\}$, consisting of configurations whose two kernel windows overlap, and has Lebesgue measure $O(h_j^{d_j})$.

\begin{assumption}[Class of integrands]\label{ass:psi}
$\Psi:\mathbb R\to\mathbb R$ is (a) Lipschitz with constant one; (b) positively homogeneous of degree one, $\Psi(su)=s\,\Psi(u)$ for $s>0$; and (c) such that $|\Psi(u)|\le|u|$.
\end{assumption}

The leading members are $\Psiabs(u):=|u|$ and $\Psimax(u):=\max\{u,0\}$, but the class excludes $\Psi_{\mathrm{sq}}(u)=u^2$. Homogeneity fixes $\Psi$ on each half-line, so every $\Psi$ satisfying (b) and (c) is of the form
\begin{equation}\label{eq:stat-psi-rep}
\Psi(u)=a_\Psi|u|+b_\Psi u,\qquad a_\Psi:=\tfrac12\{\Psi(1)+\Psi(-1)\},\quad b_\Psi:=\tfrac12\{\Psi(1)-\Psi(-1)\},
\end{equation}
with $|a_\Psi\pm b_\Psi|\le1$, and (a) is automatic; $\Psiabs$ has $(a_\Psi,b_\Psi)=(1,0)$ and $\Psimax$ has $(\frac12,\frac12)$. With $Z,Z_1,Z_2$ independent standard normals, the constants $m_\Psi:=\E[\Psi(Z)]$ and $g_\Psi(\rho):=\Cov(\Psi(\sqrt{1-\rho^2}Z_1+\rho Z_2),\Psi(Z_2))$ are, 
\begin{equation}\label{eq:stat-g}
m_\Psi=a_\Psi\sqrt{2/\pi},\quad g_\Psi(\rho)=a_\Psi^2\,g_{\mathrm{abs}}(\rho)+b_\Psi^2\,\rho,\quad g_{\mathrm{abs}}(\rho)=\tfrac{2}{\pi}\big\{\rho\arcsin\rho+\sqrt{1-\rho^2}-1\big\},
\end{equation}
since $|Z|$ and $Z$ are uncorrelated. The closed form of $g_{\mathrm{abs}}(\rho)$ is due to \citet{nabeya1951absolute}. 

For the class of integrands specified in Assumption~\ref{ass:psi}, our general statistic of interest and its population target are
\begin{equation}\label{eq:Dgen}
\Dh_j^{\Psi}(\mathcal C_j):=\sqrt N\int_{\mathcal C_j}\Psi\big(\Gammah_j(c)\big)\,dc,\qquad D_j^{\Psi}(\mathcal C_j):=\int_{\mathcal C_j}\Psi\big(\Gamma_j(c)\big)\,dc,\qquad j=1,2 .
\end{equation}
The leading forms are the absolute statistics on the symmetric sets, 
\begin{equation}\label{eq:D1}
\Dh_1:=\tfrac12\Dh_1^{\Psiabs}(\Csym{1})=\tfrac12\int_{\Xtwo}\int_{\Xone^2}\sqrt N\,\big|\tauh(x_1,x_2)-\tauh(x_1',x_2)\big|\,dx_1\,dx_1'\,dx_2,
\end{equation}
\eqlabelprev{eq:D2}its mirror image $\Dh_2$, with targets $D_j:=\tfrac12D_j^{\Psiabs}(\Csym{j})$, and the positive-part statistics on the ordered sets, $\Dh_j^{\max}:=\Dh_j^{\Psimax}(\Cord{j})$\label{eq:Dmax}. As $\tau$ is continuous under the regularity conditions of Section~\ref{sec:asym} below, $D_j=0$ if and only if $H_0^j$ holds, and $D_j^{\Psimax}(\Cord{j})=0$ if and only if $H_0^{j,\ge}$ holds. In addition, since $(u)_+:=\max\{u,0\}$ satisfies $(u)_++(-u)_+=|u|$, symmetry implies $\Dh_j^{\Psimax}(\Csym{j})=\Dh_j$. Thus, the positive-part form carries distinct information only on ordered sets.\label{rem:symmetric-max}

\begin{remark}[Weighting of pairwise differences]\label{rem}
Existing one-point \(L_1\) statistics commonly use the pointwise inverse-standard-error of CATE estimators as weights to achieve asymptotic pivotality \citep[cf.][]{chang2015nonparametric,shi2019testing}. For our proposed pairwise-difference statistics, however, this weight is undefined on the diagonal \(c_1=c_2\), where the standard error of \(\widehat\Gamma_j(c)=\widehat\tau(c_1)-\widehat\tau(c_2)\) equals zero, and may diverge near it. We therefore focus on the unweighted statistics above, which preserve the interpretation of heterogeneity in outcome units but do not account for differences in estimation precision across configurations. The centering and variance expressions below account for this heterogeneity in precision and the dependence across configurations.
\qed
\end{remark}

As noted above, the population parameter $D_1$ measures treatment effect heterogeneity in outcome units. With $\lambda$ denoting Lebesgue measure, $D_1$ equals $\frac12\lambda(\Xone)^2$ times the integral over $x_2\in\Xtwo$ of the Gini mean difference $\E|U-U'|$ \citep{yitzhaki2003gini}, where, for each fixed $x_2$, $U=\tau(X_1,x_2)$ and $U'=\tau(X_1',x_2)$, with $X_1$ and $X_1'$ independently and uniformly distributed on $\Xone$. Its density-weighted analog is defined on the evaluation subset $\mathcal R \subseteq \mathcal{X}_1\times \mathcal{X}_2$ used in practice. Let $X^{\mathcal R}$ have the law of $X$ given $X\in\mathcal R$ and $X_1'$ the law of $X_1^{\mathcal R}$ given $X_2^{\mathcal R}$, independently of $X_1^{\mathcal R}$, and
\begin{equation}\label{eq:gini-index}
\Delta_1:=\E\big|\tau(X^{\mathcal R}_1,X^{\mathcal R}_2)-\tau(X_1',X^{\mathcal R}_2)\big|,\qquad
G_1:=\frac{\Delta_1}{2\,\E|\tau(X^{\mathcal R})|}\in[0,1],
\end{equation}
with analogous definitions for $D_2$, $\Delta_2$ and $G_2$. By the triangle inequality, $G_j\le1$. When $\tau$ has the same sign throughout $\mathcal R$, $G_j$ is a mean-weighted average, over the other block, of the Gini coefficients of the conditional treatment effect distributions, the Gini coefficient of a distribution being half its mean absolute difference divided by its mean \citep{dorfman1979formula}. It measures the relative dispersion of CATEs across values of \(X_j\), holding the other covariate block fixed, before aggregating over the distribution of the latter.

\section{Asymptotic theory}\label{sec:asym}

\subsection{Regularity Conditions}\label{sec:asym-ass}
Results are stated for a direction $j$, $j'\neq j$, configuration set $\mathcal C_j$ and $\Psi$ satisfying Assumption~\ref{ass:psi}; those for $j=2$ follow upon exchanging $(x_1,\Xone,d_1,h_1,K_1)$ and $(x_2,\Xtwo,d_2,h_2,K_2)$. Let $\bar h:=\max\{h_1, h_2\}$. We require the following regularity conditions for the asymptotic results that follow.  

\begin{assumption}[Regularity]\label{ass:reg}
(a) The joint law of $(Y,X_1,X_2)$ is absolutely continuous; (b) $f$ is bounded on a neighborhood of $\dom$; (c) $\varrho$ is continuous on $\dom$ with $0<\inf_{\dom}\varrho\le\sup_{\dom}\varrho<\infty$; (d) $K_1$ and $K_2$ are products of univariate kernels of orders $s_1$ and $s_2$, supported on $[-\frac12,\frac12]$, symmetric, bounded, Lipschitz, of bounded variation and integrating to one; (e) $\bar\mu_t(x):=\E[Y\mid X=x,T=t]$, $\varsigma_t^2$, $f$ and $P_t$, $t=0,1$, are $s_1$ times continuously differentiable in $x_1$ and $s_2$ times in $x_2$ on a neighborhood of $\dom$, with bounded derivatives, $\min\{s_1, s_2\}\ge2$;\\ (f) $\E[|Y|^6\mid X=x,T=t]$ is bounded uniformly in $t$ and in $x$ in a neighborhood of $\dom$; (g) $h_1,h_2\to0$, $N(h_1^{2s_1}+h_2^{2s_2})\to0$, $N\eta^3\to\infty$ and $(N\eta^2)^{1/2}/\log N\to\infty$; (h) $h_j^{3d_j}=O(h_{j'}^{d_{j'}})$, $j=1,2$, and $\bar h^2\eta^{-1/2}+\bar hh_j^{d_j}\eta^{-1/2}\to0$.
\end{assumption}
The conditions in Assumption \ref{ass:reg} extend standard regularity assumptions for $L_1$-type functionals of kernel estimators. Thus, we discuss only those requiring further clarification and refer readers to \citet{gine2003bm} and \citet{chang2015nonparametric} for the remaining conditions.
With \(h_j\asymp N^{-\gamma}\) and \(s_1=s_2=s\), condition (g) requires \(1/(2s)<\gamma<1/\{3(d_1+d_2)\}\). This interval is nonempty if and only if \(s>3(d_1+d_2)/2\). When \(d_1=d_2=1\), fourth-order kernels permit \(\gamma=1/7\). When \(d_1+d_2=3\), sixth-order kernels permit, for example, \(\gamma=1/10\), with condition (h) also satisfied under the stated assumptions. Condition (h) adds a separate limit: with $h_1\asymp h_2$, $\bar h^2\eta^{-1/2}\to0$ requires $d_1+d_2\le3$, whatever the kernel order. These restrictions make the method suitable for low-dimensional covariate blocks. Conditions (d) and (g) are those of \citet[Assumption~4.1(e),(i)]{chang2015nonparametric} for a one-point statistic, with $d=d_1+d_2$ in place of the dimension of their single covariate: with $d=1$ their second-order kernel $3/2\cdot(1-4u^2)\ind\{|u|\le1/2\}$ suffices, whereas smoothing jointly over both blocks requires $s\ge4$, and the kernel of Section~\ref{sec:impl} is its fourth-order counterpart on the same support.

\begin{proposition}[Transfer]\label{prop:transfer}
If Assumptions~\ref{ass:sampling}--\ref{ass:overlap} hold, $r_t$ and $\mu_t$ are known, and Assumption~\ref{ass:reg} holds with $\zeta$ for $Y$, $(\tau,\varsigma_\zeta^2,f)$ for $(\bar\mu_t,\varsigma_t^2,P_t)$ in (e) and $\varrho_\zeta$ for $\varrho$, all results of this section hold for \eqref{eq:tauhat-psi}, with $\varrho_\zeta$ for $\varrho$, $\varsigma_\zeta^2/f$ for $\sum_t\varsigma_t^2/P_t$ and the score $\xi(W;x):=\{\zeta-\tau(x)\}/f(x)$ below.
\end{proposition}
\subsection{Centering}\label{sec:asym-center}
Under a constancy null, $\Dh_j^\Psi(\mathcal C_j)$ diverges at the rate $\eta^{-1/2}$ and must be centered (bias corrected) at a diverging sequence. Let $\theta_N(W;x):=\eta^{-1}\xi(W;x)\mathcal K(x;X)-\E[\eta^{-1}\xi(W;x)\mathcal K(x;X)]$ be the influence function of $\tauh$ at $x$, with $\xi(W;x):=T\{Y-\bar\mu_1(x)\}/P_1(x)-(1-T)\{Y-\bar\mu_0(x)\}/P_0(x)$ for \eqref{eq:tauhat}, and $\Theta_j(W;c):=\theta_N(W;c_1)-\theta_N(W;c_2)$ that of $\Gammah_j(c)$. The variance $\kappa_{j,N}(c):=\E[\Theta_j(W;c)^2]$\label{eq:kappa} is of order $\eta^{-1}$ off the band of $\mathcal C_j$ and equal to zero on the diagonal $c_1=c_2$.  In Section~\ref{osec:aux-center} of the Online Appendix, we show that the  centering term (bias) simplifies as
\begin{equation}\label{eq:center}
m^{\kappa}_{j,N}(\mathcal C_j):=m_\Psi\int_{\mathcal C_j}\kappa_{j,N}(c)^{1/2}\,dc,
\end{equation}
which  is of order $\eta^{-1/2}$. Since $m^{\kappa}_{j,N}(\mathcal C_j)$ depends on the unknown variance, $\kappa_{j,N}(c)$, we propose a feasible plug-in estimator based on realized kernel weights and corrected fitted residuals. Details are provided in Sections~\ref{osec:from-main} and \ref{osec:centering-details}. Recall that the CATE estimator  is $\tauh(x)=\sum_i\ell_i(x)\zeta_i$, with $\ell_i$ of \eqref{eq:tauhat-psi}, or $\ell_i(x):=\mathcal K_i(x)\{T_i/\widehat P_1(x)-(1-T_i)/\widehat P_0(x)\}/(N\eta)$, $\zeta_i:=Y_i$ and residuals formed within each arm for \eqref{eq:tauhat}. With $L_{il}:=\ell_l(X_i)$ and $\widehat e_i:=\zeta_i-\sum_lL_{il}\zeta_l$, let
\begin{equation}\label{eq:center-feasible}
q_i:=(1-L_{ii})^2+\sum_{l\neq i}L_{il}^2,\qquad
\widetilde\kappa_j(c):=N\sum_{i=1}^N\{\ell_i(c_1)-\ell_i(c_2)\}^2\frac{\widehat e_i^{\,2}}{q_i},\qquad
\widetilde m_j(\mathcal C_j):=m_\Psi\int_{\mathcal C_j}\widetilde\kappa_j^{1/2}.
\end{equation}
The factor $q_i>0$ corrects for the shrinkage of fitted residuals caused by each observation's contribution to its own fitted value. This is analogous to the leverage correction in the OLS estimation literature. Under local smoothness of the conditional variance, $\E[\widehat e_i^2\mid X]\approx q_i\varsigma_\zeta^2(X_i)$, up to smoothing terms. Although $q_i=1+O_p((N\eta)^{-1})$, the correction can matter because the centering term diverges at rate $\eta^{-1/2}$. Let $\kappa^X_{j,N}(c):=N\sum_i\{\ell_i(c_1)-\ell_i(c_2)\}^2\varsigma_\zeta^2(X_i)$ and $m^X_{j,N}:=m_\Psi\int_{\mathcal C_j}(\kappa^X_{j,N})^{1/2}$. The following lemma shows that, using $\Tilde{\kappa}_{j}(c)$, the plug-in estimator of bias is consistent.

\begin{lemma}[Estimated centering]\label{prop:centering-corr}
Under the conditions of Proposition~\ref{prop:transfer}, $\widetilde m_j(\mathcal C_j)-m^X_{j,N}(\mathcal C_j)=o_p(1)$, also with $q_i$ replaced by $1$ (uncorrected), $1-L_{ii}$ (HC2) or $(1-L_{ii})^2$ (leave-one-out), and $m^X_{j,N}(\mathcal C_j)-m^\kappa_{j,N}(\mathcal C_j)=o_p(1)$.
\end{lemma}

At second order, $\widetilde m_j-m^X_{j,N}$ has two terms of order $(N\eta^{3/2})^{-1}$. The first arises from the shrinkage of fitted residuals, with coefficients $4.8$, $0.8$, $-3.1$, and $0$ for uncorrected, HC2, leave-one-out, and $q_i$-corrected residuals, respectively. Thus, the $q_i$ correction eliminates this term. The second arises from the concavity of the square root and is nonpositive. No divisor correction can eliminate it, and its magnitude increases with the pseudo-outcome's kurtosis (see Lemma~\ref{olem:centering-bias}). This centering error shifts the null distribution of the standardized statistic, explaining the location distortion observed under uncorrected calibration in Section~\ref{sec:mc}. Lemmas~\ref{prop:centering-corr} and \ref{prop:sigmaN} (below) are proved for \eqref{eq:tauhat-psi}; for \eqref{eq:tauhat} the arguments carry over with the arm-wise weights.

\subsection{Main results}\label{sec:asym-main}\label{sec:asym-proj}
For all the results that follow, suppose Assumptions~\ref{ass:sampling}--\ref{ass:reg} hold. The proofs involve \textit{linearizing} the centered estimated pairwise differences, $\Gammah_j(c)-\Gamma_j(c)=N^{-1}\sum_i\Theta_j(W_i;c)+B_{j, N}(c)+o_p(N^{-1/2})$ uniformly, with $B_{j, N}$ the smoothing bias (Lemma~\ref{olem:pf-linearization} in Section~\ref{osec:proofs}), and then \textit{Poissonizing} the sample size, so that estimators with disjoint windows are independent. Unlike one-point $L_1$ statistics, whose integrands depend on kernel estimates at individual evaluation points, our statistics compare estimates at pairs of potentially distant points. As a result, comparisons involving distant regions can remain dependent when they share an evaluation point, preventing a direct partition of the integration domain into locally dependent contributions as in existing one-point $L_1$ theory (e.g., see \cite{gine2003bm}). We overcome this difficulty by applying a \textit{Hájek-type projection} to the Poissonized statistic, which separates its leading first-order component from a remainder with variance \(O(h_j^{d_j})\). It is Hájek-type rather than an exact Hájek decomposition because the Poissonized CATE estimates at the two points of a pair are dependent when their windows overlap; such pairs form a band of configurations whose contribution to the remainder vanishes. The projected component depends on individual evaluation points and can be represented as a sum of locally dependent contributions, allowing us to establish asymptotic normality (Lemmas~\ref{olem:pf-projection} and~\ref{olem:pf-clt} in Section~\ref{osec:proofs}); the result is then \textit{de-Poissonized} as in \citet{beirlant1995asymptotic}. To the best of our knowledge, this is the first application of a Hájek-type projection within the modern Poissonization framework of \citet{gine2003bm}.
In summary, the proof for asymptotic normality proceeds through linearization, Poissonization, a Hájek-type projection that restores local dependence, a central limit theorem for the projected statistic, and de-Poissonization to recover the fixed-sample result.
\par
Let $\rho_N(c,c'):=\E[\Theta_j(W;c)\Theta_j(W;c')]\{\kappa_{j,N}(c)\kappa_{j,N}(c')\}^{-1/2}$ and $w_{\mathcal C_j}(x):=\lambda\{c\in\mathcal C_j:c_1=x\}-\lambda\{c\in\mathcal C_j:c_2=x\}$, which is zero on symmetric sets. The following theorem states the asymptotic variance of the kernel-based integrated pairwise difference statistic.

\begin{theorem}[Asymptotic variances]\label{thm:variance}
(a) The limit
\begin{equation}\label{eq:sigma-gen}
\sigma^2_{j,\Psi}(\mathcal C_j):=\lim_{N\to\infty}\int_{\mathcal C_j}\int_{\mathcal C_j}g_\Psi\big(\rho_N(c,c')\big)\big\{\kappa_{j,N}(c)\kappa_{j,N}(c')\big\}^{1/2}dc\,dc'
\end{equation}
exists, is finite, and equals $a_\Psi^2\sigma^2_{j,\mathrm{abs}}(\mathcal C_j)+b_\Psi^2\sigma^2_{j,\mathrm{lin}}(\mathcal C_j)$, where $\sigma^2_{j,\mathrm{abs}}$ is \eqref{eq:sigma-gen} with $g_\Psi(\cdot)=g_{\mathrm{abs}}(\cdot)$ and $\sigma^2_{j,\mathrm{lin}}(\mathcal C_j):=\int_{\dom}w_{\mathcal C_j}^2(x)\sum_{t}\varsigma_t^2(x)/P_t(x)$ is the limit variance of $\sqrt N\int_{\mathcal C_j}\{\Gammah_j(c)-\Gamma_j(c)\}$.\\ (b) The limit variance $\sigma_j^2$ of $\Dh_j$ is positive and has the closed form \eqref{oeq:pf-sigma2} of Section~\ref{osec:proofs}.
\end{theorem}
Although the limiting variance is well defined, it omits contributions from overlapping kernel neighborhoods and expansion terms of order \(h_1+h_2\), which may remain substantial at practical bandwidths. In our simulations, the overlap band contains \(55\%\) to \(77\%\) of the configurations, and plugging estimated quantities into the limiting variance overstates the Monte Carlo dispersion of \(\widehat D_j-\widetilde m_j\) by \(10\%\) to \(25\%\). We therefore estimate the variance using its pre-limit representation, which retains the dependence induced by overlapping kernel neighborhoods through the realized kernel weights and residuals. Specifically,  with $\widetilde\kappa_j(c,c'):=N\sum_i\{\ell_i(c_1)-\ell_i(c_2)\}\{\ell_i(c_1')-\ell_i(c_2')\}\widehat e_i^{\,2}/q_i$ and $\widetilde\rho_N:=\widetilde\kappa_j(c,c')\{\widetilde\kappa_j(c)\widetilde\kappa_j(c')\}^{-1/2}$, the pre-limit estimator of the variance is
\begin{equation}\label{eq:sigmaN}
\widetilde\sigma^2_{j,\Psi,N}(\mathcal C_j):=\int_{\mathcal C_j}\int_{\mathcal C_j}g_\Psi\big(\widetilde\rho_N(c,c')\big)\big\{\widetilde\kappa_j(c)\widetilde\kappa_j(c')\big\}^{1/2}dc\,dc' ,
\end{equation}
whose grid form is in Section~\ref{osec:grid-formulas} of the Online Appendix. Based on the foregoing bias and variance estimators, the studentized statistics are
\begin{equation}\label{eq:S}
\Sh_j^\Psi(\mathcal C_j):=\frac{\Dh_j^\Psi(\mathcal C_j)-\widetilde m_j(\mathcal C_j)}{\widetilde\sigma_{j,\Psi,N}(\mathcal C_j)},\qquad
\Sh_j:=\frac{\Dh_j-\widetilde m_j}{\widetilde\sigma_{j,N}},
\end{equation}
with $\widetilde m_j$ and $\widetilde\sigma_{j,N}$ formed on $\Csym{j}$ with the factor $1/2$ of \eqref{eq:D1}.

\begin{lemma}[Finite-sample variance]\label{prop:sigmaN}
Under the conditions of Proposition~\ref{prop:transfer}, $\widetilde\sigma^2_{j,\Psi,N}(\mathcal C_j)-\sigma^2_{j,\Psi}(\mathcal C_j)=o_p(1)$ if $\Gamma_j\equiv0$ on $\mathcal C_j$; otherwise, $\widetilde\sigma^2_{j,\Psi,N}(\mathcal C_j)$ converges in probability to the limit variance of the statistic \eqref{eq:Dgen} of the noise field $\Gammah_j-\Gamma_j$.
\end{lemma}

\begin{theorem}[Asymptotic normality]\label{thm:normality}
If $a_\Psi>0$ and $\Gamma_j\equiv0$ on $\mathcal C_j$, $\Sh_j^\Psi(\mathcal C_j)\dto N(0,1)$; in particular $\Sh_j\dto N(0,1)$ under $H_0^j$.
\end{theorem}

 Theorem \ref{thm:normality} therefore motivates the testing procedure where we reject when $p_j:=1-\Phi(\Sh_j^\Psi(\mathcal C_j))<\alpha$, with $\alpha\in (0,1)$ and $\Phi(\cdot)$ denotes the standard normal distribution function.

\begin{proposition}[Asymptotic size and consistency]\label{prop:size}\label{prop:consistency}\label{prop:monotone}
Let $a_\Psi>0$ and $\alpha\in(0,1)$. (a) If $\Gamma_j\equiv0$ on $\mathcal C_j$, $\Pr(p_j<\alpha)\to\alpha$. (b) If $D^\Psi_j(\mathcal C_j)>0$, $\Pr(p_j<\alpha)\to1$; in particular the tests based on $\Sh_1$ and $\Sh_2$ are consistent against every fixed alternative to $H_0^1$ and $H_0^2$. (c) Let $p_j^{\max}:=1-\Phi(\Sh_j^{\Psimax}(\Cord{j}))$. Under $H_0^{j,\ge}$, $\limsup_N\Pr(p_j^{\max}<\alpha)\le\alpha$, with equality when $\tau$ is constant in the $j$th block; and $\Pr(p_j^{\max}<\alpha)\to1$ if $\tau$ is strictly increasing in the $j$th block on a set of configurations of positive measure.
\end{proposition}

All statements are pointwise in the law of the data.  By part (c), the test of the monotone null is asymptotically conservative. For $d_j=1$, its centering is half that of $\Dh_j$ and its limit variance is $1/4\cdot(\sigma_j^2+\sigma^2_{j,\mathrm{lin}}(\Cord{j}))$ (Proposition~\ref{oprop:maxvar}). The least-favorable-configuration argument behind (c) is standard (\citealp{chetverikov2019testing}, and \citealp{lee2013testing}); related tests of monotonicity are discussed in Section~\ref{osec:lit-mono}. When both blocks are coarsely discrete, the limit is a non-Gaussian functional of finitely many normals; when only one is, the integral over the continuous block keeps it normal, with a centering and variance given by sums over the atoms rather than by the formulas of Theorem~\ref{thm:variance}. This is analogous to the treatment of discrete covariates by \citet{shi2019testing}, whose limit is non-normal with simulated critical values when all covariates are discrete, and who replace integrals with sums over the atoms when at least one covariate is continuous \label{rem:discrete}.
\begin{remark}[Local Power]
   Although we do not formally derive the tests' local power functions,
Lemma~\ref{lem:drift} characterizes the drift that governs
their behavior under local alternatives. For the absolute-value statistics $b_\Psi=0$, so a deviation $\delta_N$ from the null with $\|\delta_N\|_2=o(N^{-1/2}\eta^{-1/4})$ leaves $\Dh_j$ unchanged to first order and cannot be detected. $N^{-1/2}\eta^{-1/4}$ is the analog of the rate in the local alternatives $n^{-1/2}h^{-d/4}\delta_D$ of the one-point $L_1$ test of \citet[Section~4.7]{chang2015nonparametric}, with $h^d$ in the role of $\eta$. Against these alternatives, their test has nontrivial local power, which depends on $\delta_D$ only through $\int\delta_D^2$ \citep[Theorem~4.3(b)]{chang2015nonparametric}, because the absolute value is quadratic in a small drift; we expect the same for $\Dh_j$. The positive-part statistics on ordered sets keep the linear term of Lemma~\ref{lem:drift} and, like the one-sided test of \citet[Theorem~4.3(a)]{chang2015nonparametric}, can detect alternatives of order $N^{-1/2}$ in the direction of $w_{\mathcal C_j}$.
\end{remark}

\subsection{Estimated nuisance parameters}\label{sec:asym-nuisance}\label{rem:firststage}
In observational settings, the CATE statistics in \eqref{eq:tauhat-psi}
are constructed using estimated pseudo-outcomes
$\widehat\zeta_i=\zeta(W_i;\widehat\chi)$, where
$\chi=(r_0,r_1,\mu_0,\mu_1)$ denotes the nuisance functions and
$\zeta=\zeta(W;\chi)$ is the pseudo-outcome evaluated at their true values.
Nuisance estimation introduces a conditional mean distortion
$\pi_N:=\E[\widehat\zeta-\zeta\mid X,V,\widehat\chi]$,
whose average
$\bar\pi_N(x):=\E[\pi_N\mid X=x,\widehat\chi]$
determines the resulting drift in the conditional mean of the
estimated pseudo-outcome. For the DR and IPW pseudo-outcomes, respectively, these
distortions take the form
\begin{equation}\label{eq:nui-drift}
\pi_N^{\mathrm{DR}}
=\sum_{t=0}^1(-1)^{1-t}
\frac{(\widehat r_t-r_t)(\widehat\mu_t-\mu_t)}
{\widehat r_t},
\qquad
\pi_N^{\mathrm{IPW}}
=\sum_{t=0}^1(-1)^{1-t}
\frac{\mu_t(r_t-\widehat r_t)}
{\widehat r_t}.
\end{equation}
Thus, the DR drift is second order in nuisance-estimation errors, whereas the IPW drift is first order in propensity-score estimation errors. Under the null, the stochastic error in the kernel estimator of the CATE difference is asymptotically symmetric about zero, so the first-stage-induced shift $\delta_N$ has no first-order effect on the expectation of the absolute-value component of $\Psi(u)=a_\Psi|u|+b_\Psi u$. The resulting drift consists of a linear term proportional to $b_\Psi$ and a quadratic term of order $N\eta^{1/2}\|\delta_N\|_2^2$, where $\|\cdot\|_2$ denotes the norm on $L_2(\dom)$, the space of square-integrable functions on $\dom$. 

%

\begin{lemma}[Drift]\label{lem:drift}
Let $\Gamma_j\equiv0$ on $\mathcal C_j$, so $\Gammah_j=\Gamma^{\mathrm{lin}}_j:=B_{j,N}+N^{-1}\sum_i\Theta_j(W_i;\cdot)$. Let  $\delta_N\in L_2(\dom)$, denote a deterministic sequence, $\delta^\Gamma_N(c):=\delta_N(c_1)-\delta_N(c_2)$ and $D^{\mathrm{lin}}_j(\delta_N):=\sqrt N\int_{\mathcal C_j}\Psi(\Gamma^{\mathrm{lin}}_j(c)+\delta^\Gamma_N(c))\,dc$. If (a) $N\eta^{1/2}\|\delta_N\|_2^2\to0$ and (b) $Nh_j^{2d_j}\|\delta_N\|_2^2\to0$, then $D^{\mathrm{lin}}_j(\delta_N )-D^{\mathrm{lin}}_j(0)-b_\Psi\sqrt N\int_{\dom}w_{\mathcal C_j}\delta_N\pto0$, also along the sequences of laws of Lemma~\ref{olem:uniform}.
\end{lemma}
The lemma shows that, under the null, sufficiently small perturbations affect the statistic only through a linear term proportional to $b_\Psi$, which equals zero for absolute-value statistics or symmetric configuration domains (i.e., $\mathcal C_j$ is symmetric).
Lemma~\ref{lem:drift} holds for any deterministic shift. The relevant shift is the linearized effect of nuisance estimation in the parametric first-stage setting (Theorem~\ref{thm:nuisance-param}) and the smoothed drift $\delta_N(x):=\E[\bar\pi_N(X)\mathcal K(x;X)\mid\widehat\chi]/\{\eta f(x)\}$ in the independent-sample setting (Theorem~\ref{thm:nuisance-split}). It is worthwhile to note that since $\delta_N$ is a kernel average of $\bar\pi_N$, Young's inequality gives $\|\delta_N\|_2\le C\|\bar\pi_N\|_{L_2(\mathcal U)}$ for $N$ large, with $C$ depending only on $f$ and the kernels (Lemma~\ref{olem:nui-lin}), so conditions (a) and (b) of Lemma~\ref{lem:drift} follow from the first two conditions of Assumption~\ref{ass:firststage}(b)(iv), which are imposed directly on $\bar\pi_N$.


We consider two classes of first-stage estimators. (a) A \textit{parametric} first stage models the nuisances as $r_t(x,v;\vartheta)$ and $\mu_t(x,v;\vartheta)$, $\vartheta\in\mathbb R^p$, so that $\chi_\vartheta:=(r_0(\cdot;\vartheta),r_1(\cdot;\vartheta),\mu_0(\cdot;\vartheta),\mu_1(\cdot;\vartheta))$ are the nuisance functions implied by $\vartheta$ and $\zeta(W;\chi_\vartheta)$ is the pseudo-outcome computed from them. The parameter is estimated on the same sample as the test; the estimated pseudo-outcomes are $\widehat\zeta_i=\zeta(W_i;\chi_{\widehat\vartheta})$, and $\widehat\vartheta$ has a probability limit $\vartheta^*$, the pseudo-true value, which exists whether or not the working models are correctly specified. (b) A \textit{flexible} first stage estimates $\chi$ by any method, for instance machine learning, on a sample independent of the test sample, and $\widehat\zeta_i=\zeta(W_i;\widehat\chi)$. The \textit{reference pseudo-outcome} is the pseudo-outcome at the limit of the first stage: $\zeta^*:=\zeta(W;\chi_{\vartheta^*})$ under (a) and $\zeta=\zeta(W;\chi)$ under (b). Under (a), $\zeta^*$ differs from $\zeta$ unless both working models are correctly specified, but it has conditional mean $\tau$ when one of them is; the tests are then calibrated by the law of the statistic computed from $\zeta^*$.

\begin{assumption}[First stage]\label{ass:firststage}
Assumptions~\ref{ass:sampling}--\ref{ass:overlap} hold,
$\E[|Y|^6\mid T,X,V]$ is bounded, and there exists an open
set $\mathcal U\supset\dom$ with compact closure on which
$f\ge\underline p$. The conditional variance of the reference
pseudo-outcome is continuous and bounded away from zero on
$\mathcal U$. For a parametric first stage, we have 
\vspace{-0.3cm}
\begin{itemize}
    \item[(a)] (i) $\zeta(W;\chi_\vartheta)$ satisfies the smoothness and moment conditions of Section~\ref{osec:proofs-nui} on a ball around $\vartheta^*$; (ii) $\sqrt N(\widehat\vartheta-\vartheta^*)=O_p(1)$; (iii) $r_t(\cdot;\vartheta^*)=r_t$, $t=0,1$, or, for $\zeta^{\mathrm{DR}}$ only, $\mu_t(\cdot;\vartheta^*)=\mu_t$, $t=0,1$; (iv) the conditions of Proposition~\ref{prop:transfer} hold for the response $\zeta^*$.
\end{itemize}
Next, for a flexible first stage, we have
\begin{itemize}   
    \item[(b)] (i) $\widehat\chi$ is a function of a sample independent of $\{W_i\}_{i\le N}$; (ii) with probability tending to one, $\underline p\le\widehat r_t\le1-\underline p$ and $|\widehat\mu_t|\le C$; (iii) $\sup_{x\in\mathcal U}\E[(\widehat\zeta-\zeta)^2\mid X=x,\widehat\chi]=o_p(1)$; (iv) $N\eta^{1/2}\|\bar\pi_N\|^2_{L_2(\mathcal U)}$, $Nh_j^{2d_j}\|\bar\pi_N\|^2_{L_2(\mathcal U)}$ and $\eta^{-1/2}\sup_{\mathcal U}|\bar\pi_N|$ are $o_p(1)$; (v) the conditions of Proposition~\ref{prop:transfer} hold for $\zeta$.
\end{itemize}
\end{assumption}
Assumption \ref{ass:firststage} provides regularity conditions for parametric and flexible first-stage estimators under which the asymptotic theory of the proposed statistics can accommodate nuisance estimation.

\begin{theorem}[Parametric first stage]\label{thm:nuisance-param}
Let Assumption~\ref{ass:firststage}(a) hold, $a_\Psi>0$, and $\Sh_j^\Psi(\mathcal C_j)$ be computed from $\widehat\zeta_i=\zeta(W_i;\chi_{\widehat\vartheta})$, $\widehat\vartheta$ estimated on the same sample. (a) If $\Gamma_j\equiv0$ on $\mathcal C_j$ and either $b_\Psi=0$ or $\mathcal C_j$ is symmetric, $\Sh_j^\Psi(\mathcal C_j)$ differs by $o_p(1)$ from the statistic computed from $\zeta(W_i;\chi_{\vartheta^*})$, and $\Sh_j^\Psi(\mathcal C_j)\dto N(0,1)$. (b) If $D_j^\Psi(\mathcal C_j)>0$, $\Pr(p_j<\alpha)\to1$.
\end{theorem}

\begin{theorem}[Flexible First stage on an independent sample]\label{thm:nuisance-split}
Let Assumption~\ref{ass:firststage}(b) hold, $a_\Psi>0$, and $\Sh_j^\Psi(\mathcal C_j)$ be computed from $\widehat\zeta_i=\zeta(W_i;\widehat\chi)$. (a) If\,\, $\Gamma_j\equiv0$ on $\mathcal C_j$ and one of (i) $b_\Psi=0$, (ii) $\mathcal C_j$ is symmetric, or (iii) $\sqrt N\int_{\dom}w_{\mathcal C_j}\delta_N=o_p(1)$ holds, $\Sh_j^\Psi(\mathcal C_j)\dto N(0,1)$. (b) If $D_j^\Psi(\mathcal C_j)>0$, $\Pr(p_j<\alpha)\to1$.
\end{theorem}
The proofs are in Section~\ref{osec:proofs-nui}.
Recall that Lemma~\ref{lem:drift} characterizes the effect of first-stage
estimation on the limiting null distribution. For absolute-value
statistics, such as $\Dh_j$, the linear drift vanishes because
$b_\Psi=0$; the same cancellation occurs for symmetric
configuration sets. Consequently, root-$N$ first-stage
estimation error has no first-order effect on these statistics.
The DR pseudo-outcome is Neyman orthogonal, i.e., its drift \eqref{eq:nui-drift} is a product of first-stage errors. By Lemma~\ref{lem:drift}, the remaining drift affects $\Dh_j$ quadratically under the null. This resembles Neyman orthogonality but stems from symmetry: $|\cdot|$ is even, and the kernel error is asymptotically symmetric, not a moment condition.
Theorem~\ref{thm:nuisance-param} therefore requires only
root-$N$ consistency of $\widehat\vartheta$, without an
asymptotically linear representation, provided at least one
DR working model is correctly specified. This contrasts with
one-point positive-part statistics, such as that of
\citet[Theorem~4.3]{shi2019testing}, for which same-sample
parametric nuisance estimation generally contributes a term
of order one unless both working models are correctly specified.
To the best of our knowledge, Theorem~\ref{thm:nuisance-split} is the first asymptotic result for $L_1$-type functionals of kernel CATE estimators that allows a flexible first stage estimated on an independent sample, with the effect of first-stage error characterized by the drift of Lemma~\ref{lem:drift}. The level result of \citet[Theorem~4.3]{shi2019testing} assumes a parametric first stage and examines a flexible first stage only in simulations; \citet{hsu2017consistent}, whose series logit propensity score is nonparametric, considers a Kolmogorov--Smirnov statistic without kernel smoothing.

Our two-point positive-part statistics on ordered configuration sets require
additional conditions. In this case, $b_\Psi=1/2$ and
$w_{\mathcal C_j}$ is generally nonzero, so first-stage
estimation contributes the linear drift
$b_\Psi\sqrt N\int_{\dom}w_{\mathcal C_j}\delta_N$.
Theorem~\ref{thm:nuisance-param} does not cover this case as
stated, but the discussion after Proposition~\ref{oprop:nui-ext} establishes
an extension when both DR working models are correctly
specified. Theorem~\ref{thm:nuisance-split} covers ordered
sets when this linear drift is $o_p(1)$, as required by
condition (a)(iii). For DR pseudo-outcomes, a sufficient
condition is the double machine learning product rate
$\|\widehat r-r\|_2\|\widehat\mu-\mu\|_2=o_p(N^{-1/2})$.
This condition generally fails for IPW pseudo-outcomes when
the propensity score is estimated on an independent sample
of comparable size. Both theorems establish the null distribution
under the equality restriction $\Gamma_j\equiv0$; validity over
the composite monotone null follows from
Proposition~\ref{oprop:nui-ext}(b) under the one-sided
version of condition (a)(iii), which (a)(iii) implies.

\begin{remark}[Rates and cross-fitting]
\label{rem:nuisance-rate}\label{rem:crossfit}
For DR pseudo-outcomes, Assumption~\ref{ass:firststage}(b)(ii)
ensures that the nuisance-induced drift is bounded by products
of propensity score and outcome regression estimation errors.
Together with Assumption~\ref{ass:reg}(h), this implies that
Assumption~\ref{ass:firststage}(b)(iv) holds if
$\|\widehat r-r\|_4\|\widehat\mu-\mu\|_4
=o_p(N^{-1/2}\eta^{-1/4})
\quad\text{and}\quad
\|\widehat r-r\|_\infty\|\widehat\mu-\mu\|_\infty
=o_p(\eta^{1/2}).$
 These conditions apply to all the statistics considered and differ from the root-$N$ product rate of double machine learning: the first permits a slower product rate but uses stronger norms, and the second requires uniform control of the estimation errors. For positive-part statistics on ordered sets, Theorem~\ref{thm:nuisance-split} additionally requires condition (a)(iii), which the root-$N$ product rate satisfies. Relative to conventional kernel estimation of conditional treatment effects \citep{fan2022estimation,kennedy2023towards}, our integrated statistics require a faster rate for the nuisance-induced drift. Finally, Theorem~\ref{thm:nuisance-split} assumes that the nuisance functions are estimated on a sample independent of the test sample; although cross-fitting has been studied in related settings \citep{semenova2021debiased,lundborg2024projected}, extending our results to cross-fitted estimators requires controlling the dependence across folds, which we leave for future research. \qed
\end{remark}

\subsection{The magnitude of the integrated absolute difference statistic}\label{sec:app-gini}

The absolute-value transformation introduces a noise-induced bias that is nonnegative, up to the smoothing bias $B_{j, N}$, of order $h_1^{s_1}+h_2^{s_2}=o(N^{-1/2})$ by Assumption~\ref{ass:reg}(g), and bounded above by the expected absolute estimation noise, whose integrated magnitude $m^\kappa_{j, N}/\sqrt N\asymp(N\eta)^{-1/2}$ is estimated by $\widetilde m_j/\sqrt N$. The bound holds with equality wherever $\Gamma_j=0$, including under homogeneity, while the bias becomes negligible where $|\Gamma_j|$ is large relative to the estimation noise. The bias-corrected and uncorrected statistics can therefore serve as lower and upper bounds for $D_j$. Let $\widetilde D_j:=(\Dh_j-\widetilde m_j)/\sqrt N$, $F_j:=\{c\in\Csym{j}:\Gamma_j(c)=0\}$, $\omega_1(x):=\int_{\Xone}\mathrm{sgn}\{\tau(x_1,x_2)-\tau(x_1',x_2)\}dx_1'$, $\omega_2$ its mirror image, and $\Omega_j:=\int_{\dom}\omega_j^2\varsigma_\zeta^2/f$, or $\int_{\dom}\omega_j^2\sum_t\varsigma_t^2/P_t$ for \eqref{eq:tauhat}.

\begin{proposition}[Magnitudes]\label{prop:gini-clt}
Under the conditions of Proposition~\ref{prop:transfer}, (a) $\widetilde D_j-O_p(N^{-1/2})\le D_j\le\Dh_j/\sqrt N+O_p(N^{-1/2})$, so both ends of the bracket $[\widetilde D_j,\Dh_j/\sqrt N]$, of width $O_p((N\eta)^{-1/2})$, are consistent for $D_j$; (b) under $H_0^j$, $\sqrt N\widetilde D_j\dto N(0,\sigma_j^2)$; (c) if, for some constants $C<\infty$ and $\epsilon_0>0$, $\lambda\{c\in\Csym{j}:|\Gamma_j(c)|\le\epsilon\}\le C\epsilon$ for all $0<\epsilon\le\epsilon_0$, then $\Dh_j-\sqrt ND_j\dto N(0,\Omega_j)$ with $\Omega_j>0$; (d) if $\lambda(F_j)>0$, $F_j$ has a Lebesgue-null boundary and, for some $C<\infty$ and $\epsilon_0>0$, $\lambda\{c\in\Csym{j}:0<|\Gamma_j(c)|\le\epsilon\}\le C\epsilon$ for all $0<\epsilon\le\epsilon_0$, then $\Dh_j-\sqrt ND_j-m^\kappa_{j,N}(F_j)=O_p(1)$, where $m^\kappa_{j,N}(F_j)\asymp\eta^{-1/2}\lambda(F_j)\to\infty$.
\end{proposition}

Proposition \ref{prop:gini-clt} shows that when the CATE varies throughout the evaluation domain,
conventional root-$N$ inference is available. When the CATE
is constant over part of the domain, root-$N$ inference
may fail because estimation noise contributes a
nonnegligible bias. Nevertheless, the proposition provides
consistent bounds and a confidence interval that accommodates
both cases. Specifically, define
\begin{equation}\label{eq:gini-interval}
\left[
\widetilde D_j-z\widehat{\mathrm{se}}_j/\sqrt N,\,
\Dh_j/\sqrt N+z\widehat{\mathrm{se}}_j/\sqrt N
\right],
\quad
\widehat{\mathrm{se}}_j
:=\max\{\widetilde\sigma_{j,N},
\widehat\Omega_{j,N}^{1/2}\},
\quad
z:=\Phi^{-1}(1-\alpha/2),
\end{equation}
where
$\widehat\Omega_{j,N}
:=N\sum_i\{\int\widehat\omega_j(x)\ell_i(x)dx\}^2
\widehat e_i^{\,2}/q_i$,
with $\widehat\omega_j$ constructed from $\tauh$.
The interval has asymptotic coverage of at least
$1-\alpha$ in each of the cases covered by the proposition.
The results extend to the directional Gini mean difference $\Delta_j$ and its normalized counterpart $G_j$. The interval construction applies to $\Delta_j$, using the density-weighted analogs of the estimators and variances,
and to $G_j$ after normalization by twice the estimated
$\E|\tau(X^{\mathcal R})|$ (Section~\ref{osec:gini-if}).
These coverage results are pointwise in the CATE function
and do not establish uniform validity.

\subsection{Implementation}\label{sec:impl}
\begin{algorithm}[Directional tests and magnitudes]\label{alg:steps}
\setcounter{bean}{0}
\leavevmode
\begin{list}{\textsc{Step} \arabic{bean}.}{\usecounter{bean}}
\item (First stage.) Fit $\widehat r_t$ and $\widehat\mu_t$ for $\zeta^{\mathrm{DR}}$, parametrically on the full sample (Theorem~\ref{thm:nuisance-param}) or flexibly on an independent subsample (Theorem~\ref{thm:nuisance-split}); report the range of the fitted propensities and a specification check, and treat trimming as a sensitivity check.
\item (Kernels and bandwidths.) For $d_1=d_2=1$ use products of the fourth-order kernel $K(u)=45/16\cdot(1-28/3\cdot u^2)(1-4u^2)\ind\{|u|\le 1/2\}$ and $h_j=k_j\widehat{\mathrm{sd}}_jN^{-1/7}$, with $\widehat{\mathrm{sd}}_j$ the standard deviation of block $j$ and $k_j\in[3,5]$; for $d_1+d_2=3$ a sixth-order kernel and, e.g., $N^{-1/10}$. No data-driven rule is available for selecting $k_j$, since bandwidths suitable for estimation need not be optimal for testing, and the choice involves a trade-off between size and power \citep{fan2003some,gao2008bandwidth}. We therefore recommend reporting results across a range of $k_j$ values, following \citet[Remark~4.1 ]{chang2015nonparametric}.
\item (Grid.) Evaluate $\tauh$ on a uniform grid between the 10th and 90th percentiles of each block, with spacing at most $h_j/6$, and compute $\Dh_j$, $\widetilde m_j$ and $\widetilde\sigma_{j, N}$ with the same equal weights. On $M$ grid points in $\Xone$ with fitted values $\tauh_k$ and order statistics $\tauh_{(k)}$, the inner double sum over the grid is computed exactly from the order statistics \citep{yitzhaki2003gini} 
\begin{equation}\label{eq:gini-sort}
\frac{1}{M^{2}}\sum_{k,l}\big|\tauh_k-\tauh_l\big|=\frac{2}{M^{2}}\sum_{k=1}^{M}(2k-M-1)\,\tauh_{(k)} .
\end{equation}
\item ($p$-values and magnitudes.) Set $p_j=1-\Phi(\Sh_j)$, apply Holm's adjustment, and report $\widetilde D_j$, $\Dh_j/\sqrt N$ and the interval \eqref{eq:gini-interval}. \qed
\end{list}
\end{algorithm}

The kernel in step 2 is the fourth-order member of the Epanechnikov family, supported on $[-1/2,1/2]$. Let $K(u)=(a-bu^2)(1-4u^2)\ind\{|u|\le1/2\}$; the normalization and zero second-moment conditions, $\int K(u)\,du=1$ and $\int u^2K(u)\,du=0$, imply $a=45/16$ and $b/a=28/3$.
 Equivalently, $K(u)=2\varphi(2u)$, where $\varphi(v)=15/32\cdot(3-10v^2+7v^4)\ind\{|v|\le1\}$ is the smooth polynomial kernel of \citet[Theorem~2.4]{muller1984smooth}, with $(\nu,k,\mu)=(0,4,1)$ in his notation, whose second-order member is the Epanechnikov kernel. The exponent of the bandwidth, $1/7$, lies in the range $(1/8,1/6)$ allowed by Assumption~\ref{ass:reg}(g) for $s=4$ and $d_1=d_2=1$ and balances its two conditions, as $Nh_j^{8}$ and $(N\eta^3)^{-1}$ are both of order $N^{-1/7}$; scaling by $\widehat{\mathrm{sd}}_j$ makes the bandwidth invariant to the units of block $j$, and $k_j$ is a tuning constant reported over a range. Larger values give windows that cover more than half of the evaluation region in the application of Section~\ref{sec:app-emp}, where they smooth out variation that is local within it (Section~\ref{osec:emp-rob}). Note that the estimated scale, the percentile region, and a grid with spacing $o(h_j)$ do not affect the test limits (Lemma~\ref{olem:tuning}). For the magnitudes, the equal-weight grid sum, which gives the endpoints of the grid the weight of interior points, estimates $D_j$ with a relative error of order $1/M$ (Section~\ref{osec:tuning}).

 For an integer-valued modifier such as age, adding to it in the second stage an independent jitter, a sum of $s_1+2$ independent $U[-1/2,1/2]$ variables \citep{machado2005quantiles}, puts the design within the assumptions\footnote{Smoothing the discrete covariate with a discrete kernel \citep{racine2004nonparametric} is an alternative approach.}, and $H_0^1$ holds for the jittered data if and only if it holds on the original atoms within $(s_1+2)/2$ of the evaluation region (Proposition~\ref{oprop:jitter}). 
 Under $H_0^j$, the average $\bar S_j$ of the statistics over jitter draws has a limit with variance at most one if the draws are jointly asymptotically normal, so that $\bar p_j:=1-\Phi(\bar S_j)$ is then conservative, as for averages of statistics over random sample splits \citep{lundborg2024projected}.

\section{Monte Carlo evidence}\label{sec:mc}\label{sec:mc-tests}

In the base designs $X_1$ and $X_2$ are independent $U[0,1]$, the propensity $\Pr(T=1\mid X)=1/2$ is known, and $Y=1+ 1/2\cdot X_1+1/2\cdot X_2+T\tau(X_1,X_2)+\epsilon$, $\epsilon\sim N(0,1)$. The CATE functions are
\begin{align*}
\text{A0:}\quad &\tau=1
 &&\text{both nulls true;}\\
\text{A1:}\quad &\tau=1+0.8\sin(2\pi x_1)
 &&H_0^{1}\ \text{false},\ H_0^{2}\ \text{true;}\\
\text{A2:}\quad &\tau=1+0.8\sin(2\pi x_2)
 &&H_0^{1}\ \text{true},\ H_0^{2}\ \text{false;}\\
\text{A3:}\quad &\tau=1+0.8\{\sin(2\pi x_1)+\sin(2\pi x_2)\}
 &&\text{both nulls false;}\\
\text{A4:}\quad &\tau=1+\beta\,(x_1-\tfrac12)(x_2-\tfrac12)
 &&\text{both nulls false, marginal nulls true;}\\
\text{A5:}\quad &\tau=1+\beta\,\{(x_1-\tfrac12)^2-\tfrac1{12}\}
 &&H_0^{1}\ \text{false},\ H_0^{2}\ \text{true,}
\end{align*}
with $\beta=13.0016$, which gives A4 and A5 the value $0.178$ of $D_1$ in A1. A4 is the interaction \eqref{eq:interaction-example}, invisible to marginal tests. A5 is symmetric in $x_1$ about $1/2$, so that $\Cov(X_1,\tau(X_1,X_2)\mid X_2)=0$ and the target of the WGCM equals zero although $H_0^1$ is false \citep[p.~2853]{lundborg2024projected}. Every method uses the IPW pseudo-outcome with the known propensity, $\zeta_i=2(2T_i-1)Y_i$. The joint $L_1$ test using our proposed integrated pairwise absolute difference statistic follows Algorithm~\ref{alg:steps} with $k_j=3.5$ and a $13\times13$ grid on $[0.1,0.9]^2$, with spacing $0.18h_j$ at $N=1000$.

\begin{figure}[h]
\centering
\includegraphics[width=\textwidth]{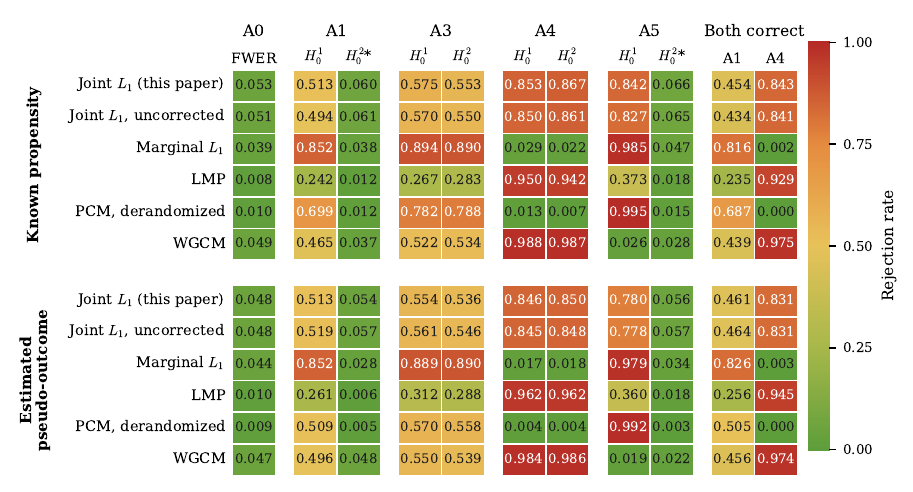}
\caption{\label{fig:mc}Size, power and attribution at $N=1000$: Holm-adjusted rejection frequencies at the $5\%$ level: individual, for each null separately in A1--A5 (true nulls marked $^*$); and joint, as the FWER in A0 (rejecting either true null) and under ``Both correct'' (both decisions right in A1 and A4).
}
\end{figure}

We report two versions of our joint $L_1$ test, differing only
in centering and variance estimation. The \textit{corrected}
version, recommended in Section~\ref{sec:impl}, uses
$\widetilde m_j$ from \eqref{eq:center-feasible}, based on
$q_i$-adjusted squared residuals, and the finite-sample
standard deviation $\widetilde\sigma_{j,N}$ from
\eqref{eq:sigmaN}.  The
\textit{uncorrected} version uses raw squared residuals
($q_i=1$).
The benchmarks are the \textit{marginal $L_1$} that uses the same
functional to a CATE estimator that smooths over $X_j$ alone, \textit{the test of \citet{lavergne2015significance} (LMP)} with asymptotic critical values,  \textit{PCM of \citet{lundborg2024projected}} and \textit{WGCM of \citet{scheidegger2022weighted}}.  We compute the PCM and WGCM using the replication code of \citet{lundborg2024projected} and its generalized additive model learners, and we run the LMP test with the kernels and bandwidths of \citet[Section~4]{lavergne2015significance}. Rejection rates reported in Figure \ref{fig:mc} are based on $1{,}000$ replications per design with N=1000.\footnote{
The number of replications is constrained by computational cost.
Our test requires numerical integration over a grid of configurations
to compute the centering and finite-sample variance, PCM refits
its learners across six sample splits, and LMP has computational
complexity of order $N^3$.} Details of the designs and further results at different sample sizes are in Section~\ref{osec:mc}.

\textit{Size.} Figure~\ref{fig:mc} shows that both versions
of the joint $L_1$ test maintain rejection frequencies close
to the nominal level. In A0, the corrected test has familywise
error rates of $0.053$ with the known propensity and $0.048$
with estimated pseudo-outcomes, compared with $0.051$ and
$0.048$ for the uncorrected version. In A1 and A5, rejection
frequencies for the true $H_0^2$ range from $0.054$ to $0.066$
for the corrected test and from $0.057$ to $0.065$ for the
uncorrected test, all within $2.3$ Monte Carlo standard errors
of $0.05$. The similar size performance of the uncorrected
test reflects offsetting centering and scaling errors, which
the corrections remove (Section~\ref{osec:mc-diag}).
Among the benchmarks, derandomized PCM and LMP are markedly
conservative, with FWERs of $0.008$--$0.010$,
whereas WGCM ($0.049$ and $0.047$) and marginal
$L_1$ ($0.039$ and $0.044$) maintain their level.

\textit{Power and attribution.} Figure~\ref{fig:mc} shows
that the corrected joint $L_1$ test rejects false nulls with
frequencies ranging from $0.513$ to $0.867$. It is the only
method with power exceeding $0.5$ against every false null
in both panels, whereas each benchmark performs poorly in
at least one design. With estimated pseudo-outcomes, our test is competitive with derandomized PCM, with close rates in A1 and A3 ($0.513$ versus $0.509$ and $0.554$
versus $0.570$); with the known propensity PCM rejects in A1 with frequency $0.699$.
PCM has little power against the pure interaction in A4
($0.013$ and $0.004$), reflecting the additive learners
used in its replication code, which cannot detect
the interaction (see Tables~\ref{otab:mc2-500}--\ref{otab:mc2-2000}).
LMP rejects with frequency $0.950$ and $0.962$ in A4, and $0.242$ and
$0.261$ in A1.
Marginal $L_1$ rejects with frequency $0.852$ in A1, where
heterogeneity survives marginalization, but has power
close to its size in A4.

WGCM rejects in A1 with frequency $0.465$ and $0.496$ but fails against
symmetric effect modification in A5. In this design,
$\Cov(X_1,\tau(X_1,X_2)\mid X_2)=0$, so its weighted
covariance target cannot detect variation in $X_1$ \citep{lundborg2024projected}.
Its rejection frequencies for the false $H_0^1$ are
$0.026$ and $0.019$, compared with $0.842$ and $0.780$
for the corrected joint test and $0.827$ and $0.778$
for the uncorrected version. PCM, whose learned projection
captures the curvature, achieves power of $0.995$ and
$0.992$. Our proposed joint test also maintains rejection frequencies
of $0.066$ and $0.056$ for the true $H_0^2$, correctly
attributing heterogeneity to $X_1$.

These differences matter for attributing effect heterogeneity. In design A4, the corrected joint test correctly identifies both
covariate blocks as sources of heterogeneity in $0.843$ and
$0.831$ of samples, compared with zero for PCM and at most
$0.003$ for marginal $L_1$. The uncorrected joint test has
similar power throughout, with rejection frequencies of
$0.434$ and $0.841$ for the two attribution outcomes in
A1 and A4 under the known propensity, compared with
$0.454$ and $0.843$ for the corrected test. Across designs,
the corrections therefore improve the theoretical and
finite-sample calibration without significantly changing power.
PCM and WGCM necessarily rely on random sample splitting,
which can make their decisions depend on the particular splits.

\textit{Nuisance estimation.} The bottom panel of
Figure~\ref{fig:mc} examines the effect of estimating nuisance
parameters on the test sample. A confounder $V\sim N(0,1)$
affects both treatment assignment,
$\Pr(T=1\mid X,V)=\mathrm{logit}^{-1}(-1+X_1+X_2+0.75V)$,
and the outcome,
$Y=1+1/2\cdot X_1+1/2\cdot X_2+V+T\tau(X_1,X_2)+2\epsilon$.
All methods use the DR pseudo-outcome \eqref{eq:psi-dr},
constructed from a correctly specified logistic propensity
model and arm-specific linear outcome regressions on
$(1,X_1,X_2,V)$, estimated on the full sample. The treated-arm
outcome model is misspecified when $\tau$ is nonlinear.

The corrected joint test remains well calibrated, with
familywise error of $0.048$ versus $0.053$ under the known
propensity and rejection rates of $0.054$ and $0.056$ for
the true $H_0^2$ in A1 and A5, respectively. Its power
changes by at most $0.02$ except in A5, where it falls
from $0.842$ to $0.780$. The uncorrected joint test is
similarly stable, with familywise error of $0.048$ and
power changes generally below $0.02$, apart from A5
($0.827$ to $0.778$). Marginal $L_1$ and LMP also barely change, whereas PCM loses substantial power in designs A1
($0.699$ to $0.509$) and A3 (about $0.78$ to $0.57$),
reducing its exact attribution rate in A1 from $0.687$
to $0.505$. WGCM retains its power and its level, with FWER of $0.047$ and rejection
of the true $H_0^2$ in A1 at $0.048$. The stable size of the joint test corroborates Theorem~\ref{thm:nuisance-param}, which establishes
asymptotic level control under same-sample parametric
nuisance estimation, even when the outcome model is
misspecified.

The Online Appendix reports the unadjusted rejection rates of $H_0^1$ in the designs of Figure~\ref{fig:mc} (Figure~\ref{ofig:unadj}); results of supplementary runs with the same methods for the base designs at $N=500$, $1000$, $2000$ and $5000$, with the null distribution of $\Sh_j$ under both calibrations (Sections~\ref{osec:mc-results} and \ref{osec:mc-diag}), for additional designs with misspecified outcome models, correlated blocks and heavy-tailed errors (Section~\ref{osec:mc-new}) and for the sensitivity to the grid and the bandwidth (Section~\ref{osec:mc-sens}); the coverage of the magnitude intervals \eqref{eq:gini-interval} (Section~\ref{osec:mc-mag}); and designs with an estimated first stage that illustrate Theorems~\ref{thm:nuisance-param} and \ref{thm:nuisance-split} (Section~\ref{osec:mc-fs}).

%

\section{Empirical illustration: 401(k) eligibility and net financial assets}\label{sec:app-emp}\label{sec:app}
In this section, we apply the proposed methods to examine heterogeneity in the effect of 401(k) eligibility on household savings by age and income. We describe the data, estimation procedure, and implementation choices, including boundary limitations.

The data are from the 1991 Survey of Income and Program Participation extract of \citet{chernozhukov2004effects}, $N=9{,}915$ households, of which $3{,}682$ are eligible (see Table~\ref{otab:emp-data} for summary statistics of all relevant variables). The treatment is eligibility for an employer-provided 401(k) plan, treated as unrelated to saving preferences given income and other observables (\citealp{abadie2003semiparametric}, \citealp{poterba1995do}), and the outcome is net financial assets in US dollars. The modifiers are age ($X_1$) and family income ($X_2$), and $V$ consists of education, family size, marital status, two-earner status, defined-benefit pension status, IRA participation, and home ownership. The first stage, fitted on the full sample, is a \textit{logistic propensity} with cubic terms in age and income, their interaction, and linear terms in $V$, with linear outcome models on the same regressors in each arm. The fitted propensities lie in $[0.035,0.955]$ and \textit{are not} trimmed. Because age is integer-valued, the second stage uses jittered age (Section~\ref{sec:impl}), and we average every statistic over $50$ jitter draws. With $k_j=3.5$, the bandwidths are $h_1=9.8$ years and
$h_2=23{,}289$ USD, corresponding to CATE estimation
windows of $\pm4.9$ years and $\pm11{,}645$ USD around
each grid point. Each joint window contains at least $153$ households, and the $31\times31$ grid spacing is below $0.1h_j$. %

Theorem~\ref{thm:nuisance-param} justifies the $p$-values if the cubic logistic propensity is correctly specified, as the linear outcome models \textit{are not} credible for a heavy-tailed outcome. 
A flexible series first stage gives practically the same results, and Theorem~\ref{thm:nuisance-split} requires no correctly specified model. The results should be read as follows. The estimand is the effect of eligibility, an intention-to-treat effect: $70.5\%$ of eligible households participate, and take-up rises with income, so an income gradient combines variation in the effect of participation with variation in take-up. As in the literature using these data, we treat eligibility as unconfounded given income and $V$. Although this assumption has been disputed \citep{engen1996illusory}, \citet{chernozhukov2026long} find in a recent paper that the average effect, and its increase across income quartiles, remain positive under plausible confounding by omitted firm characteristics. The conclusions do not depend on including IRA participation and home ownership, which may respond to eligibility, in $V$ (see the robustness checks in Section~\ref{osec:emp}). A rejection of $H_0^1$ establishes that age modifies the effect at fixed income, possibly through covariates in $V$ or job tenure, rather than a causal interaction \citep{vanderweele2009distinction}. This covariate-specific
attribution is not provided by the omnibus test of
\citet{lapenta2026machine} applied to the same data.

\begin{table}[h]
\caption{\label{tab:empirical}Attribution of heterogeneity in the effect of 401(k) eligibility on net financial assets.}
\begin{center}
\setlength{\tabcolsep}{4pt}
\begin{tabular*}{\textwidth}{@{\extracolsep{\fill}}lccccccc@{}}
\hline\hline
& \multicolumn{4}{c}{Test of $H_0^{j}$} & \multicolumn{3}{c}{Magnitude (USD)} \\
\cline{2-5}\cline{6-8}
Direction & $\bar S_j$ & $\bar p_j$ & $\tilde p_j$ & Rej. & $\widetilde D_j$ & $\Dh_j/\sqrt N$ & $\widetilde\Delta_j$ \\
\hline
Age ($j=1$) & 2.15 & 0.016 & 0.016 & 0.96 & 1{,}045 & 5{,}143 & 1{,}369 \\
 & (0.27) &  &  &  & (690) &  & (1{,}332) \\
Income ($j=2$) & 3.68 & $<$0.001 & $<$0.001 & 1.00 & 1{,}835 & 5{,}989 & 2{,}704 \\
 & (0.26) &  &  &  & (736) &  & (1{,}315) \\
\hline\hline
\end{tabular*}
\end{center}
\footnotesize
\renewcommand{\baselineskip}{11pt}
\textbf{Note:} Doubly robust pseudo-outcome, cubic logistic propensity and linear outcome models; $k_j=3.5$; averages over $50$ jitter draws of age. $\bar S_j$: average of $\Sh_j$ of \eqref{eq:S} (standard deviation over draws); $\bar p_j=1-\Phi(\bar S_j)$; $\tilde p_j$: Holm-adjusted, $\tilde p_{(2)}=\max\{2\bar p_{(1)},\bar p_{(2)}\}$; Rej.: share of draws whose own Holm-adjusted $p$-value is at most $0.05$. Valid if the propensity model is correct (Theorem~\ref{thm:nuisance-param}). $\widetilde D_j$ and $\Dh_j/\sqrt N$ bracket $D_j$ on the evaluation box, standardized to the unit square; $\widetilde\Delta_j$: noise-corrected density-weighted magnitude (uncorrected $7{,}380$ and $8{,}708$); standard errors include the first-stage term and assume the margin condition. With the flexible first stage, $\bar S=(2.08,3.85)$ and $\tilde p=(0.019,<0.001)$.
\end{table}

Both nulls are rejected at the $5\%$ level after Holm's adjustment (Table~\ref{tab:empirical}): income decisively, and age with $\tilde p_1=0.016$, $96\%$ of the individual jitter draws rejecting. Neither decision depends on the bandwidth. Over the recommended range $k_j\in[3,5]$, under both pseudo-outcomes and both first stages, every draw rejects $H_0^2$ ($\bar S_2$ between $3.45$ and $4.02$), and the jitter-averaged test rejects $H_0^1$ at every $k_j$, with $\tilde p_1$ between $0.009$ and $0.045$ and at least half of the individual jitter draws rejecting. On the split-sample route of Theorem~\ref{thm:nuisance-split}, both halves confirm income, whereas age is rejected in one half at $k_j=3.5$ and in neither at $k_j=5$, as with the cubic first stage on the same halves: the cost of halving the sample. Net financial assets are heavy-tailed (excess kurtosis $184$; Remark~\ref{orem:heavytails}); winsorizing the outcome at the 1st and 99th percentiles, which changes the estimand, gives $\Sh_1=4.30$ and $\Sh_2=4.93$ for one draw, so a small number of extreme households attenuate, rather than drive, the age-related result.
The eligibility effect thus varies with income at fixed age and, with age at fixed income, across the recommended bandwidths. For targeting, neither modifier is redundant given the other: a rule that assigns eligibility, or promotes access to a plan, on income alone leaves gains from conditioning on age unexploited; conversely, a welfare-maximizing assignment rule should use both.
\begin{figure}
\centering
\includegraphics[width=1\textwidth]{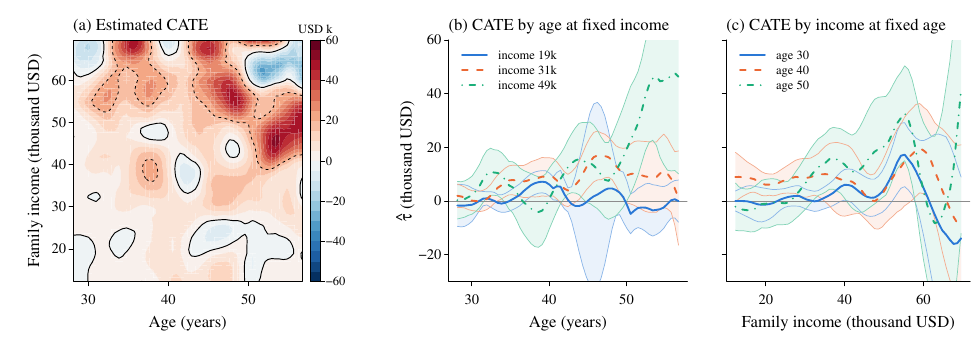}
\caption{\label{fig:cate}Estimated CATE (thousand USD); doubly robust pseudo-outcome, $k_j=3.5$, one jitter draw. (a) CATE function over the evaluation Cartesian plane (box), contours at zero (solid) and 20,000 USD (dashed). (b) CATE by age at fixed income, at the income quartiles. (c) CATE by income at fixed age, at ages 30, 40 and 50. Shaded: pointwise 95\% bands.}
\end{figure}

The scaled estimated bias $\widetilde m_j/\sqrt N$ is about $80\%$ of $\Dh_1/\sqrt N$ and $70\%$ of $\Dh_2/\sqrt N$. Net of it, two households of the same age but independently drawn incomes in the box differ in their eligibility effect by $\widetilde\Delta_2=2{,}704$ USD on average (standard error $1{,}315$; upper end of the bracket $8{,}708$), and the interval \eqref{eq:gini-interval} for $D_2$ is $[392,7{,}433]$. For age, $\widetilde\Delta_1=1{,}369$ ($1{,}332$), and the interval for $D_1$, $[-307,6{,}496]$, includes zero although the test rejects: the interval is built to be valid in all three cases of Proposition~\ref{prop:gini-clt} and is not an inversion of the test. Over $k_j\in[3,5]$ the corrected magnitudes are stable ($\widetilde D_1$ between $759$ and $1{,}078$ USD, $\widetilde D_2$ between $1{,}729$ and $1{,}835$), whereas $\Dh_1/\sqrt N$ falls from $5{,}891$ to $3{,}571$ with the estimated bias. As the fitted CATE function is negative on $16\%$ of the box, $G_j$ is not a Gini coefficient here (Section~\ref{osec:emp}). The pointwise bands in Figure~\ref{fig:cate} are wide, so the figure describes the point estimate rather than a tested shape.

Section~\ref{osec:emp} of the Online Appendix reports additional results and robustness checks of the application. It covers the data and the tails of the outcome, the debate on the exogeneity of eligibility, the specification tests and covariate balance of the first stage, and the evaluation region, its edge windows and the jittered age; the jitter-averaged statistics over $k_j\in[3,5]$ under both pseudo-outcomes and both first stages (Table~\ref{otab:emp-sens}); single-draw checks with integer-valued age, finer grids, a winsorized outcome, an adjustment set without IRA participation and home ownership, trimmed propensities and the sample-split route of Theorem~\ref{thm:nuisance-split} (Table~\ref{otab:emp-sens2}); and the magnitudes and the index $G_j$ over the bandwidth range.
\section{Conclusion}\label{sec:concl}

This paper develops asymptotic theory for integrated pairwise-difference functionals of kernel estimators, which integrate absolute or positive-part differences between kernel estimates at pairs of points and fall outside the one-point $L_1$ theory. Poissonization and a Hájek-type projection onto single-point terms yield asymptotic normality. Furthermore, under some regularity conditions, we show that estimating the nuisance parameters of pseudo-outcomes does not affect the limit theory.  We develop two applications of our asymptotic results: tests that attribute CATE heterogeneity to a covariate block conditional on another,
and root-$N$ inference for the directional Gini mean difference,
which measures effect heterogeneity in outcome units.
Using  401(k) eligibility data, we show that treatment effects
vary with income at fixed age and with age at fixed income
across all recommended bandwidths, suggesting that targeting
rules should account for both covariates.

Several questions remain open. Cross-fitting, permitted by
competing procedures \citep{lapenta2026machine,semenova2021debiased}, induces dependence among second-stage summands across folds. We conjecture that the same rate conditions suffice, but leave a formal proof for future work. Extending our theory to higher-dimensional covariate blocks requires a different centering and is left for future work. The numerical evidence in Section~\ref{osec:mc-blocks} supports the feasibility of such an extension. Moreover, interpreting non-rejection as evidence of no effect modification would require uniformly valid inference, such as an upper confidence bound for $D_j$ with uniform coverage.

\section*{Acknowledgments}
This paper builds on portions of my working paper, \cite{owusu2024nonparametric}, particularly the statistic and its application to treatment effect heterogeneity testing. The research benefited from my time at the University of Bristol, UK, as a Lecturer, and I am grateful to Youngki Shin, Jeffrey S.\ Racine, Michael Veall, Saraswata Chaudhuri, Emmanuel S.\ Tsyawo, Monika Avila M\'arquez, Augustine Denteh, and conference and seminar participants for their helpful comments. I acknowledge the use of  Claude Opus 5.5, GPT-6 Astra, and Grammarly for language editing and  Claude Opus 5.5 for R-coding assistance. I reviewed and verified all AI-assisted text and R code.  All errors are mine.
This research is supported by the Social Sciences and Humanities Research Council of Canada (SSHRC), Grant No.\ 430-2025-01562. I have no conflicts of interest to declare. Replication code is available from the author on request.

\onehalfspacing
\setlength{\bibsep}{0pt plus 0.3ex}
\bibliographystyle{chicago}
{\small
\bibliography{reference-gen}
}
\normalsize
\onehalfspacing

\pagebreak
\setcounter{page}{1}
\begin{center}
\Large{Online Appendix to ``Asymptotic Theory  of Integrated Pairwise-Difference Functionals of Kernel Estimators, with Applications to  Treatment Effect Heterogeneity''}\\
\vspace{0.5cm}
\large{Julius Owusu}
\end{center}

\noindent This Online Appendix collects material that supports the main paper. A table of contents and a table of notation precede Section~S1. Section~S1 proves the identification results. Section~S2 proves the main asymptotic results, Section~S3 the results for estimated nuisance parameters, and Section~S4 the results on the magnitudes. Section~S5 gives the auxiliary lemmas and routine derivations behind the proofs of Section~S2, and Section~S6 gives the auxiliary results for estimated nuisance parameters and the uniformity of the oracle results along sequences of laws. Section~S7 states the implementation results that the main text uses in detail (among them the scale of the max form, data-dependent tuning and integer-valued modifiers), proves them together with the sharp uniform rate and the auxiliary parts of the results on the corrected centering and on the magnitudes, and gives the grid formulas used in the computations. Section~\ref{osec:mc} reports the full Monte Carlo study, including the size diagnostics of the corrected test, the additional designs and designs with an estimated first stage, and Section~\ref{osec:emp} the details and robustness checks of the empirical application. Section~\ref{osec:lit} collects further discussion of the literature.

\addtocontents{toc}{\protect\setcounter{tocdepth}{2}}
{\singlespacing\small\tableofcontents}
\clearpage
\StartSupplementCitations

\setcounter{section}{0}
\setcounter{equation}{0}
\setcounter{figure}{0}
\setcounter{table}{0}
\setcounter{footnote}{0}
\renewcommand{\thesection}{S\arabic{section}}
\renewcommand{\thefigure}{S\arabic{figure}}
\renewcommand{\thetable}{S\arabic{table}}
\makeatletter\def\@currentlabel{}\makeatother
\section*{Notation}\label{osec:notation}
\addcontentsline{toc}{section}{\protect\numberline{}Notation}
{\small\singlespacing
\renewcommand{\arraystretch}{1.08}
\begin{tabular*}{\textwidth}{@{\extracolsep{\fill}}p{0.27\textwidth}p{0.69\textwidth}@{}}
\hline\hline
Symbol & Meaning \\
\hline
$W=(Y,T,X_1,X_2,V)$ & outcome, binary treatment, the two blocks $X=(X_1,X_2)$ of dimensions $d_1,d_2$ ($d:=d_1+d_2$), adjustment covariates $V$\\
$\tau$, $\zeta$, $\zeta^{\mathrm{IPW}}$, $\zeta^{\mathrm{DR}}$ & CATE on $(X_1,X_2)$; pseudo-outcome, IPW (IPW) or doubly robust\\
$r_t$, $\mu_t$; $\chi$ & propensity and outcome regression given $(X,V)$; the nuisances $(r_0,r_1,\mu_0,\mu_1)$\\
$\bar\mu_t$, $\varsigma_t^2$, $P_t$, $f$ & mean and variance of $Y$ given $(X,T=t)$, $P_t:=\Pr(T=t\mid X)f$, density of $X$\\
$\varsigma_\zeta^2$, $\varrho$, $\varrho_\zeta$ & $\Var(\zeta\mid X)$; scaled pointwise variances of $\tauh$\\
$K_j$, $h_j$, $k_j$, $\widehat{\mathrm{sd}}_j$, $s_j$, $\eta$ & kernel and bandwidth of block $j$, bandwidth constant and estimated standard deviation of block $j$ ($h_j=k_j\widehat{\mathrm{sd}}_jN^{-1/7}$), kernel order, $\eta:=h_1^{d_1}h_2^{d_2}$\\
$\mathcal K_i(x)$, $\ell_i(x)$, $L_{il}$, $\widehat e_i$, $q_i$ & product kernel at $X_i$, regression weights, $L_{il}:=\ell_l(X_i)$, fitted residual, own-weight divisor\\
$\mathcal W(x)$, $\mathcal R$, $\mathcal B$ & kernel window at $x$; evaluation box; box containing all windows (proofs)\\
$c=(c_1,c_2)$; $\mathcal C_j$, $\Csym{j}$, $\Cord{j}$ & configuration and its two evaluation points; configuration sets, symmetric and ordered\\
$\Gamma_j$, $\Gammah_j$; $\Psi$, $a_\Psi$, $b_\Psi$, $m_\Psi$, $g_\Psi$ & differenced CATE estimands and estimate; integrand and its constants\\
$\Dh_j^\Psi(\mathcal C_j)$, $D_j^\Psi(\mathcal C_j)$; $\Dh_j$, $D_j$ & statistic and target; absolute-value statistic and target on $\Csym{j}$\\
$\theta_N$, $\Theta_j$, $\xi$, $\kappa_{j,N}$ & influence function of $\tauh$, its difference at $c_1,c_2$, score, variance of $\Theta_j$\\
$m^\kappa_{j,N}$, $m^X_{j,N}$, $\widetilde m_j$, $\widetilde\kappa_j$ & centering, its conditional version given the covariates, its estimate; estimated variance $\widetilde\kappa_j(c)$ of the differenced field, the diagonal of the covariance estimate $\widetilde\kappa_j(c,c')$\\
$\sigma^2_{j,\Psi}$, $\sigma_j^2$, $\widetilde\sigma^2_{j,\Psi,N}$, $\Sh_j$ & limit variances, finite-sample variance, studentized statistic\\
$\widetilde D_j$, $\Delta_j$, $\widetilde\Delta_j$, $\widehat f_{\mathcal C_j}$, $G_j$ & noise-corrected magnitude, density-weighted magnitude and its correction, estimated density of the pair of evaluation points, index\\
$F_j$, $\omega_j$, $\Omega_j$, $\widehat\Omega_{j,N}$ & flat set, sign weight, variance of the magnitude and its estimate\\
$\vartheta$, $\vartheta^*$; $\pi_N$, $\bar\pi_N$, $\delta_N$ & first-stage parameter and its limit; drift, its average given $X$, smoothed drift\\
$D^{\mathrm{lin}}_j$, $D^{\mathrm{lin}}_j(\delta_N)$, $\delta^\Gamma_N$ & linearized statistic; the same with the field shifted by $\delta^\Gamma_N(c):=\delta_N(c_1)-\delta_N(c_2)$\\
$\mathcal N$, $\Gamma_{\mathcal N}$, $D^P_j$, $\projfield_{\mathcal N}$, $R_{\mathcal N}$ & Poisson sample size, Poissonized field and statistic, projection field, remainder\\
$B_{j,N}$, $B_N$, $\epsilon_N$ & smoothing bias of $\Gammah_j$ and of $\tauh$, uniform rate\\
$p_{\mathcal B}$, $U_{\mathcal N}$, $U'_{\mathcal N}$, $Q_{\mathcal N}$ & probability outside $\mathcal B$, standardized counts in and outside $\mathcal B$, normalized projection\\
\hline\hline
\end{tabular*}}

\section{Proofs of the identification results}\label{osec:id}

This section proves Proposition~\ref{prop:id} of the main text: its second statement, identification of $\tau$ by the difference of the arm-wise regressions when there is no adjustment block $V$, in Section~\ref{osec:id-cate}, and its first statement, $\E[\zeta\mid X=x]=\tau(x)$ for the two pseudo-outcomes, in Sections~\ref{osec:id-pseudo} and \ref{osec:id-reduced}, where it is split into a statement given the full adjustment set (part (a)) and its consequence given $X$ (part (b)).\ Throughout, $W=(Y,T,X_1,X_2)$ in the baseline case and $W=(Y,T,X_1,X_2,V)$ when an additional adjustment block $V$ is present; the sample is i.i.d.\ by Assumption~\ref{ass:sampling}. All statements are restated so that this section can be read independently of the main text.

\subsection{Identification of the conditional average treatment effect}\label{osec:id-cate}

\textit{Proposition~\ref{prop:id} of the main text, second statement (no adjustment block $V$).} Under Assumptions~\ref{ass:sampling}, \ref{ass:unconf} and \ref{ass:overlap}, for every $(x_1,x_2)\in\dom$,
\begin{align}
\tau(x_1,x_2)&:=\E\big[Y(1)-Y(0)\mid X_1=x_1,X_2=x_2\big]\notag\\
&=\E[Y\mid T=1,X_1=x_1,X_2=x_2]-\E[Y\mid T=0,X_1=x_1,X_2=x_2].\label{oeq:id-claim}
\end{align}

\begin{proof}[Proof of the second statement of Proposition~\ref{prop:id}]
By Assumption~\ref{ass:unconf}(a) the realized outcome satisfies
\begin{equation}\label{oeq:id-consistency}
Y=Y(0)+\big[Y(1)-Y(0)\big]\,T .
\end{equation}
Assumption~\ref{ass:overlap} guarantees that $\Pr(T=t\mid X_1=x_1,X_2=x_2)\ge\underline p>0$ for $t\in\{0,1\}$ and every $(x_1,x_2)$ in a neighborhood of $\dom$, so the conditional expectations $\E[Y\mid T=t,X_1=x_1,X_2=x_2]$ are well defined there. Using \eqref{oeq:id-consistency} and then Assumption~\ref{ass:unconf}(b),
\begin{align}
&\E[Y\mid T=1,X_1=x_1,X_2=x_2]\nonumber\\
&\quad=\E\big[Y(0)+\{Y(1)-Y(0)\}T\;\big|\;T=1,X_1=x_1,X_2=x_2\big]\nonumber\\
&\quad=\E\big[Y(1)\mid T=1,X_1=x_1,X_2=x_2\big]
=\E\big[Y(1)\mid X_1=x_1,X_2=x_2\big],
\label{oeq:id-mean1}
\end{align}
the last equality by the conditional independence $(Y(1),Y(0))\perp\!\!\!\perp T\mid(X_1,X_2)$ of Assumption~\ref{ass:unconf}(b). The same argument with $T=0$ gives
\begin{equation}\label{oeq:id-mean0}
\E[Y\mid T=0,X_1=x_1,X_2=x_2]=\E\big[Y(0)\mid X_1=x_1,X_2=x_2\big].
\end{equation}
Differencing \eqref{oeq:id-mean1} and \eqref{oeq:id-mean0} and using the linearity of the conditional expectation gives \eqref{oeq:id-claim}.
\end{proof}

Two remarks emerge. First, no restriction is placed on the dependence between the two covariate blocks: the partition of $X$ into $(X_1,X_2)$ plays no role in identification and affects the analysis only through the hypotheses. Second, the argument is pointwise in $(x_1,x_2)$; the continuity of $\tau$ used in the asymptotic theory is supplied separately by Assumption~\ref{ass:reg}.

\subsection{Pseudo-outcome identification}\label{osec:id-pseudo}

Let $Z:=(X_1,X_2,V)$, let $r_t(z):=\Pr(T=t\mid Z=z)$ and $\mu_t(z):=\E[Y\mid T=t,Z=z]$, and let $\tau^{\mathrm{full}}(z):=\E[Y(1)-Y(0)\mid Z=z]$. The two pseudo-outcomes of the main text, equation~\eqref{eq:psi-ipw}, are
\begin{gather*}
\zeta^{\mathrm{IPW}}(W):=\frac{\ind(T=1)\,Y}{r_1(Z)}-\frac{\ind(T=0)\,Y}{r_0(Z)},\\
\zeta^{\mathrm{DR}}(W):=\mu_1(Z)-\mu_0(Z)+\frac{\ind(T=1)\{Y-\mu_1(Z)\}}{r_1(Z)}-\frac{\ind(T=0)\{Y-\mu_0(Z)\}}{r_0(Z)} .
\end{gather*}

\textit{Proposition~\ref{prop:id} of the main text, first statement, in two parts.} Suppose Assumptions~\ref{ass:sampling}, \ref{ass:unconf} and \ref{ass:overlap} hold on the full adjustment set $(X_1,X_2,V)$, and let $\zeta\in\{\zeta^{\mathrm{IPW}},\zeta^{\mathrm{DR}}\}$.\\ (a) For every $z=(x_1,x_2,v)$, $\E[\zeta(W)\mid Z=z]=\tau^{\mathrm{full}}(z)$.\\ (b) For every $(x_1,x_2)\in\dom$,
\[
\E[\zeta(W)\mid X_1=x_1,X_2=x_2]=\tau(x_1,x_2):=\E\big[\tau^{\mathrm{full}}(x_1,x_2,V)\mid X_1=x_1,X_2=x_2\big].
\]

\begin{proof}[Proof of part (a)]
By Assumption~\ref{ass:overlap}, $r_t(z)\in[\underline p,1-\underline p]$, so every ratio below is well defined and bounded.

\textit{IPW.} On the event $\{T=1\}$, Assumption~\ref{ass:unconf}(a) gives $Y=Y(1)$. Hence, using Assumption~\ref{ass:unconf}(b) in the form $Y(1)\perp\!\!\!\perp T\mid Z$,
\begin{align*}
\E\Big[\frac{\ind(T=1)}{r_1(Z)}Y\;\Big|\;Z=z\Big]
&=\frac{1}{r_1(z)}\E\big[\ind(T=1)Y(1)\mid Z=z\big]\\
&=\frac{r_1(z)}{r_1(z)}\E\big[Y(1)\mid Z=z\big]
=\E\big[Y(1)\mid Z=z\big],
\end{align*}
where the middle equality factorizes the expectation by conditional independence. The same computation with $T=0$ gives $\E[\ind(T=0)Y/r_0(Z)\mid Z=z]=\E[Y(0)\mid Z=z]$. Subtracting yields $\E[\zeta^{\mathrm{IPW}}(W)\mid Z=z]=\E[Y(1)-Y(0)\mid Z=z]=\tau^{\mathrm{full}}(z)$.

\textit{DR (augmented inverse probability weighted, AIPW) pseudo-outcome.} The first term of $\zeta^{\mathrm{DR}}$ has conditional expectation $\mu_1(z)-\mu_0(z)$, which equals $\tau^{\mathrm{full}}(z)$ by \eqref{oeq:id-claim} with $Z$ in place of $(X_1,X_2)$. For the residual terms, the same factorization gives, for $t=1$,
\[
\E\Big[\frac{\ind(T=1)\{Y-\mu_1(Z)\}}{r_1(Z)}\;\Big|\;Z=z\Big]
=\frac{1}{r_1(z)}\,r_1(z)\,\E\big[Y-\mu_1(z)\mid T=1,Z=z\big]=0
\]
by the definition of $\mu_1$, and identically for $t=0$. Hence $\E[\zeta^{\mathrm{DR}}(W)\mid Z=z]=\tau^{\mathrm{full}}(z)$.
\end{proof}

\subsection{Dimension-reduced identification}\label{osec:id-reduced}

\begin{proof}[Proof of part (b)]
By the law of iterated expectations, conditioning first on $(X_1,X_2,V)$ and then on $(X_1,X_2)$,
\begin{align*}
\E[\zeta(W)\mid X_1=x_1,X_2=x_2]
&=\E\Big[\E\big[\zeta(W)\mid X_1,X_2,V\big]\;\Big|\;X_1=x_1,X_2=x_2\Big]\\
&=\E\big[\tau^{\mathrm{full}}(x_1,x_2,V)\mid X_1=x_1,X_2=x_2\big],
\end{align*}
the second equality by part (a). The right-hand side equals $\E[Y(1)-Y(0)\mid X_1=x_1,X_2=x_2]=\tau(x_1,x_2)$ by a second application of iterated expectations, so the two expressions for $\tau(x_1,x_2)$ in part (b) agree. Integrability of $\zeta$ follows from $\E[|Y|]<\infty$ and the bounds $r_t\ge\underline p$.
\end{proof}

The estimand $\tau(x_1,x_2)$ produced by part (b){} is the treatment effect averaged over the conditional law of the nuisance block $V$ within the stratum $\{X_1=x_1,X_2=x_2\}$. As noted in the main text, the hypotheses tested are hypotheses about $\tau$, not about $\tau^{\mathrm{full}}$: constancy of $\tau^{\mathrm{full}}$ in $x_1$ implies constancy of $\tau$ in $x_1$, but the converse fails whenever variation of $\tau^{\mathrm{full}}$ across $v$ cancels upon averaging within a stratum. Consequently, a rejection transfers upward to the finer null, while a non-rejection does not transfer downward.
\section{Proofs of the asymptotic results}\label{osec:proofs}
This section proves Theorems~\ref{thm:variance} and \ref{thm:normality} and Proposition~\ref{prop:size} of the main text; the auxiliary lemmas are in Section~\ref{osec:aux}. The proof has four steps: linearization (Section~\ref{osec:pf-lin}), Poissonization (Section~\ref{osec:pf-poisson}), a H\'ajek-type projection (Section~\ref{osec:pf-proj}), and a central limit theorem for the projected statistic followed by de-Poissonization (Section~\ref{osec:pf-clt}); Section~\ref{osec:pf-main} combines them.

Throughout, $j$ is fixed, $j'\neq j$ is the other block, $\mathcal C:=\mathcal C_j$, $\kappa:=\kappa_{j,N}$, $\Theta:=\Theta_j$, $\mathcal W(x):=\{z:\|x_1-z_1\|_\infty\le h_1/2,\|x_2-z_2\|_\infty\le h_2/2\}$ is the window at $x$, and two windows overlap if they intersect. The extension from compact product sets inside the box $\mathcal B$ of Lemma~\ref{olem:pf-clt} to $\dom$, Lemmas~\ref{olem:poisson}, \ref{olem:centering}, \ref{olem:variance-gen} and \ref{olem:depoisson}, and routine computations are in Section~\ref{osec:aux}. 

\subsection{Linearization}\label{osec:pf-lin}
The first step replaces the estimated pairwise differences by sums of i.i.d.\ influence-function terms plus the smoothing bias. Let $\epsilon_N:=\{\log N/(N\eta)\}^{1/2}+h_1^{s_1}+h_2^{s_2}$. The linearization holds for both estimators, without any null hypothesis; its proof, with that of Proposition~\ref{prop:transfer}, is in Section~\ref{osec:from-main}.

\begin{lemma}[Linearization]\label{olem:pf-linearization}
(a) $\sup_{\dom}|\widehat P_t-P_t|$ and $\sup_{\dom}|\tauh-\tau|$ are $O_p(\epsilon_N)$, $\sup_{\dom}|\rhoh-\varrho|=o_p(\eta^{1/2})$ for the plug-in kernel estimator $\rhoh$ of $\varrho$ in \eqref{oeq:rhodef}, $\sqrt N \epsilon_N^2\to0$ and $\epsilon_N=o(\eta^{1/2})$. (b) $\Gammah_j(c)-\Gamma_j(c)=N^{-1}\sum_{i=1}^N\Theta_j(W_i;c)+B_{j,N}(c)+R_{j,N}(c)$, where $\sup_c|B_{j,N}(c)|=O(h_1^{s_1}+h_2^{s_2})$ and $\sup_{c}|R_{j,N}(c)|=O_p(\epsilon_N^2)=o_p(N^{-1/2})$. (c) $\Dh_j^\Psi(\mathcal C_j)-D^{\mathrm{lin}}_j=o_p(1)$, where $D^{\mathrm{lin}}_j$ is \eqref{eq:Dgen} with $\Gamma_j+B_{j,N}+N^{-1}\sum_i\Theta_j(W_i;\cdot)$ in place of $\Gammah_j$.
\end{lemma}
\noindent\hyperref[oprf:pf-linearization]{\emph{Proof}} (Section~\ref{osec:from-main}).

\subsection{Poissonization}\label{osec:pf-poisson}
Let $\mathcal N$ be a Poisson variable with mean $N$ independent of the data, and define the Poissonized version of the CATE estimator as $$\tau_{\mathcal N}(x):=\tau(x)+N^{-1}\sum_{i\le\mathcal N}\eta^{-1}\xi(W_i;x)\mathcal K_i(x)$$ with mean $\tau(x)+B_N(x)$, where  $B_N(x):=\E[\eta^{-1}\xi(W;x)\mathcal K(x;X)]$ is the smoothing bias, so that $B_{j,N}(c)=B_N(c_1)-B_N(c_2)$, and
\begin{equation}\label{oeq:pf-poisson-field}
\Gamma_{\mathcal N}(c):=\tau_{\mathcal N}(c_1)-\tau_{\mathcal N}(c_2),\qquad
D^P_j:=\sqrt N\int_{\mathcal C}\Psi\big(\Gamma_{\mathcal N}(c)\big)\,dc .
\end{equation}
By the definition of the influence function, $\eta^{-1}\xi(W_i;x)\mathcal K_i(x)=\theta_N(W_i;x)+B_N(x)$, so that the Poissonized CATE estimator (field), which adds the uncentered summands, becomes
\begin{equation*}
\tau_{\mathcal N}(x)=\tau(x)+\frac{\mathcal N}{N}\,B_N(x)+\frac1N\sum_{i\le\mathcal N}\theta_N(W_i;x).
\end{equation*}
On $\{\mathcal N=N\}$ the factor $\mathcal N/N$ equals one and the sum runs over $i=1,\dots,N$; differencing at $c_1$ and $c_2$ then gives the Poissonized  CATE difference estimator
\begin{align*}
\Gamma_{\mathcal N}(c)&=\{\tau(c_1)-\tau(c_2)\}+\{B_N(c_1)-B_N(c_2)\}+\frac1N\sum_{i=1}^N\{\theta_N(W_i;c_1)-\theta_N(W_i;c_2)\}\\
&=\Gamma_j(c)+B_{j,N}(c)+\frac1N\sum_{i=1}^N\Theta_j(W_i;c),
\end{align*}
the linearized field of Lemma~\ref{olem:pf-linearization}(b) without its remainder, so that $D^P_j=D^{\mathrm{lin}}_j$ on $\{\mathcal N=N\}$.

The proofs rely on three properties of the Poissonized field, established in Lemma~\ref{olem:poisson}. First, $\mathcal K_i(x)=0$ unless $X_i\in\mathcal W(x)$, so $\tau_{\mathcal N}(x)$ depends on the sample only through the points in $\mathcal W(x)$, and $\Gamma_{\mathcal N}(c)$ only through those in $\mathcal W(c_1)\cup\mathcal W(c_2)$. As a Poisson process has independent counts on disjoint sets (independent increments), $\Gamma_{\mathcal N}(c)$ and $\Gamma_{\mathcal N}(c')$ are independent whenever the windows of $c$ do not overlap those of $c'$. Second, differencing the display for $\tau_{\mathcal N}$ at $c_1$ and $c_2$ gives, for every value of $\mathcal N$,
\begin{equation*}
\Gamma_{\mathcal N}(c)=\Gamma_j(c)+\frac1N\sum_{i\le\mathcal N}\big\{\Theta_j(W_i;c)+B_{j,N}(c)\big\},\qquad \E\Gamma_{\mathcal N}(c)=\Gamma_j(c)+B_{j,N}(c),
\end{equation*}
the mean by Wald's identity. The influence-function term $\Theta_j(W;c)$ is centered, but the summand $f_c(W):=\Theta_j(W;c)+B_{j,N}(c)$ has mean $B_{j,N}(c)$. The variance of a compound-Poisson sum is governed by the second moment of its summand rather than by its variance: as $\E\mathcal N=\Var\mathcal N=N$,
\begin{equation*}
\Var\Big(\sum_{i\le\mathcal N}f_c(W_i)\Big)=\E\mathcal N\,\Var f_c(W)+\Var(\mathcal N)\{\E f_c(W)\}^2=N\,\E[f_c(W)^2].
\end{equation*}
With $\kappa_{j,N}(c):=\E[\Theta_j(W;c)^2]$, the variance of $\Theta_j$ defined in Section~\ref{sec:asym-center} of the main text, this gives
\begin{equation*}
N\Var\big(\Gamma_{\mathcal N}(c)\big)=\E\big[\{\Theta_j(W;c)+B_{j,N}(c)\}^2\big]=\kappa_{j,N}(c)+B_{j,N}(c)^2=:\kappa_{\mathcal N}(c),
\end{equation*}
the cross term vanishing as $\E\Theta_j=0$, and the standardized field is
\begin{equation*}
S_{\mathcal N}(c):=\frac{\sqrt N\{\Gamma_{\mathcal N}(c)-\E\Gamma_{\mathcal N}(c)\}}{\kappa_{\mathcal N}(c)^{1/2}} .
\end{equation*}
Third, $S_{\mathcal N}(c)$ is a normalized sum of $N$ i.i.d.\ blocks. By the infinite divisibility of the Poisson distribution, we may take $\mathcal N=\sum_{k=1}^N\mathcal N_k$ without loss of generality, with $\mathcal N_1,\dots,\mathcal N_N$ i.i.d.\ Poisson with mean one and independent of the data, since only the law of $\Gamma_{\mathcal N}$ matters. Let $I_k:=\{i:\mathcal N_1+\dots+\mathcal N_{k-1}<i\le\mathcal N_1+\dots+\mathcal N_k\}$ index the observations of the $k$th block, so that $\sum_{i\le\mathcal N}=\sum_{k=1}^N\sum_{i\in I_k}$. Since also $\sqrt N B_{j,N}(c)=N^{-1/2}\sum_{k=1}^NB_{j,N}(c)$, the two displays above give
\begin{equation*}
S_{\mathcal N}(c)=\frac1{\sqrt N}\sum_{k=1}^NQ_k(c),\qquad
Q_k(c):=\frac{1}{\kappa_{\mathcal N}(c)^{1/2}}\Big[\sum_{i\in I_k}\big\{\Theta_j(W_i;c)+B_{j,N}(c)\big\}-B_{j,N}(c)\Big].
\end{equation*}
The blocks $Q_k(c)$ are i.i.d.\ because the pairs $(\mathcal N_k,\{W_i\}_{i\in I_k})$ are. Each is a centered and scaled compound-Poisson sum: its mean is zero by Wald's identity, as $\E\mathcal N_k=1$ and $\E\Theta_j=0$, and its variance is one, by the compound-Poisson variance formula above with $\E\mathcal N_k=\Var\mathcal N_k=1$, which gives $\Var(\sum_{i\in I_k}f_c(W_i))=\E[f_c(W)^2]=\kappa_{\mathcal N}(c)$. Its third absolute moment is $O(\eta^{-1/2})$, because $\Theta_j$ is of order $\eta^{-1}$ on a window of probability of order $\eta$, so that $\E|\Theta_j|^3=O(\eta^{-2})$ while $\kappa_{\mathcal N}$ is of order $\eta^{-1}$. The same holds jointly for $(Q_k(c),Q_k(c'))$.

The normal approximations below compare the moments of $\Psi(S_{\mathcal N}(c))$ with those of $\Psi$ at standard normals. With $Z,Z_1,Z_2$ independent standard normals, let
\begin{equation*}
m_\Psi:=\E[\Psi(Z)],\qquad g_\Psi(\rho):=\Cov\big(\Psi(\sqrt{1-\rho^2}Z_1+\rho Z_2),\Psi(Z_2)\big),\quad \rho\in[-1,1],
\end{equation*}
the mean of $\Psi$ at a standard normal and the covariance of $\Psi$ at two standard normals with correlation $\rho$, as in \eqref{eq:stat-g} of the main text. By the representation $\Psi(u)=a_\Psi|u|+b_\Psi u$ of \eqref{eq:stat-psi-rep}, $m_\Psi=a_\Psi\sqrt{2/\pi}$ and $g_\Psi(\rho)=a_\Psi^2g_{\mathrm{abs}}(\rho)+b_\Psi^2\rho$, with $g_{\mathrm{abs}}(\rho)=\tfrac2\pi\{\rho\arcsin\rho+\sqrt{1-\rho^2}-1\}$. We use the Berry--Esseen inequalities of \citet{sweeting1977speeds}, in the form of Fact~6.1 of \citet{gine2003bm}. That is, for i.i.d.\ standardized bivariate blocks $(Q_i,Q_i')$ with correlation $\rho$, $|\rho|\le1-\delta$, $S:=N^{-1/2}\sum_iQ_i$ and $S':=N^{-1/2}\sum_iQ_i'$, we have
\begin{equation}\label{oeq:pf-BE}
\big|\E\Psi(S)-m_\Psi\big|\le C_1\frac{\E|Q_1|^3}{\sqrt N},\qquad
\big|\Cov(\Psi(S),\Psi(S'))-g_\Psi(\rho)\big|\le C_\delta\frac{\E|Q_1|^3+\E|Q_1'|^3}{\sqrt N},
\end{equation}
which for general $\Psi$ follow from \eqref{eq:stat-psi-rep}. Applied with $Q_i=Q_i(c)$ and $Q_i'=Q_i(c')$, \eqref{oeq:pf-BE} approximates the means and covariances of $\Psi(S_{\mathcal N}(c))$ and $\Psi(S_{\mathcal N}(c'))$ with errors $O((N\eta)^{-1/2})$.

\subsection{H\'ajek-type projection}\label{osec:pf-proj}
This step is the H\'ajek-type projection outlined in Section~\ref{sec:asym-main} of the main text. With the centered integrand $\bar\Psi_{\mathcal N}(c):=\sqrt N\Psi(\Gamma_{\mathcal N}(c))-\E[\sqrt N\Psi(\Gamma_{\mathcal N}(c))]$, the Poissonized statistic is $D^P_j-\E[D^P_j]=\int_{\mathcal C}\bar\Psi_{\mathcal N}(c)\,dc$. By the first property in Section~\ref{osec:pf-poisson}, the one-point field $x\mapsto\tau_{\mathcal N}(x)$ is locally dependent: $\tau_{\mathcal N}(x)$ and $\tau_{\mathcal N}(x')$ are independent once $\mathcal W(x)$ and $\mathcal W(x')$ do not overlap. The pairwise integrand does not inherit this property. As $\bar\Psi_{\mathcal N}(c)$ depends on the points in both $\mathcal W(c_1)$ and $\mathcal W(c_2)$, the integrands $\bar\Psi_{\mathcal N}(c)$ and $\bar\Psi_{\mathcal N}(c')$ are dependent whenever a window of $c$ overlaps a window of $c'$, for instance when $c_1'$ is close to $c_1$ while $c_2'$ is far from $c_2$. Configurations far apart in $\mathcal C$ can therefore be dependent, so $\mathcal C$ cannot be partitioned into cells that are dependent only on neighboring cells, a partition that the blocking argument of \citet{gine2003bm} for one-point statistics requires. 

We therefore project $\bar\Psi_{\mathcal N}(c)$ onto the two single-point fields $\tau_{\mathcal N}(c_1)$ and $\tau_{\mathcal N}(c_2)$. Off the band these are independent, and $\sum_{k=1}^2\E[\bar\Psi_{\mathcal N}(c)\mid\tau_{\mathcal N}(c_k)]$ is the H\'ajek projection of $\bar\Psi_{\mathcal N}(c)$ onto sums of a function of $\tau_{\mathcal N}(c_1)$ and a function of $\tau_{\mathcal N}(c_2)$. On the band, where $\mathcal W(c_1)$ and $\mathcal W(c_2)$ overlap, $\tau_{\mathcal N}(c_1)$ and $\tau_{\mathcal N}(c_2)$ are dependent and the sum is not an orthogonal projection; there it serves only as a definition. The decomposition below is an identity at every $c$, and the band, of measure $O(h_j^{d_j})$ in $\mathcal C$, contributes $O(h_j^{d_j})$ to the variance of the remainder (proof of Lemma~\ref{olem:pf-projection}(c), case (III)). Collecting, for each point $x$, the projections of all configurations that have $x$ as one of their two points gives the projection field $\Pi_{\mathcal N}$, and a remainder $R_{\mathcal N}$ defined respectively  as 
\begin{align}
\Pi_{\mathcal N}(x)&:=\int_{\mathcal C}\sum_{k=1}^{2}\E\big[\bar\Psi_{\mathcal N}(c)\,\big|\,\tau_{\mathcal N}(c_k)\big]\,\delta_x(c_k)\,dc ,\label{oeq:pf-Delta-proj}\\
R_{\mathcal N}(c)&:=\bar\Psi_{\mathcal N}(c)-\sum_{k=1}^{2}\E\big[\bar\Psi_{\mathcal N}(c)\,\big|\,\tau_{\mathcal N}(c_k)\big],\label{oeq:pf-R-rem}
\end{align}
where $\delta_x$ is the Dirac measure at $x$, so that $\delta_x(c_k)$ restricts the integral to the configurations whose $k$th point is $x$ and $\Pi_{\mathcal N}(x)$, a function of $\tau_{\mathcal N}(x)$, collects the projections of every configuration with $x$ in either slot. For $j=2$, with $c=(x_1,x_2,x_2')$, $c_1=(x_1,x_2)$ and $c_2=(x_1,x_2')$,
\begin{align*}
\Pi_{\mathcal N}(x_1,x_2)&=\int\E\big[\bar\Psi_{\mathcal N}(x_1,x_2,x_2')\mid\tau_{\mathcal N}(x_1,x_2)\big]\ind_{\mathcal C}(x_1,x_2,x_2')\,dx_2'\\
&\quad+\int\E\big[\bar\Psi_{\mathcal N}(x_1,x_2',x_2)\mid\tau_{\mathcal N}(x_1,x_2)\big]\ind_{\mathcal C}(x_1,x_2',x_2)\,dx_2' ;
\end{align*}%
for $j=1$, with $c=(x_1,x_1',x_2)$, the integral runs over $x_1'$ instead. Integrating over $x$ and applying Fubini's theorem removes the restriction,
\begin{equation*}
\int_{\dom}\Pi_{\mathcal N}(x)\,dx=\int_{\mathcal C}\sum_{k=1}^{2}\E\big[\bar\Psi_{\mathcal N}(c)\,\big|\,\tau_{\mathcal N}(c_k)\big]\,dc ,
\end{equation*}
so that, by the definition of $R_{\mathcal N}$, $\int_{\dom}\Pi_{\mathcal N}(x) dx+\int_{\mathcal C}R_{\mathcal N}(c) dc=\int_{\mathcal C}\bar\Psi_{\mathcal N}(c)dc=D^P_j-\E[D^P_j]$. Lemma~\ref{olem:pf-projection} below shows that the remainder is asymptotically negligible. As a function of $\tau_{\mathcal N}(x)$ alone, $\Pi_{\mathcal N}(x)$ depends only on the points in $\mathcal W(x)$, so that $\int\Pi_{\mathcal N}(x) dx$ is locally dependent; for $\Dh_2$ the configurations are written out in Lemma~\ref{olem:projection} of Section~\ref{osec:aux-proj}.

\begin{lemma}[Projection]\label{olem:pf-projection}
(a) $D^P_j-\E[D^P_j]=\int_{\dom}\Pi_{\mathcal N}(x)\,dx+\int_{\mathcal C}R_{\mathcal N}(c)\,dc$; (b) for $c$ off the band, $\E[R_{\mathcal N}(c)\mid\mathcal G]=0$ for every $\sigma$-field $\mathcal G$ generated by Poisson increments over a region disjoint from $\mathcal W(c_2)$, and symmetrically with $c_1$; (c) $\Var(\int_{\mathcal C}R_{\mathcal N}(c)dc)=O(h_j^{d_j})$; (d) $\Var(\int_{\dom}\Pi_{\mathcal N}(x)dx)=\Var(D^P_j)+O(h_j^{d_j/2})$ and $\sup_{x}\E|\Pi_{\mathcal N}(x)|^3=O(\eta^{-3/2})$.
\end{lemma}

\begin{proof}[Proof of Lemma~\ref{olem:pf-projection}]
Of the conditions on $\Psi$, the proof uses only those of Assumption~\ref{ass:psi} of the main text.\\ \emph{(a)} Integrating the two slot projections in \eqref{oeq:pf-R-rem} over the points $\{c_1=x\}$ and $\{c_2=x\}$ gives $\int\Pi_{\mathcal N}$; the two-slot definition dispenses with the evenness of $\Psi$.\\ \emph{(b)} Off the band $\tau_{\mathcal N}(c_1)$ and $\tau_{\mathcal N}(c_2)$ are independent; if $\mathcal G$ is generated by increments over a region disjoint from $\mathcal W(c_2)$, $\tau_{\mathcal N}(c_2)$ is independent of $\mathcal G$, so $\E[\sqrt N\Psi(\Gamma_{\mathcal N}(c))\mid\mathcal G]$ is the first slot projection, the second has conditional mean zero, and the three terms of \eqref{oeq:pf-R-rem} cancel.\\ \emph{(c)} \emph{Size of the remainder.} We show that $\sup_{c,c'\in\mathcal C}|\E[R_{\mathcal N}(c)R_{\mathcal N}(c')]|=O(\eta^{-1})$, in four steps. First, by the definition \eqref{oeq:pf-R-rem} of $R_{\mathcal N}(c)$ and Minkowski's inequality,
\begin{equation*}
\|R_{\mathcal N}(c)\|_2\le\|\bar\Psi_{\mathcal N}(c)\|_2+\sum_{k=1}^{2}\big\|\E[\bar\Psi_{\mathcal N}(c)\mid\tau_{\mathcal N}(c_k)]\big\|_2\le3\|\bar\Psi_{\mathcal N}(c)\|_2 .
\end{equation*}
The second inequality is conditional Jensen's inequality: for a square-integrable $Y$ and a $\sigma$-field $\mathcal F$, $\{\E[Y\mid\mathcal F]\}^2\le\E[Y^2\mid\mathcal F]$, and taking expectations gives $\|\E[Y\mid\mathcal F]\|_2\le\|Y\|_2$. Second, $\bar\Psi_{\mathcal N}(c)$ is centered, so $\|\bar\Psi_{\mathcal N}(c)\|_2^2=N\Var\Psi(\Gamma_{\mathcal N}(c))$. Let $\Gamma'$ be an independent copy of $\Gamma_{\mathcal N}(c)$. As $\Psi$ is $1$-Lipschitz (Assumption~\ref{ass:psi}(a)),
\begin{align*}
N\Var\Psi(\Gamma_{\mathcal N}(c))&=\frac N2\E\big[\{\Psi(\Gamma_{\mathcal N}(c))-\Psi(\Gamma')\}^2\big]\\
&\le\frac N2\E\big[\{\Gamma_{\mathcal N}(c)-\Gamma'\}^2\big]=N\Var\Gamma_{\mathcal N}(c)=\kappa_{\mathcal N}(c),
\end{align*}
where the last equality is the variance identity of Section~\ref{osec:pf-poisson}. Third, by the same identity, $\kappa_{\mathcal N}(c)=\E[\{\Theta_j(W;c)+B_{j,N}(c)\}^2]=\kappa_{j, N}(c)+B_{j, N}(c)^2$: the influence-function variance $\kappa_{j,N}(c)=\E[\Theta_j(W;c)^2]$ plus the squared smoothing bias, the cross term vanishing as $\E\Theta_j=0$. By Lemma~\ref{olem:poisson}, $\eta\kappa_{j,N}(c)\le 4\sup\varrho+o(1)$ uniformly on $\mathcal C$ and $\sup B_{j,N}^2=O((h_1^{s_1}+h_2^{s_2})^2)=o(1)$, so $\sup_{c\in\mathcal C}\kappa_{\mathcal N}(c)=O(\eta^{-1})$. Combining the three steps, $\sup_{c\in\mathcal C}\|R_{\mathcal N}(c)\|_2\le3\sup_{c\in\mathcal C}\kappa_{\mathcal N}(c)^{1/2}=O(\eta^{-1/2})$. Fourth, by the Cauchy--Schwarz inequality,
\begin{equation*}
|\E[R_{\mathcal N}(c)R_{\mathcal N}(c')]|\le\|R_{\mathcal N}(c)\|_2\|R_{\mathcal N}(c')\|_2\le9\sup_{c\in\mathcal C}\kappa_{\mathcal N}(c)=O(\eta^{-1}).
\end{equation*} \emph{Classification of pairs.} By the law of iterated expectations and the centering of $\bar\Psi_{\mathcal N}$, at every $c\in\mathcal C$, on and off the band,
\begin{equation*}
\E R_{\mathcal N}(c)=\E\bar\Psi_{\mathcal N}(c)-\sum_{k=1}^{2}\E\big\{\E[\bar\Psi_{\mathcal N}(c)\mid\tau_{\mathcal N}(c_k)]\big\}=\E\bar\Psi_{\mathcal N}(c)-2\,\E\bar\Psi_{\mathcal N}(c)=0 .
\end{equation*}
Hence, by Fubini's theorem, $\Var(\int_{\mathcal C}R_{\mathcal N})=\E[(\int_{\mathcal C}R_{\mathcal N})^2]=\iint_{\mathcal C^2}\E[R_{\mathcal N}(c)R_{\mathcal N}(c')]\,dc\,dc'$. The uniform bound alone gives only $O(\eta^{-1})$, so we show that most pairs contribute zero. 

We partition all pairs \((c,c')\in\mathcal C^2\) into three mutually exclusive cases.
In particular, let $n_{cc'}:=\sum_{k=1}^2\sum_{l=1}^2
\mathbf 1\left\{
\mathcal W(c_k)\cap\mathcal W(c'_l)\neq\emptyset
\right\}$ be the number of overlapping cross pairs of windows $(\mathcal W(c_k),\mathcal W(c'_l))$, $k,l\in\{1,2\}$. Recall that $R_{\mathcal N}(c)$ depends only on the Poisson increments over $\mathcal W(c_1)\cup\mathcal W(c_2)$ and that increments over disjoint regions are independent (Section~\ref{osec:pf-poisson}).\\ (I) $n_{cc'}=0$. The remainders depend on increments over disjoint regions, so they are independent and $\E[R_{\mathcal N}(c)R_{\mathcal N}(c')]=\E R_{\mathcal N}(c)\,\E R_{\mathcal N}(c')=0$.\\ (II) $n_{cc'}=1$, say $\mathcal W(c_1)\cap\mathcal W(c_1')\neq\emptyset$, with $c$ and $c'$ off the band. Then $A:=\mathcal W(c_1)\cup\mathcal W(c_1')$, $\mathcal W(c_2)$ and $\mathcal W(c_2')$ are pairwise disjoint. Let $\mathcal G$ be generated by the increments over $A$. Given $\mathcal G$, the remaining randomness of $R_{\mathcal N}(c)$ and $R_{\mathcal N}(c')$ lies in $\mathcal W(c_2)$ and $\mathcal W(c_2')$ respectively, so the two are conditionally independent. Since $A$ is disjoint from $\mathcal W(c_2)$ and $\mathcal W(c_2')$, part (b) gives $\E[R_{\mathcal N}(c)\mid\mathcal G]=\E[R_{\mathcal N}(c')\mid\mathcal G]=0$, so $\E[R_{\mathcal N}(c)R_{\mathcal N}(c')]=\E\{\E[R_{\mathcal N}(c)\mid\mathcal G]\,\E[R_{\mathcal N}(c')\mid\mathcal G]\}=0$ exactly. This is the case that requires the projection. An overlap constrains the shared block $j'$ to a set of measure $O(h_{j'}^{d_{j'}})$ and one block-$j$ point to a set of measure $O(h_j^{d_j})$, so these pairs have measure $O(\eta)$. Without the degeneracy (b), they would contribute $O(\eta\cdot\eta^{-1})=O(1)$, the order of $\Var(D^P_j)$.\\ (III) All other pairs, with $n_{cc'}\ge2$, or $n_{cc'}=1$ and $c$ or $c'$ on the band. Besides the shared-block constraint, these pairs carry two block-$j$ constraints of measure $O(h_j^{d_j})$ each: the overlap, and a second overlap or the band condition. For example, for $j=2$, $c=(x_1,x_2,x_2')$, $c'=(y_1,y_2,y_2')$ and the overlap $\mathcal W(c_1)\cap\mathcal W(c_1')\neq\emptyset$, these are $|x_1-y_1|=O(h_1)$, $|x_2-y_2|=O(h_2)$, and $|x_2'-y_2'|=O(h_2)$ or $|x_2-x_2'|=O(h_2)$. Their measure is therefore $O(\eta h_j^{d_j})$, and against the uniform bound they contribute $O(h_j^{d_j})$. Summing the three cases, $\Var(\int_{\mathcal C}R_{\mathcal N}(c) dc)=0+0+O(h_j^{d_j})$.\\
\emph{(d)} \emph{Variance.} By (a), $\int_{\dom}\Pi_{\mathcal N}(x)dx=(D^P_j-\E D^P_j)-\int_{\mathcal C}R_{\mathcal N}(c)dc$, so
\begin{equation*}
\Var\Big(\int_{\dom}\Pi_{\mathcal N}(x) dx\Big)=\Var(D^P_j)-2\Cov\Big(D^P_j,\int_{\mathcal C}R_{\mathcal N}(c)dc\Big)+\Var\Big(\int_{\mathcal C}R_{\mathcal N}(c)dc\Big).
\end{equation*}
By Lemma~\ref{olem:variance-gen}, $\Var(D^P_j)=O(1)$, and by (c), $\Var(\int_{\mathcal C}R_{\mathcal N}(c)dc)=O(h_j^{d_j})$. By Cauchy--Schwarz, the covariance is at most $\{\Var(D^P_j)\Var(\int_{\mathcal C}R_{\mathcal N}(c)dc)\}^{1/2}=O(h_j^{d_j/2})$, which proves the first claim.\\
\emph{Third moment:} To show this result, 
we express $\Pi_{\mathcal N}(x)$ as an integral of conditional
expectations, partition the domain according to kernel-window
overlap, and use the Lipschitz property of $\Psi$ to bound
both contributions without distributional assumptions.

\emph{Setup:} We show that $\|\Pi_{\mathcal N}(x)\|_3=O(\eta^{-1/2})$ uniformly in $x$. Fix $x$ and a slot $k$. A configuration $c$ with $c_k=x$ is determined by the block-$j$ coordinate $y$ of its other point $\tilde x(y):=c_{3-k}$; write $c(y)$ for it. By \eqref{oeq:pf-Delta-proj}, $\Pi_{\mathcal N}(x)=\sum_{k=1}^2\int\E[\bar\Psi_{\mathcal N}(c(y))\mid\tau_{\mathcal N}(x)]\ind_{\mathcal C}(c(y))\,dy$, as in the expression for $j=2$ above. The $y$-domain of each integral has measure bounded by a constant $L$, as $\mathcal C$ is bounded. Split it into the off-band slice $\{\|y-x_j\|_\infty>h_j\}$, where $\mathcal W(x)$ and $\mathcal W(\tilde x(y))$ are disjoint, and the band slice $\{\|y-x_j\|_\infty\le h_j\}$, of measure at most $(2h_j)^{d_j}$. Let
\begin{gather*}
t:=\sqrt N\{\tau_{\mathcal N}(x)-\E\tau_{\mathcal N}(x)\},\qquad V(y):=\sqrt N\{\tau_{\mathcal N}(\tilde x(y))-\E\tau_{\mathcal N}(\tilde x(y))\},\\
\mu(y):=\sqrt N\,\E\Gamma_{\mathcal N}(c(y)),
\end{gather*}
so that $\sqrt N\Gamma_{\mathcal N}(c(y))=\pm\{t-V(y)\}+\mu(y)$, with sign $+$ for $k=1$ and $-$ for $k=2$. By \eqref{oeq:thirdmom} (Rosenthal's inequality for Poissonized sums, with $N\eta\to\infty$), $\sup_x\|t\|_3=O(\eta^{-1/2})$, and the same holds for $V(y)$.\\

\emph{Off-band slice.} Here $V(y)$ is independent of $\tau_{\mathcal N}(x)$ by independent increments, so conditioning on $\tau_{\mathcal N}(x)$ fixes $t$ and averages over $V(y)$:
\begin{equation*}
\E[\sqrt N\Psi(\Gamma_{\mathcal N}(c(y)))\mid\tau_{\mathcal N}(x)]=\varphi_y(t),\qquad\varphi_y(s):=\E\,\Psi\big(\pm\{s-V(y)\}+\mu(y)\big).
\end{equation*}
As $\Psi$ is $1$-Lipschitz (Assumption~\ref{ass:psi}(a)), $|\varphi_y(s)-\varphi_y(s')|\le|s-s'|$. Hence, with $t'$ an independent copy of $t$, $|\E[\bar\Psi_{\mathcal N}(c(y))\mid\tau_{\mathcal N}(x)]|=|\varphi_y(t)-\E\varphi_y(t')|\le\E_{t'}|t-t'|\le|t|+\E|t|$ for every $y$. Integrating over the slice, the off-band part of the $k$th integral is bounded in absolute value by $L(|t|+\E|t|)$, whose $L_3$-norm is at most $2L\|t\|_3=O(\eta^{-1/2})$.\\

\emph{Band slice.} Here $\tau_{\mathcal N}(x)$ and $V(y)$ share Poisson points and $\varphi_y$ is unavailable. Let $Y(y):=\sqrt N\Psi(\Gamma_{\mathcal N}(c(y)))$ and $Y'(y)$ be an independent copy. By Minkowski's integral inequality and conditional Jensen's inequality,
\begin{equation*}
\Big\|\int_{\{\|y-x_j\|_\infty\le h_j\}}\E[\bar\Psi_{\mathcal N}(c(y))\mid\tau_{\mathcal N}(x)]\,dy\Big\|_3\le(2h_j)^{d_j}\sup_y\|Y(y)-\E Y(y)\|_3 .
\end{equation*}
As $Y-\E Y=\E[Y-Y'\mid Y]$, Jensen's inequality and the Lipschitz property give $\|Y-\E Y\|_3\le\|Y-Y'\|_3\le2\sqrt N\|\Gamma_{\mathcal N}-\E\Gamma_{\mathcal N}\|_3\le2(\|t\|_3+\|V(y)\|_3)$, so the band part is $O(h_j^{d_j}\eta^{-1/2})$. The deterministic shift $\mu(y)$ cancels, so no null hypothesis is needed.\\

\emph{Conclusion.} By Minkowski's inequality over the two slots and the two slices, $\sup_x\|\Pi_{\mathcal N}(x)\|_3\le2\{O(\eta^{-1/2})+O(h_j^{d_j}\eta^{-1/2})\}=O(\eta^{-1/2})$, that is, $\sup_x\E|\Pi_{\mathcal N}(x)|^3=O(\eta^{-3/2})$; for $\Dh_2$ this is \eqref{oeq:delta3}. This bound gives the Lyapunov condition in the proof of the central limit theorem in Lemma~\ref{olem:pf-clt}.
\end{proof}

\subsection{Central limit theorem and de-Poissonization}\label{osec:pf-clt}
By Lemma~\ref{olem:pf-projection}, $D^P_j-\E[D^P_j]$ equals the locally dependent integral $\int_{\dom}\Pi_{\mathcal N}(x) dx$ up to a negligible remainder, so a central limit theorem for locally dependent fields applies to it. Stated jointly with the Poisson count in a box, which the de-Poissonization requires, it reads as follows.

\begin{lemma}[Central limit theorem for the projected field]\label{olem:pf-clt}
Let $\mathcal B\supset\dom$ be a box containing every window centered in $\dom$, $p_{\mathcal B}:=\Pr(X\notin\mathcal B)\in(0,1)$; such a box with $p_{\mathcal B}>0$ exists only if the support of $X$ extends beyond $\dom$ enlarged by half a window, and otherwise the lemma is applied, as in Section~\ref{osec:aux-sets}, on compact product sets whose windows lie in a box inside the support, with the extension to $\dom$ in Section~\ref{osec:aux-depoisson}, $U_{\mathcal N}:=\{\mathcal N(\mathcal B)-N(1-p_{\mathcal B})\}/\sqrt N$ the standardized count in $\mathcal B$ and $U'_{\mathcal N}{}:=\{\mathcal N(\mathcal B^c)-Np_{\mathcal B}\}/\sqrt N$ the standardized count outside it, where $\mathcal N(A):=\#\{i\le\mathcal N:X_i\in A\}$ is the Poisson count in $A$. If $\sigma^2:=\sigma^2_{j,\Psi}(\mathcal C)>0$ and $Q_{\mathcal N}:=\sigma^{-1}\int_{\dom}\Pi_{\mathcal N}$, then (a) $(Q_{\mathcal N},U_{\mathcal N})\dto N(0,\mathrm{diag}(1,1-p_{\mathcal B}))$ and (b) $(Q_{\mathcal N},U_{\mathcal N})$ is independent of $U'_{\mathcal N}$ for every $N$.
\end{lemma}
\noindent\hyperref[oprf:pf-clt]{\emph{Proof}} (Section~\ref{osec:aux-clt}).

In the proof of Lemma~\ref{olem:pf-clt}, we express $Q_{\mathcal N}$ and $U_{\mathcal N}$ as sums of integrals over cells of sides $h_1$ and $h_2$, a one-dependent array with Lyapunov sum $O(\eta^{1/2})$ by Lemma~\ref{olem:pf-projection}(d), and apply the central limit theorem of \citet{shergin1990central} for one-dependent fields (Section~\ref{osec:aux-clt}).

The law of $D^{\mathrm{lin}}_j$ is that of $D^P_j$ given $\{\mathcal N=N\}$ (Section~\ref{osec:pf-poisson}), and the de-Poissonization Lemma~\ref{olem:depoisson} of Section~\ref{osec:aux} transfers the limit of Lemma~\ref{olem:pf-clt} to it.

\subsection{Proofs of the main results}\label{osec:pf-main}
With the rescaled kernel autocorrelations $\bar\rho_1(u_1):=\int K_1(z)K_1(z+u_1)dz/\int K_1^2$, $u_1\in[-1,1]^{d_1}$, and $\bar\rho_2(u_2):=\int K_2(z)K_2(z+u_2)dz/\int K_2^2$, $u_2\in[-1,1]^{d_2}$, the limit variance of $\Dh_2$ in Theorem~\ref{thm:variance}(b) is
\begin{align}\label{oeq:pf-sigma2}
\sigma^2_{2}=\int_{\Xone}\int_{\Xtwo^3}\int_{[-1,1]^{d_2}}\int_{[-1,1]^{d_1}}
&g_{\mathrm{abs}}\big(\rho^*_2(x_1,x_2,x_2',x_2'',u_2,u_1)\big)\{\varrho(x_1,x_2)+\varrho(x_1,x_2')\}^{1/2}\\
&\times\{\varrho(x_1,x_2)+\varrho(x_1,x_2'')\}^{1/2}\;du_1\,du_2\,dx_2''\,dx_2'\,dx_2\,dx_1,\nonumber
\end{align}
\begin{equation}\label{oeq:pf-rhostar}
\rho^*_2:=\varrho(x_1,x_2)\,\bar\rho_1(u_1)\,\bar\rho_2(u_2)\big\{[\varrho(x_1,x_2)+\varrho(x_1,x_2')][\varrho(x_1,x_2)+\varrho(x_1,x_2'')]\big\}^{-1/2},
\end{equation}
and $\sigma^2_1$\label{oeq:pf-sigma1} is its mirror image; configurations interact only via the coordinate of an overlap.
\begin{proof}[Proof of Theorem~\ref{thm:variance}]
By homogeneity $\sqrt N\Psi(\Gamma_{\mathcal N}(c))=\kappa_{\mathcal N}(c)^{1/2}\Psi(S_{\mathcal N}(c)+o(1))$ uniformly under $\Gamma_j\equiv0$, so $$\Var(D^P_j)=\iint_{\mathcal C^2}\{\kappa_{\mathcal N}(c)\kappa_{\mathcal N}(c')\}^{1/2}\Cov(\Psi(S_{\mathcal N}(c)),\Psi(S_{\mathcal N}(c')))+o(1)$$ by the Lipschitz property. The covariance vanishes unless the window unions intersect and is $O(\eta^{-1})$ uniformly; pairs with two or more overlapping cross pairs, or with one and some other pair of $j$-coordinates within $2h_j$, have measure $O(\eta h_j^{d_j})$ and contribute $o(1)$, which leaves the four single-overlap events $\mathcal E_{kl}$, $k,l\in\{1,2\}$, in which exactly $\mathcal W(c_k)$ and $\mathcal W(c'_l)$ overlap. On $\mathcal E_{kl}$ the substitution $c'_l=c_k+(u_1h_1,u_2h_2)$, with Jacobian $\eta$, and the kernel-convolution expansion give, for the correlation $\rho_N$ of $\Theta_j$ defined before Theorem~\ref{thm:variance} of the main text, $\rho_N(c,c')=(-1)^{k+l}\rho^*_{kl}+O(h_1+h_2)$ uniformly, where $$\rho^*_{kl}:=\varrho(c_k)\bar\rho_1(u_1)\bar\rho_2(u_2)\{[\varrho(c_1)+\varrho(c_2)][\varrho(c_1')+\varrho(c_2')]\}^{-1/2}$$ is \eqref{oeq:pf-rhostar} for $(k,l)=(1,1)$, $j=2$. Since $\sup|\rho_N|\le1-\delta<1$ on these events, the second inequality in \eqref{oeq:pf-BE} replaces the covariance by $g_\Psi(\rho_N)$ at cost $O((N\eta)^{-1/2})$; this is Lemma~\ref{olem:variance-gen}. Bounded convergence gives equation \eqref{eq:sigma-gen} of the main text as the sum over $(k,l)$ of $\int g_\Psi((-1)^{k+l}\rho^*_{kl})\{[\varrho(c_1)+\varrho(c_2)][\varrho(c_1')+\varrho(c_2')]\}^{1/2}$ over the limiting domain of $\mathcal E_{kl}$ and $(u_1,u_2)\in[-1,1]^d$. For $\Psiabs$ on $\Csym{j}$, the four domains coincide after relabeling, and the sum, with the factor $1/4$, is \eqref{oeq:pf-sigma2}, positive as $g_{\mathrm{abs}}\ge0$ with $g_{\mathrm{abs}}(\rho^*)>0$ near $(u_1,u_2)=0$. In general $g_\Psi(\pm\rho)=a_\Psi^2g_{\mathrm{abs}}(\rho)\pm b_\Psi^2\rho$ splits the limit into $a_\Psi^2\sigma^2_{j,\mathrm{abs}}(\mathcal C)$ and $b_\Psi^2\lim\iint_{\mathcal C^2}\E[\Theta(W;c)\Theta(W;c')]=b_\Psi^2\lim\E[(\int_{\dom}w_{\mathcal C}\theta_N)^2]=b_\Psi^2\sigma^2_{j,\mathrm{lin}}(\mathcal C)$, with $w_{\mathcal C}$ the signed slot multiplicity defined before Theorem~\ref{thm:variance} of the main text.
\end{proof}

\begin{proof}[Proof of Theorem~\ref{thm:normality}]
Fix $\mathcal C$ with $\Gamma_j\equiv0$ on it. By Lemma~\ref{olem:pf-linearization}(c) it suffices to treat $D^{\mathrm{lin}}_j$, whose law is that of $D^P_j$ given $\mathcal N=N$. By Lemma~\ref{olem:pf-projection}(a),(c), $D^P_j-\E D^P_j=\int\Pi_{\mathcal N}+o_p(1)$, and $\Var(D^P_j)\to\sigma^2:=\sigma^2_{j,\Psi}(\mathcal C)\ge a_\Psi^2\sigma^2_{j,\mathrm{abs}}(\mathcal C)>0$ by Theorem~\ref{thm:variance}. Lemma~\ref{olem:pf-clt} and the de-Poissonization Lemma~\ref{olem:depoisson} \citep{beirlant1995asymptotic} yield $(D^{\mathrm{lin}}_j-\E D^P_j)/\sigma\dto N(0,1)$. Finally $\E D^P_j-m^\kappa_{j,N}(\mathcal C)=o(1)$ by Lemma~\ref{olem:centering}, $\widetilde m_j(\mathcal C)-m^\kappa_{j,N}(\mathcal C)=o_p(1)$ by Lemma~\ref{prop:centering-corr}, $\widetilde\sigma^2_{j,\Psi,N}(\mathcal C)\pto\sigma^2$ by Lemma~\ref{prop:sigmaN}, and Slutsky's theorem completes the proof.
\end{proof}

\begin{proof}[Proof of Proposition~\ref{prop:size}]
Part (a) follows from Theorem~\ref{thm:normality}. For part (b), Lemma~\ref{olem:pf-linearization}(a) and the Lipschitz property give $N^{-1/2}\Dh_j^\Psi(\mathcal C)=D_j^\Psi(\mathcal C)+O_p(\epsilon_N)$, while $\widetilde m_j=O_p(\eta^{-1/2})$ and $\widetilde\sigma_{j,\Psi,N}=O_p(1)$ by Lemmas~\ref{prop:centering-corr} and \ref{prop:sigmaN} for the noise field, so $\Sh_j^\Psi(\mathcal C)\ge\sqrt N\{D_j^\Psi(\mathcal C)+o_p(1)\}/\widetilde\sigma_{j,\Psi,N}-O_p(\eta^{-1/2})\pto\infty$, as $\sqrt N\eta^{1/2}\to\infty$. For part (c), let $\mathcal C:=\Cord{j}$ and $\Gamma:=\Gamma_j$. Under $H_0^{j,\ge}$, $\Gamma\le0$ on $\mathcal C$, as $c_1\ge c_2$ in block $j$, so $(u+v)_+\le u_+$ for $v\le0$ gives $\Dh_j^{\max}\le\check D_j^{\max}:=\sqrt N\int_{\mathcal C}(\Gammah_j-\Gamma)_+$, the statistic \eqref{eq:Dgen} of the noise field $\Gammah_j-\Gamma$. As the null only removes $\sqrt N\Gamma_j$ from the integrand, the proof of Theorem~\ref{thm:normality}, with Lemmas~\ref{prop:centering-corr} and~\ref{prop:sigmaN} for the noise field and $a_{\Psimax}=\frac12$, gives $\{\check D_j^{\max}-\widetilde m_j(\mathcal C)\}/\widetilde\sigma_{j,\Psimax,N}(\mathcal C)\dto N(0,1)$, whence $\limsup_N\Pr(p_j^{\max}<\alpha)\le\alpha$. If $\tau$ is constant in the $j$th block, $\Gamma\equiv0$ on $\mathcal C$ and part (a) gives equality; if it strictly increases in that block on a set of positive measure, $D_j^{\Psimax}(\mathcal C)>0$ and part (b) applies.
\end{proof}

\section{Proofs for estimated nuisance parameters}\label{osec:proofs-nui}
This section proves Theorems~\ref{thm:nuisance-param} and \ref{thm:nuisance-split} of the main text; the auxiliary results are in Section~\ref{osec:nui}.
The conditions of Assumption~\ref{ass:firststage}(a)(i) are: for a closed ball $\mathbb B_0$ around $\vartheta^*$, $\vartheta\mapsto\zeta(W;\chi_\vartheta)$ is twice continuously differentiable on $\mathbb B_0$, with derivatives written $\partial_\vartheta\zeta(W;\chi_\vartheta)$ and $\partial^2_\vartheta\zeta(W;\chi_\vartheta)$ and evaluated at $\vartheta^*$ as $\partial_\vartheta\zeta(W;\chi_{\vartheta^*})$, $\|\partial_\vartheta\zeta(W;\chi_{\vartheta^*})\|$ and $\sup_{\mathbb B_0}\|\partial^2_\vartheta\zeta(W;\chi_\vartheta)\|$ are bounded by $M(W)$ with $\sup_{x\in\mathcal U}\E[M(W)^6\mid X=x]<\infty$, and $\tau_\vartheta(x)=\E[\partial_\vartheta\zeta(W;\chi_{\vartheta^*})\mid X=x]$ is continuous on $\bar{\mathcal U}$, where $\mathcal U\supset\dom$ is the open set of Assumption~\ref{ass:firststage} and $\bar{\mathcal U}$ its (compact) closure; norms on $\mathcal U$ are Lebesgue. Throughout, $j$ is fixed, $j'\neq j$, $\mathcal C:=\mathcal C_j$, $C$ does not depend on $N$ or on the first stage, and $\theta_N$, $\Theta_j$, $\kappa_{j,N}$ are formed with the score $\xi(W;x):=\{\zeta-\tau(x)\}/f(x)$, as in Proposition~\ref{prop:transfer}. The proof of Lemma~\ref{lem:drift} is in Section~\ref{osec:from-main}.

\begin{proof}[Proof of Theorem~\ref{thm:nuisance-param}]
Put $\delta_\vartheta:=\widehat\vartheta-\vartheta^*$, $\zeta^*_i:=\zeta(W_i;\chi_{\vartheta^*})$, $\zeta_{\vartheta,i}:=\partial_\vartheta\zeta(W_i;\chi_{\vartheta^*})$ and $\Gamma_\vartheta(c):=\tau_\vartheta(c_1)-\tau_\vartheta(c_2)$; a subscript $*$ marks statistics computed from $\zeta^*_i$, to which Theorem~\ref{thm:normality} and Lemma~\ref{olem:pf-linearization} apply by Proposition~\ref{prop:transfer} and Assumption~\ref{ass:firststage}(a)(iii),(iv). Fix $\epsilon>0$ and $C_\epsilon$ such that $\mathcal E_N:=\{\widehat\vartheta\in \mathbb B_0,\sqrt N\|\delta_\vartheta\|\le C_\epsilon,\inf_{\dom}\widehat f\ge\inf f/2\}$, with $\widehat f(x):=(N\eta)^{-1}\sum_i\mathcal K_i(x)$ the kernel density estimate of $f$, has probability at least $1-\epsilon$ for $N$ large; all bounds below are on $\mathcal E_N$.

\emph{Step 1 (expansion).} By Taylor's theorem on the ball $\mathbb B_0$, $\widehat\zeta_i=\zeta^*_i+\delta_\vartheta'\zeta_{\vartheta,i}+R_i$, $|R_i|\le\|\delta_\vartheta\|^2M(W_i)$, and by linearity $\tauh=\tauh_*+\delta_\vartheta'\widehat\tau_\vartheta+\tauh^{(R)}$, where $\tauh^{(Z)}$ is the regression of $Z_i$ and $\widehat\tau_\vartheta:=\tauh^{(\zeta_\vartheta)}$. Here $\int|\tauh^{(R)}|=O_p(N^{-1})$, as $\int\eta^{-1}|\mathcal K_i|\le\|K_1K_2\|_{L_1}$, and $\E\int|\widehat\tau_\vartheta-\tau_\vartheta|=o(1)$ by the uniform continuity of $\tau_\vartheta$. Hence $\Gammah_j=\Gammah_{j*}+\delta_\vartheta'\Gamma_\vartheta+R^\vartheta_N$ with $\sqrt N\int_{\mathcal C}|R^\vartheta_N|=o_p(1)$, and by the Lipschitz property and Lemma~\ref{olem:pf-linearization}(b),(c) for $\zeta^*$, $\Dh_j^\Psi(\mathcal C)=D^{\mathrm{lin}}_j(\delta_\vartheta'\tau_\vartheta)+o_p(1)$, with $D^{\mathrm{lin}}_j(\cdot)$ of Lemma~\ref{lem:drift} for $\zeta^*$.

\emph{Step 2 (uniformity in $\delta_\vartheta$).} Let $F_N(\delta_\vartheta):=D^{\mathrm{lin}}_j(\delta_\vartheta'\tau_\vartheta)-D^{\mathrm{lin}}_j(0)$ for deterministic $\delta_\vartheta$. For fixed $u$, $\delta_N:=N^{-1/2}u'\tau_\vartheta$ satisfies (a) and (b) of Lemma~\ref{lem:drift}, whose linear term vanishes as $b_\Psi=0$ or $w_{\mathcal C}\equiv0$, so $F_N(N^{-1/2}u)=o_p(1)$. As $|F_N(\delta_\vartheta)-F_N(\widetilde\delta_\vartheta)|\le C_\vartheta\sqrt N\|\delta_\vartheta-\widetilde\delta_\vartheta\|$, $C_\vartheta:=2\lambda(\mathcal X_j)\int\|\tau_\vartheta\|$, a finite $\epsilon$-net $\mathcal A_\epsilon$ of $\{\|u\|\le C_\epsilon\}$ gives $\sup_{\sqrt N\|\delta_\vartheta\|\le C_\epsilon}|F_N(\delta_\vartheta)|\le\max_{u\in\mathcal A_\epsilon}|F_N(N^{-1/2}u)|+C_\vartheta\epsilon$. Hence $F_N(\widehat\vartheta-\vartheta^*)=o_p(1)$ and $\Dh_j^\Psi(\mathcal C)=\Dh^\Psi_{j*}(\mathcal C)+o_p(1)$.

\emph{Step 3.} By Lemma~\ref{olem:nui-scale}(b), $\widetilde m_j-\widetilde m_{j*}=o_p(1)$ and $\widetilde\sigma^2_{j,\Psi,N}-\widetilde\sigma^2_{j,\Psi,N*}=o_p(1)$. Under (a), with Step~2, $\Sh_j^\Psi(\mathcal C)-\Sh^\Psi_{j*}(\mathcal C)=o_p(1)$, and $\Sh^\Psi_{j*}(\mathcal C)\dto N(0,1)$ by Proposition~\ref{prop:transfer}. For (b), Step~1 without the null gives $N^{-1/2}\Dh_j^\Psi(\mathcal C)=D_j^\Psi(\mathcal C)+O_p(\epsilon_N+N^{-1/2})$, and Proposition~\ref{prop:size}(b) follows as before.
\end{proof}

\begin{proof}[Proof of Theorem~\ref{thm:nuisance-split}]
\emph{Step 1 (deterministic first stage).} For a deterministic $\chi_N$ satisfying Assumption~\ref{ass:firststage}(b)(ii) and deterministic versions of (b)(iii),(iv) and of part (a), the law $P_N$ of $(\zeta_N,X)$, $\zeta_N:=\zeta(W;\chi_N)$, satisfies the hypotheses of Lemma~\ref{olem:uniform}, with $\sup_{\mathcal U}|\bar\pi_N|=o(\eta^{1/2})$ for $\bar\pi_N:=\E[\zeta_N-\zeta\mid X]$ (Lemma~\ref{olem:nui-seq}). By Lemma~\ref{olem:nui-lin}, whose hypothesis follows from (b)(iv) and $\|\delta_N\|_2\le C\|\bar\pi_N\|_{L_2(\mathcal U)}$, $\Dh_j^\Psi(\mathcal C)=D^{\mathrm{lin}}_j(\delta_N)+o_p(1)$, and Lemma~\ref{lem:drift} along $P_N$ gives $D^{\mathrm{lin}}_j(\delta_N)=D^{\mathrm{lin}}_j(0)+b_\Psi\sqrt N\int w_{\mathcal C}\delta_N+o_p(1)$, the last term being $o_p(1)$ under each condition of part (a). Lemma~\ref{olem:uniform}(b),(c) gives $\{D^{\mathrm{lin}}_j(0)-m^\kappa_{j, N}(\mathcal C)\}/\sigma\dto N(0,1)$, $\sigma:=\sigma_{j,\Psi}(\mathcal C)$ computed from $\varrho_\zeta$, $\widetilde m_j-m^\kappa_{j,N}(\mathcal C)=o_p(1)$ and $\widetilde\sigma_{j,\Psi,N}\pto\sigma$ along $P_N$, the uniformity of the estimated centering and variance being part (vii) of the proof of that lemma. Hence $\Pr_{P_N}(\Sh_j^\Psi(\mathcal C)\le s)\to\Phi(s)$ for every $s$.

\emph{Step 2 (estimated first stage).} By (b)(i), given $\widehat\chi$ the second stage is i.i.d., so $F_N(s):=\Pr(\Sh_j^\Psi(\mathcal C)\le s\mid\widehat\chi)=H_N(s;\widehat\chi)$, $H_N(s;\chi)$ the probability of Step~1 at $\chi$. Along a further subsequence of every subsequence, the failure of (b)(ii) and the $o_p(1)$ quantities of (b)(iii),(iv) and part (a) vanish almost surely, so $F_N(s)\to\Phi(s)$ almost surely by Step~1. Thus $F_N(s)\pto\Phi(s)$ and $\Pr(\Sh_j^\Psi(\mathcal C)\le s)=\E F_N(s)\to\Phi(s)$, proving (a); (b) follows by the same reduction and the proof of Proposition~\ref{prop:size}(b), as $\sup_{\dom}|\tauh-\tau|=o_p(1)$ along $P_N$ (Lemma~\ref{olem:uniform}(c)).
\end{proof}
\section{Proof of the magnitude results}\label{osec:proofs-gini}
This section proves Proposition~\ref{prop:gini-clt} of the main text; the auxiliary parts of the proof are in Section~\ref{osec:gini-if}.

\begin{proof}[Proof of Proposition~\ref{prop:gini-clt}]
Let $\Gamma:=\Gamma_j$, $\mathcal C:=\Csym{j}$, $B:=B_{j,N}$ and $\bar\Theta_N(c):=N^{-1}\sum_i\Theta_j(W_i;c)$, so that $D^{\mathrm{lin}}_j=\tfrac12\sqrt N\int_{\mathcal C}|\Gamma+B+\bar\Theta_N|$.

\emph{(a)} Replacing $W_i$ by an independent copy $W_i'$ changes $D^{\mathrm{lin}}_j$ by at most $N^{-1/2}\{A(W_i)+A(W_i')\}$, $A(W):=\tfrac12\int_{\mathcal C}|\Theta_j(W;c)|dc\le C(1+|\zeta|)$, as the windows integrate to $O(1)$; the Efron--Stein inequality \citep{efron1981jackknife} gives $\Var(D^{\mathrm{lin}}_j)\le C$ under no null or margin condition (the margin condition of Proposition~\ref{prop:gini-clt}(c)), so $\Dh_j/\sqrt N=\E D^{\mathrm{lin}}_j/\sqrt N+O_p(N^{-1/2})$ by Lemma~\ref{olem:pf-linearization}(c). As $\E \bar\Theta_N(c)=0$, Jensen's and the triangle inequalities give $|\Gamma|-|B|\le\E|\Gamma+B+\bar\Theta_N|\le|\Gamma|+|B|+\E|\bar\Theta_N|$. The Berry--Esseen step of Lemma~\ref{olem:centering} for the fixed-sample sum $\bar\Theta_N$ gives $\tfrac12\sqrt N\int_{\mathcal C}\E|\bar\Theta_N|=m^\kappa_{j,N}+O((N\eta^2)^{-1/2})$ with or without the null. With $\sqrt N\sup|B|\to0$, $D_j-o(N^{-1/2})\le\E D^{\mathrm{lin}}_j/\sqrt N\le D_j+\{m^\kappa_{j,N}+o(1)\}/\sqrt N$, and (a) follows from $\widetilde m_j-m^\kappa_{j,N}=o_p(1)$ (Lemma~\ref{prop:centering-corr}) and $m^\kappa_{j,N}=O(\eta^{-1/2})$.

\emph{(b)} This is Lemma~\ref{prop:sigmaN} with Theorem~\ref{thm:normality}, since $\sqrt N\widetilde D_j=\widetilde\sigma_{j,N}\Sh_j$.

\emph{(c)} Let $u_N:=\Gammah_j-\Gamma$ and $\upsilon_N:=|\Gamma+u_N|-|\Gamma|-\mathrm{sgn}(\Gamma)u_N$, $\mathrm{sgn}(0):=0$, so that $\Dh_j-\sqrt ND_j=\mathcal L_N+R_N$ with $\mathcal L_N:=\tfrac12\sqrt N\int_{\mathcal C}\mathrm{sgn}(\Gamma)u_N$ and $R_N:=\tfrac12\sqrt N\int_{\mathcal C}\upsilon_N$. Step~1: since $||x+y|-|x|-\mathrm{sgn}(x)y|\le2|y|\ind\{|y|\ge|x|\}$, the margin condition gives $|R_N|\le C\sqrt N\bar u_N^2$ on $\{\bar u_N\le\epsilon_0\}$, $\bar u_N:=\sup_{\mathcal C}|u_N|=O_p(\epsilon_N)$, and $\sqrt N\epsilon_N^2\to0$ by Lemma~\ref{olem:pf-linearization}(a). Step~2: by Lemma~\ref{olem:pf-linearization}(b), the bias contributes $O(\sqrt N(h_1^{s_1}+h_2^{s_2}))=o(1)$ and the remainder $o_p(1)$, so $\mathcal L_N=N^{-1/2}\sum_iZ_{N,i}+o_p(1)$ with the i.i.d.\ centered $Z_{N,i}:=\tfrac12\int_{\mathcal C}\mathrm{sgn}(\Gamma)\Theta_j(W_i;c)dc=\int_{\dom}\omega_j(x)\theta_N(W_i;x)dx$, as $\mathrm{sgn}(\Gamma)$ is antisymmetric in the two evaluation points. Step~3: $|Z_{N,i}|\le C(1+|\zeta_i|)$, and $Z_{N,i}\to\omega_j(X_i)\xi(W_i;X_i)$ in $L_2$, the kernel being an approximate identity on the bounded $\omega_j\xi(W;\cdot)$; so $\Var Z_{N,i}\to \Omega_j=\E[\omega_j(X)^2\xi(W;X)^2]$, positive when $D_j>0$ as $\inf\varrho>0$ and $\omega_j\not\equiv0$; Lyapunov's theorem concludes. Section~\ref{osec:from-main} shows the consistency of the plug-in $\widehat\Omega_{j,N}$.

\emph{(d)} Split $\Dh_j/\sqrt N-D_j=\tfrac12\int_{F_j}|\Gammah_j|+\tfrac12\int_{\mathcal C\setminus F_j}(|\Gammah_j|-|\Gamma|)$. As $F_j$ is invariant under the exchange $c\mapsto(c_2,c_1)$ and $\Gamma\equiv0$ on it, $\sqrt N$ times the first term is the statistic of Theorem~\ref{thm:normality} on $F_j$, equal to $m^\kappa_{j,N}(F_j)+O_p(1)$, with $m^\kappa_{j,N}(F_j)\asymp\eta^{-1/2}\lambda(F_j)$ as $\eta\kappa_{j,N}\in[2\inf\varrho,2\sup\varrho]+o(1)$ off the band; on $\mathcal C\setminus F_j$ the margin condition holds, and Steps~1 and~2 of (c) make $\sqrt N$ times the second term $O_p(1)$.
\end{proof}

\providecommand{\projfield}{\Delta}
\providecommand{\bwc}{c}
\providecommand{\pext}{p_{\mathcal B}}\providecommand{\Uext}{U'_{\mathcal N}}\providecommand{\Qproj}{Q_{\mathcal N}}
\section{Auxiliary lemmas and routine derivations}\label{osec:aux}

This section collects the auxiliary lemmas and the routine derivations behind the proofs of Theorems~\ref{thm:variance} and \ref{thm:normality} of the main text. Section~\ref{osec:proofs} states Lemmas~\ref{olem:pf-linearization}--\ref{olem:pf-clt} and gives the proof of the projection lemma, the inclusion--exclusion computation of the variance over the four overlap events, the assembly of the normality theorem and the short proofs of the size, consistency and monotone-null results; the proof of Lemma~\ref{olem:pf-linearization} is in Section~\ref{osec:from-main}, the full proof of the central limit theorem of Lemma~\ref{olem:pf-clt} is in Section~\ref{osec:aux-clt}, and everything else is here.\ 

The organization mirrors the lemma list of Section~\ref{sec:asym-proj} of the main text: Section~\ref{osec:aux-lin} sets out the linearization behind Lemma~\ref{olem:pf-linearization}, whose proof are  is in Section~\ref{osec:from-main}, Section~\ref{osec:aux-poisson} Lemma~\ref{olem:poisson} (Poissonization facts), Section~\ref{osec:aux-center} Lemma~\ref{olem:centering} (bias and centering), Section~\ref{osec:aux-var} the routine bounds behind Lemma~\ref{olem:variance-gen}, Section~\ref{osec:aux-proj} the configuration bookkeeping and the moment bounds attached to Lemma~\ref{olem:pf-projection}, Section~\ref{osec:aux-clt} Lemma~\ref{olem:pf-clt} (the joint central limit theorem) and Section~\ref{osec:aux-depoisson} Lemma~\ref{olem:depoisson} (de-Poissonization). Section~\ref{osec:aux-remarks} documents two remarks, Section~\ref{osec:aux-mirror} the transfer to the mirror statistic $\Dh_1$, and Section~\ref{osec:aux-moved} the routine computations referred to from Section~\ref{osec:proofs}, in the notation of that section. Each main-text lemma is restated before its proof, so that this document can be read on its own.{} The results for estimated nuisance parameters and the uniformity of the oracle results along sequences of laws are in Section~\ref{osec:nui}, and the proofs and computational details of the implementation results (the sharp uniform rate, the corrected centering, the finite-sample variance, data-dependent tuning, the magnitudes and the jittered modifier) in Section~\ref{osec:impl-proofs}.

\subsection{Setting, notation and the class of integrands}\label{osec:aux-setting}

The two covariate blocks take values in compact sets $\Xone\subset\mathbb R^{d_1}$ and $\Xtwo\subset\mathbb R^{d_2}$ with nonempty interior, and $\tau(x_1,x_2):=\E[Y(1)-Y(0)\mid X_1=x_1,X_2=x_2]$ is estimated by smoothing jointly over both blocks. Let $K_2(\tilde z_2)=\prod_{j=1}^{d_2}K_{2j}(\tilde z_{2j})$ and $K_1(\tilde z_1)=\prod_{l=1}^{d_1}K_{1l}(\tilde z_{1l})$ be product kernels with bandwidths $h_2$ and $h_1$. The estimator of the main text, equation~\eqref{eq:tauhat}, is
\begin{equation}\label{oeq:tauhat}
\tauh(x_1,x_2):=\frac{1}{N\eta}\sum_{i=1}^{N}Y_i
\Big[\frac{T_i}{\widehat P_1(x_1,x_2)}-\frac{1-T_i}{\widehat P_0(x_1,x_2)}\Big]
K_2\Big(\frac{x_2-X_{2i}}{h_2}\Big)K_1\Big(\frac{x_1-X_{1i}}{h_1}\Big),
\end{equation}
where $\widehat P_1(x_1,x_2):=(N\eta)^{-1}\sum_iT_iK_2(\tfrac{x_2-X_{2i}}{h_2})K_1(\tfrac{x_1-X_{1i}}{h_1})$, similarly $\widehat P_0$ with $1-T_i$, are kernel estimators of the density-scale objects
\begin{equation}\label{oeq:Pt}
P_t(x_1,x_2):=\Pr(T=t\mid X_1=x_1,X_2=x_2)\,f(x_1,x_2),\qquad t\in\{0,1\},
\end{equation}
$f$ being the joint Lebesgue density of $(X_1,X_2)$. Throughout, for $x=(x_1,x_2)$,
\[
\mathcal K(x;X):=K_2\Big(\frac{x_2-X_{2}}{h_2}\Big)K_1\Big(\frac{x_1-X_{1}}{h_1}\Big),\qquad \mathcal K_i(x):=\mathcal K(x;X_i),
\]
abbreviates the joint kernel weight, and
\begin{equation}\label{oeq:eta}
\eta:=h_1^{d_1}h_2^{d_2}
\end{equation}
is the smoothing volume. A configuration in direction $2$ is $c=(x_1,x_2,x_2')$ with evaluation points $c_1:=(x_1,x_2)$ and $c_2:=(x_1,x_2')$, and $\Gamma(c):=\Gamma_2(c)=\tau(c_1)-\tau(c_2)$; the general statistic is $\Dh^\Psi_2(\mathcal C_2):=\sqrt N\int_{\mathcal C_2}\Psi(\Gammah_2(c))\,dc$, equation~\eqref{eq:Dgen}. The theory is developed for the second-direction statistic in its symmetric absolute form,
\begin{equation}\label{oeq:D2}
\Dh_2:=\tfrac12\Dh_2^{\Psiabs}(\Csym{2})=\frac12\int_{\Xone}\int_{\Xtwo}\int_{\Xtwo}
\sqrt N\,\big|\tauh(x_1,x_2)-\tauh(x_1,x_2')\big|\,dx_2\,dx_2'\,dx_1 ,
\end{equation}
under $H_0^{2}$: $\tau(x_1,x_2)=\tau(x_1,x_2')$ for all $x_1\in\Xone$ and $x_2,x_2'\in\Xtwo$, that is, $\Gamma\equiv0$ on $\Csym2$. The pointwise scaled variance of the main text, equation~\eqref{eq:rho}, is
\begin{equation}\label{oeq:rhodef}
\varrho(x_1,x_2):=\Big[\sum_{t\in\{0,1\}}\frac{\varsigma_t^2(x_1,x_2)}{P_t(x_1,x_2)}\Big]\int K_2(\tilde z_2)^2d\tilde z_2\int K_1(\tilde z_1)^2d\tilde z_1 ,
\end{equation}
with $\bar\mu_t(x_1,x_2):=\E[Y\mid X_1=x_1,X_2=x_2,T=t]$ and $\varsigma_t^2(x_1,x_2):=\Var(Y\mid X_1=x_1,X_2=x_2,T=t)$. Throughout, Assumptions~\ref{ass:sampling}--\ref{ass:overlap}, \ref{ass:psi} and \ref{ass:reg} of the main text are in force; the parts of Assumption~\ref{ass:reg} are referred to by letter as in the main text: (b) $f$ is bounded, twice continuously differentiable and bounded below by $\underline p$ on a neighborhood of $\dom$; (c) $0<\inf\varrho\le\sup\varrho<\infty$; (d) each $K_{2j}$ and $K_{1l}$ is a bounded, even, Lipschitz kernel of order $s_2$, respectively $s_1$, supported on $[-1/2,1/2]$; (e) $\bar\mu_t,\varsigma_t^2,f,P_t$ are $s_1$ times continuously differentiable in $x_1$ and $s_2$ times in $x_2$ on a neighborhood of $\dom$, with $s_1\wedge s_2\ge2$; (f) $\E[|Y|^6\mid X_1=x_1, X_2=x_2, T=t]$ is bounded; (g) $h_1\to0$, $h_2\to0$, $N(h_1^{2s_1}+h_2^{2s_2})\to0$, $N\eta^3\to\infty$ and $(N\eta^2)^{1/2}/\log N\to\infty$. Overlap, Assumption~\ref{ass:overlap}, bounds $\Pr(T=t\mid X_1=x_1,X_2=x_2)$ and $f$ below by $\underline p$ on a neighborhood of $\dom$. As in the main text, $\epsilon_N:=\{\log N/(N\eta)\}^{1/2}+h_1^{s_1}+h_2^{s_2}$, the sharp uniform rate of Lemma~\ref{olem:sharp}. The statistic \eqref{oeq:D2} contains no estimated weight, and we impose no high-level uniform-consistency condition: the uniform rates for $\tauh$, $\widehat P_t$, and $\rhoh$ are derived in Section~\ref{osec:aux-lin}.

The rate conditions in (g) are feasible: with $h_2\asymp N^{-\gamma_2}$ and $h_1\asymp N^{-\gamma_1}$ all three displays hold simultaneously if and only if $2s_2\gamma_2>1$, $2s_1\gamma_1>1$ and $3(d_2\gamma_2+d_1\gamma_1)<1$, a nonempty region whenever $s_1\wedge s_2>\tfrac32(d_1+d_2)$ in the equal-bandwidth case and under weaker order requirements when the bandwidths differ. The condition $N\eta^3\to\infty$ controls the Lyapunov and Rosenthal terms, and $(N\eta^2)^{1/2}/\log N\to\infty$ controls the Bernstein rate of the feasible centering. For parts (a) and (b) of Lemma~\ref{olem:pf-linearization}, the truncation step of Lemma~\ref{olem:sharp} needs only $N^2\eta^3/(\log N)^3\to\infty$; the stronger $N\eta^3\to\infty$ is used for the truncation of $Y^2$ in $\rhoh$ and for the Lyapunov and Rosenthal bounds, and it implies $(N\eta^2)^{1/2}/\log N\to\infty$, since $N\eta^2\ge N^{1/3}(N\eta^3)^{2/3}$.

\textit{Influence functions.} As in Section~\ref{sec:asym-center} of the main text, the score and the centered influence function of $\tauh$ are
\begin{gather}
\xi(W;x):=\frac{T\{Y-\bar\mu_1(x)\}}{P_1(x)}-\frac{(1-T)\{Y-\bar\mu_0(x)\}}{P_0(x)},\notag\\
\theta_N(W;x):=\frac{1}{\eta}\,\xi(W;x)\,\mathcal K(x;X)-B_N(x),\qquad
B_N(x):=\E\Big[\frac1\eta\,\xi(W;x)\,\mathcal K(x;X)\Big],\label{oeq:theta}
\end{gather}
so that $\E[\theta_N(W;x)]=0$ and $B_N(x)$, the smoothing bias, is $O(h_1^{s_1}+h_2^{s_2})$ uniformly on $\dom$ by the kernel orders in Assumption~\ref{ass:reg}(d), the smoothness in (e) and the standard bias expansion of the joint smoother. For a configuration $c=(x_1,x_2,x_2')$ put
\begin{equation}\label{oeq:Theta}
\Theta(W;c):=\theta_N(W;c_1)-\theta_N(W;c_2),\quad B_{2,N}(c):=B_N(c_1)-B_N(c_2),\quad \kappa_{N}(c):=\E\big[\Theta(W;c)^2\big],
\end{equation}
the influence function $\Theta_2$, bias $B_{2,N}$ and exact variance $\kappa_{2,N}$ defined in Section~\ref{sec:asym-center}{} of the main text, from which the Poissonized second moment $\kappa_{\mathcal N}=\kappa_N+B_{2,N}^2$ of Section~\ref{osec:pf-poisson} differs by the squared bias; $\Theta(W,x_1,x_2,x_2')$ is also written for $\Theta(W;c)$. Both terms of $\Theta$ carry the same $X_1$ kernel factor: the two CATE estimators being differenced share their $X_1$ window and differ only in the $X_2$ window.

\textit{The class of integrands.} The main text states the asymptotic results for integrands $\Psi$ in the class of Assumption~\ref{ass:psi}: $\Psi$ is Lipschitz with constant one, positively homogeneous of degree one, and satisfies $|\Psi(u)|\le|u|$. The Gaussian constants are $m_\Psi:=\E[\Psi(Z)]$ and $g_\Psi(\rho):=\Cov\big(\Psi(\sqrt{1-\rho^2}\,Z_1+\rho Z_2),\Psi(Z_2)\big)$ for independent standard normals $Z,Z_1,Z_2$. The following observation, equations~\eqref{eq:stat-psi-rep}--\eqref{eq:stat-g} in the main text, reduces every statement about the class to statements about the absolute value and the identity; we provide a short verification for completeness.

\begin{remark}\label{orem:psi-structure}
\textit{Structure of the class.} Positive homogeneity gives $\Psi(0)=0$, $\Psi(u)=\Psi(1)u$ for $u>0$ and $\Psi(u)=\Psi(-1)|u|$ for $u<0$, so that every member of the class is of the form
\begin{equation}\label{oeq:psi-structure}
\Psi(u)=a_\Psi\,|u|+b_\Psi\,u,\qquad
a_\Psi:=\tfrac12\{\Psi(1)+\Psi(-1)\},\quad b_\Psi:=\tfrac12\{\Psi(1)-\Psi(-1)\},
\end{equation}
and the Lipschitz and domination conditions both reduce to $|\Psi(1)|\vee|\Psi(-1)|\le1$, which is $|a_\Psi\pm b_\Psi|\le1$. For $\Psiabs$, $(a_\Psi,b_\Psi)=(1,0)$; for $\Psimax$, $(a_\Psi,b_\Psi)=(\tfrac12,\tfrac12)$. Consequently $m_\Psi=a_\Psi\sqrt{2/\pi}$, and, since $\Cov(|Z_1'|,Z_2)=\E[|Z_1'|\,\E[Z_2\mid Z_1']]=\rho\,\E[|Z_1'|Z_1']=0$ for a standardized jointly normal pair $(Z_1',Z_2)$ with correlation $\rho$,
\begin{equation}\label{oeq:gpsi}
g_\Psi(\rho)=a_\Psi^2\,g_{\mathrm{abs}}(\rho)+b_\Psi^2\,\rho,\qquad
g_{\mathrm{abs}}(\rho)=\tfrac2\pi\big\{\rho\arcsin\rho+\sqrt{1-\rho^2}-1\big\},
\end{equation}
the closed form of $g_{\mathrm{abs}}$ being due to \citet{nabeya1951absolute}; in particular $g_{\Psimax}(\rho)=\{\rho+g_{\mathrm{abs}}(\rho)\}/4$, and $g_\Psi$ is even if and only if $b_\Psi=0$. Also $\Var(\Psi(Z))=a_\Psi^2(1-2/\pi)+b_\Psi^2\le1$, so $|g_\Psi|\le1$.
\end{remark}

In each subsection, we write the proofs for $\Psiabs$ and the symmetric statistic \eqref{oeq:D2}. A closing paragraph headed ``Extension to general $\Psi$'' states precisely which properties of $|\cdot|$ the argument used and why, by \eqref{oeq:psi-structure}, the step holds for the class; where a step uses the symmetry $|-u|=|u|$, which the class does not share, this is said explicitly. Standard normal variables are denoted $Z,Z_1,Z_2$; $\lambda$ denotes Lebesgue measure; $\|\cdot\|_\infty$ the maximum norm; and $\|\cdot\|_p$ the $L_p$-norm of a random variable.

\subsection{Uniform asymptotic approximation and consistency}\label{osec:aux-lin}

Write
\begin{align}
\tauh(x_1,x_2)&=\tau(x_1,x_2)+\big(\tau_{N0}(x_1,x_2)-\E[\tau_{N0}(x_1,x_2)]\big)\notag\\
&\quad+\big(\E[\tau_{N0}(x_1,x_2)]-\tau(x_1,x_2)\big)+R_N(x_1,x_2),\label{oeq:lin-decomp}
\end{align}
where $\tau_{N0}$ is \eqref{oeq:tauhat} with $\widehat P_t$ replaced by $P_t$,
\[
\tau_{N0}(x_1,x_2):=\frac{1}{N\eta}\sum_{i=1}^{N}Y_i\Big[\frac{T_i}{P_1(x_1,x_2)}-\frac{1-T_i}{P_0(x_1,x_2)}\Big]\mathcal K_i(x_1,x_2),
\]
and
\begin{align*}
R_N(x_1,x_2):=&\frac{1}{N\eta}\sum_{i=1}^{N}Y_i\Big[\frac{T_i}{P_1(x_1,x_2)}-\frac{1-T_i}{P_0(x_1,x_2)}\Big]\\
&\times\Big(T_i\frac{P_1(x_1,x_2)-\widehat P_1(x_1,x_2)}{\widehat P_1(x_1,x_2)}
+(1-T_i)\frac{P_0(x_1,x_2)-\widehat P_0(x_1,x_2)}{\widehat P_0(x_1,x_2)}\Big)\mathcal K_i(x_1,x_2).
\end{align*}
Define the linearization of the denominator correction,
\begin{align*}
\tau_{N1}(x_1,x_2):=&\ \bar\mu_1(x_1,x_2)-\bar\mu_0(x_1,x_2)
-\frac{\bar\mu_1(x_1,x_2)}{N\eta\,P_1(x_1,x_2)}\sum_{i=1}^{N}T_i\,\mathcal K_i(x_1,x_2)\\
&+\frac{\bar\mu_0(x_1,x_2)}{N\eta\,P_0(x_1,x_2)}\sum_{i=1}^{N}(1-T_i)\,\mathcal K_i(x_1,x_2),
\end{align*}
and $\tau_N(x_1,x_2):=\tau_{N0}(x_1,x_2)+\tau_{N1}(x_1,x_2)$. Collecting terms and using $\tau=\bar\mu_1-\bar\mu_0$,
\begin{gather}
\tau_N(x)=\tau(x)+\frac1N\sum_{i=1}^{N}\frac1\eta\,\xi(W_i;x)\,\mathcal K_i(x),\qquad
\E[\tau_N(x)]=\tau(x)+B_N(x),\notag\\
\tau_N(x)-\E[\tau_N(x)]=\frac1N\sum_{i=1}^{N}\theta_N(W_i;x),\label{oeq:tauN}
\end{gather}
with $\xi$, $\theta_N$ and $B_N$ as in \eqref{oeq:theta}. The fixed-sample linearized field is
\begin{equation}\label{oeq:GammaN}
\Gamma_N(c):=\tau_N(c_1)-\tau_N(c_2)=\Gamma(c)+B_{2,N}(c)+\frac1N\sum_{i=1}^{N}\Theta(W_i;c),
\end{equation}
and $D^{\mathrm{lin}}_2$ of Lemma~\ref{olem:pf-linearization}(c) is $\tfrac12\int_{\Csym2}\sqrt N|\Gamma_N(c)|\,dc$.

\begin{lemma}[Uniform rates]\label{olem:consis}
Under Assumptions~\ref{ass:sampling}--\ref{ass:reg},
(a) $\sup_{\dom}|\widehat P_t-P_t|=O_p(\epsilon_N)$ for $t\in\{0,1\}$;
(b) $\sup_{\dom}|\tauh-\tau|=O_p(\epsilon_N)$;
(c) $\sup_{\dom}|\rhoh-\varrho|=o_p(\eta^{1/2})$, where $\rhoh$ is the plug-in kernel estimator of \eqref{oeq:rhodef}; the kernel regressions $\widehat{\bar\mu}_t$ of $Y$ on $(X_1,X_2)$ within each arm converge uniformly at the rate $\epsilon_N$.
Consequently, under Assumption~\ref{ass:reg}(g), $\epsilon_N=o(\eta^{1/2})$ and $\sqrt N\epsilon_N^2\to0$.
\end{lemma}

\begin{proof}
Parts (a) and (b) and the rate for $\widehat{\bar\mu}_t$ follow from Lemma~\ref{olem:sharp} applied to $Z\in\{\ind(T=t),\,Y\ind(T=t),\,1\}$, the bias being $O(h_1^{s_1}+h_2^{s_2})$ by the kernel orders, and from the ratio expansion with denominators bounded below by Assumption~\ref{ass:overlap}; part (c) is the truncation argument stated after Lemma~\ref{olem:sharp}, which needs only three conditional moments of $Y^2$. Since $\{\log N/(N\eta)\}^{1/2}=\eta^{1/2}\{\log N/(N\eta^2)\}^{1/2}$ and $h_1^{s_1}+h_2^{s_2}=o(N^{-1/2})$, $\epsilon_N=o(\eta^{1/2})$; and $\sqrt N\epsilon_N^2\le2\log N/(N\eta^2)^{1/2}+2\sqrt N(h_1^{s_1}+h_2^{s_2})^2\to0$ under Assumption~\ref{ass:reg}(g), with no logarithmic strengthening.
\end{proof}

\begin{lemma}[Linearization of the denominator correction]\label{olem:three}
Under Assumptions~\ref{ass:sampling}--\ref{ass:reg},
\[
\sup_{(x_1,x_2)\in\dom}\big|R_N(x_1,x_2)-\tau_{N1}(x_1,x_2)\big|=o_p(N^{-1/2}),
\]
and hence
\[
\sup_{(x_1,x_2,x_2')\in\Xone\times\Xtwo^2}|\{R_N(x_1,x_2)-R_N(x_1,x_2')\}-\{\tau_{N1}(x_1,x_2)-\tau_{N1}(x_1,x_2')\}|=o_p(N^{-1/2}).
\]
\end{lemma}

\begin{proof}
This is the elementary ratio expansion in the proof of Lemma~\ref{olem:pf-linearization} in Section~\ref{osec:from-main}: with $\widehat{\bar\mu}_t:=\widehat A_t/\widehat P_t$ the kernel regression of $Y$ within arm $t$, the definitions of $R_N$ and $\tau_{N1}$ give the exact identity $R_N-\tau_{N1}=-\sum_{t}(-1)^{1-t}(\widehat P_t-P_t)(\widehat{\bar\mu}_t-\bar\mu_t)/P_t$, a sum of products of two uniform errors over the deterministic $P_t$, so that $\sup_{\dom}|R_N-\tau_{N1}|=O_p(\epsilon_N^2)=o_p(N^{-1/2})$ by Lemma~\ref{olem:consis}(a),(c) and Assumption~\ref{ass:reg}(g); the second claim follows from the triangle inequality.
\end{proof}

Lemma~\ref{olem:pf-linearization} combines the decomposition \eqref{oeq:lin-decomp}, the identity \eqref{oeq:tauN} and Lemmas~\ref{olem:consis} and \ref{olem:three}; its statement is in Section~\ref{osec:proofs} and its proof is in Section~\ref{osec:from-main}. Its part (b) gives, uniformly in $c\in\mathcal C_j$, $\Gammah_j(c)-\Gamma_j(c)=N^{-1}\sum_i\Theta_j(W_i;c)+B_{j,N}(c)+O_p(\epsilon_N^2)$, and its part (c) that $\Dh_j^\Psi(\mathcal C_j)-D^{\mathrm{lin}}_j=o_p(1)$, where $D^{\mathrm{lin}}_j$ is \eqref{eq:Dgen} with the linearized field \eqref{oeq:GammaN} in place of $\Gammah_j$. No null hypothesis is used.

By Lemma~\ref{olem:pf-linearization}(c) it suffices to analyze $D^{\mathrm{lin}}_2$, built from the linearized field $\Gamma_N$. No condition on any estimated weight is needed, because the integrand of \eqref{oeq:D2} contains no estimated quantity beyond the CATE estimators themselves.

\textit{Extension to general $\Psi$.} Lemmas~\ref{olem:consis} and \ref{olem:three} and part (b) do not involve $\Psi$. Part (c) used the absolute value only through its Lipschitz property, which is Assumption~\ref{ass:psi}(a); the argument is verbatim on any configuration set $\mathcal C_j$ of finite measure.

\subsection{Preliminary facts and Poissonization}\label{osec:aux-poisson}

\subsubsection{The integration sets}\label{osec:aux-sets}

Let $\epsilon>0$ be arbitrary. For constants $\{M_l>0:\,l=1,\dots,d_1+d_2\}$ let $\mathcal B(M)=\prod_{l=1}^{d_1+d_2}[-M_l,M_l]\subset\dom$ denote a box with nonempty interior, and for $h_0>0$ let $\mathcal B(M,h_0):=\{(x_1,x_2)\in\mathcal B(M):\inf_{y\notin\mathcal B(M)}\|(x_1,x_2)-y\|\ge h_0\}$ be its $h_0$-contraction (in Sections~\ref{osec:aux-poisson}--\ref{osec:aux-depoisson} the evaluation region is $\mathcal B_1\times\mathcal B_2\subset\mathcal B(M,h_0)$ rather than $\dom$, so $\mathcal B(M)$, which contains every window centered in $\mathcal B_1\times\mathcal B_2$, plays the role of the box $\mathcal B\supset\dom$ of Lemma~\ref{olem:pf-clt}, and $\pext$ below is its exterior probability). Choose $M,h_0>0$ and Borel sets $\mathcal B_2\subset\Xtwo$, $\mathcal B_1\subset\Xone$ with Lebesgue-null boundaries such that
\begin{align}
&\mathcal B_1\times \mathcal B_2\subset\mathcal B(M,h_0),\label{oeq:M1}\\
&\int_{\mathbb R^{d_1+d_2}\setminus\mathcal B(M)}f(x_1,x_2)\,dx_2\,dx_1=:\pext>0,\label{oeq:M2}\\
&\int_{\dom}f(x_1,x_2)\,dx_2\,dx_1-\int_{\mathcal B_1\times \mathcal B_2}f(x_1,x_2)\,dx_2\,dx_1<\epsilon .\label{oeq:M3}
\end{align}
Conditions \eqref{oeq:M1}--\eqref{oeq:M3} are the joint-smoothing counterparts of (6.4)--(6.6) of \citet{gine2003bm}; \eqref{oeq:M3} requires $\mathcal B_1\times \mathcal B_2$ to capture all but $\epsilon$ of the mass, which makes the extension $\mathcal B_2\uparrow\Xtwo$, $\mathcal B_1\uparrow\Xone$ of Section~\ref{osec:aux-depoisson} possible. Such $M,h_0,\mathcal B_2,\mathcal B_1$ exist: $\dom$ is compact and $f$ is continuous and bounded below on it, so it suffices to take $\mathcal B(M)$ a box exhausting all but a small part of the mass and $\mathcal B_1\times \mathcal B_2$ a product of closed rectangles inside its $h_0$-contraction. The contraction serves the kernel windows: for $h_2\vee h_1<h_0$ (in particular for all $N$ large) every window $\mathcal W(x_1,x_2)$, defined in \eqref{oeq:window} below, centered in $\mathcal B_1\times \mathcal B_2$ is contained in $\mathcal B(M)$, and $f\ge\underline p$ on $\mathcal B(M)$ by Assumption~\ref{ass:overlap}. This containment is used essentially in Section~\ref{osec:aux-clt}: it makes the Poissonized statistic measurable with respect to the increments inside $\mathcal B(M)$, hence independent of the exterior count required by the de-Poissonization principle. The theory is developed on $\mathcal B_1\times \mathcal B_2$ and extended to $\dom$ in Section~\ref{osec:aux-depoisson} by letting $\epsilon\downarrow0$.

\subsubsection{Two analytic tools}

\begin{lemma}[Uniform kernel-convolution expansion, with a rate]\label{olem:gine}
Let $\mathcal J$ be an arbitrary index set and let $\{H_\varpi:\varpi\in\mathcal J\}$ be a family of Borel functions on $\mathbb R^{d_1+d_2}$ that are uniformly bounded, $\sup_\varpi\|H_\varpi\|_\infty\le\bar H<\infty$, and vanish outside a common compact set $\mathcal C_H\subset\mathbb R^{d_1+d_2}$. Let $\Xi:\mathbb R^{d_1+d_2}\times\mathcal J\to\mathbb R$ be such that $\Xi(\cdot,\cdot;\varpi)$ is continuously differentiable on a fixed neighborhood $\mathcal U$ of $\mathcal B_1\times \mathcal B_2$ for each $\varpi$, with
\[
C_\Xi:=\sup_{\varpi\in\mathcal J}\ \sup_{(z,p)\in\mathcal U}\big\{|\Xi(z,p;\varpi)|+\|\nabla_{(z,p)}\Xi(z,p;\varpi)\|\big\}<\infty .
\]
Then, for all $h_2,h_1$ small enough that $(x_2-h_2\tilde z_2,x_1-h_1\tilde z_1)\in\mathcal U$ for every $(x_1,x_2)\in \mathcal B_1\times \mathcal B_2$ and $(\tilde z_2,\tilde z_1)\in\mathcal C_H$,
\begin{align}
&\sup_{\varpi\in\mathcal J}\ \sup_{(x_1,x_2)\in \mathcal B_1\times \mathcal B_2}
\bigg|\frac{1}{\eta}\iint \Xi(z,p;\varpi)\,H_\varpi\Big(\tfrac{x_2-z}{h_2},\tfrac{x_1-p}{h_1}\Big)\,dz\,dp\notag\\
&\qquad-\Xi(x_2,x_1;\varpi)\iint H_\varpi(\tilde z_2,\tilde z_1)\,d\tilde z_2\,d\tilde z_1\bigg|=O(h_2+h_1),\label{oeq:gine1}
\end{align}
with a constant depending only on $C_\Xi$, $\bar H$ and $\mathcal C_H$. If, in addition, $\Xi(\cdot,\cdot;\varpi)$ is twice continuously differentiable on $\mathcal U$ with second derivatives bounded uniformly in $\varpi$, and the family has vanishing first moments,
\begin{equation}\label{oeq:Hmoments}
\iint\tilde z_{2j}\,H_\varpi(\tilde z_2,\tilde z_1)\,d\tilde z_2\,d\tilde z_1=\iint\tilde z_{1l}\,H_\varpi(\tilde z_2,\tilde z_1)\,d\tilde z_2\,d\tilde z_1=0\qquad\text{for all }j\le d_2,\ l\le d_1,\ \varpi\in\mathcal J,
\end{equation}
then the right-hand side of \eqref{oeq:gine1} improves to $O(h_2^2+h_1^2)$.
\end{lemma}

Here and below the first argument of $\Xi$ is the $X_2$-coordinate $z\in\mathbb R^{d_2}$ and the second the $X_1$-coordinate $p\in\mathbb R^{d_1}$ of the conditioning point, so $\Xi(x_2,x_1;\varpi)$ is $\Xi$ evaluated at the window center.

\phantomsection\label{oprf:gine}
\begin{proof}
Substitute $z=x_2-h_2\tilde z_2$, $p=x_1-h_1\tilde z_1$, so that $dz\,dp=\eta\,d\tilde z_2\,d\tilde z_1$ and, the support condition confining the integral to $(\tilde z_2,\tilde z_1)\in\mathcal C_H$, the quantity inside the supremum equals
\[
\bigg|\iint_{\mathcal C_H}\big\{\Xi(x_2-h_2\tilde z_2,x_1-h_1\tilde z_1;\varpi)-\Xi(x_2,x_1;\varpi)\big\}\,H_\varpi(\tilde z_2,\tilde z_1)\,d\tilde z_2\,d\tilde z_1\bigg| .
\]
By the mean value theorem along the segment joining $(x_2-h_2\tilde z_2,x_1-h_1\tilde z_1)$ to $(x_2,x_1)$, which lies in $\mathcal U$ for $h_2,h_1$ small,
\[
\big|\Xi(x_2-h_2\tilde z_2,x_1-h_1\tilde z_1;\varpi)-\Xi(x_2,x_1;\varpi)\big|\le C_\Xi\,c_H\,(h_2+h_1),\qquad c_H:=\sup_{(\tilde z_2,\tilde z_1)\in\mathcal C_H}\big(\|\tilde z_2\|+\|\tilde z_1\|\big),
\]
uniformly in $\varpi\in\mathcal J$ and $(x_1,x_2)\in \mathcal B_1\times \mathcal B_2$; multiplying by $|H_\varpi|\le\bar H$ and integrating over $\mathcal C_H$ gives \eqref{oeq:gine1} with constant $C_\Xi c_H\bar H\lambda(\mathcal C_H)$. Under the additional hypotheses, expand to second order instead: the first-order term is
\[
-h_2\sum_{j\le d_2}\partial_{z_j}\Xi(x_2,x_1;\varpi)\iint\tilde z_{2j}H_\varpi-h_1\sum_{l\le d_1}\partial_{p_l}\Xi(x_2,x_1;\varpi)\iint\tilde z_{1l}H_\varpi ,
\]
which vanishes identically by \eqref{oeq:Hmoments}, and the second-order remainder is bounded by
\[
\tfrac12\sup_\varpi\sup_{\mathcal U}\|\nabla^2\Xi\|\,c_H^2(h_2+h_1)^2
\]
uniformly, whence the $O(h_2^2+h_1^2)$ rate.
\end{proof}

Lemma~\ref{olem:gine} is the tool behind every ``change of variables and first-order expansion'' step below, namely the uniform second-moment facts in part (d) of the proof of Lemma~\ref{olem:poisson} below, on which the bias computation rests, and Part~2 of the proof of Lemma~\ref{olem:variance}. We state it in the form those steps require rather than in the form of Lemma~6.1 of \citet{gine2003bm}, which covers a finite class of kernels and delivers only $o(1)$. Two features matter. First, the index set $\mathcal J$ is a continuum: the families actually used are
\begin{align}
H_{u_2,u_1}(\tilde z_2,\tilde z_1)&:=K_2(\tilde z_2)K_2(\tilde z_2+u_2)K_1(\tilde z_1)K_1(\tilde z_1+u_1),\qquad(u_2,u_1)\in[-1,1]^{d_2}\times[-1,1]^{d_1},\notag\\
H_{*}(\tilde z_2,\tilde z_1)&:=K_2(\tilde z_2)^2K_1(\tilde z_1)^2,\label{oeq:Hfamilies}
\end{align}
both uniformly bounded by $\|K_2\|_\infty^2\|K_1\|_\infty^2$ and supported in the common compact set $\mathcal C_H=[-\tfrac12,\tfrac12]^{d_1+d_2}$ (for $H_{u_2,u_1}$ the factor $K_2(\tilde z_2)K_2(\tilde z_2+u_2)$ is supported in the intersection of $[-\tfrac12,\tfrac12]^{d_2}$ with its translate, whatever $u_2$), so that the hypotheses hold with $\mathcal J=[-1,1]^{d_1+d_2}$ (respectively a singleton) and the $O(h_2+h_1)$ bound is uniform in $(u_2,u_1)$, as every application below needs. Second, $H_*$ is even in each coordinate, by the evenness of $K_2$ and $K_1$ in Assumption~\ref{ass:reg}(d), hence satisfies \eqref{oeq:Hmoments}: this upgrades the rate to $O(h_2^2+h_1^2)$ in the diagonal second-moment expansions, and only there.

The second tool is the Berry--Esseen-type comparison of \citet{gine2003bm}, Fact~6.1 there, which they obtain from Theorem~1 of \citet{sweeting1977speeds}.

\begin{fact}\label{ofact:BE}
Let $\{(Q_i,Q_i')':i=1,\dots,N\}$ be i.i.d.\ random vectors in $\mathbb R^2$ such that each component has mean zero, variance one and a finite third absolute moment, and let $(Z_1,Z_2)'$ be bivariate normal with standard normal marginals and $\Cov(Z_1,Z_2)=\Cov(Q_i,Q_i')=\rho$. Write $S:=N^{-1/2}\sum_{i\le N}Q_i$ and $S':=N^{-1/2}\sum_{i\le N}Q_i'$. There exist universal positive constants $C_1,C_2,C_2'$ such that
\begin{equation}\label{oeq:BEfact1}
\big|\E[|S|]-\E[|Z_1|]\big|\le\frac{C_1}{\sqrt N}\E[|Q_i|^3]
\end{equation}
and, whenever $\rho^2<1$,
\begin{align}
\big|\E[|SS'|]-\E[|Z_1Z_2|]\big|&\le\frac{C_2}{(1-\rho^2)^{3/2}\sqrt N}\big(\E[|Q_i|^3]+\E[|Q_i'|^3]\big),\label{oeq:BEfact2}\\
\big|\E[S\,|S'|]-\E[Z_1|Z_2|]\big|&\le\frac{C_2'}{(1-\rho^2)^{3/2}\sqrt N}\big(\E[|Q_i|^3]+\E[|Q_i'|^3]\big).\label{oeq:BEfact3}
\end{align}
\end{fact}

Inequalities \eqref{oeq:BEfact1}--\eqref{oeq:BEfact3} are (6.18)--(6.20) of \citet{gine2003bm}, with their $(\omega_i,\zeta_i)$ in the role of $(Q_i,Q_i')$ and (6.18) applied to either component; their (6.20) bounds $|\E[S\,|S'|]|$ itself, which is \eqref{oeq:BEfact3} because $\E[Z_1|Z_2|]=0$ by the symmetry $(Z_1,Z_2)\mapsto(-Z_1,-Z_2)$. The absolute form uses only \eqref{oeq:BEfact1}--\eqref{oeq:BEfact2}; \eqref{oeq:BEfact3}, which they need for one-sided functionals, enters here only through the extension to general $\Psi$.

The argument below needs the comparison in covariance form. Put
\begin{equation}\label{oeq:gdef}
g(\rho):=\Cov\big(|Z_1|,|Z_2|\big)=\E[|Z_1Z_2|]-\E[|Z_1|]\,\E[|Z_2|],
\end{equation}
so that $g=g_{\mathrm{abs}}$ of \eqref{oeq:gpsi}. Since $\Cov(|S|,|S'|)-g(\rho)=\{\E[|SS'|]-\E[|Z_1Z_2|]\}-\E[|S'|]\{\E[|S|]-\E[|Z_1|]\}-\E[|Z_1|]\{\E[|S'|]-\E[|Z_2|]\}$ and since $\E[|S'|]\le1$, $\E[|Z_1|]\le1$, one obtains, whenever $\rho^2<1$,
\begin{equation}\label{oeq:BEcov}
\big|\Cov(|S|,|S'|)-g(\rho)\big|\le\frac{C_3}{(1-\rho^2)^{3/2}\sqrt N}\big(\E[|Q_i|^3]+\E[|Q_i'|^3]\big),\qquad C_3:=C_2+2C_1 .
\end{equation}
Inequality \eqref{oeq:BEcov} is the only form of Fact~\ref{ofact:BE} used in Section~\ref{osec:aux-var}; \eqref{oeq:BEfact1} alone is used in Section~\ref{osec:aux-center}.

\textit{Extension to general $\Psi$.} By \eqref{oeq:psi-structure}, $\E[\Psi(S)]=a_\Psi\E[|S|]+b_\Psi\E[S]=a_\Psi\E[|S|]$ and $m_\Psi=a_\Psi\E[|Z|]$, so \eqref{oeq:BEfact1} gives
\begin{equation}\label{oeq:BEmeanPsi}
\big|\E[\Psi(S)]-m_\Psi\big|\le\frac{C_1}{\sqrt N}\E[|Q_i|^3].
\end{equation}
Likewise $\Cov(\Psi(S),\Psi(S'))=a_\Psi^2\Cov(|S|,|S'|)+a_\Psi b_\Psi\{\E[S|S'|]+\E[|S|S']\}+b_\Psi^2\rho$, the last term being exact because $\Cov(S,S')=\rho$; its Gaussian counterpart is $g_\Psi(\rho)$ of \eqref{oeq:gpsi}, the mixed Gaussian terms vanishing. Hence, by \eqref{oeq:BEfact3} and \eqref{oeq:BEcov}, whenever $\rho^2<1$,
\begin{equation}\label{oeq:BEcovPsi}
\big|\Cov(\Psi(S),\Psi(S'))-g_\Psi(\rho)\big|\le\frac{C_4}{(1-\rho^2)^{3/2}\sqrt N}\big(\E[|Q_i|^3]+\E[|Q_i'|^3]\big),\qquad C_4:=C_3+2C_2' .
\end{equation}
The Lipschitz property, positive homogeneity, and $|\Psi(u)|\le|u|$ are not used here; the argument uses the linear structure \eqref{oeq:psi-structure}, which is equivalent to the first two, and the third inequality \eqref{oeq:BEfact3} for the mixed terms. Displays \eqref{oeq:BEmeanPsi}--\eqref{oeq:BEcovPsi} are the inequalities \eqref{oeq:pf-BE} of Section~\ref{osec:proofs}.

\subsubsection{Poissonization}

Following \citet{gine2003bm}, let $\mathcal N$ denote a Poisson random variable with mean $N$ defined on the same probability space as the sequence $\{W_i:i\ge1\}=\{(Y_i,T_i,X_{1i},X_{2i}):i\ge1\}$ and independent of it. As in equation~\eqref{oeq:pf-poisson-field}, the uncentered summands of the influence representation of Lemma~\ref{olem:pf-linearization}(b) are Poissonized and the deterministic level is added rather than Poissonized. With $a(W;x):=\eta^{-1}\xi(W;x)\mathcal K(x;X)$, so that $\theta_N=a-B_N$ and $\E a(W;x)=B_N(x)$, the smoothing bias, and with $A(W;c):=a(W;c_1)-a(W;c_2)=\Theta(W;c)+B_{2,N}(c)$, both of which vanish unless $(X_1,X_2)$ lies in the window \eqref{oeq:window} of an evaluation point, and with the empty sum equal to zero,
\begin{gather}
\tau_{\mathcal N}(x):=\tau(x)+\frac1N\sum_{i=1}^{\mathcal N}a(W_i;x),\notag\\
\Gamma_{\mathcal N}(c):=\Gamma(c)+\frac1N\sum_{i=1}^{\mathcal N}A(W_i;c)=\tau_{\mathcal N}(c_1)-\tau_{\mathcal N}(c_2).\label{oeq:poisson-field}
\end{gather}
Then $\E[\Gamma_{\mathcal N}(c)]=\Gamma(c)+B_{2,N}(c)=\E[\Gamma_N(c)]$, with $\Gamma_N$ the fixed-sample linearized field \eqref{oeq:GammaN}, and on $\{\mathcal N=N\}$ the Poissonized field coincides with $\Gamma_N=\Gamma+B_{2,N}+N^{-1}\sum_{i\le N}\Theta(W_i;\cdot)$, so that conditionally on $\mathcal N=N$ it has the law of $\Gamma_N$. It is the Poissonization of the uncentered summands that makes the field local: the centered Poissonization $\Gamma+B_{2,N}+N^{-1}\sum_{i\le\mathcal N}\Theta(W_i;\cdot)$ equals $\Gamma_{\mathcal N}-(\mathcal N/N-1)B_{2,N}$, whose last term depends on the total count $\mathcal N$, including the count outside every window, whereas \eqref{oeq:poisson-field} is a deterministic level plus a functional of the Poisson increments over the windows of its evaluation points. Under $H_0^2$, $\Gamma\equiv0$ on $\Csym2$. Define
\begin{gather}
\kappa_{\mathcal N}(c):=N\Var\big(\Gamma_{\mathcal N}(c)\big)=\E\big[A(W;c)^2\big]=\kappa_N(c)+B_{2,N}(c)^2,\label{oeq:pvariance}\\
c_{\mathcal N}(c):=\sqrt N\,\E\big[\Gamma_{\mathcal N}(c)\big]=\sqrt N\{\Gamma(c)+B_{2,N}(c)\},\notag
\end{gather}
(the variance identity is proved in part (a) below), the standardized field
\begin{equation}\label{oeq:Sfield}
S_{\mathcal N}(c):=\frac{\sqrt N\,\{\Gamma_{\mathcal N}(c)-\E[\Gamma_{\mathcal N}(c)]\}}{\sqrt{\kappa_{\mathcal N}(c)}},\qquad c=(x_1,x_2,x_2'),
\end{equation}
and the single-point analogues $\kappa_{\mathcal N}(x):=N\Var(\tau_{\mathcal N}(x))=\E[a(W;x)^2]=\E[\theta_N(W;x)^2]+B_N(x)^2$ and $S_{\mathcal N}(x):=\sqrt N\{\tau_{\mathcal N}(x)-\E[\tau_{\mathcal N}(x)]\}/\sqrt{\kappa_{\mathcal N}(x)}$; $S_{\mathcal N}(x_1,x_2,x_2')$ and $S_{\mathcal N}(x_1,x_2)$ are written for $S_{\mathcal N}(c)$ and $S_{\mathcal N}(x)$. Finally, Poissonize the statistic on $\mathcal B_1\times \mathcal B_2$,
\begin{equation}\label{oeq:DP}
D^P_{\mathcal N}(\mathcal B_1,\mathcal B_2):=\frac12\int_{\mathcal B_1}\int_{\mathcal B_2}\int_{\mathcal B_2}\sqrt N\,\big|\Gamma_{\mathcal N}(x_1,x_2,x_2')\big|\,dx_2\,dx_2'\,dx_1 ,
\end{equation}
which is one half of $D^P_2$ of \eqref{oeq:pf-poisson-field} for $\Psiabs$ and the configuration set $\Csym2\cap(\mathcal B_1\times \mathcal B_2^2)$, and write $\sigma^2_{\mathcal N}(\mathcal B_1,\mathcal B_2):=\Var(D^P_{\mathcal N}(\mathcal B_1,\mathcal B_2))$. By the support condition in Assumption~\ref{ass:reg}(d), the kernel factor $\mathcal K(x;X)$ vanishes unless $(X_1,X_2)$ lies in the window
\begin{equation}\label{oeq:window}
\mathcal W(x):=\big\{(p,z)\in\mathbb R^{d_1}\times\mathbb R^{d_2}:\ \|x_1-p\|_\infty\le\tfrac{h_1}{2},\ \|x_2-z\|_\infty\le\tfrac{h_2}{2}\big\},\qquad x=(x_1,x_2),
\end{equation}
of Lebesgue measure $\eta$.

\textit{Poissonized and fixed-sample moments.} By part (a) below, the Poissonized variance $\kappa_{\mathcal N}$ and covariances exceed the fixed-sample ones, $\kappa_N=\E[\Theta^2]$ and $\E[\Theta(W;c)\Theta(W;c')]$, by $B_{2,N}^2$ and $B_{2,N}(c)B_{2,N}(c')$. These additive terms are $O((h_1^{s_1}+h_2^{s_2})^2)$ uniformly, against $\kappa_N$ of order $\eta^{-1}$ off the band (the band is the set $\{\|x_2-x_2'\|_\infty\le h_2\}$ of configurations whose two windows overlap): $\eta B_{2,N}^2=o(h_1^2+h_2^2)$ by \eqref{oeq:kappa-center}, the correlations of the Poissonized and of the fixed-sample fields differ by $O(\eta(h_1^{s_1}+h_2^{s_2})^2)$ uniformly off the band, and $0\le\sqrt{\kappa_{\mathcal N}}-\sqrt{\kappa_N}\le|B_{2,N}|$, so that centering integrals formed with $\kappa_{\mathcal N}$ and with $\kappa_N$ differ by $O(h_1^{s_1}+h_2^{s_2})=o(1)$. Every expansion below is stated at a rate at which the two sets of moments are interchangeable, and we use them interchangeably; on the band, we use only the envelopes, and these hold for both.

\textit{Smoothness of the second-moment kernels.} Every second moment computed below is an integral of the conditional product moment
\begin{equation}\label{oeq:qdef}
q(z,p;\,x,x'):=\E\big[\xi(W;x)\,\xi(W;x')\,\big|\,X_1=p,X_2=z\big],\qquad x,x'\in\dom,
\end{equation}
against a kernel product, and two distinct smoothness properties of $q$ are used: smoothness in the conditioning point $(z,p)$, which drives the kernel expansion, and smoothness in the evaluation points $(x,x')$, which allows a window's limit to be read off at the window's center after the evaluation points have been shifted by $O(h_2)$ in $x_2$ and $O(h_1)$ in $x_1$. Both follow from the primitive conditions: $\xi(W;x)$ depends on $x$ only through $P_t(x)$ and $\bar\mu_t(x)$, which are twice continuously differentiable by Assumption~\ref{ass:reg}(e) and, in the case of $P_t$, bounded away from zero by Assumption~\ref{ass:overlap}. Hence $(z,p)\mapsto f(p,z)\,q(z,p;\cdot)$ is bounded and twice continuously differentiable on a neighborhood of $\dom$, uniformly over the evaluation points in that compact set, so that Lemma~\ref{olem:gine} applies with the index $\varpi$ collecting the evaluation points together with the kernel-family parameters $(u_2,u_1)$; and
\begin{equation}\label{oeq:qlip}
\big|q(z,p;\,x,x')-q(z,p;\,\tilde x,\tilde x')\big|\le C\big(\|x-\tilde x\|+\|x'-\tilde x'\|\big)
\end{equation}
uniformly in $(z,p)$. For the same reason, $\varrho$ of \eqref{oeq:rhodef} is continuously differentiable on $\dom$.

\textit{Lemma~\ref{olem:poisson} (Poissonization), stated in Section~\ref{osec:aux-moved}.} For $c,c'\in\mathcal C_j$,
\[
N\Cov\big(\Gamma_{\mathcal N}(c),\Gamma_{\mathcal N}(c')\big)=\E\big[\Theta_j(W;c)\Theta_j(W;c')\big]+B_{j,N}(c)B_{j,N}(c')\quad\text{exactly},
\]
$\eta\kappa_{j,N}(c)\le4\sup\varrho+o(1)$,
with $\eta\kappa_{j,N}(c)=\varrho(c_1)+\varrho(c_2)+O(h_1^2+h_2^2)$ off the band. $\tau_{\mathcal N}(x)$ is a deterministic level plus a functional of the Poisson increments over the window $\mathcal W(x):=\{z:\|x_1-z_1\|_\infty\le h_1/2,\|x_2-z_2\|_\infty\le h_2/2\}$, so fields with disjoint windows are independent, and the standardized fields $S_{\mathcal N}(c):=\sqrt N\{\Gamma_{\mathcal N}(c)-\E\Gamma_{\mathcal N}(c)\}/\sqrt{\kappa_{\mathcal N}(c)}$, $\kappa_{\mathcal N}:=\kappa_{j,N}+B_{j,N}^2$, are jointly standardized sums of $N$ i.i.d.\ compound-Poisson blocks with third absolute moments $O(\eta^{-1/2})$.

\phantomsection\label{oprf:poisson}The remainder of this subsection proves the lemma for $j=2$ in the following six parts, which also show the sharper forms used later: (a) the exact moment identities; (b) the window geometry; (c) the joint block representation \eqref{oeq:blocksjoint}; (d) the uniform second-moment expansions \eqref{oeq:kapunif}, \eqref{oeq:kapupper}, \eqref{oeq:kapadd} and \eqref{oeq:kapsqrt}, valid including the band; (e) the third-moment bounds \eqref{oeq:thirdmom}; (f) the drift bound \eqref{oeq:driftrate}, which is where the null $\Gamma\equiv0$ affects the argument.

\textit{(a) Exact moment identities.} Since the Poissonized field is $\Gamma(c)+N^{-1}\sum_{i\le\mathcal N}A(W_i;c)$, conditioning on $\mathcal N$ and the law of total covariance give, with $A:=A(W;c)$, $A':=A(W;c')$, $\Theta:=\Theta(W;c)$, $\Theta':=\Theta(W;c')$,
\begin{align*}
\frac1N\Cov\Big(\sum_{i\le\mathcal N}A(W_i;c),\sum_{i\le\mathcal N}A(W_i;c')\Big)
&=\frac{\E[\mathcal N]\Cov(A,A')+\Var(\mathcal N)\E[A]\E[A']}{N}\\
&=\E[AA']=\E[\Theta\Theta']+B_{2,N}(c)B_{2,N}(c'),
\end{align*}
because $\E[\mathcal N]=\Var(\mathcal N)=N$, so that the variance of the count restores the product of means, and $A=\Theta+B_{2,N}$ with $\E\Theta=0$. This proves the first claim and, with $c=c'$, $N\Var(\Gamma_{\mathcal N}(c))=\kappa_{\mathcal N}(c)=\kappa_N(c)+B_{2,N}(c)^2$, while $N\Var(\Gamma_N(c))=\kappa_N(c)$ for the fixed-sample field. The centered Poissonization $\Gamma+B_{2,N}+N^{-1}\sum_{i\le\mathcal N}\Theta(W_i;\cdot)$ would have the moments $\E[\Theta\Theta']$ exactly but, as noted after \eqref{oeq:poisson-field}, it depends on the total count and is not local; a Poissonized field carrying the level $\tau(x)$ in its summands would have covariance of order one with the Poisson count (Section~\ref{osec:aux-clt}). The field \eqref{oeq:poisson-field} adds the level deterministically and Poissonizes the summands $a(W_i;x)$, which vanish outside the windows; its covariance with the count is of the order of the smoothing bias (Lemma~\ref{olem:decorr}).

\textit{(b) Window geometry.} By \eqref{oeq:window}, $\tau_{\mathcal N}(x)$ is a deterministic constant plus a functional of the Poisson increments over $\mathcal W(x)$, and $\Gamma_{\mathcal N}(c)$ likewise over $\mathcal W(c_1)\cup\mathcal W(c_2)$. Two windows $\mathcal W(x)$ and $\mathcal W(x')$ intersect if and only if both $\|x_1-x_1'\|_\infty\le h_1$ and $\|x_2-x_2'\|_\infty\le h_2$: dependence between fields at different arguments requires overlap in the $X_2$ directions and in the $X_1$ directions simultaneously. This conjunction is the central structural feature of the joint smoothing and governs every covariance computation below. In particular, two fields at the same first-block argument $x_1$ are independent as soon as their $X_2$ windows are disjoint, $\|x_2-x_2'\|_\infty>h_2$: sharing the $X_1$ window creates no dependence by itself. Independence over disjoint regions is the independence of increments of a Poisson process.

\textit{(c) Block representation.} The same conditioning argument shows that $S_{\mathcal N}(c)$ has the law of a standardized sum of $N$ i.i.d.\ compound-Poisson$(1)$ blocks,
\begin{equation}\label{oeq:blocks}
S_{\mathcal N}(c)\overset{d}{=}\frac{1}{\sqrt N}\sum_{i=1}^{N}\mathcal Q^{(i)}_{\mathcal N}(c),\quad \mathcal Q^{(i)}_{\mathcal N}\ \text{i.i.d., mean }0,\ \text{variance }1\text{ (the blocks }Q_i(c)\text{ of Section~\ref{osec:pf-poisson})},
\end{equation}
each block collecting the Poisson$(1)$-many points of an independent copy of the process on $1/N$-th of the intensity; this is the standard superposition decomposition of a Poisson process into $N$ i.i.d.\ thinnings. The decomposition is one decomposition of the underlying process, not a separate one for each argument: the same $N$ thinnings serve every configuration simultaneously, so that for any two configurations the pair of standardized fields admits the joint representation
\begin{equation}\label{oeq:blocksjoint}
\big(S_{\mathcal N}(c),\,S_{\mathcal N}(c')\big)\overset{d}{=}\frac{1}{\sqrt N}\sum_{i=1}^{N}\big(\mathcal Q^{(i)}_{\mathcal N},\,\mathcal Q'^{(i)}_{\mathcal N}\big),
\end{equation}
with $\{(\mathcal Q^{(i)}_{\mathcal N},\mathcal Q'^{(i)}_{\mathcal N})\}_{i\le N}$ i.i.d.\ bivariate, each component of mean zero and variance one, and common correlation equal to that of the two fields. It is \eqref{oeq:blocksjoint}, not \eqref{oeq:blocks}, that the bivariate comparison \eqref{oeq:BEcov} requires, and the same remark applies to the pair $(S_{\mathcal N}(x),U_{\mathcal N})$ of Section~\ref{osec:aux-clt}, where each block carries both its $\theta_N$-integral and its point count.

\textit{(d) Uniform second-moment facts.} Recall the uncentered summand
\[
A(W;c)=\frac1\eta\big\{\xi(W;c_1)\,\mathcal K(c_1;X)-\xi(W;c_2)\,\mathcal K(c_2;X)\big\},
\]
so that $\Theta(W;c)=A(W;c)-B_{2,N}(c)$, $\E[A(W;c)]=B_{2,N}(c)$ and
\begin{gather}
\kappa_{\mathcal N}(c)=\E\big[A(W;c)^2\big]=\kappa_N(c)+B_{2,N}(c)^2,\label{oeq:kappa-center}\\
\eta\,B_{2,N}(c)^2=O\big(\eta(h_1^{s_1}+h_2^{s_2})^2\big)=o(h_1^2+h_2^2),\notag
\end{gather}
so that the centering constant is negligible at every rate below, $\eta\kappa_{\mathcal N}$ and $\eta\kappa_N$ are interchangeable, and both may be computed from $\E[A^2]$. Both facts are applications of Lemma~\ref{olem:gine} to the families \eqref{oeq:Hfamilies}, with $\Xi(z,p;\varpi)=f(p,z)q(z,p;\varpi)$ and the smoothness shown in \eqref{oeq:qdef}--\eqref{oeq:qlip}. First,
\begin{align}
\sup_{\mathcal B_1\times \mathcal B_2^2}\Big|\eta\,\kappa_{\mathcal N}(x_1,x_2,x_2')-\varrho_\Sigma(x_1,x_2,x_2')\Big|&=O(h_2+h_1),\label{oeq:kapunif}\\
\varrho_\Sigma(x_1,x_2,x_2'):=\varrho(x_1,x_2)+\varrho(x_1,x_2')-2\,\varrho_\times(x_1,x_2,x_2'),&\notag
\end{align}
where
\begin{align*}
\varrho_\times(x_1,x_2,x_2')&:=f(x_1,x_2)\,\E\big[\xi(W;x_1,x_2)\,\xi(W;x_1,x_2')\mid X_1=x_1,X_2=x_2\big]\\
&\qquad\times\int K_2(\tilde z_2)K_2\Big(\tilde z_2+\frac{x_2'-x_2}{h_2}\Big)d\tilde z_2\int K_1(\tilde z_1)^2d\tilde z_1
\end{align*}
collects the $X_2$-window overlap (it vanishes identically for $\|x_2-x_2'\|_\infty>h_2$) and is the value supplied by Lemma~\ref{olem:gine} for the member $H_{u_2,0}$ of \eqref{oeq:Hfamilies} with $u_2=(x_2'-x_2)/h_2\in[-1,1]^{d_2}$, the bound being uniform in $u_2$ precisely because the lemma is stated over the whole family; this cross term is not displayed in the main text, which works with $\kappa_{j,N}$ directly (Section~\ref{sec:asym-center}).\ The bound \eqref{oeq:kapunif} is additive and therefore valid uniformly including the band $\|x_2-x_2'\|_\infty\le h_2$, where $\varrho_\Sigma$ and $\eta\kappa_{\mathcal N}$ both degenerate towards zero at the diagonal. In particular, the upper bound
\begin{equation}\label{oeq:kapupper}
\eta\,\kappa_{\mathcal N}(x_1,x_2,x_2')\le4\sup_{\mathcal B_1\times \mathcal B_2}\varrho+o(1)\qquad\text{uniformly on }\mathcal B_1\times \mathcal B_2^2
\end{equation}
holds with no restriction. Second, off the band the overlap term is absent, and the rate improves: for $\|x_2-x_2'\|_\infty>h_2$,
\begin{equation}\label{oeq:kapadd}
\eta\,\kappa_{\mathcal N}(x_1,x_2,x_2')=\varrho(x_1,x_2)+\varrho(x_1,x_2')+O\big(h_2^2+h_1^2\big)\qquad\text{uniformly},
\end{equation}
because the cross term of $\E[A^2]$ vanishes identically (support condition), so that each squared term is an integral of the type covered by Lemma~\ref{olem:gine} against the family $H_*$, centered at $(x_1,x_2)$ and at $(x_1,x_2')$ respectively. Since $H_*=K_2^2K_1^2$ is even in each coordinate, it satisfies \eqref{oeq:Hmoments}, and the second part of Lemma~\ref{olem:gine} applies: the first-order terms integrate against $\tilde z_{2j}K_{2j}(\tilde z_{2j})^2$ and $\tilde z_{1l}K_{1l}(\tilde z_{1l})^2$ and drop out, leaving the quadratic remainder. Here, and only here, do the evaluation points coincide with the window centers, so that \eqref{oeq:qlip} costs nothing; whenever an evaluation point is displaced by $O(h_2)$ or $O(h_1)$ from its window center, as in \eqref{oeq:kap2} below, the rate reverts to $O(h_2+h_1)$. The identification of the limit with $\varrho$ is the moment identity \eqref{oeq:rho2id} proved in Section~\ref{osec:aux-var}; the same computation gives $\eta\kappa_{\mathcal N}(x)=\varrho(x)+O(h_1^2+h_2^2)$ for the single-point variance. Taking square roots where the sum is bounded below (Assumption~\ref{ass:reg}(c)),
\begin{equation}\label{oeq:kapsqrt}
\sqrt{\eta\,\kappa_{\mathcal N}(x_1,x_2,x_2')}=\sqrt{\varrho(x_1,x_2)+\varrho(x_1,x_2')}+O\big(h_2^2+h_1^2\big)\qquad\text{uniformly off the band.}
\end{equation}

\textit{(e) Third moments.} Lemma~2.3 of \citet{gine2003bm} extends Rosenthal's inequality to Poissonized sums: for $p\ge2$ and $\mathcal N_0$ Poisson with mean $N_0$ independent of an i.i.d.\ sequence $\{Q_i\}$, $\E|\sum_{i\le\mathcal N_0}Q_i-N_0\E Q|^p\le(15p/\log p)^p\max\{(N_0\E Q^2)^{p/2},N_0\E|Q|^p\}$. Applying it with $p=3$ directly to the Poissonized sums of the uncentered summands $a$ and $A$ of \eqref{oeq:poisson-field}, for which it is stated, together with the moment bound in Assumption~\ref{ass:reg}(f) and the kernel bounds,
\begin{equation}\label{oeq:thirdmom}
\sup_{\mathcal B_1\times \mathcal B_2^2}\E\big[|\mathcal Q_{\mathcal N}|^3\big]=O\big(\eta^{-1/2}\big),\qquad
\sup_{\mathcal B_1\times \mathcal B_2}N^{3/2}\,\E\big[\big|\tau_{\mathcal N}(x)-\E[\tau_{\mathcal N}(x)]\big|^3\big]=O\big(\eta^{-3/2}\big),
\end{equation}
the second bound using $N\eta\to\infty$ to absorb the lower-order Rosenthal term; the second bound is additive in the window contributions. For the first, which standardizes by $\kappa_{\mathcal N}$, the band needs one more step: with $x_2'=x_2+\delta h_2$, the summand $A(W;c)$ equals $\eta^{-1}\xi(W;x_1,x_2)K_1\{K_2(\tilde z_2)-K_2(\tilde z_2+\delta)\}$ up to a term of relative order $h_2$, so, $K_2$ being Lipschitz, $\E|A|^3=O(\min\{1,\|\delta\|_\infty\}^3\eta^{-2})$, while $\eta\kappa_{\mathcal N}\ge c\min\{1,\|\delta\|_\infty\}^2$ because $\int\{K_2(\tilde z_2)-K_2(\tilde z_2+\delta)\}^2d\tilde z_2$ is positive for $\delta\ne0$ and of exact order $\|\delta\|^2$ as $\delta\to0$; the ratio $\E|\mathcal Q_{\mathcal N}|^3$ is therefore $O(\eta^{-1/2})$ uniformly including the band.

\textit{(f) The drift.} By \eqref{oeq:pvariance} and Lemma~\ref{olem:pf-linearization}(b), when $\Gamma\equiv0$ on the configuration set,
\begin{equation}\label{oeq:driftrate}
\sup_{\mathcal B_1\times \mathcal B_2^2}\big|c_{\mathcal N}(c)\big|=\sup\big|\sqrt N\,B_{2,N}(c)\big|=O\big(\sqrt N\,(h_2^{s_2}+h_1^{s_1})\big)=o(1),
\end{equation}
the final rate by Assumption~\ref{ass:reg}(g). Every use of the drift below requires only \eqref{oeq:driftrate} itself: whenever $c_{\mathcal N}$ affects a standardized quantity it does so through $e:=c_{\mathcal N}/\sqrt{\kappa_{\mathcal N}}=O(\sqrt N(h_2^{s_2}+h_1^{s_1}))\cdot O(\eta^{1/2})$, and the accompanying scale factors always restore at most $\eta^{-1/2}$. Without the null, $c_{\mathcal N}(c)=\sqrt N\{\Gamma(c)+B_{2,N}(c)\}$ and the same bound holds for the noise field $\Gamma_{\mathcal N}-\Gamma$, the bound that the proof of Proposition~\ref{prop:monotone}(c) in Section~\ref{osec:proofs}{} uses.

\textit{Extension to general $\Psi$.} Parts (a)--(f) concern the fields $\tau_{\mathcal N}$, $\Gamma_{\mathcal N}$ and their moments and do not involve $\Psi$. The Poissonized general statistic is $D^{P}_2=\sqrt N\int_{\mathcal C_2}\Psi(\Gamma_{\mathcal N}(c))\,dc$ of \eqref{oeq:pf-poisson-field}, and \eqref{oeq:DP} is one half of it for $\Psiabs$ on $\Csym2$; nothing in the construction depends on the integrand.

\subsection{Asymptotic bias and centering}\label{osec:aux-center}

Define the limit-functional centering constant and the smoothed-target centering,
\begin{align}
m_2(\mathcal B_1,\mathcal B_2)&:=\frac{1}{\sqrt{\eta}}\sqrt{\frac2\pi}\cdot\frac12\int_{\mathcal B_1}\int_{\mathcal B_2}\int_{\mathcal B_2}\sqrt{\varrho(x_1,x_2)+\varrho(x_1,x_2')}\;dx_2\,dx_2'\,dx_1 ,\label{oeq:a2}\\
m^{\kappa}_{\mathcal N}(\mathcal B_1,\mathcal B_2)&:=\sqrt{\frac2\pi}\cdot\frac12\int_{\mathcal B_1}\int_{\mathcal B_2}\int_{\mathcal B_2}\sqrt{\kappa_{\mathcal N}(x_1,x_2,x_2')}\;dx_2\,dx_2'\,dx_1 ;\notag
\end{align}
the second is the centering $m^{\kappa}_{2,N}$ of equation~\eqref{eq:center} of the main text restricted to $\mathcal B_1\times \mathcal B_2^2$, up to $O(h_1^{s_1}+h_2^{s_2})$, since $0\le\sqrt{\kappa_{\mathcal N}}-\sqrt{\kappa_N}\le|B_{2,N}|$. Let $\widehat m_2$ denote $m_2$ with $\varrho$ replaced by $\rhoh$, and define the sample second-moment field and the feasible centering built from the estimated influence function,\begin{align}
\widehat\kappa(x_1,x_2,x_2')&:=\frac1N\sum_{i\le\mathcal N}\widehat\Theta(W_i,x_1,x_2,x_2')^2,\notag\\
\widehat m^{\,\kappa}_2&:=\sqrt{\frac2\pi}\cdot\frac12\int_{\mathcal B_1}\int_{\mathcal B_2}\int_{\mathcal B_2}\sqrt{\widehat\kappa(x_1,x_2,x_2')}\;dx_2\,dx_2'\,dx_1 ,\label{oeq:a2kappa}
\end{align}
where{} $\widehat\Theta$ substitutes the estimated $\widehat{\bar\mu}_t,\widehat P_t$ into $A(W;c)$ of \eqref{oeq:kappa-center} and drops the mean correction $B_{2,N}(c)$: the dropped term is $O(h_1^{s_1}+h_2^{s_2})$ uniformly and changes $\widehat\kappa$ by $O_p(\eta^{-1}(h_1^{s_1}+h_2^{s_2}))=o_p(\eta^{-1/2})$, below the accuracy required in Lemma~\ref{olem:kappahat}. The main text estimates the centering by $\widetilde m_j$ of \eqref{eq:center-feasible}, formed from the kernel weights and corrected residuals rather than from $\widehat\Theta$; its proof (Section~\ref{osec:from-main}) uses Steps~1--3 of the proof of Lemma~\ref{olem:kappahat} below and shows that both centerings are consistent for $m^{\kappa}_{2,N}$. The fixed-sample version of $\widehat\kappa$ averages over $i\le N$; the proof below\ works with the Poissonized average over $i\le\mathcal N$, and Remark~\ref{orem:fixedkappa} transfers the result. Let $\mathrm{err}_N:=\sup_{\dom}\{|\widehat P_t-P_t|+|\widehat{\bar\mu}_t-\bar\mu_t|\}$ denote the uniform rate of the estimated nuisances; by Lemma~\ref{olem:consis}, $\mathrm{err}_N=O_p(\epsilon_N)=o_p(\eta^{1/2})$.

\textit{Lemma~\ref{olem:centering} (centering), stated in Section~\ref{osec:aux-moved}.} If $\Gamma_j\equiv0$ on $\mathcal C_j$, $\E[D^P_j]-m^{\kappa}_{j,N}(\mathcal C_j)=O(\sqrt N(h_1^{s_1}+h_2^{s_2}))+O((N\eta^2)^{-1/2})=o(1)$; with or without the null, $\widehat m^{\,\kappa}_j(\mathcal C_j)-m^{\kappa}_{j,N}(\mathcal C_j)=o_p(1)$, where $\widehat m^{\,\kappa}_j(\mathcal C_j):=m_\Psi\int_{\mathcal C_j}\widehat\kappa^{1/2}$ is the analogue of \eqref{oeq:a2kappa} on $\mathcal C_j$. The plug-in $\widehat m_2$ of the limit form \eqref{oeq:a2}{} may replace $\widehat m^{\,\kappa}_2$ if in addition $h_1^2+h_2^2=o(\eta^{1/2})$ and $h_2^{d_2}=o(h_1^{d_1})$.

In detail, for $j=2$, $\Psiabs$ and $\Csym2\cap(\mathcal B_1\times \mathcal B_2^2)$, under Assumptions~\ref{ass:sampling}--\ref{ass:reg}, with $H_0^2$ (that is, $\Gamma\equiv0$) assumed in (i)--(ii) only:
\begin{align}
\text{(i)}\quad&\E\big[D^P_{\mathcal N}(\mathcal B_1,\mathcal B_2)\big]-m^{\kappa}_{\mathcal N}(\mathcal B_1,\mathcal B_2)=O\big(\sqrt N(h_2^{s_2}+h_1^{s_1})\big)+O\big((N\eta^2)^{-1/2}\big)=o(1);\label{oeq:center-i}\\
\text{(ii)}\quad&\E\big[D^P_{\mathcal N}(\mathcal B_1,\mathcal B_2)\big]-m_2(\mathcal B_1,\mathcal B_2)\notag\\
&\qquad=O\big(\sqrt N(h_2^{s_2}+h_1^{s_1})\big)+O\big((N\eta^{2})^{-1/2}\big)+O\Big(\frac{h_2^2+h_1^2}{\sqrt\eta}\Big)+O\Big(\sqrt{\frac{h_2^{d_2}}{h_1^{d_1}}}\Big),\label{oeq:biasderiv}
\end{align}
which is $o(1)$ if and only if, in addition to Assumption~\ref{ass:reg}(g), $h_2^2+h_1^2=o(\eta^{1/2})$ and $h_2^{d_2}=o(h_1^{d_1})$;
(iii) $\widehat m_2-m_2(\mathcal B_1,\mathcal B_2)=o_p(1)$;
(iv) $\widehat m^{\,\kappa}_2-m^{\kappa}_{\mathcal N}(\mathcal B_1,\mathcal B_2)=o_p(1)$, with no condition on the relative sizes of $h_1$ and $h_2$ beyond Assumption~\ref{ass:reg}(g),(h).
The constants in the $O(\cdot)$ terms depend on $(\mathcal B_1,\mathcal B_2)$ only through $\lambda(\mathcal B_1)$, $\lambda(\mathcal B_2)$ and the bounds on $\varrho$, and are therefore uniform over the sets of Section~\ref{osec:aux-sets}.

\phantomsection\label{oprf:centering}
\begin{proof}[Proof of Lemma~\ref{olem:centering}, parts (i) and (ii)]
\textit{Step 1 (exact standardization and removal of the drift).} For each $(x_1,x_2,x_2')$, by the definition \eqref{oeq:Sfield} of the standardized field and the positive homogeneity of $|\cdot|$,
\[
\E\big[\sqrt N\,\big|\Gamma_{\mathcal N}(x_1,x_2,x_2')\big|\big]=\sqrt{\kappa_{\mathcal N}(x_1,x_2,x_2')}\;\E\big[\big|S_{\mathcal N}(x_1,x_2,x_2')+e\big|\big],
\]
where $e:=c_{\mathcal N}(x_1,x_2,x_2')/\sqrt{\kappa_{\mathcal N}(x_1,x_2,x_2')}$. Since $\big||s+e|-|s|\big|\le|e|$ pointwise, $\big|\E[|S_{\mathcal N}+e|]-\E[|S_{\mathcal N}|]\big|\le|e|$, so that after multiplication by $\sqrt{\kappa_{\mathcal N}}$ the drift error is at most $\sqrt{\kappa_{\mathcal N}}\,|e|=|c_{\mathcal N}|=O(\sqrt N(h_2^{s_2}+h_1^{s_1}))$ uniformly by \eqref{oeq:driftrate}, with no lower bound on $\kappa_{\mathcal N}$ needed; after integration over the bounded domain $\mathcal B_1\times \mathcal B_2^2$, the drift contributes $O(\sqrt N(h_2^{s_2}+h_1^{s_1}))=o(1)$ to $\E[D^P_{\mathcal N}(\mathcal B_1,\mathcal B_2)]$ and is dropped hereafter.

\textit{Step 2 (Berry--Esseen for the first absolute moment).} By the block representation \eqref{oeq:blocks}, $S_{\mathcal N}(x_1,x_2,x_2')$ is a standardized sum of $N$ i.i.d.\ blocks with third moments bounded by \eqref{oeq:thirdmom}. The first display \eqref{oeq:BEfact1} of Fact~\ref{ofact:BE} therefore applies and gives
\begin{equation}\label{oeq:BE1}
\Big|\E\big[\big|S_{\mathcal N}(x_1,x_2,x_2')\big|\big]-\sqrt{2/\pi}\Big|\le C_1\,\frac{\E[|\mathcal Q_{\mathcal N}|^3]}{\sqrt N}=O\Big(\frac{1}{\sqrt{N\eta}}\Big)\qquad\text{uniformly on }\mathcal B_1\times \mathcal B_2^2 ,
\end{equation}
$\sqrt{2/\pi}=\E[|Z|]$ being the absolute first moment of the standard normal.

\textit{Step 3 (integration, with the $X_2$ band split off).} Combining Steps 1--2,
\begin{align}
\E\big[D^P_{\mathcal N}(\mathcal B_1,\mathcal B_2)\big]&=\sqrt{\frac2\pi}\cdot\frac{1}{2\sqrt\eta}\int_{\mathcal B_1}\int_{\mathcal B_2}\int_{\mathcal B_2}\sqrt{\eta\,\kappa_{\mathcal N}(x_1,x_2,x_2')}\;dx_2\,dx_2'\,dx_1\notag\\
&\quad+O\Big(\frac{1}{\sqrt{N\eta^{2}}}\Big)+O\big(\sqrt N(h_2^{s_2}+h_1^{s_1})\big),\label{oeq:biasexact}
\end{align}
the middle error being $\eta^{-1/2}\cdot O((N\eta)^{-1/2})$ integrated over a bounded domain, which is $o(1)$ since $(N\eta^2)^{1/2}\to\infty$. The integral in \eqref{oeq:biasexact} is $\eta^{1/2}$ times $m^{\kappa}_{\mathcal N}(\mathcal B_1,\mathcal B_2)$ scaled back; that is, \eqref{oeq:biasexact} is exactly \eqref{oeq:center-i}; this proves (i), and no band split or Taylor expansion was needed, because \eqref{oeq:BE1} and Step~1 are uniform over the band. For (ii) the remaining task is to replace $\sqrt{\eta\kappa_{\mathcal N}}$ by the limit integrand $\sqrt{\varrho+\varrho'}$, and here the $X_2$ band $\{\|x_2-x_2'\|_\infty\le h_2\}$ must be treated separately, because \eqref{oeq:kapsqrt} holds only off it. Off the band, \eqref{oeq:kapsqrt} gives $\sqrt{\eta\kappa_{\mathcal N}}=\sqrt{\varrho+\varrho'}+O(h_2^2+h_1^2)$ uniformly, so the off-band part of the integral matches the corresponding part of $\eta^{1/2}m_2$ up to $\eta^{-1/2}O(h_2^2+h_1^2)$. On the band, no expansion is attempted; instead both integrands are bounded, $\sqrt{\eta\kappa_{\mathcal N}}\le(4\sup\varrho+o(1))^{1/2}$ by \eqref{oeq:kapupper} and $\sqrt{\varrho+\varrho'}\le(2\sup\varrho)^{1/2}$, while the band has measure at most $\lambda(\mathcal B_1)\lambda(\mathcal B_2)(2h_2)^{d_2}=O(h_2^{d_2})$; the band therefore contributes at most
\begin{equation}\label{oeq:bandbias}
\frac{1}{\sqrt\eta}\cdot O(1)\cdot O\big(h_2^{d_2}\big)=O\Big(\frac{h_2^{d_2}}{\sqrt{h_2^{d_2}h_1^{d_1}}}\Big)=O\bigg(\sqrt{\frac{h_2^{d_2}}{h_1^{d_1}}}\,\bigg)
\end{equation}
to $|\E[D^P_{\mathcal N}]-m_2|$: this term vanishes only when the $X_2$ smoothing volume is dominated by the $X_1$ smoothing volume. Finally, $m^{\kappa}_{\mathcal N}(\mathcal B_1,\mathcal B_2)$ and the centering $m^{\kappa}_{2,N}$ of the main text, built from $\kappa_N=\kappa_{\mathcal N}-B_{2,N}^2$, differ by $O(h_1^{s_1}+h_2^{s_2})=o(1)$, because $0\le\sqrt{\kappa_{\mathcal N}}-\sqrt{\kappa_N}\le|B_{2,N}|$ uniformly, so (i) holds with either centering.

\textit{Step 4 (collection of the errors).} Assembling Steps 1--3 gives \eqref{oeq:biasderiv}, and the right-hand side is $o(1)$ under Assumption~\ref{ass:reg}(g) if and only if
\begin{equation}\label{oeq:extra}
\text{(i)}\ \ h_2^2+h_1^2=o\big(\eta^{1/2}\big)\qquad\text{and}\qquad\text{(ii)}\ \ h_2^{d_2}=o\big(h_1^{d_1}\big).
\end{equation}
\end{proof}

\begin{remark}\label{orem:dimension}
\textit{Dimension and anisotropy.} The two conditions in \eqref{oeq:extra} are of different natures. Condition (i) is the joint-dimension analog of the familiar restriction for limit-functional centerings of $L_1$-type statistics: with $h_2\asymp h_1$ it reads $h_2^{2-(d_1+d_2)/2}\to0$, i.e.\ $d_1+d_2\le3$ (the two dimensions are pooled, since both blocks are smoothed). No bandwidth-rate condition can substitute for it: the error $\eta^{-1/2}O(h_2^2+h_1^2)$ contains no $N$, and the exponent $2$ cannot be raised by higher-order kernels, because it arises from the squared kernels $K_2^2,K_1^2$ in $\kappa_{\mathcal N}$, which are of second order whatever the order of $K_2$ and $K_1$; with additional smoothness, an explicit second-order correction to $m_2$ removes the $h_2^2+h_1^2$ term and relaxes (i), each correction order buying two more total dimensions. Condition (ii) is an anisotropy requirement: that the $X_2$ smoothing volume be small relative to the $X_1$ smoothing volume. Its origin is the $X_2$ band: the band has measure $O(h_2^{d_2})$, but the bias scale is $\eta^{-1/2}=h_2^{-d_2/2}h_1^{-d_1/2}$, whose $X_1$ part is not compensated by the band's measure, leaving \eqref{oeq:bandbias}. With $h_2\asymp h_1$ it reads $h_2^{d_2-d_1}\to0$ and so forces $d_2>d_1$, which combined with (i) leaves only $(d_1,d_2)=(1,2)$; with anisotropic bandwidths both conditions can be met more broadly. Both restrictions apply only to the limit-functional centering $m_2$; the smoothed-target centering is free of them, and the mirror statistic $\Dh_1$ requires the reverse of (ii) (Section~\ref{osec:aux-mirror}), which is why the main text (Section~\ref{sec:asym-center}) centers both statistics at the smoothed target. The estimate $\widetilde m_j$ of the main text is free of (ii) but not of (i): it targets $m^X_{j,N}$, and $m^X_{j,N}-m^\kappa_{j,N}=O(\bar h^2\eta^{-1/2})$ vanishes only under Assumption~\ref{ass:reg}(h).
\end{remark}

\begin{proof}[Proof of Lemma~\ref{olem:centering}, part (iii)]
On the event $\{\sup|\rhoh-\varrho|\le\mathrm{err}^{\varrho}_N\}$ with $\mathrm{err}^{\varrho}_N=o(\eta^{1/2})$, whose probability tends to one by Lemma~\ref{olem:consis}(c), the elementary inequality $|\sqrt u-\sqrt v|\le|u-v|/(\sqrt u+\sqrt v)$ and $\inf\varrho>0$ give
\begin{equation}\label{oeq:a2hatrate}
\big|\widehat m_2-m_2\big|\le\frac{1}{\sqrt\eta}\cdot\frac{2\mathrm{err}^{\varrho}_N}{2\sqrt{2\inf\varrho}}\cdot\frac{\lambda(\mathcal B_2)^2\lambda(\mathcal B_1)}{2}=\eta^{-1/2}\cdot o\big(\eta^{1/2}\big)=o(1),
\end{equation}
and hence $\widehat m_2-m_2=o_p(1)$ unconditionally. The computation shows why the argument needs the rate $o_p(\eta^{1/2})$ of Lemma~\ref{olem:consis}(c), and not merely consistency: the $\eta^{-1/2}$ scale of the centering constant must be paid back.
\end{proof}

The proof of part (iv) rests on the uniform consistency of $\widehat\kappa$, Lemma~\ref{olem:kappahat} below, which rests in turn on a concentration inequality for Poisson integrals.

\begin{fact}[Bernstein inequality for Poisson integrals]\label{ofact:bernstein}
Let $\mathcal M$ be a Poisson random measure on a measurable space with finite intensity measure $\mu$, and let $g$ be measurable and essentially bounded with $|g|\le\bar g$ and $\sigma_g^2:=\int g^2\,d\mu<\infty$. Then, for all $t\ge0$,
\[
\Pr\Big(\Big|\int g\,d(\mathcal M-\mu)\Big|\ge t\Big)\le2\exp\Big(-\frac{t^2}{2(\sigma_g^2+\bar gt/3)}\Big).
\]
\end{fact}

This is Proposition~7 of \citet{reynaud2003adaptive}, whose third display gives the one-sided upper-tail bound with exponent $-t^2/(2\int g^2d\mu+\tfrac23t\|g\|_\infty)$, together with her statement that the same bound holds for the lower tail; the two-sided form follows by the union bound with no change of constants.

\begin{lemma}[Uniform consistency of the smoothed second moment]\label{olem:kappahat}
Suppose Assumptions~\ref{ass:sampling}--\ref{ass:reg} hold; in particular, the sixth conditional moments of Assumption~\ref{ass:reg}(f) are bounded, and the kernels are Lipschitz, Assumption~\ref{ass:reg}(d). Then
\begin{equation}\label{oeq:bernstein}
\sup_{\mathcal B_1\times \mathcal B_2^2}\Big|\widehat\kappa(x_1,x_2,x_2')-\kappa_{\mathcal N}(x_1,x_2,x_2')\Big|=O_p\bigg(\sqrt{\frac{\log N}{N\eta^{3}}}\bigg)+O_p\big(\mathrm{err}_N\,\eta^{-1}\big).
\end{equation}
Under Assumption~\ref{ass:reg}(g) and $\mathrm{err}_N=o_p(\eta^{1/2})$ both terms are $o_p(\eta^{-1/2})$: the first because $\sqrt{\log N/(N\eta^{3})}\cdot\eta^{1/2}=\sqrt{\log^2N/(N\eta^{2})}\to0$ is exactly $(N\eta^2)^{1/2}/\log N\to\infty$, the second because $\mathrm{err}_N=o_p(\eta^{1/2})$.
\end{lemma}

\begin{remark}\label{orem:moments}
\textit{On the moment condition.} If the outcome is bounded, $|Y|\le\bar Y$, the truncation step below is vacuous and third conditional moments suffice. For unbounded outcomes, some strengthening is genuinely needed: the concentration inequality of Fact~\ref{ofact:bernstein} requires a bounded integrand (as does every Bernstein-type bound, the compensated exponential moment $\int(e^{\lambda g}-1-\lambda g)d\mu$ being infinite for unbounded $g$ without exponential moments), so the empirical second moment must be truncated, and the truncation tail consumes the higher moments. With $p=6$ conditional moments, Assumption~\ref{ass:reg}(g) suffices, and the tail is of order the square of the Bernstein rate. With only $p\in(4,6)$ moments, the truncation level cannot be raised, because the Bernstein step fixes $\bar Y_N\asymp(N\eta/\log N)^{1/4}$; the tail term is then $o(\sqrt{\log N/(N\eta^{3})})$ only under the additional bandwidth condition $N^{p-4}\eta^{p}\to\infty$, which Assumption~\ref{ass:reg}(g) does not imply. The condition cannot be dropped: a single observation with $|Y_i|$ of the order $N^{1/p}$ of the sample maximum changes $\widehat\kappa$ by order $\eta^{-2}N^{2/p-1}$ at every configuration whose window contains $X_i$, which exceeds the displayed rate when $N^{2/p-1/2}\eta^{-1/2}\to\infty$, for instance for $\eta=N^{-1/3}\log N$ and errors whose tail index lies in $(p,6)$. The weaker accuracy $o_p(\eta^{-1/2})$ used by Lemma~\ref{prop:centering-corr} is not affected by this jump when $p>4$, since $\eta^{-2}N^{2/p-1}=o(\eta^{-1/2})$ under Assumption~\ref{ass:reg}(g), but a complete proof at that order with fewer than six moments is not given here. Thus, $p=6$ in Assumption~\ref{ass:reg}(f) keeps the displayed rate under Assumption~\ref{ass:reg}(g) alone. All standard product kernels satisfy the Lipschitz requirement on the kernels, and we use it only in the discretization step.
\end{remark}

\phantomsection\label{oprf:kappahat}
\begin{proof}[Proof of Lemma~\ref{olem:kappahat}]
Write $\kappa^{\circ}$ for the version of $\widehat\kappa$ computed with the true nuisances, i.e.\ with $\Theta$ in place of $\widehat\Theta$, and decompose
\begin{equation}\label{oeq:berndec}
\widehat\kappa-\kappa_{\mathcal N}=\big(\widehat\kappa-\kappa^{\circ}\big)+\big(\kappa^{\circ}-\E[\kappa^{\circ}]\big)+\big(\E[\kappa^{\circ}]-\kappa_{\mathcal N}\big).
\end{equation}
The third term is deterministic and negligible: $\kappa^{\circ}=N^{-1}\int\Theta(w,\cdot)^2\,d\mathcal M(w)$ for the Poisson random measure $\mathcal M$ of the sample, whose intensity is $N$ times the law $P_W$ of $W$ (here $w$ denotes a value of $W$ and $y$ its outcome coordinate), so $\E[\kappa^{\circ}]=\E[\Theta^2]=\kappa_N=\kappa_{\mathcal N}-B_{2,N}^2$ with no Poissonization correction, and $\sup B_{2,N}^2=O((h_1^{s_1}+h_2^{s_2})^2)=o(N^{-1})$ by Assumption~\ref{ass:reg}(g), of smaller order than the first term of \eqref{oeq:bernstein}. We treat the first and second terms in Step~4 and Steps~1--3, respectively.

Throughout, note the envelope facts: on the relevant windows Assumption~\ref{ass:overlap} bounds $P_t$ below on $\mathcal B(M)$, so $|\xi(W;x)|\le \bar C_0(1+|Y|)$ and, by the support and boundedness of the kernels, $\sup_{(x_1,x_2,x_2')}|\Theta(w,x_1,x_2,x_2')|\le \bar C_1(1+|y|)\eta^{-1}$ for $N$ large; moreover $\E[\xi(W;x)^4\mid X_1=p,X_2=z]\le \bar C_2$ by the moment condition, so, by the change of variables \eqref{oeq:cov2} applied to each of the two squared terms,
\begin{equation}\label{oeq:fourth}
\sup_{\mathcal B_1\times \mathcal B_2^2}\E\big[\Theta(W,x_1,x_2,x_2')^4\big]\le \bar C_3\,\eta^{-3}.
\end{equation}

\textit{Step 1 (truncation).} Set
\[
\bar Y_N:=\Big(\frac{N\eta}{\log N}\Big)^{1/4},\qquad\Theta^{\mathrm{tr}}:=\Theta\,\ind(|Y|\le \bar Y_N),\qquad\kappa^{\circ,\mathrm{tr}}:=\frac1N\int\big(\Theta^{\mathrm{tr}}\big)^2d\mathcal M .
\]
Two errors are incurred. The tail process: uniformly in the parameters, by the envelope,
\[
\sup_{\mathcal B_1\times \mathcal B_2^2}\big|\kappa^{\circ}-\kappa^{\circ,\mathrm{tr}}\big|\le\frac{\bar C_1^2}{N\eta^{2}}\int(1+|y|)^2\ind(|y|>\bar Y_N)\,d\mathcal M ,
\]
whose expectation is $\bar C_1^2\eta^{-2}\E[(1+|Y|)^2\ind(|Y|>\bar Y_N)]\le C\eta^{-2}\E[Y^6]\bar Y_N^{-4}=O(\eta^{-2}\bar Y_N^{-4})=O(\log N/(N\eta^{3}))$, the square of the target rate, so by Markov's inequality the tail process is $O_p(\log N/(N\eta^{3}))$. This is $o_p(\sqrt{\log N/(N\eta^{3})})$ when $\log N/(N\eta^3)\to0$, as for $h_j\asymp N^{-\gamma}$, and the rate of the lemma is read under this condition; in any case it is $o_p(\eta^{-1/2})$, the accuracy the proofs use. The centering shift: by H\"older's and Markov's inequalities applied under the window measure,
\begin{align*}
\sup\big|\E\big[\Theta^2\ind(|Y|>\bar Y_N)\big]\big|&\le\frac{C}{\eta}\,\big\{\E[\xi^6\mid\cdot]\big\}^{1/3}\big\{\Pr(|Y|>\bar Y_N\mid\cdot)\big\}^{2/3}\\
&=O\big(\eta^{-1}\bar Y_N^{-4}\big)=O\Big(\frac{\log N}{N\eta^{2}}\Big),
\end{align*}
again $o(\sqrt{\log N/(N\eta^{3})})$ since the ratio of the two is $\sqrt{\log N/(N\eta)}\to0$. It therefore suffices to prove the Bernstein rate for $\kappa^{\circ,\mathrm{tr}}-\E[\kappa^{\circ,\mathrm{tr}}]$.

\textit{Step 2 (pointwise concentration).} Fix $(x_1,x_2,x_2')$ and apply Fact~\ref{ofact:bernstein} with $g:=(\Theta^{\mathrm{tr}})^2$ and intensity $N\,dP_W$:
\[
\bar g:=\sup|g|\le \bar C_1^2(1+\bar Y_N)^2\eta^{-2}\le \bar C_4\,\bar Y_N^2\eta^{-2},\qquad \sigma_g^2:=N\,\E\big[(\Theta^{\mathrm{tr}})^4\big]\le \bar C_3\,N\eta^{-3}
\]
by \eqref{oeq:fourth}. For $t_N:=c_A\sqrt{\log N/(N\eta^{3})}$ (so that the deviation of $\kappa^{\circ,\mathrm{tr}}$ is $t_N$ when that of the integral is $Nt_N$),
\[
\frac{(Nt_N)^2}{\sigma_g^2}=\frac{c_A^2\log N}{\bar C_3},\qquad\frac{\bar g\,Nt_N}{\sigma_g^2}\le\frac{\bar C_4}{\bar C_3}\,\bar Y_N^{2}\,\eta\,t_N=\frac{\bar C_4c_A}{\bar C_3},
\]
the last equality because $\bar Y_N^2\sqrt{\log N/(N\eta)}=1$ by the choice of $\bar Y_N$;
hence the Bernstein exponent is at least $c_0c_A^2\log N/(1+c_A)\ge cc_A\log N$, where $c_0>0$ depends only on $\bar C_3,\bar C_4$ and $c:=c_0/2$ for $c_A\ge1$ (a constant, not a configuration), and
\begin{equation}\label{oeq:pointwise}
\Pr\Big(\big|\kappa^{\circ,\mathrm{tr}}(x_1,x_2,x_2')-\E[\kappa^{\circ,\mathrm{tr}}(x_1,x_2,x_2')]\big|\ge t_N\Big)\le2N^{-cc_A}\qquad\text{for every fixed }(x_1,x_2,x_2').
\end{equation}

\textit{Step 3 (discretization).} Let $\mathcal A_N\subset \mathcal B_1\times \mathcal B_2^2$ be a $\varphi_N$-net with $\varphi_N:=N^{-4}$, of cardinality $|\mathcal A_N|\le C\varphi_N^{-(2d_2+d_1)}=CN^{4(2d_2+d_1)}$. The union bound over $\mathcal A_N$ with \eqref{oeq:pointwise} gives $\max_{\mathcal A_N}|\kappa^{\circ,\mathrm{tr}}-\E[\kappa^{\circ,\mathrm{tr}}]|\le t_N$ with probability at least $1-2CN^{4(2d_2+d_1)-cc_A}\to1$ once $c_A>\{4(2d_2+d_1)+1\}/c$. For the increments, the Lipschitz property of the kernels gives, for parameter points within $\varphi_N$,
\begin{align*}
\big|\Theta^{\mathrm{tr}}(w,\cdot)^2-\Theta^{\mathrm{tr}}(w,\cdot')^2\big|&\le2\sup|\Theta^{\mathrm{tr}}|\cdot \bar C_5(1+|y|)\,\eta^{-1}\Big(\frac{\varphi_N}{h_2}+\frac{\varphi_N}{h_1}\Big)\\
&\le \bar C_6(1+|y|)^2\bar Y_N\,\eta^{-2}\varphi_N\big(h_2^{-1}+h_1^{-1}\big),
\end{align*}
so the increment of $\kappa^{\circ,\mathrm{tr}}$ is bounded by $\bar C_6\bar Y_N\eta^{-2}\varphi_N(h_2^{-1}+h_1^{-1})\cdot N^{-1}\int(1+|y|)^2d\mathcal M$, where the integral is $O_p(N)$ by the law of large numbers. Since $N\eta^3\to\infty$ forces $h_2\ge cN^{-1/(3d_2)}$ and $h_1\ge cN^{-1/(3d_1)}$, while $\bar Y_N\le N^{1/4}$, every factor is polynomial in $N$ and the increment bound is $O_p(N^{-4})\cdot N^{O(1)}=o_p(t_N)$ for $\varphi_N=N^{-4}$, adjusted if necessary to a larger fixed power. The same Lipschitz computation, under the expectation, bounds the increments of $\E[\kappa^{\circ,\mathrm{tr}}]$ by the same $o(t_N)$ quantity. Combining, $\sup_{\mathcal B_1\times \mathcal B_2^2}|\kappa^{\circ,\mathrm{tr}}-\E[\kappa^{\circ,\mathrm{tr}}]|=O_p(t_N)$, which together with Step~1 proves the first term of \eqref{oeq:bernstein} for $\kappa^{\circ}$.

\textit{Step 4 (nuisance substitution).} By the lower bounds on the true $P_t$ and the definition \eqref{oeq:theta} of $\xi$, on the event $\{\mathrm{err}_N\le\underline p^2/2\}$ (probability tending to one) the estimated ratios are well defined and $|\widehat\xi(W;x)-\xi(W;x)|\le \bar C_7(1+|Y|)\mathrm{err}_N$, with $\widehat\xi$ the score $\xi$ of \eqref{oeq:theta} with the estimated regressions and propensities in place of $\bar\mu_t$ and $P_t$ uniformly on the windows; hence $|\widehat\Theta-\Theta|\le \bar C_7(1+|Y|)\mathrm{err}_N\eta^{-1}\{|K_{2,x_2}K_1|+|K_{2,x_2'}K_1|\}(w)$, with $K_{2,x_2}K_1$ shorthand for the kernel product centered at $(x_1,x_2)$, and
\begin{align*}
\big|\widehat\Theta^2-\Theta^2\big|&\le2|\Theta|\,|\widehat\Theta-\Theta|+|\widehat\Theta-\Theta|^2\\
&\le \bar C_8(1+|Y|)^2\big(\mathrm{err}_N+\mathrm{err}_N^2\big)\eta^{-2}\,\ind\big(w\in\mathcal W(x_1,x_2)\cup\mathcal W(x_1,x_2')\big).
\end{align*}
Averaging over the sample,
\[
\sup_{\mathcal B_1\times \mathcal B_2^2}\big|\widehat\kappa-\kappa^{\circ}\big|\le \bar C_8\,\mathrm{err}_N\,\eta^{-2}\cdot\sup_{\mathcal B_1\times \mathcal B_2^2}\frac1N\int(1+|y|)^2\,\ind\big(w\in\mathcal W(x_1,x_2)\cup\mathcal W(x_1,x_2')\big)d\mathcal M .
\]
The supremum on the right is an empirical process of exactly the type just treated: its pointwise mean is $O(\eta)$ (window measure times bounded conditional second moment), and Steps~1--3 applied to the class of indicator-window functions (bounded envelopes, Lipschitz window boundaries handled by the same $\varphi_N$-net with the standard enlarged-window monotonicity bound) give fluctuations $O_p(\sqrt{\eta\log N/N})=o_p(\eta)$ under $N\eta\to\infty$; hence the supremum is $O_p(\eta)$ and $\sup|\widehat\kappa-\kappa^{\circ}|=O_p(\mathrm{err}_N\eta^{-1})$, the second term of \eqref{oeq:bernstein}. Assembling \eqref{oeq:berndec} completes the proof.
\end{proof}

\begin{remark}\label{orem:fixedkappa}
\textit{Fixed-sample version.} Lemma~\ref{olem:kappahat} is stated for the Poissonized average $\widehat\kappa=N^{-1}\sum_{i\le\mathcal N}\widehat\Theta^2$. Conditionally on $\mathcal N=N$ this is the fixed-sample average $\widehat\kappa_2:=N^{-1}\sum_{i\le N}\widehat\Theta^2$, and since $\Pr(\mathcal N=N)\ge cN^{-1/2}$, an event of Poissonized probability $O(N^{-c_A'})$, for any $c_A'>0$, has fixed-sample probability $O(N^{-c_A'+1/2})$. The bounds of Steps~1--4 are either of this polynomial type with a free exponent (the union bound over the net, \eqref{oeq:pointwise}, holds with $c_A$ as large as desired) or are statements about fixed-sample i.i.d.\ averages that hold under the fixed-sample law directly (the Markov and law-of-large-numbers steps and the nuisance rates of Lemma~\ref{olem:consis}); hence \eqref{oeq:bernstein} holds for the fixed-sample $\widehat\kappa_2$ as well, and every statement about $\widehat\kappa$ below applies to it.
\end{remark}

\begin{proof}[Proof of Lemma~\ref{olem:centering}, part (iv)]
The comparison of $\widehat m^{\,\kappa}_2$ with $m^{\kappa}_{\mathcal N}(\mathcal B_1,\mathcal B_2)$ is split at the $X_2$ band. Off the band, $\kappa_{\mathcal N}\ge\{2\inf\varrho-o(1)\}\eta^{-1}$ by \eqref{oeq:kapadd}, so $|\sqrt{\widehat\kappa}-\sqrt{\kappa_{\mathcal N}}|\le|\widehat\kappa-\kappa_{\mathcal N}|/\sqrt{\kappa_{\mathcal N}}=O(\eta^{1/2})\cdot o_p(\eta^{-1/2})=o_p(1)$ uniformly by Lemma~\ref{olem:kappahat}, and the off-band integral contributes $o_p(1)$. On the band, where $\kappa_{\mathcal N}$ may vanish, the elementary $|\sqrt u-\sqrt v|\le\sqrt{|u-v|}$ and the band measure $O(h_2^{d_2})$ give a contribution $O(h_2^{d_2})\cdot o_p(\eta^{-1/4})=o_p(h_2^{d_2}\eta^{-1/4})=o_p(1)$, since $h_2^{d_2}\eta^{-1/4}=(h_2^{3d_2}/h_1^{d_1})^{1/4}=O(1)$ by Assumption~\ref{ass:reg}(h). Hence $|\widehat m^{\,\kappa}_2-m^{\kappa}_{\mathcal N}(\mathcal B_1,\mathcal B_2)|=o_p(1)$. The centering $\widehat m^{\,\kappa}_2$ is therefore valid for all $(d_1,d_2)$ and all bandwidth shapes compatible with Assumption~\ref{ass:reg}(g),(h), and is a valid default; \eqref{oeq:extra} is invoked only when the limit-functional form $\widehat m_2$ is preferred for interpretability.
\end{proof}

\textit{Extension to general $\Psi$.} The absolute value entered the proof of parts (i)--(ii) in exactly two places. In Step~1, positive homogeneity was used to write $\E[\Psi(\sqrt N\Gamma_{\mathcal N})]=\sqrt{\kappa_{\mathcal N}}\,\E[\Psi(S_{\mathcal N}+e)]$ (since $\sqrt{\kappa_{\mathcal N}}>0$), and the Lipschitz property was used to remove the drift, $|\E[\Psi(S_{\mathcal N}+e)]-\E[\Psi(S_{\mathcal N})]|\le|e|$. In Step~2, we replace the first-moment comparison \eqref{oeq:BE1} with \eqref{oeq:BEmeanPsi}, which yields the same rate with $m_\Psi$ in place of $\sqrt{2/\pi}$. Hence, for a configuration set $\mathcal C_2$ of finite measure, $\E[D^P_2]=m_\Psi\int_{\mathcal C_2}\sqrt{\kappa_{\mathcal N}(c)}\,dc+o(1)$, which is the centering $m^{\kappa}_{2,N}$ of the main text for general $\Psi$, and the feasible centering is $m_\Psi\int_{\mathcal C_2}\sqrt{\widehat\kappa}$. Parts (iii)--(iv) and Lemma~\ref{olem:kappahat} do not involve $\Psi$; the band is now $\{\|x_2-x_2'\|_\infty\le h_2\}\cap\mathcal C_2$, of measure $O(h_2^{d_2})$ as before. In particular $m_{\Psimax}=1/\sqrt{2\pi}$, and on the symmetric set the $\Psimax$ centering is one half of the $\Psiabs$ centering, as it must be. The symmetry $|-u|=|u|$ is not used.

\subsection{Routine bounds for the asymptotic variance}\label{osec:aux-var}

Define the $X_2$- and $X_1$-kernel autocorrelations
\[
\bar\rho_2(u_2):=\frac{\int K_2(\tilde z_2)K_2(\tilde z_2+u_2)\,d\tilde z_2}{\int K_2(\tilde z_2)^2d\tilde z_2},\qquad
\bar\rho_1(u_1):=\frac{\int K_1(\tilde z_1)K_1(\tilde z_1+u_1)\,d\tilde z_1}{\int K_1(\tilde z_1)^2d\tilde z_1},
\]
for $u_2\in[-1,1]^{d_2}$ and $u_1\in[-1,1]^{d_1}$, the scale factor
\begin{equation}\label{oeq:omega}
\kappa^*_2(x_1,x_2,x_2',x_2''):=\sqrt{\big[\varrho(x_1,x_2)+\varrho(x_1,x_2')\big]\big[\varrho(x_1,x_2)+\varrho(x_1,x_2'')\big]},
\end{equation}
and, as in equation~\eqref{oeq:pf-rhostar},
\begin{equation}\label{oeq:rho2star}
\rho^*_2(x_1,x_2,x_2',x_2'',u_2,u_1):=\frac{\varrho(x_1,x_2)}{\kappa^*_2(x_1,x_2,x_2',x_2'')}\;\bar\rho_2(u_2)\,\bar\rho_1(u_1).
\end{equation}
Section~\ref{osec:proofs} gives the inclusion--exclusion computation over the four overlap events that assembles the limit. This subsection restates the lemma and supplies the routine ingredients of that computation: the exact standardization identity, the removal of the drift, the Cauchy--Schwarz envelope that is uniform over the diagonal bands, the trimming of the dependent configuration set by measure, the limits of the second moments under the two changes of variables, and the Berry--Esseen comparison; the assembly is then recalled in a few lines.

\textit{Lemma~\ref{olem:variance-gen} (variance), stated in Section~\ref{osec:aux-moved}.} With
\begin{gather*}
\rho_N(c,c'):=\frac{\E[\Theta_j(W;c)\Theta_j(W;c')]}{\sqrt{\kappa_{j,N}(c)\kappa_{j,N}(c')}},\\
\Var(D^P_j)=\int_{\mathcal C_j}\int_{\mathcal C_j}g_\Psi\big(\rho_N(c,c')\big)\sqrt{\kappa_{j,N}(c)\kappa_{j,N}(c')}\,dc\,dc'+o(1),
\end{gather*}
and the double integral converges, its limit $\sigma^2_{j,\Psi}(\mathcal C_j)$ of \eqref{eq:sigma-gen} being $a_\Psi^2\sigma^2_{j,\mathrm{abs}}(\mathcal C_j)+b_\Psi^2\sigma^2_{j,\mathrm{lin}}(\mathcal C_j)$.

The following lemma is the explicit form for $\Dh_2$, that is, for $j=2$, $\Psiabs$ and $\Csym2\cap(\mathcal B_1\times \mathcal B_2^2)$, where $\sigma^2_{2,\mathcal B_1\times \mathcal B_2}=\tfrac14\sigma^2_{2,\mathrm{abs}}(\Csym2\cap(\mathcal B_1\times \mathcal B_2^2))$ because of the factor $\tfrac12$ in \eqref{oeq:D2}; its proof contains the display of Lemma~\ref{olem:variance-gen} as step (3b), and the general case is treated in the extension paragraph and in Section~\ref{osec:aux-moved}.

\begin{lemma}[Explicit form of Lemma~\ref{olem:variance-gen} for $\Dh_2$]\label{olem:variance}
Suppose Assumptions~\ref{ass:sampling}--\ref{ass:reg} hold, $\Gamma\equiv0$ on $\Csym2$, and $(\mathcal B_1,\mathcal B_2)$ satisfy \eqref{oeq:M1}--\eqref{oeq:M3}. Then, with $g=g_{\mathrm{abs}}$ of \eqref{oeq:gdef},
\begin{align}
\lim_{N\to\infty}\sigma^2_{\mathcal N}(\mathcal B_1,\mathcal B_2)&=\sigma^2_{2,\mathcal B_1\times \mathcal B_2},\label{oeq:sigma2}\\
\sigma^2_{2,\mathcal B_1\times \mathcal B_2}&:=\int_{\mathcal B_1}\int_{\mathcal B_2}\int_{\mathcal B_2}\int_{\mathcal B_2}\int_{[-1,1]^{d_2}}\int_{[-1,1]^{d_1}}
g\big(\rho^*_2(x_1,x_2,x_2',x_2'',u_2,u_1)\big)\notag\\
&\qquad\times\kappa^*_2(x_1,x_2,x_2',x_2'')
\;du_1\,du_2\,dx_2''\,dx_2'\,dx_2\,dx_1 .\notag
\end{align}
\end{lemma}

\begin{remark}\label{orem:variance-features}
Two features of the limit deserve notice before the proof. First, although the variance computation involves two $X_1$ arguments $x_1,x_1'$, dependence requires $\|x_1-x_1'\|_\infty\le h_1$, and the change of variables $x_1'=x_1+u_1h_1$ collapses the pair into one free $X_1$ argument and one local coordinate $u_1$, leaving a single integral over $x_1\in \mathcal B_1$. Second, the limiting correlation \eqref{oeq:rho2star} carries the product $\bar\rho_2(u_2)\bar\rho_1(u_1)$: the joint kernel being a product, its autocorrelation factorizes into an $X_2$ part and an $X_1$ part; $X_1$ separation beyond $h_1$ yields exactly zero correlation, while $X_1$ proximity within $h_1$ contributes the continuum of partial correlations $\bar\rho_1(u_1)$.
\end{remark}

\phantomsection\label{oprf:variance-gen}
\begin{proof}
Since the deterministic centering does not affect covariances, Fubini's theorem gives
\begin{equation}\label{oeq:varquad}
\sigma^2_{\mathcal N}(\mathcal B_1,\mathcal B_2)=\frac14\int_{(\mathcal B_1\times \mathcal B_2^2)^2}\Cov\Big(\sqrt N\big|\Gamma_{\mathcal N}(x_1,x_2,x_2')\big|,\;\sqrt N\big|\Gamma_{\mathcal N}(x_1',x_2'',x_2''')\big|\Big)\,dc\,dc' ,
\end{equation}
where $dc\,dc'$ abbreviates $dx_2\,dx_2'\,dx_1\,dx_2''\,dx_2'''\,dx_1'$.

\textit{Part 1 (reduction).} By \eqref{oeq:Sfield},
\[
\sqrt N\,\Gamma_{\mathcal N}(x_1,x_2,x_2')-\sqrt N\,\E[\Gamma_{\mathcal N}(x_1,x_2,x_2')]=\sqrt{\kappa_{\mathcal N}(x_1,x_2,x_2')}\,S_{\mathcal N}(x_1,x_2,x_2')
\]
with $\kappa_{\mathcal N}$ deterministic and nonnegative, so, exactly,
\begin{align}
\Cov\Big(\sqrt N\big|\Gamma_{\mathcal N}(x_1,x_2,x_2')\big|,\;&\sqrt N\big|\Gamma_{\mathcal N}(x_1',x_2'',x_2''')\big|\Big)\notag\\
&=\sqrt{\kappa_{\mathcal N}(x_1,x_2,x_2')\,\kappa_{\mathcal N}(x_1',x_2'',x_2''')}\notag\\
&\qquad\times\Cov\Big(\big|S_{\mathcal N}(x_1,x_2,x_2')+e\big|,\,\big|S_{\mathcal N}(x_1',x_2'',x_2''')+e'\big|\Big),\label{oeq:standardize}
\end{align}
where $e,e'$ collect the drifts as in Section~\ref{osec:aux-center}. Since $|\Cov(|Q+e|,|Q'+e'|)-\Cov(|Q|,|Q'|)|\le2|e|\,\E[|Q'|]+2|e'|\,\E[|Q|]+4|e||e'|$ for standardized $Q,Q'$, and the covariance vanishes on independent configurations for both the drifted and undrifted pairs, the drift correction acts only on the dependent-configuration set, whose measure is $O(\eta)$ (below); multiplied by the scale $O(\eta^{-1})$ and by $\sup(|e|+|e'|)=O(\sqrt N(h_2^{s_2}+h_1^{s_1}))\cdot O(\eta^{1/2})$ it contributes $o(1)$ to \eqref{oeq:varquad}, and the drifts are dropped hereafter.

By the window geometry, part (b) of Lemma~\ref{olem:poisson}, $\Gamma_{\mathcal N}(x_1,x_2,x_2')$ and $\Gamma_{\mathcal N}(x_1',x_2'',x_2''')$ are functionals of Poisson increments over $\mathcal W(x_1,x_2)\cup\mathcal W(x_1,x_2')$ and $\mathcal W(x_1',x_2'')\cup\mathcal W(x_1',x_2''')$ respectively; the covariance in \eqref{oeq:varquad} vanishes unless the two unions intersect, which requires
\begin{equation}\label{oeq:depcond}
\ind^{(1)}:=\ind\big(\|x_1-x_1'\|_\infty\le h_1\big)=1\qquad\text{and}\qquad n_{cc'}:=\ind_1+\ind_2+\ind_3+\ind_4\ge1,
\end{equation}
where $\ind_1:=\ind(\|x_2-x_2''\|_\infty\le h_2)$, $\ind_2:=\ind(\|x_2-x_2'''\|_\infty\le h_2)$, $\ind_3:=\ind(\|x_2'-x_2''\|_\infty\le h_2)$, $\ind_4:=\ind(\|x_2'-x_2'''\|_\infty\le h_2)$: $X_1$ overlap and at least one $X_2$ cross overlap. By the Cauchy--Schwarz inequality, $\Var(\sqrt N|\Gamma_{\mathcal N}|)\le N\E[\Gamma_{\mathcal N}^2]=\kappa_{\mathcal N}+c_{\mathcal N}^2=\kappa_{\mathcal N}+o(1)$, so, by the upper bound \eqref{oeq:kapupper},
\begin{equation}\label{oeq:csb}
\Big|\Cov\Big(\sqrt N\big|\Gamma_{\mathcal N}(x_1,x_2,x_2')\big|,\,\sqrt N\big|\Gamma_{\mathcal N}(x_1',x_2'',x_2''')\big|\Big)\Big|\le\eta^{-1}\big(4\sup\nolimits_{\mathcal B_1\times \mathcal B_2}\varrho+o(1)\big)
\end{equation}
uniformly on all of $(\mathcal B_1\times \mathcal B_2^2)^2$: only the upper bound in \eqref{oeq:kapunif} is used, so the vanishing of $\varrho_\Sigma$ on the $X_2$ bands is harmless. This is the analytical payoff of the unweighted form of \eqref{oeq:D2}.

Now trim the dependent set $\{\ind^{(1)}=1,n_{cc'}\ge1\}$. Let $\mathcal P$ denote the six $X_2$ pairs of a configuration, namely the four cross pairs $(x_2,x_2'')$, $(x_2,x_2''')$, $(x_2',x_2'')$, $(x_2',x_2''')$ indexing $\ind_1,\dots,\ind_4$, together with the two within-pair differences $(x_2,x_2')$ and $(x_2'',x_2''')$, and, for each $\iota\in\{1,\dots,4\}$, define the separation indicator
\begin{equation}\label{oeq:sep}
\ind^{\mathrm{sep}}_\iota:=\prod_{(z,z')\in\mathcal P\setminus\{\iota\text{-th cross pair}\}}\ind\big(\|z-z'\|_\infty>2h_2\big),
\end{equation}
which requires every $X_2$ pair other than the active one to be separated by more than $2h_2$ rather than merely by more than $h_2$ (in particular it forces $n_{cc'}=1$). The margin $2h_2$ makes the vanishing statements of Part~2 exact rather than asymptotic: a center displaced by $\|u_2h_2\|_\infty\le h_2$ from $x_2$ is still more than $h_2$ away from any point more than $2h_2$ from $x_2$. We discard three classes of configurations, and each is bounded by \eqref{oeq:csb} against its measure. Every configuration carries the $X_1$ constraint $\|x_1-x_1'\|_\infty\le h_1$, of $x_1'$-measure $(2h_1)^{d_1}$, and each active $X_2$ constraint pins one $X_2$ argument to a cube of side at most $4h_2$. (a) Configurations with $n_{cc'}\ge2$ carry at least two $X_2$ constraints on distinct $X_2$ pairs, hence (by the two-case count shown after \eqref{oeq:buckets} below: the two constraints either share an $X_2$ argument or involve disjoint pairs) have $X_2$ measure $O(h_2^{2d_2})$ and total measure $O(h_2^{2d_2}h_1^{d_1})$; against \eqref{oeq:csb} they contribute $O(h_2^{2d_2}h_1^{d_1}\eta^{-1})=O(h_2^{d_2})=o(1)$. (b) Configurations with $n_{cc'}=1$ in which a within-pair difference is small, $\|x_2-x_2'\|_\infty\le2h_2$ or $\|x_2''-x_2'''\|_\infty\le2h_2$, carry a second $X_2$ constraint and are bounded identically by $O(h_2^{d_2})$. (c) Configurations with $n_{cc'}=1$ in which some inactive cross pair lies in the annulus $(h_2,2h_2]$ likewise carry a second $X_2$ constraint and contribute $O(h_2^{d_2})$. Finally, the exact dependence indicator and the sum of the cross indicators differ only on $\{n_{cc'}\ge2\}$, $0\le\sum_{\iota=1}^{4}\ind_\iota-\ind(n_{cc'}\ge1)\le3\sum_{\iota<\iota'}\ind_\iota\ind_{\iota'}$, whose contribution is the $O(h_2^{d_2})$ of (a), so the indicator may be replaced by the sum. Collecting,
\begin{align}
\sigma^2_{\mathcal N}(\mathcal B_1,\mathcal B_2)&=\bar\sigma^2_{\mathcal N}+O(h_2^{d_2}),\qquad\bar\sigma^2_{\mathcal N}:=\sum_{\iota=1}^{4}\bar\sigma^2_{\mathcal N,\iota},\label{oeq:sigmabar}\\
\bar\sigma^2_{\mathcal N,\iota}&:=\frac14\int_{(\mathcal B_1\times \mathcal B_2^2)^2}\sqrt{\kappa_{\mathcal N}(x_1,x_2,x_2')\,\kappa_{\mathcal N}(x_1',x_2'',x_2''')}\notag\\
&\qquad\times\Cov\big(|S_{\mathcal N}(x_1,x_2,x_2')|,\,|S_{\mathcal N}(x_1',x_2'',x_2''')|\big)\,
\ind_\iota\,\ind^{(1)}\,\ind^{\mathrm{sep}}_\iota\;dc\,dc' .\notag
\end{align}
On the support of the trimming indicator all within-pair $X_2$ separations exceed $2h_2$, so \eqref{oeq:kapsqrt} applies to both scale factors and, uniformly there,
\begin{align}
\eta\,\sqrt{\kappa_{\mathcal N}(x_1,x_2,x_2')\,\kappa_{\mathcal N}(x_1',x_2'',x_2''')}
&=\sqrt{\big[\varrho(x_1,x_2)+\varrho(x_1,x_2')\big]\big[\varrho(x_1',x_2'')+\varrho(x_1',x_2''')\big]}\notag\\
&\quad+O\big(h_2^2+h_1^2\big).\label{oeq:scale}
\end{align}

\textit{Part 2 (limits of the second moments under the changes of variables).} Consider the term with $\ind_1$ active and apply the two changes of variables
\begin{equation}\label{oeq:cov1}
x_2''=x_2+u_2h_2,\quad dx_2''=h_2^{d_2}\,du_2;\qquad x_1'=x_1+u_1h_1,\quad dx_1'=h_1^{d_1}\,du_1,
\end{equation}
with $u_2\in[-1,1]^{d_2}$ and $u_1\in[-1,1]^{d_1}$,
which convert the $(x_1',x_2'')$-integrals into $(u_2,u_1)$-integrals and cancel the factor $\eta^{-1}$ implicit in $\sqrt{\kappa\kappa'}=\eta^{-1}\cdot\eta\sqrt{\kappa\kappa'}$, where $\kappa:=\kappa_{\mathcal N}(x_1,x_2,x_2')$ and $\kappa':=\kappa_{\mathcal N}(x_1',x_2'',x_2''')$ are the variances of the two configurations. The free coordinates $(x_2,x_2',x_2''')\in \mathcal B_2^3$ are pairwise separated by more than $2h_2$ (the content of the trimming indicator) and $x_1\in \mathcal B_1$ is free. We compute the limit of
\begin{align}
\rho^*_{\mathcal N}&:=\E\big[S_{\mathcal N}(x_1,x_2,x_2')\,S_{\mathcal N}(x_1+u_1h_1,x_2+u_2h_2,x_2''')\big]\notag\\
&=\frac{\eta\,\E\big[\Theta(W,x_1,x_2,x_2')\,\Theta(W,x_1+u_1h_1,x_2+u_2h_2,x_2''')\big]}{\sqrt{\eta\,\kappa_{\mathcal N}(x_1,x_2,x_2')}\,\sqrt{\eta\,\kappa_{\mathcal N}(x_1+u_1h_1,x_2+u_2h_2,x_2''')}}\label{oeq:rhostarN}
\end{align}
by treating the numerator and the two denominator factors in turn, using the exact moment identities of part (a) of Lemma~\ref{olem:poisson}.

\textit{(2a) A vanishing principle.} Because each $K_{2j}$ and each $K_{1l}$ vanishes outside $[-\tfrac12,\tfrac12]$, for any centers and any bounded measurable $\Xi(\cdot,\cdot)$,
\begin{equation}\label{oeq:disjoint}
\iint \Xi(z,p)\,K_2\Big(\tfrac{\bar z-z}{h_2}\Big)K_2\Big(\tfrac{\bar z'-z}{h_2}\Big)K_1\Big(\tfrac{\bar p-p}{h_1}\Big)K_1\Big(\tfrac{\bar p'-p}{h_1}\Big)dz\,dp=0
\end{equation}
whenever $\|\bar z-\bar z'\|_\infty>h_2$ or $\|\bar p-\bar p'\|_\infty>h_1$,
the integrand is supported on the intersection of the corresponding boxes. Equivalently, substituting $z=\bar z-h_2\tilde z_2$, $p=\bar p-h_1\tilde z_1$ turns the kernel product into $K_2(\tilde z_2)K_2(\tilde z_2+\tfrac{\bar z'-\bar z}{h_2})K_1(\tilde z_1)K_1(\tilde z_1+\tfrac{\bar p'-\bar p}{h_1})$, which vanishes pointwise once either shift exceeds $1$ in maximum norm.

\textit{(2b) The numerator.} By \eqref{oeq:kappa-center}, $\E[\Theta(W;c)\Theta(W;c')]=\E[A(W;c)A(W;c')]-B_{2,N}(c)B_{2,N}(c')$, and $\eta\,B_{2,N}(c)B_{2,N}(c')=o(h_1+h_2)$ is dropped; the remaining factor is $A(W,x_1,x_2,x_2')=\{\xi(W;x_1,x_2)K_2(\tfrac{x_2-X_2}{h_2})-\xi(W;x_1,x_2')K_2(\tfrac{x_2'-X_2}{h_2})\}\eta^{-1}K_1(\tfrac{x_1-X_1}{h_1})$, and $\Theta$ is written for $A$ in the rest of Part~2. Conditioning on $(X_1,X_2)$ and using the notation \eqref{oeq:qdef}, the product $\eta\,\E[\Theta(W,x_1,x_2,x_2')\Theta(W,x_1+u_1h_1,x_2+u_2h_2,x_2''')]$ expands into four terms indexed by the $X_2$ center pairs $\{x_2,x_2'\}\times\{x_2+u_2h_2,x_2'''\}$, each of the form
\begin{equation}\label{oeq:numfour}
\pm\frac{1}{\eta}\iint f(p,z)\,q(z,p;\cdot)\,K_2\Big(\tfrac{c_1-z}{h_2}\Big)K_2\Big(\tfrac{c_2-z}{h_2}\Big)K_1\Big(\tfrac{x_1-p}{h_1}\Big)K_1\Big(\tfrac{x_1+u_1h_1-p}{h_1}\Big)\,dz\,dp ,
\end{equation}
with $X_2$ centers $(c_1,c_2)\in\{x_2,x_2'\}\times\{x_2+u_2h_2,x_2'''\}$ and the common $X_1$ centers $x_1$, $x_1+u_1h_1$ ($\|u_1h_1\|_\infty\le h_1$, so the $X_1$ kernels always overlap). By \eqref{oeq:disjoint}, the three terms with $X_2$ center pairs $(x_2,x_2''')$, $(x_2',x_2''')$ and $(x_2',x_2+u_2h_2)$ vanish identically on the support of $\ind^{\mathrm{sep}}$: the first two because $\|x_2-x_2'''\|_\infty>2h_2>h_2$ and $\|x_2'-x_2'''\|_\infty>2h_2>h_2$; the third because $\|x_2'-(x_2+u_2h_2)\|_\infty\ge\|x_2'-x_2\|_\infty-\|u_2h_2\|_\infty>2h_2-h_2=h_2$, which is exactly the gain from the margin $2h_2$ in \eqref{oeq:sep}. In the surviving term, with $X_2$ centers $(x_2,x_2+u_2h_2)$ and sign $+$, substitute
\begin{equation}\label{oeq:cov2}
z=x_2-h_2\tilde z_2,\qquad p=x_1-h_1\tilde z_1,\qquad dz\,dp=\eta\,d\tilde z_2\,d\tilde z_1,
\end{equation}
so that $K_2(\tfrac{x_2-z}{h_2})=K_2(\tilde z_2)$, $K_2(\tfrac{x_2+u_2h_2-z}{h_2})=K_2(\tilde z_2+u_2)$, $K_1(\tfrac{x_1-p}{h_1})=K_1(\tilde z_1)$, $K_1(\tfrac{x_1+u_1h_1-p}{h_1})=K_1(\tilde z_1+u_1)$: the term is exactly the convolution integral of Lemma~\ref{olem:gine} with kernel family $H_{u_2,u_1}$ of \eqref{oeq:Hfamilies} and integrand $\Xi(z,p;\varpi)=f(p,z)\,q(z,p;\,x_1,x_2,x_1+u_1h_1,x_2+u_2h_2)$, the index $\varpi=(u_2,u_1,x_1,x_2,h_1,h_2)$ ranging over a set on which the hypotheses of that lemma hold uniformly by \eqref{oeq:qdef}--\eqref{oeq:qlip}. Lemma~\ref{olem:gine} therefore gives the value of the integral at the center, $f(x_1,x_2)q(x_2,x_1;\,x_1,x_2,x_1+u_1h_1,x_2+u_2h_2)\int K_2(\tilde z_2)K_2(\tilde z_2+u_2)d\tilde z_2\int K_1(\tilde z_1)K_1(\tilde z_1+u_1)d\tilde z_1$, up to $O(h_2+h_1)$ uniformly in $\varpi$; and \eqref{oeq:qlip}, applied with $\|u_2h_2\|_\infty\le h_2$ and $\|u_1h_1\|_\infty\le h_1$, replaces the evaluation points $(x_1+u_1h_1,x_2+u_2h_2)$ by $(x_1,x_2)$ at a further uniform cost $O(h_2+h_1)$. Hence, uniformly over the surviving configurations and over $(u_2,u_1)\in[-1,1]^{d_1+d_2}$,
\begin{align}
\eta\,\E\big[\Theta(W,x_1,x_2,x_2')\,&\Theta(W,x_1+u_1h_1,x_2+u_2h_2,x_2''')\big]\notag\\
&=f(x_1,x_2)\,\E\big[\xi(W;x_1,x_2)^2\,\big|\,X_1=x_1,X_2=x_2\big]\notag\\
&\qquad\times\int K_2(\tilde z_2)K_2(\tilde z_2+u_2)d\tilde z_2\int K_1(\tilde z_1)K_1(\tilde z_1+u_1)d\tilde z_1+O(h_2+h_1).\label{oeq:num}
\end{align}

\textit{(2c) The moment identity.} By the definition \eqref{oeq:theta} of $\xi$,
\[
\xi(W;x_1,x_2)=\frac{Y-\bar\mu_1(x_1,x_2)}{P_1(x_1,x_2)}\ \text{on }\{T=1\},\qquad
\xi(W;x_1,x_2)=-\frac{Y-\bar\mu_0(x_1,x_2)}{P_0(x_1,x_2)}\ \text{on }\{T=0\}.
\]
Decomposing the conditional second moment over the two events and using the joint-density convention \eqref{oeq:Pt},
\begin{align*}
\E\big[\xi(W;x_1,x_2)^2\,\big|\,X_1=x_1,X_2=x_2\big]&=\sum_{t\in\{0,1\}}\Pr(T=t\mid X_1=x_1,X_2=x_2)\,\frac{\varsigma_t^2(x_1,x_2)}{P_t(x_1,x_2)^2}\\
&=\frac{1}{f(x_1,x_2)}\sum_{t\in\{0,1\}}\frac{\varsigma_t^2(x_1,x_2)}{P_t(x_1,x_2)}\, ,
\end{align*}
whence
\begin{equation}\label{oeq:rho2id}
f(x_1,x_2)\,\E\big[\xi(W;x_1,x_2)^2\,\big|\,X_1=x_1,X_2=x_2\big]\int K_2(\tilde z_2)^2d\tilde z_2\int K_1(\tilde z_1)^2d\tilde z_1=\varrho(x_1,x_2),
\end{equation}
the definition \eqref{oeq:rhodef}. Dividing and multiplying \eqref{oeq:num} by $\int K_2^2\int K_1^2$ and using the definitions of $\bar\rho_2,\bar\rho_1$, the numerator of \eqref{oeq:rhostarN} equals $\varrho(x_1,x_2)\,\bar\rho_2(u_2)\,\bar\rho_1(u_1)+O(h_2+h_1)$ uniformly.

\textit{(2d) The first denominator factor.} Expanding $\Theta^2(W,x_1,x_2,x_2')$ and conditioning on $(X_1,X_2)$ produces two squared terms and a cross term; the cross term is identically zero for $h_2<\|x_2-x_2'\|_\infty$ by \eqref{oeq:disjoint} (the $X_1$ kernels coincide, but the $X_2$ windows are disjoint), which holds on the support of $\ind^{\mathrm{sep}}$. Each squared term is an integral of the type covered by Lemma~\ref{olem:gine} against the single kernel $H_*=K_2^2K_1^2$: the first centered at $(x_1,x_2)$, by the substitution \eqref{oeq:cov2}, and the second at the  of the box $\mathcal W(x_1,x_2')$ where its mass lies, by
\begin{equation}\label{oeq:cov3}
z=x_2'-h_2\tilde z_2',\qquad p=x_1-h_1\tilde z_1,\qquad K_2\Big(\tfrac{x_2'-z}{h_2}\Big)=K_2(\tilde z_2').
\end{equation}
In both terms the evaluation points coincide with the window center, so \eqref{oeq:qlip} is not called upon and the second part of Lemma~\ref{olem:gine} applies, $H_*$ being even; the two terms are $f(x_1,x_2)\E[\xi(W;x_1,x_2)^2\mid X_1=x_1,X_2=x_2]\int K_2^2\int K_1^2+O(h_2^2+h_1^2)$ and $f(x_1,x_2')\E[\xi(W;x_1,x_2')^2\mid X_1=x_1,X_2=x_2']\int K_2^2\int K_1^2+O(h_2^2+h_1^2)$. Combining with \eqref{oeq:rho2id} at $(x_1,x_2)$ and at $(x_1,x_2')$,
\begin{equation}\label{oeq:kap}
\eta\,\kappa_{\mathcal N}(x_1,x_2,x_2')=\varrho(x_1,x_2)+\varrho(x_1,x_2')+O(h_2^2+h_1^2)\qquad\text{uniformly over }\|x_2-x_2'\|_\infty>h_2 ,
\end{equation}
which is \eqref{oeq:kapadd}, established here in the form in which it was announced.

\textit{(2e) The second denominator factor.} The pair $(x_2+u_2h_2,x_2''')$ at $X_1$-argument $x_1+u_1h_1$ is treated identically. The cross term vanishes because, on the support of $\ind^{\mathrm{sep}}$, $\|(x_2+u_2h_2)-x_2'''\|_\infty\ge\|x_2-x_2'''\|_\infty-\|u_2h_2\|_\infty>h_2$; this is the second place where the margin $2h_2$ of \eqref{oeq:sep} is consumed. Applying Lemma~\ref{olem:gine} with $H_*$ centered at $(x_1+u_1h_1,x_2+u_2h_2)$ and at $(x_1+u_1h_1,x_2''')$ gives $\eta\kappa_{\mathcal N}(x_1+u_1h_1,x_2+u_2h_2,x_2''')=\varrho(x_1+u_1h_1,x_2+u_2h_2)+\varrho(x_1+u_1h_1,x_2''')+O(h_2^2+h_1^2)$; the arguments are then moved to $(x_1,x_2)$ and $(x_1,x_2''')$ using the continuous differentiability of $\varrho$ in both arguments (shown after \eqref{oeq:qlip}) together with $\|u_2h_2\|_\infty\le h_2$, $\|u_1h_1\|_\infty\le h_1$, at a cost $O(h_2+h_1)$ which now dominates:
\begin{equation}\label{oeq:kap2}
\eta\,\kappa_{\mathcal N}(x_1+u_1h_1,x_2+u_2h_2,x_2''')=\varrho(x_1,x_2)+\varrho(x_1,x_2''')+O(h_2+h_1)\qquad\text{uniformly.}
\end{equation}
The asymmetry between the rates in \eqref{oeq:kap} and \eqref{oeq:kap2} is not an oversight: \eqref{oeq:kap} is a statement about a window read at its own center, the form that the bias computation of Section~\ref{osec:aux-center} needs, whereas \eqref{oeq:kap2} involves a displacement of the evaluation points and cannot be better than $O(h_2+h_1)$. Only $o(1)$ is used in Part~3.

\textit{(2f) The limiting correlation, with a uniform rate.} The denominators \eqref{oeq:kap}--\eqref{oeq:kap2} are bounded below by $2\inf\varrho-O(h_2+h_1)>0$, so assembling (2b)--(2e) in \eqref{oeq:rhostarN},
\begin{equation}\label{oeq:rhostar}
\sup\big|\rho^*_{\mathcal N}-\rho^*_2(x_1,x_2,x_2',x_2''',u_2,u_1)\big|=O(h_2+h_1),
\end{equation}
the supremum over all surviving configurations and $(u_2,u_1)\in[-1,1]^{d_1+d_2}$; all constants below may therefore be chosen uniformly. Since $|\bar\rho_2(u_2)|\le1$ and $|\bar\rho_1(u_1)|\le1$ by the Cauchy--Schwarz inequality applied to each autocorrelation integral,
\begin{equation}\label{oeq:rhobar}
\sup|\rho^*_2|\le\frac{\sup_{\mathcal B_1\times \mathcal B_2}\varrho}{\sup_{\mathcal B_1\times \mathcal B_2}\varrho+\inf_{\mathcal B_1\times \mathcal B_2}\varrho}=:\bar\rho<1,
\end{equation}
and by \eqref{oeq:rhostar} there are $h_{2,0},h_{1,0},\delta>0$ with $\sup|\rho^*_{\mathcal N}|\le1-\delta$ for all $h_2<h_{2,0}$, $h_1<h_{1,0}$. The bound \eqref{oeq:rhobar} is the step at which the present argument parts company with the corresponding arguments for one-sided functionals of a single field, Lemma~6.4 of \citet{gine2003bm} or Lemma~B.10 of \citet{chang2015nonparametric}: there the standardized fields at a point and at its $h$-displacement have correlation exactly one at zero displacement, on a set of positive limiting mass, the Berry--Esseen pre-factor $(1-\rho^2)^{-3/2}$ blows up, and those proofs must regularize by an independent perturbation calibrated against the bandwidth. Here no such device is needed: the numerator of $\rho^*_2$ carries only the shared window's $\varrho$ while each denominator carries the sum of two, and $t\mapsto t/(t+\inf\varrho)$ is increasing; and the configurations at which the correlation would degenerate to one, those with $n_{cc'}\ge2$, were removed in Part~1 by \eqref{oeq:csb} against a measure $O(h_2^{2d_2}h_1^{d_1})$. The unweighted form of the statistic makes \eqref{oeq:csb} available uniformly, including on the $X_2$ bands where $\varrho_\Sigma$ degenerates; an inverse-standard-error weight would reinstate the difficulty.

\textit{(2g) The remaining three pairings.} Second term ($\ind_2$ active): substitute $x_2'''=x_2+u_2^{(2)}h_2$ and $x_1'=x_1+u_1h_1$; the surviving $X_2$ product carries the sign of $(+\xi(W;x_1,x_2)K_{2,x_2})\cdot(-\xi(W;x_1+u_1h_1,x_2+u_2^{(2)}h_2)K_{2,x_2+u_2^{(2)}h_2})$, a minus; the substitution centered at $(x_1,x_2)$ and (2c) give numerator $-\varrho(x_1,x_2)\bar\rho_2(u_2^{(2)})\bar\rho_1(u_1)+O(h_2+h_1)$, denominators $\varrho(x_1,x_2)+\varrho(x_1,x_2')+O(h_2+h_1)$ and $\varrho(x_1,x_2'')+\varrho(x_1,x_2)+O(h_2+h_1)$: correlation $-\rho^*_2(x_1,x_2,x_2',x_2'',u_2^{(2)},u_1)+O(h_2+h_1)$, scale limit $\sqrt{[\varrho(x_1,x_2)+\varrho(x_1,x_2')][\varrho(x_1,x_2)+\varrho(x_1,x_2'')]}$. Third term ($\ind_3$ active): substitute $x_2''=x_2'+u_2^{(3)}h_2$, $x_1'=x_1+u_1h_1$; the shared $X_2$ center is now $x_2'$; the surviving product $(-\xi(W;x_1,x_2')K_{2,x_2'})\cdot(+\xi(W;x_1+u_1h_1,x_2'+u_2^{(3)}h_2)K_{2,x_2'+u_2^{(3)}h_2})$ is again a minus; the substitution centered at $(x_1,x_2')$ gives numerator $-\varrho(x_1,x_2')\bar\rho_2(u_2^{(3)})\bar\rho_1(u_1)+O(h_2+h_1)$ by \eqref{oeq:rho2id} at $(x_1,x_2')$, with denominators $\varrho(x_1,x_2)+\varrho(x_1,x_2')+O(h_2+h_1)$ and $\varrho(x_1,x_2')+\varrho(x_1,x_2''')+O(h_2+h_1)$: correlation $-\rho^*_2(x_1,x_2',x_2,x_2''',u_2^{(3)},u_1)+O(h_2+h_1)$. Fourth term ($\ind_4$ active): substitute $x_2'''=x_2'+u_2^{(4)}h_2$, $x_1'=x_1+u_1h_1$; the product $(-\xi(W;x_1,x_2')K_{2,x_2'})\cdot(-\xi(W;x_1+u_1h_1,x_2'+u_2^{(4)}h_2)K_{2,x_2'+u_2^{(4)}h_2})$ is a plus: correlation $+\rho^*_2(x_1,x_2',x_2,x_2'',u_2^{(4)},u_1)+O(h_2+h_1)$. In every case the shared $X_2$ point's $\varrho$ occupies the numerator, the two free points the denominators, the factor $\bar\rho_1(u_1)$ is common to all four (the $X_1$ window is always the shared one), and the uniform rate and the bound \eqref{oeq:rhobar} hold verbatim.

\textit{Part 3 (Gaussian comparison and convergence).} For each configuration define
\[
(Z_{1\mathcal N},Z_{2\mathcal N}):=\big(\sqrt{1-\rho^{*2}_{\mathcal N}}\,Z_1+\rho^*_{\mathcal N}Z_2,\;Z_2\big),
\]
matching the mean, variance and correlation of the standardized pair; directly from the definition of $g$, $\Cov(|Z_{1\mathcal N}|,|Z_{2\mathcal N}|)=g(\rho^*_{\mathcal N})$ exactly. Three properties of $g$ are used: boundedness, $|g(\rho)|\le\{\Var(|Z_{1\mathcal N}|)\Var(|Z_{2\mathcal N}|)\}^{1/2}=1-2/\pi<1$ by the Cauchy--Schwarz inequality and $\Var(|Z|)=1-2/\pi$; evenness, $g(-\rho)=g(\rho)$, because replacing $Z_1$ by $-Z_1$ preserves the joint law while mapping the correlation to $-\rho$; and continuity on $[-1,1]$, from the closed form in \eqref{oeq:gpsi}. Define $\check\sigma^2_{\mathcal N}=\sum_{\iota=1}^4\check\sigma^2_{\mathcal N,\iota}$ by replacing, in each term $\bar\sigma^2_{\mathcal N,\iota}$ of \eqref{oeq:sigmabar}, the covariance of the standardized fields by $g(\rho^*_{\mathcal N})$.

\textit{(3a) Limit of the Gaussian version.} In the first term, after the changes of variables \eqref{oeq:cov1}, the integrand is $g(\rho^*_{\mathcal N})\cdot\eta\sqrt{\kappa\kappa'}\cdot\ind(x_2+u_2^{(1)}h_2\in \mathcal B_2)\ind(x_1+u_1h_1\in \mathcal B_1)$ over $(x_1,x_2,x_2',x_2''',u_2^{(1)},u_1)$; as $h_2,h_1\to0$ the correlation converges by \eqref{oeq:rhostar} and the continuity of $g$, the scale by \eqref{oeq:kap} and \eqref{oeq:kap2}, the boundary indicators to one a.e.\ (Lebesgue-null boundaries), and the integrand is bounded by $(1-2/\pi)(2\sup\varrho+o(1))$; by bounded convergence,
\begin{align}
\check\sigma^2_{\mathcal N,1}\longrightarrow\frac14\int_{\mathcal B_1}\int_{\mathcal B_2^3}\int_{[-1,1]^{d_2}}\int_{[-1,1]^{d_1}}
&g\big(\rho^*_2(x_1,x_2,x_2',x_2''',u_2^{(1)},u_1)\big)\,\kappa^*_2(x_1,x_2,x_2',x_2''')\notag\\
&\qquad\times du_1\,du_2^{(1)}\,dx_2'''\,dx_2'\,dx_2\,dx_1=:\tfrac14\sigma^2_\circ .\label{oeq:tau1}
\end{align}
The identical argument in the remaining three terms yields the integrals of
\begin{gather*}
g\big(-\rho^*_2(x_1,x_2,x_2',x_2'',u_2^{(2)},u_1)\big),\quad g\big(-\rho^*_2(x_1,x_2',x_2,x_2''',u_2^{(3)},u_1)\big)\\
\text{and}\quad g\big(\rho^*_2(x_1,x_2',x_2,x_2'',u_2^{(4)},u_1)\big)
\end{gather*}
against the corresponding scale limits. By the evenness of $g$ the signs are immaterial; renaming $x_2''\mapsto x_2'''$ in the second, interchanging $x_2\leftrightarrow x_2'$ in the third, and doing both in the fourth reproduces $\sigma^2_\circ/4$ each time. Hence $\check\sigma^2_{\mathcal N}\to4\cdot\sigma^2_\circ/4=\sigma^2_{2,\mathcal B_1\times \mathcal B_2}$. This is the inclusion--exclusion step of Section~\ref{osec:proofs}.

\textit{(3b) Vanishing of the difference.} The pair $(S_{\mathcal N},S'_{\mathcal N})$ of standardized fields entering each surviving configuration admits the joint block representation \eqref{oeq:blocksjoint}, with third moments bounded by \eqref{oeq:thirdmom}; and $|\rho^*_{\mathcal N}|\le1-\delta<1$ by (2f), so the covariance form \eqref{oeq:BEcov} of Fact~\ref{ofact:BE} applies and gives, for each such configuration,
\begin{equation}\label{oeq:BEapply}
\Big|\Cov\big(|S_{\mathcal N}|,|S'_{\mathcal N}|\big)-g\big(\rho^*_{\mathcal N}\big)\Big|\le\frac{C_3\big(1-\rho^{*2}_{\mathcal N}\big)^{-3/2}}{\sqrt N}\Big(\E[|\mathcal Q_{\mathcal N}|^3]+\E[|\mathcal Q'_{\mathcal N}|^3]\Big)=O\Big(\frac{1}{\sqrt{N\eta}}\Big)
\end{equation}
uniformly, the pre-factor $(1-\rho^{*2}_{\mathcal N})^{-3/2}\le(2\delta-\delta^2)^{-3/2}$ being a fixed constant by (2f) and the third moments being $O(\eta^{-1/2})$ by \eqref{oeq:thirdmom}. Since $\check\sigma^2_{\mathcal N}$ and $\bar\sigma^2_{\mathcal N}$ differ only in the replacement of the covariance by $g(\rho^*_{\mathcal N})$, in each of the four terms and after the changes of variables \eqref{oeq:cov1}, whose Jacobian $\eta$ cancels the $\eta^{-1}$ carried by $\sqrt{\kappa\kappa'}$, the integrand is $\eta\sqrt{\kappa\kappa'}$, which is $O(1)$ by \eqref{oeq:kap}--\eqref{oeq:kap2}, times $\Cov(|S_{\mathcal N}|,|S'_{\mathcal N}|)-g(\rho^*_{\mathcal N})$, which is $O((N\eta)^{-1/2})$ by \eqref{oeq:BEapply}, integrated over a domain of measure at most $2^{d_1+d_2}\lambda(\mathcal B_2)^3\lambda(\mathcal B_1)<\infty$. Hence $|\check\sigma^2_{\mathcal N}-\bar\sigma^2_{\mathcal N}|=O((N\eta)^{-1/2})=o(1)$. Combining with \eqref{oeq:sigmabar} and (3a) completes the proof.
\end{proof}

\textit{Extension to general $\Psi$.} Part~1 used positive homogeneity in \eqref{oeq:standardize}, which for $\Psi$ reads $\Cov(\Psi(\sqrt N\Gamma_{\mathcal N}),\Psi(\sqrt N\Gamma'_{\mathcal N}))=\sqrt{\kappa\kappa'}\Cov(\Psi(S_{\mathcal N}+e),\Psi(S'_{\mathcal N}+e'))$, with $\Gamma'_{\mathcal N}:=\Gamma_{\mathcal N}(c')$; the Lipschitz property and $\E[|\Psi(Q)|]\le\E[|Q|]\le1$ in the removal of the drift; and $|\Psi(u)|\le|u|$ in the Cauchy--Schwarz envelope, $\Var(\Psi(\sqrt N\Gamma_{\mathcal N}))\le N\E[\Gamma_{\mathcal N}^2]$, so that \eqref{oeq:csb} and the trimming by measure hold verbatim. Part~2 concerns the second moments of the fields and does not involve $\Psi$. In Part~3, boundedness and continuity of $g_\Psi$ follow from \eqref{oeq:gpsi} ($|g_\Psi|\le\Var(\Psi(Z))\le1$), and the comparison \eqref{oeq:BEapply} is replaced by \eqref{oeq:BEcovPsi}, with the same rate, so (3b) holds for the class and gives the display of Lemma~\ref{olem:variance-gen}. The only step that uses the symmetry of $|\cdot|$ is the merging of the four single-overlap terms in (3a), which rests on $g(-\rho)=g(\rho)$; for general $\Psi$, $g_\Psi(-\rho)=g_\Psi(\rho)-2b_\Psi^2\rho$, so the two pairings with a minus sign ($\iota=2,3$) carry $g_\Psi(-\rho^*_2)$ and the two with a plus sign ($\iota=1,4$) carry $g_\Psi(+\rho^*_2)$, each over its own domain. On the symmetric set $\Csym2$ the four domains coincide after relabeling and the linear terms cancel in pairs (consistently with $\int_{\Csym2}\Gamma_{\mathcal N}=0$), so the limit is $a_\Psi^2$ times the absolute limit; on an ordered set $\Cord2$ the domains differ and the four terms must be kept separate. In either case the limit is the one obtained by carrying the general covariance $g_\Psi(\rho_{\mathcal N}(c,c'))\sqrt{\kappa_{\mathcal N}(c)\kappa_{\mathcal N}(c')}$, with $\rho_{\mathcal N}(c,c'):=N\Cov(\Gamma_{\mathcal N}(c),\Gamma_{\mathcal N}(c'))/\sqrt{\kappa_{\mathcal N}(c)\kappa_{\mathcal N}(c')}$ the correlation of the Poissonized fields through Parts~1--3 without the merge, that is, the limit of $\int_{\mathcal C_2}\int_{\mathcal C_2}g_\Psi(\rho_{\mathcal N}(c,c'))\sqrt{\kappa_{\mathcal N}(c)\kappa_{\mathcal N}(c')}\,dc\,dc'$, which is the general form, equation~\eqref{eq:sigma-gen} of the main text, with the explicit absolute form \eqref{oeq:sigma2} as the special case. Since $g_\Psi(\pm\rho)=a_\Psi^2g_{\mathrm{abs}}(\rho)\pm b_\Psi^2\rho$, the general limit splits as $\sigma^2_{j,\Psi}(\mathcal C_j)=a_\Psi^2\sigma^2_{j,\mathrm{abs}}(\mathcal C_j)+b_\Psi^2\sigma^2_{j,\mathrm{lin}}(\mathcal C_j)$, the cross term vanishing because $\Cov(|S|,S')=0$ for centered jointly normal pairs, where $\sigma^2_{j,\mathrm{lin}}(\mathcal C_j)=\int_{\dom}w_{\mathcal C_j}(x)^2\sum_{t}\{\varsigma_t^2(x)/P_t(x)\}\,dx$ is the limit variance of the linear functional $\sqrt N\int_{\mathcal C_j}\{\Gammah_j-\Gamma_j\}$ and $w_{\mathcal C_j}$ the signed slot multiplicity (Theorem~\ref{thm:variance}(a) of the main text, with the computation of $\sigma^2_{j,\mathrm{lin}}$ in Section~\ref{osec:aux-moved}); hence $\sigma^2_{j,\Psi}(\mathcal C_j)\ge a_\Psi^2\sigma^2_{j,\mathrm{abs}}(\mathcal C_j)>0$ whenever $a_\Psi>0$, which is the positivity assumed in Theorem~\ref{thm:normality}.

\subsection{The projection: configuration bookkeeping and moment bounds}\label{osec:aux-proj}

The proof of the projection lemma is in Section~\ref{osec:proofs}. This subsection restates the lemma, documents the configuration bookkeeping behind the variance bound for the remainder, and derives the three consequences that the central limit theorem of Section~\ref{osec:aux-clt} consumes: the variance of the projection, a third-moment bound for the projected field, and an $L_2$-bound for its band slice.

\textit{Lemma~\ref{olem:pf-projection}.} With $\projfield_{\mathcal N}$ the two-slot projection field of \eqref{oeq:pf-Delta-proj} and $R_{\mathcal N}$ the remainder of \eqref{oeq:pf-R-rem}: (a) $D^P_j-\E[D^P_j]=\int_{\dom}\projfield_{\mathcal N}(x)\,dx+\int_{\mathcal C_j}R_{\mathcal N}(c)\,dc$; (b) for $c$ off the band, $\E[R_{\mathcal N}(c)\mid\mathcal G]=0$ for every $\sigma$-field $\mathcal G$ generated by Poisson increments over a region disjoint from $\mathcal W(c_2)$, and symmetrically with $c_1$; (c) $\Var(\int_{\mathcal C_j}R_{\mathcal N})=O(h_j^{d_j})$; (d) $\Var(\int_{\dom}\projfield_{\mathcal N})=\Var(D^P_j)+O(h_j^{d_j/2})$ and $\sup_{x}\E|\projfield_{\mathcal N}(x)|^3=O(\eta^{-3/2})$.

For $j=2$, $\Psiabs$ and $\Csym2\cap(\mathcal B_1\times \mathcal B_2^2)$ the two slot projections in \eqref{oeq:pf-Delta-proj} coincide after the relabeling $(x_2,x_2')\mapsto(x_2',x_2)$ (this is shown in the proof below), so that $\projfield_{\mathcal N}(x)=2\projfield^{\circ}_{\mathcal N}(x)$, where, for $x_1\in \mathcal B_1$ and $x_2\in \mathcal B_2$, the one-slot projection field is
\begin{multline}\label{oeq:Delta}
\projfield^{\circ}_{\mathcal N}(x_1,x_2):=\int_{\mathcal B_2}\Big(\E\big[\sqrt N\,|\Gamma_{\mathcal N}(x_1,x_2,x_2')|\,\big|\,\tau_{\mathcal N}(x_1,x_2)\big]\\
-\E\big[\sqrt N\,|\Gamma_{\mathcal N}(x_1,x_2,x_2')|\big]\Big)\,dx_2' ,
\end{multline}
where the conditional expectation is with respect to the $\sigma$-field generated by the scalar $\tau_{\mathcal N}(x_1,x_2)$;
and, for each $(x_1,x_2,x_2')$, the remainder of equation~\eqref{oeq:pf-R-rem},
\begin{align}
R_{\mathcal N}(x_1,x_2,x_2'):=\sqrt N\,|\Gamma_{\mathcal N}(x_1,x_2,x_2')|&-\E\big[\sqrt N\,|\Gamma_{\mathcal N}|\big]\notag\\
&-\Big(\E\big[\sqrt N\,|\Gamma_{\mathcal N}|\,\big|\,\tau_{\mathcal N}(x_1,x_2)\big]-\E\big[\sqrt N\,|\Gamma_{\mathcal N}|\big]\Big)\notag\\
&-\Big(\E\big[\sqrt N\,|\Gamma_{\mathcal N}|\,\big|\,\tau_{\mathcal N}(x_1,x_2')\big]-\E\big[\sqrt N\,|\Gamma_{\mathcal N}|\big]\Big),\label{oeq:R}
\end{align}
where $\Gamma_{\mathcal N}$ abbreviates $\Gamma_{\mathcal N}(x_1,x_2,x_2')$.
These conditional expectations are well defined as functions of the scalars $\tau_{\mathcal N}(x_1,x_2)$ and $\tau_{\mathcal N}(x_1,x_2')$: for $\|x_2-x_2'\|_\infty>h_2$ the two estimators are functionals of disjoint Poisson increments (at a common $X_1$, box disjointness is decided by the $X_2$ windows alone), hence independent, and each conditional expectation integrates out the other window's field; on the $X_2$ band $\|x_2-x_2'\|_\infty\le h_2$ the conditional expectations remain well defined (a conditional expectation given $\sigma(\tau_{\mathcal N}(x_1,x_2))$ exists whatever the dependence between the windows, and is a measurable function of that scalar), but the representation through an independent $x_2'$-field is lost; every statement below therefore specifies whether it holds on the band, and only crude bounds are used there.

With the factor $\tfrac12$ of \eqref{oeq:DP}, $\tfrac12\int_{\dom}\projfield_{\mathcal N}=\int_{\mathcal B_1}\int_{\mathcal B_2}\projfield^{\circ}_{\mathcal N}$ and $D^P_2=2D^P_{\mathcal N}(\mathcal B_1,\mathcal B_2)$, so that parts (a), (c) and (d) of Lemma~\ref{olem:pf-projection} for $\Dh_2$ read as follows, and part (b) is the degeneracy \eqref{oeq:degen} below.

\begin{lemma}[Lemma~\ref{olem:pf-projection} for $\Dh_2$]\label{olem:projection}
Under the conditions of Lemma~\ref{olem:variance},
\begin{align}
D^P_{\mathcal N}(\mathcal B_1,\mathcal B_2)-\E\big[D^P_{\mathcal N}(\mathcal B_1,\mathcal B_2)\big]&=\int_{\mathcal B_1}\int_{\mathcal B_2}\projfield^{\circ}_{\mathcal N}(x_1,x_2)\,dx_2\,dx_1\notag\\
&\quad+\frac12\int_{\mathcal B_1}\int_{\mathcal B_2}\int_{\mathcal B_2}R_{\mathcal N}(x_1,x_2,x_2')\,dx_2\,dx_2'\,dx_1 ,\label{oeq:decomp}
\end{align}
and
\begin{equation}\label{oeq:Rvar}
\Var\Big(\frac12\int_{\mathcal B_1}\int_{\mathcal B_2}\int_{\mathcal B_2}R_{\mathcal N}(x_1,x_2,x_2')\,dx_2\,dx_2'\,dx_1\Big)=O\big(h_2^{d_2}\big)=o(1).
\end{equation}
\end{lemma}

\begin{proof}
\textit{Decomposition.} By \eqref{oeq:R}, for each $(x_1,x_2,x_2')$ the centered integrand equals the sum of the two pointwise projections and $R_{\mathcal N}(x_1,x_2,x_2')$. Integrating $\tfrac12\int_{\mathcal B_1}\iint_{\mathcal B_2^2}(\cdot)$, the $x_2$-window projection gives $\tfrac12\int_{\mathcal B_1}\int_{\mathcal B_2}\projfield^{\circ}_{\mathcal N}$ by \eqref{oeq:Delta} and Fubini's theorem. For the $x_2'$-window projection, relabel the dummy variables $(x_2,x_2')\mapsto(x_2',x_2)$, a measure-preserving bijection of $\mathcal B_2\times \mathcal B_2$ at each fixed $x_1$, and use the antisymmetry $\Gamma_{\mathcal N}(x_1,x_2',x_2)=-\Gamma_{\mathcal N}(x_1,x_2,x_2')$, so that $|\Gamma_{\mathcal N}(x_1,x_2',x_2)|=|\Gamma_{\mathcal N}(x_1,x_2,x_2')|$ as random variables; the $x_2'$-window projection integrates to the same $\tfrac12\int_{\mathcal B_1}\int_{\mathcal B_2}\projfield^{\circ}_{\mathcal N}$. The two projection integrals, although they project onto different windows pointwise, are therefore the same integral after integration, and their sum is $\int_{\mathcal B_1}\int_{\mathcal B_2}\projfield^{\circ}_{\mathcal N}$: the statistic's $\tfrac12$ is consumed exactly by the merge, the same $4\cdot\tfrac14$ bookkeeping as in the proof of Lemma~\ref{olem:variance}, which is why \eqref{oeq:projvar} below reproduces $\sigma^2_{2,\mathcal B_1\times \mathcal B_2}$ with no factor discrepancy. The relabeling also identifies the two slot integrals of \eqref{oeq:pf-Delta-proj}, giving $\projfield_{\mathcal N}=2\projfield^{\circ}_{\mathcal N}$ as claimed; for general $\Psi$ the two slot integrals are kept separate in \eqref{oeq:pf-Delta-proj} and no relabeling is needed.

\textit{Degeneracy.} Fix $x_1$ and let $x_2,x_2'$ with $\|x_2-x_2'\|_\infty>h_2$. Conditionally on any $\sigma$-field generated by Poisson increments over a region disjoint from $\mathcal W(x_1,x_2')$ (in particular, on the increments over $\mathcal W(x_1,x_2)$, or over any union of boxes disjoint from $\mathcal W(x_1,x_2')$),
\begin{equation}\label{oeq:degen}
\E\big[R_{\mathcal N}(x_1,x_2,x_2')\,\big|\,\text{increments independent of the $(x_1,x_2')$-box}\big]=0 ,
\end{equation}
because, given such a $\sigma$-field, the conditional expectation of $\sqrt N|\Gamma_{\mathcal N}|$ reduces to the conditional expectation given $\tau_{\mathcal N}(x_1,x_2)$ (the $(x_1,x_2')$-field being independent of the conditioning), the $x_2'$-projection is independent of the conditioning and integrates to its mean, and the four terms of \eqref{oeq:R} cancel. The symmetric statement holds with $x_2$ and $x_2'$ interchanged. The hypothesis $\|x_2-x_2'\|_\infty>h_2$ is essential: on the $X_2$ band the two boxes share increments, the reduction fails, and no degeneracy is claimed or used there. The unconditional mean-zero property holds everywhere, band included: by the tower property, $\E[R_{\mathcal N}(x_1,x_2,x_2')]=0$ for every $(x_1,x_2,x_2')$.

\textit{Variance of the remainder.} Expand \eqref{oeq:Rvar} into $\tfrac14\int\E[R_{\mathcal N}(x_1,x_2,x_2')R_{\mathcal N}(x_1',x_2'',x_2''')]$ over $(\mathcal B_1\times \mathcal B_2^2)^2$. We first show a moment bound valid on all of $\mathcal B_1\times \mathcal B_2^2$, band included: by the triangle inequality in $L_2$ and the conditional Jensen inequality,
\begin{equation}\label{oeq:RL2}
\big\|R_{\mathcal N}(x_1,x_2,x_2')\big\|_2\le3\,\Big\{\Var\big(\sqrt N|\Gamma_{\mathcal N}(x_1,x_2,x_2')|\big)\Big\}^{1/2}\le3\,\big\{\kappa_{\mathcal N}+o(1)\big\}^{1/2}=O\big(\eta^{-1/2}\big),
\end{equation}
using only the upper bound \eqref{oeq:kapupper} and \eqref{oeq:driftrate}, uniform over the band. Now classify the configurations $(x_1,x_2,x_2',x_1',x_2'',x_2''')$. With the indicators $\ind^{(1)}$, $\ind_1,\dots,\ind_4$ of \eqref{oeq:depcond}, $n_{cc'}$ their $X_2$ count, and the own-pair band indicators $\ind_{x_2x_2'}:=\ind(\|x_2-x_2'\|_\infty\le h_2)$, $\ind_{x_2''x_2'''}:=\ind(\|x_2''-x_2'''\|_\infty\le h_2)$, define
\begin{align}
\mathrm{(I)}&:=\{\ind^{(1)}=0\}\;\cup\;\big(\{\ind^{(1)}=1\}\cap\{n_{cc'}=0\}\big),\notag\\
\mathrm{(II)}&:=\{\ind^{(1)}=1\}\cap\{n_{cc'}=1\}\cap\{\ind_{x_2x_2'}=\ind_{x_2''x_2'''}=0\},\notag\\
\mathrm{(III)}&:=\{\ind^{(1)}=1\}\cap\Big(\{n_{cc'}\ge2\}\,\cup\,\big(\{n_{cc'}=1\}\cap\{\ind_{x_2x_2'}+\ind_{x_2''x_2'''}\ge1\}\big)\Big).\label{oeq:buckets}
\end{align}
These three sets partition the configuration space. The routing can be pictured as a decision tree whose first gate is the $X_1$ overlap: $X_1$-disjoint configurations ($\|x_1-x_1'\|_\infty>h_1$) go to (I) whatever their $X_2$ geometry; below the gate the $X_2$ count $n_{cc'}$ takes exactly one value, $\{n_{cc'}=0\}$ joins (I), $\{n_{cc'}\ge2\}$ joins (III), and $\{n_{cc'}=1\}$ is split by the own-band question into (II), when both own pairs are off their bands, and the rest of (III). Nothing falls through and nothing is counted twice; (II) is the only terminal with a hypothesis to protect, the off-band requirement of \eqref{oeq:degen}.

The $X_2$-measure count used for (III), and for the discarded classes (a)--(c) of Part~1 of the proof of Lemma~\ref{olem:variance}, is the following. A configuration in (III) carries at least two active constraints among the six $X_2$ indicators $\ind_1,\dots,\ind_4,\ind_{x_2x_2'},\ind_{x_2''x_2'''}$, each pinning a distinct $X_2$ pair into a cube of side $2h_2$ (side $4h_2$ for the trimming classes of Lemma~\ref{olem:variance}, which changes only the constant). Two distinct constrained pairs cover either three of the four $X_2$ coordinates, when they share a coordinate, or all four, when they are disjoint; in the first case one coordinate is free, the shared one is free, and the two others are pinned to it, giving $X_2$-measure $\lambda(\mathcal B_2)^2(2h_2)^{2d_2}$; in the second case each pair has one free and one pinned coordinate, giving again $\lambda(\mathcal B_2)^2(2h_2)^{2d_2}$. Either way exactly two coordinates are pinned and the $X_2$-measure is $O(h_2^{2d_2})$; the $X_1$ constraint $\|x_1-x_1'\|_\infty\le h_1$ multiplies this by $(2h_1)^{d_1}$, and the integrand bound $O(\eta^{-1})=O(h_2^{-d_2}h_1^{-d_1})$ then leaves $O(h_2^{d_2})\to0$.

(I): on $\{\ind^{(1)}=0\}$ the two box-unions are disjoint outright (every cross box pair is $X_1$-separated); on $\{\ind^{(1)}=1,n_{cc'}=0\}$ they are disjoint because every cross pair is $X_2$-separated. In both cases the two remainders are functionals of increments over disjoint regions, hence independent, and $\E[RR']=\E[R]\E[R']=0$ by the band-valid mean-zero property, with no restriction on the geometry within each pair.

(II): by relabeling it suffices to treat $\ind_1=1$; the definition of (II) makes the five remaining $X_2$ indicators zero, and each is used exactly once. Let $\mathcal G$ be the $\sigma$-field of the increments over $\mathcal W(x_1,x_2)\cup\mathcal W(x_1',x_2'')$. Given $\mathcal G$, the residual randomness of $R_{\mathcal N}(x_1,x_2,x_2')$ is the increments over $\mathcal W(x_1,x_2')$, disjoint from $\mathcal W(x_1,x_2)$ (by $\ind_{x_2x_2'}=0$) and from $\mathcal W(x_1',x_2'')$ (by $\ind_3=0$); symmetrically for $R_{\mathcal N}(x_1',x_2'',x_2''')$ and $\mathcal W(x_1',x_2''')$ (by $\ind_2=0$, $\ind_{x_2''x_2'''}=0$). The two residual boxes are disjoint (by $\ind_4=0$), so the remainders are conditionally independent given $\mathcal G$. Since $\mathcal G$ is generated by increments independent of the $(x_1,x_2')$-box and the own pair is off its band, \eqref{oeq:degen} gives $\E[R\mid\mathcal G]=0$, and symmetrically $\E[R'\mid\mathcal G]=0$; this is where the off-band restriction of (II) is consumed. Hence $\E[RR']=\E[\E[R\mid\mathcal G]\E[R'\mid\mathcal G]]=0$ exactly.

(III): by the measure count above the set has total measure $O(h_2^{2d_2}h_1^{d_1})$; against $|\E[RR']|\le\|R\|_2\|R'\|_2=O(\eta^{-1})$ from \eqref{oeq:RL2}, band-valid exactly where the degeneracy is not, it contributes $O(h_2^{2d_2}h_1^{d_1}\eta^{-1})=O(h_2^{d_2})$. Since (I)--(III) exhaust the configuration space, only (III) contributes, proving \eqref{oeq:Rvar}.
\end{proof}

\begin{remark}\label{orem:whydegen}
The degeneracy \eqref{oeq:degen} is the load-bearing step: the set (II) activates only one $X_2$ indicator, so its measure is of order $h_2^{d_2}h_1^{d_1}$, and the crude bound would give $O(h_2^{d_2}h_1^{d_1})\cdot O(\eta^{-1})=O(1)$, bounded but not vanishing, and exactly the order of $\Var(D^P_{\mathcal N}(\mathcal B_1,\mathcal B_2))$, which by Part~1 of the proof of Lemma~\ref{olem:variance} is itself carried by the one-shared-window configurations and converges to $\sigma^2_{2,\mathcal B_1\times \mathcal B_2}>0$. The content of the H\'ajek-type projection is that the projections absorb all of this first-order dependence; \eqref{oeq:degen} is the exact statement of the absorption, and converts (II) from $O(1)$ to zero.
\end{remark}

Three consequences are shown for use in Section~\ref{osec:aux-clt}. First, the variance of the projection. By \eqref{oeq:decomp}, the projection equals the centered statistic minus the remainder; expanding the variance of the difference and bounding the cross term by the Cauchy--Schwarz inequality with Lemma~\ref{olem:variance} and \eqref{oeq:Rvar},
\begin{equation}\label{oeq:projvar}
\Var\Big(\int_{\mathcal B_1}\int_{\mathcal B_2}\projfield^{\circ}_{\mathcal N}(x_1,x_2)\,dx_2\,dx_1\Big)=\sigma^2_{\mathcal N}(\mathcal B_1,\mathcal B_2)+O\big(h_2^{d_2/2}\big)\;\longrightarrow\;\sigma^2_{2,\mathcal B_1\times \mathcal B_2}>0 ,
\end{equation}
positivity holding because $g\ge0$ with $g(\rho^*_2)>0$ on a neighborhood of $(u_2,u_1)=0$ and the scale factors are bounded below by $2\inf\varrho>0$. No orthogonality between projection and remainder is asserted or needed for \eqref{oeq:projvar}; in fact the classical orthogonality of the Hoeffding decomposition holds here exactly off the doubly-constrained sets: $\E[\projfield^{\circ}_{\mathcal N}(x_1'',y)R_{\mathcal N}(x_1,x_2,x_2')]=0$ when the $(x_1'',y)$-box is disjoint from both pair boxes (independence and mean zero), and when it meets exactly one of them while $\|x_2-x_2'\|_\infty>h_2$ (condition on the union increments; $\projfield^{\circ}_{\mathcal N}(x_1'',y)$ is a function of $\tau_{\mathcal N}(x_1'',y)$ and $\E[R_{\mathcal N}\mid\cdot]=0$ by \eqref{oeq:degen}), so that the covariance between projection and remainder is $O(h_2^{d_2})$; only the cruder Cauchy--Schwarz $O(h_2^{d_2/2})$ is used above.

Second, a moment bound for $\projfield^{\circ}_{\mathcal N}(x_1,x_2)$, obtained by splitting the inner $x_2'$-integral of \eqref{oeq:Delta} into the off-band slice $\{\|x_2'-x_2\|_\infty>h_2\}$ and the band slice $\{\|x_2'-x_2\|_\infty\le h_2\}$. On the off-band slice the $(x_1,x_2')$-box is independent of $\tau_{\mathcal N}(x_1,x_2)$, so the conditional expectation is the deterministic $1$-Lipschitz function $t\mapsto\E[|t-\sqrt N(\tau_{\mathcal N}(x_1,x_2')-\E[\tau_{\mathcal N}(x_1,x_2')])+\sqrt N\E[\Gamma_{\mathcal N}]|]$ of $t=\sqrt N(\tau_{\mathcal N}(x_1,x_2)-\E[\tau_{\mathcal N}(x_1,x_2)])$; integrating over the slice, of measure at most $\lambda(\mathcal B_2)$,
\begin{align}
\bigg|\int_{\{\|x_2'-x_2\|_\infty>h_2\}}\big(\E[\,\cdot\mid\tau_{\mathcal N}(x_1,x_2)]&-\E[\,\cdot\,]\big)dx_2'\bigg|\notag\\
&\le\lambda(\mathcal B_2)\Big(\big|\sqrt N\big(\tau_{\mathcal N}(x_1,x_2)-\E[\tau_{\mathcal N}(x_1,x_2)]\big)\big|+\E\big[|\cdots|\big]\Big),\label{oeq:lip}
\end{align}
of $L_3$-norm $O(\eta^{-1/2})$ by \eqref{oeq:thirdmom}. On the band slice, Minkowski's integral inequality and conditional Jensen give
\begin{align}
\bigg\|\int_{\{\|x_2'-x_2\|_\infty\le h_2\}}\big(\E[\,\cdot\mid\tau_{\mathcal N}(x_1,x_2)]&-\E[\,\cdot\,]\big)dx_2'\bigg\|_3\notag\\
&\le(2h_2)^{d_2}\sup_{\mathcal B_1\times \mathcal B_2^2}\big\|\sqrt N|\Gamma_{\mathcal N}|-\E[\cdot]\big\|_3=O\big(h_2^{d_2}\eta^{-1/2}\big),\label{oeq:lipband}
\end{align}
the supremum being $O(\eta^{-1/2})$ uniformly over the band by the additive triangle bound $\|\sqrt N\Gamma_{\mathcal N}\|_3\le\sqrt N\|\tau_{\mathcal N}(x_1,x_2)-\E[\cdot]\|_3+\sqrt N\|\tau_{\mathcal N}(x_1,x_2')-\E[\cdot]\|_3+|c_{\mathcal N}|$ and \eqref{oeq:thirdmom}. The band slice is dominated by the off-band slice (the ratio of \eqref{oeq:lipband} to $O(\eta^{-1/2})$ is $O(h_2^{d_2})$, with no anisotropy condition), and Minkowski gives
\begin{equation}\label{oeq:delta3}
\sup_{(x_1,x_2)\in \mathcal B_1\times \mathcal B_2}\E\big[|\projfield^{\circ}_{\mathcal N}(x_1,x_2)|^3\big]=O\big(\eta^{-3/2}\big).
\end{equation}

Third, for the covariance step of Section~\ref{osec:aux-clt}, the band slice's $L_2$-norm is shown: by the same conditional-Jensen argument in $L_2$, it is $O(h_2^{d_2}\eta^{-1/2})$, uniformly over $(x_1,x_2)$.

\textit{Extension to general $\Psi$.} The degeneracy \eqref{oeq:degen} and the classification (I)--(III) use only the independence structure of the Poisson increments and the tower property, not $\Psi$. The $L_2$-bound \eqref{oeq:RL2} uses $\Var(\Psi(\sqrt N\Gamma_{\mathcal N}))\le\E[\Psi(\sqrt N\Gamma_{\mathcal N})^2]\le N\E[\Gamma_{\mathcal N}^2]$, i.e.\ $|\Psi(u)|\le|u|$; the moment bound \eqref{oeq:lip} uses that $t\mapsto\E[\Psi(t-\tilde\tau)]$ is $1$-Lipschitz for any random $\tilde\tau$ independent of $t$, i.e.\ the Lipschitz property; and \eqref{oeq:lipband}, \eqref{oeq:delta3} then follow as before. The one step that uses the symmetry $|-u|=|u|$ is the merging of the two slot projections into the one-slot field $\projfield^{\circ}_{\mathcal N}$: it relies on $\Psiabs(\Gamma_{\mathcal N}(x_1,x_2',x_2))=\Psiabs(\Gamma_{\mathcal N}(x_1,x_2,x_2'))$ together with the symmetry of $\mathcal B_2\times \mathcal B_2$ under exchange of the two $X_2$ arguments. Section~\ref{osec:proofs} avoids the merge: its projection field \eqref{oeq:pf-Delta-proj} integrates over both fibres $\{c_1=x\}$ and $\{c_2=x\}$, so that $D^P_j-\E[D^P_j]=\int_{\dom}\projfield_{\mathcal N}+\int_{\mathcal C_j}R_{\mathcal N}$ holds exactly for every $\Psi$ and every configuration set by Fubini's theorem slot by slot, with no factor bookkeeping; for $\Psiabs$ on $\Csym2$ the two fibre integrals coincide and $\projfield_{\mathcal N}=2\projfield^{\circ}_{\mathcal N}$, which with the factor $\tfrac12$ of $\Dh_2$ recovers \eqref{oeq:decomp}. With the two-slot field every statement of this subsection holds for the class: the degeneracy and (I)--(III) are unchanged, \eqref{oeq:Rvar} follows with the general variance limit, \eqref{oeq:projvar} reads $\Var(\int_{\dom}\projfield_{\mathcal N})=\Var(D^P_j)+O(h_j^{d_j/2})$, and the moment bounds \eqref{oeq:lip}--\eqref{oeq:delta3} hold for each fibre integral separately, the band in \eqref{oeq:lipband} being $\{\|x_2'-x_2\|_\infty\le h_2\}$ intersected with the relevant fibre of $\mathcal C_2$.

\subsection{The central limit theorem for the projected field}\label{osec:aux-clt}

Let $\mathcal N(A)$ denote the number of Poisson points $(X_{1i},X_{2i})$, $i\le\mathcal N$, falling in a Borel set $A\subset\mathbb R^{d_1+d_2}$, and split the sample count at the box $\mathcal B(M)$ of Section~\ref{osec:aux-sets}:
\begin{gather}
\mathcal S_{\mathcal N}:=\frac{D^P_{\mathcal N}(\mathcal B_1,\mathcal B_2)-\E\big[D^P_{\mathcal N}(\mathcal B_1,\mathcal B_2)\big]}{\sigma_{\mathcal N}(\mathcal B_1,\mathcal B_2)},\label{oeq:counts}\\
U_{\mathcal N}:=\frac{\mathcal N\big(\mathcal B(M)\big)-N\mu\big(\mathcal B(M)\big)}{\sqrt N},\qquad
\Uext:=\frac{\mathcal N\big(\mathcal B(M)^c\big)-N\mu\big(\mathcal B(M)^c\big)}{\sqrt N},\notag
\end{gather}
where $\mu(A):=\Pr((X_1,X_2)\in A)$, so that $U_{\mathcal N}+\Uext=(\mathcal N-N)/\sqrt N$. By \eqref{oeq:M2}, $\mu(\mathcal B(M)^c)=\pext\in(0,1)$: positivity is \eqref{oeq:M2} itself, and $\pext<1$ because $\mathcal B(M)$ has nonempty interior on which $f\ge\underline p$. Consequently $\mathcal N(\mathcal B(M))$ and $\mathcal N(\mathcal B(M)^c)$ are independent Poisson variables with means $N(1-\pext)$ and $N\pext$ (increments of a Poisson process over disjoint sets), and
\begin{equation}\label{oeq:countvar}
\Var(U_{\mathcal N})=1-\pext,\qquad\Var(\Uext)=\pext,\qquad\text{both exactly.}
\end{equation}
This is the two-block form required by the de-Poissonization principle of Section~\ref{osec:aux-depoisson}: it is the interior count $U_{\mathcal N}$ that affects the central limit theorem, while the exterior count $\Uext$ affects it only through the exact independence shown in Lemma~\ref{olem:clt}(ii).

\textit{Lemma~\ref{olem:pf-clt}.} Let $\mathcal B\supset\dom$ be a box containing every window centered in $\dom$, $\pext:=\Pr((X_1,X_2)\notin\mathcal B)\in(0,1)$; such a box with $\pext>0$ exists only if the support of $X$ extends beyond $\dom$ enlarged by half a window, and otherwise the lemma is applied, as in Section~\ref{osec:aux-sets}, on compact product sets whose windows lie in a box inside the support, with the extension to $\dom$ in Section~\ref{osec:aux-depoisson}, $U_{\mathcal N}:=\{\mathcal N(\mathcal B)-N(1-\pext)\}/\sqrt N$ the standardized interior count and $\Uext$ the exterior count. If $\sigma^2_{j,\Psi}(\mathcal C_j)>0$ then, with $\Qproj:=\int_{\dom}\projfield_{\mathcal N}/\sigma_{j,\Psi}(\mathcal C_j)$, (a) $(\Qproj,U_{\mathcal N})\dto N(0,\mathrm{diag}(1,1-\pext))$ and (b) $(\Qproj,U_{\mathcal N})$ is independent of $\Uext$ for every $N$.

The lemma is proved on the sets $\mathcal B_1\times \mathcal B_2$ of Section~\ref{osec:aux-sets}, with the box $\mathcal B(M)$ in the role of $\mathcal B$ and the exact standard deviation $\sigma_{\mathcal N}(\mathcal B_1,\mathcal B_2)$ in the role of the limit $\sigma_{j,\Psi}(\mathcal C_j)$; since $\sigma_{\mathcal N}(\mathcal B_1,\mathcal B_2)\to\sigma_{2,\mathcal B_1\times \mathcal B_2}>0$ by Lemma~\ref{olem:variance}, the two normalizations are interchangeable by Slutsky's theorem, and the passage from $\mathcal B_1\times \mathcal B_2$ to $\dom$ is in Section~\ref{osec:aux-depoisson}. For $\Dh_2$, $\Qproj=\int_{\mathcal B_1}\int_{\mathcal B_2}\projfield^{\circ}_{\mathcal N}/\sigma_{\mathcal N}(\mathcal B_1,\mathcal B_2)$, the factors $2$ in $\projfield_{\mathcal N}=2\projfield^{\circ}_{\mathcal N}$ and in $D^P_2=2D^P_{\mathcal N}(\mathcal B_1,\mathcal B_2)$ cancelling.

\begin{lemma}[Lemma~\ref{olem:pf-clt} for $\Dh_2$ on $\mathcal B_1\times \mathcal B_2$]\label{olem:clt}
Under the conditions of Lemma~\ref{olem:variance}:
(i) $(\mathcal S_{\mathcal N},U_{\mathcal N})\dto\big(Z_1,\sqrt{1-\pext}\,Z_2\big)$ with $Z_1,Z_2$ independent standard normals, and the same holds with $\Qproj$ in place of $\mathcal S_{\mathcal N}$;
(ii) for each $N$, the vector $(\mathcal S_{\mathcal N},U_{\mathcal N})$ is independent of $\Uext$.
\end{lemma}

\phantomsection\label{oprf:pf-clt}
\begin{proof}
\textit{Part (ii).} By \eqref{oeq:M1} and $h_2\vee h_1<h_0$, every window $\mathcal W(x_1,x_2)$ with $(x_1,x_2)\in \mathcal B_1\times \mathcal B_2$ is contained in $\mathcal B(M)$; hence $D^P_{\mathcal N}(\mathcal B_1,\mathcal B_2)$, being, by the locality of the Poissonized field \eqref{oeq:poisson-field}, whose summands vanish outside the windows and whose deterministic part does not involve the count, a functional of the Poisson increments over $\bigcup_{(x_1,x_2)\in \mathcal B_1\times \mathcal B_2}\mathcal W(x_1,x_2)\subset\mathcal B(M)$, and $U_{\mathcal N}$, a functional of the increments over $\mathcal B(M)$, are both measurable with respect to the increments over $\mathcal B(M)$, while $\Uext$ is a functional of the increments over the disjoint set $\mathcal B(M)^c$. Independence of Poisson increments over disjoint sets gives (ii). This is the exact role of the contraction \eqref{oeq:M1}: it buys the room the kernel windows need for the statistic to live strictly inside $\mathcal B(M)$, and is the counterpart of the argument on p.~756 of \citet{gine2003bm}.

\textit{Part (i).} By \eqref{oeq:decomp}--\eqref{oeq:Rvar} and Chebyshev's inequality, the remainder contributes $o_p(1)$ after division by $\sigma_{\mathcal N}(\mathcal B_1,\mathcal B_2)$, which is bounded away from zero for large $N$ by \eqref{oeq:projvar}; by Slutsky's theorem it suffices to prove the joint CLT for
\[
\Qproj:=\frac{1}{\sigma_{\mathcal N}(\mathcal B_1,\mathcal B_2)}\int_{\mathcal B_1}\int_{\mathcal B_2}\projfield^{\circ}_{\mathcal N}(x_1,x_2)\,dx_2\,dx_1\qquad\text{in place of }\mathcal S_{\mathcal N}.
\]
The proof proceeds through an anisotropic grid, in five steps.

\textit{(a) The grid and the cell variables.} Partition $\mathbb R^{d_1+d_2}$ into the half-open anisotropic cells
\[
C_{\mathbf i}:=\prod_{l=1}^{d_1}\big(i_lh_1,(i_l+1)h_1\big]\times\prod_{j=1}^{d_2}\big(i_{d_1+j}h_2,(i_{d_1+j}+1)h_2\big],\qquad\mathbf i\in\mathbb Z^{d_1+d_2},
\]
of side $h_1$ in the $X_1$ directions and $h_2$ in the $X_2$ directions, hence of volume $\eta$ each. Index the array by $\mathcal I_{\mathcal N}:=\{\mathbf i\in\mathbb Z^{d_1+d_2}:C_{\mathbf i}\cap\mathcal B(M)\ne\emptyset\}$, a finite set by the compactness of $\mathcal B(M)$, of cardinality $|\mathcal I_{\mathcal N}|\le C\eta^{-1}$ (the cells meeting $\mathcal B(M)$ lie in its one-cell enlargement, of Lebesgue measure $O(1)$, and each has volume $\eta$). Write $R_{\mathbf i}:=C_{\mathbf i}\cap(\mathcal B_1\times \mathcal B_2)$, so that $\lambda(R_{\mathbf i})\le\eta$ and $R_{\mathbf i}=\emptyset$ for those cells that miss $\mathcal B_1\times \mathcal B_2$. Define, for each $\mathbf i\in\mathcal I_{\mathcal N}$,
\begin{equation}\label{oeq:cellvars}
\alpha_{\mathbf i}:=\frac{1}{\sigma_{\mathcal N}(\mathcal B_1,\mathcal B_2)}\int_{R_{\mathbf i}}\projfield^{\circ}_{\mathcal N}(x_1,x_2)\,dx_2\,dx_1 ,\qquad
u_{\mathbf i}:=\frac{\mathcal N\big(C_{\mathbf i}\cap\mathcal B(M)\big)-N\mu\big(C_{\mathbf i}\cap\mathcal B(M)\big)}{\sqrt N},
\end{equation}
with the convention $\alpha_{\mathbf i}=0$ when $R_{\mathbf i}=\emptyset$. Since $\mathcal B_1\times \mathcal B_2\subset\mathcal B(M)$ and the cells $\{C_{\mathbf i}\}_{\mathbf i\in\mathcal I_{\mathcal N}}$ partition $\mathcal B(M)$,
\begin{equation}\label{oeq:tile}
\Qproj=\sum_{\mathbf i\in\mathcal I_{\mathcal N}}\alpha_{\mathbf i},\qquad U_{\mathcal N}=\sum_{\mathbf i\in\mathcal I_{\mathcal N}}u_{\mathbf i},\qquad\text{both exactly, with no leftover term.}
\end{equation}
It is to obtain the second identity that the array is indexed by the cells meeting $\mathcal B(M)$ rather than by those meeting $\mathcal B_1\times \mathcal B_2$: the latter would omit the cells of $\mathcal B(M)\setminus(\mathcal B_1\times \mathcal B_2)$ and lose a slab of interior count. The cells vanishing there do no harm, since $\alpha_{\mathbf i}=0$ on them, exactly as the factor $I_C$ in the corresponding array (6.42) of \citet{gine2003bm} annihilates the cells outside their set $C$. For the same reason the cell sides need no adjustment analogous to their $h^*_n=M/m_n$: the identities \eqref{oeq:tile} require only that the cells partition $\mathcal B(M)$, and irregular boundary cells are harmless, all that is used below being $\lambda(R_{\mathbf i})\le\eta$ and $\lambda(C_{\mathbf i}\cap\mathcal B(M))\le\eta$.

\textit{(b) One-dependence.} $\projfield^{\circ}_{\mathcal N}(x_1,x_2)$ is by construction a measurable function of the scalar $\tau_{\mathcal N}(x_1,x_2)$ (this covers the band slice of its inner integral as well), and $\tau_{\mathcal N}(x_1,x_2)$ is a functional of the increments over the box $\mathcal W(x_1,x_2)$, whose half-widths $h_1/2$ and $h_2/2$ are strictly smaller than the respective cell sides. Hence for $(x_1,x_2)\in C_{\mathbf i}$, $\mathcal W(x_1,x_2)$ is contained in the union of $C_{\mathbf i}$ and its $3^{d_1+d_2}-1$ neighbors, and $\alpha_{\mathbf i}$ (and trivially $u_{\mathbf i}$) is a functional of the Poisson increments over that neighborhood. More precisely, $\alpha_{\mathbf i}$ and $u_{\mathbf i}$ are functionals of the increments over the enlargement of $C_{\mathbf i}$ by $h_1/2$ in the $X_1$ directions and $h_2/2$ in the $X_2$ directions: in one coordinate of side $h$, the closed windows centered in $(ih,(i+1)h]$ lie in $(ih-h/2,(i+1)h+h/2]$, and those centered in $((i+2)h,(i+3)h]$ lie in $((i+1)h+h/2,(i+3)h+h/2]$. Two cells at lattice distance at least $2$ in the maximum norm are two sides apart in some coordinate, so their enlargements are disjoint, and families of cells pairwise at distance at least $2$ depend on increments over disjoint sets. Hence the array $\{(\alpha_{\mathbf i},u_{\mathbf i})\}$ is $1$-dependent on $\mathbb Z^{d_1+d_2}$, with exact independence beyond distance one. (The exterior count $\Uext$ plays no part here: it is independent of the whole array by part (ii), and by \eqref{oeq:tile} it is not needed to reconstitute $U_{\mathcal N}$.)

\textit{(c) Variances and the covariance.} $\Var(\sum\alpha_{\mathbf i})=\Var(\Qproj)\to1$ by \eqref{oeq:projvar}; $\Var(\sum u_{\mathbf i})=\Var(U_{\mathcal N})=1-\pext$ exactly, by \eqref{oeq:tile} and \eqref{oeq:countvar}. The cross term $\Cov(\Qproj,U_{\mathcal N})\to0$ is the content of Lemma~\ref{olem:decorr} below. Hence for any fixed $(a,c)\in\mathbb R^2$, $\Var(a\Qproj+cU_{\mathcal N})\to a^2+c^2(1-\pext)$.

\textit{(d) Lyapunov.} By Jensen's inequality over $R_{\mathbf i}$, the third-moment bound \eqref{oeq:delta3}, and $\lambda(R_{\mathbf i})\le\eta$,
\[
\E[|\alpha_{\mathbf i}|^3]\le\frac{\lambda(R_{\mathbf i})^2}{\sigma^3_{\mathcal N}}\int_{R_{\mathbf i}}\E\big[|\projfield^{\circ}_{\mathcal N}(x_1,x_2)|^3\big]dx_2\,dx_1=O\big(\eta^{2}\cdot\eta\cdot\eta^{-3/2}\big)=O\big(\eta^{3/2}\big)
\]
uniformly in $\mathbf i\in\mathcal I_{\mathcal N}$, whence $\sum_{\mathbf i}\E[|\alpha_{\mathbf i}|^3]=O(\eta^{-1}\cdot\eta^{3/2})=O(\eta^{1/2})=o(1)$, using $|\mathcal I_{\mathcal N}|=O(\eta^{-1})$. For the counts, the central third absolute moment of a Poisson variable with mean $N_0$ is $O(N_0+N_0^{3/2})$, and $N\mu(C_{\mathbf i}\cap\mathcal B(M))=O(N\eta)$, so $\E[|u_{\mathbf i}|^3]=O((N\eta)^{3/2}+N\eta)/N^{3/2}=O(\eta^{3/2})+O(\eta/\sqrt N)$ and $\sum_{\mathbf i}\E[|u_{\mathbf i}|^3]=O(\eta^{1/2})+O(1/\sqrt N)=o(1)$.

\textit{(e) Conclusion.} For fixed $(a,c)$, the array $\{a\alpha_{\mathbf i}+cu_{\mathbf i}:\mathbf i\in\mathcal I_{\mathcal N}\}$ is a $1$-dependent triangular array on $\mathbb Z^{d_1+d_2}$ whose total, by \eqref{oeq:tile}, is exactly $a\Qproj+cU_{\mathcal N}$, with variance converging to $a^2+c^2(1-\pext)$ by (c) and Lyapunov ratios vanishing by (d); the central limit theorem for $m$-dependent random fields of \citet{shergin1990central}{} therefore yields $a\Qproj+cU_{\mathcal N}\dto N(0,a^2+c^2(1-\pext))$. (Where the index set is one-dimensional this is the form used as Fact~6.2 of \citealp{gine2003bm}; the lattice $\mathbb Z^{d_1+d_2}$ requires the field version, and any local-dependence normal approximation with a Lyapunov condition serves equally.) The Cram\'er--Wold device gives $(\Qproj,U_{\mathcal N})\dto(Z_1,\sqrt{1-\pext}Z_2)$, and Slutsky's theorem transfers the statement to $(\mathcal S_{\mathcal N},U_{\mathcal N})$, proving (i).
\end{proof}

\begin{lemma}[Decorrelation from the count]\label{olem:decorr}
Under the conditions of Lemma~\ref{olem:variance}, $\Cov(\Qproj,U_{\mathcal N})\longrightarrow0$.
\end{lemma}

\begin{proof}
Fix $(x_1,x_2)\in \mathcal B_1\times \mathcal B_2$; it suffices to show $\sup_{(x_1,x_2)}|\Cov(\projfield^{\circ}_{\mathcal N}(x_1,x_2),U_{\mathcal N})|=o(1)$, since $\Qproj$ integrates $\projfield^{\circ}_{\mathcal N}$ over a bounded domain and $\sigma_{\mathcal N}$ is bounded away from zero. Because $\Var(U_{\mathcal N})=1-\pext$ is a fixed constant in $(0,1)$, it is equivalent, and notationally lighter, to work with the standardized interior count $U_{\mathcal N}/\sqrt{1-\pext}$; every $O(\cdot)$ and $o(\cdot)$ statement below is unaffected by that constant factor, and $U_{\mathcal N}$ is written for the standardized count throughout parts (b)--(d). The proof introduces a decomposition of the count that renders every step exact or explicitly quantified.

\textit{(a) Local and far counts.} Split
\[
U_{\mathcal N}=U^{\mathrm{loc}}_{(x_1,x_2)}+U^{\mathrm{far}}_{(x_1,x_2)},\qquad
U^{\mathrm{loc}}_{(x_1,x_2)}:=\frac{\mathcal N\big(\mathcal W(x_1,x_2)\big)-N\mu\big(\mathcal W(x_1,x_2)\big)}{\sqrt N},
\]
the standardized centered count in the kernel box; the split is legitimate for the interior count because $\mathcal W(x_1,x_2)\subset\mathcal B(M)$ for $(x_1,x_2)\in \mathcal B_1\times \mathcal B_2$ by \eqref{oeq:M1}, so that $U^{\mathrm{far}}_{(x_1,x_2)}$ is the standardized centered count over $\mathcal B(M)\setminus\mathcal W(x_1,x_2)$. The far part is a functional of the increments outside $\mathcal W(x_1,x_2)$, hence independent of $\tau_{\mathcal N}(x_1,x_2)$, which by \eqref{oeq:poisson-field} is a deterministic level plus a functional of the increments over $\mathcal W(x_1,x_2)$, and of every measurable function of it; since $\projfield^{\circ}_{\mathcal N}(x_1,x_2)$ is such a function and is centered,
\begin{equation}\label{oeq:farzero}
\Cov\big(\projfield^{\circ}_{\mathcal N}(x_1,x_2),\,U^{\mathrm{far}}_{(x_1,x_2)}\big)=0\qquad\text{exactly.}
\end{equation}
Moreover $\Var(U^{\mathrm{loc}}_{(x_1,x_2)})=\mu(\mathcal W(x_1,x_2))\le\sup f\,\eta$, so
\begin{equation}\label{oeq:uloc}
\big\|U^{\mathrm{loc}}_{(x_1,x_2)}\big\|_2=O\big(\eta^{1/2}\big)\qquad\text{uniformly.}
\end{equation}
The two displays already dispose of the band slice of $\projfield^{\circ}_{\mathcal N}$: its $L_2$-norm is $O(h_2^{d_2}\eta^{-1/2})$ (end of Section~\ref{osec:aux-proj}), so by \eqref{oeq:farzero}, the Cauchy--Schwarz inequality and \eqref{oeq:uloc} its covariance with $U_{\mathcal N}$ is $O(h_2^{d_2}\eta^{-1/2})\cdot O(\eta^{1/2})=O(h_2^{d_2})$, uniformly; the crude bound $\|\cdot\|_2\|U_{\mathcal N}\|_2=O(h_2^{d_2}\eta^{-1/2})$, without the local/far split, would instead require the anisotropy condition $h_2^{d_2}=o(h_1^{d_1})$; the split makes the $X_1$ smoothing volume factor cancel. It remains to treat the off-band part $\Lambda(\sqrt{\kappa_{\mathcal N}(x_1,x_2)}\,S_{\mathcal N}(x_1,x_2))$ (in the rest of this proof $\kappa:=\kappa_{\mathcal N}(x_1,x_2)$), where
\begin{align*}
\Lambda(w)&:=\int_{\{x_2'\in \mathcal B_2:\|x_2'-x_2\|_\infty>h_2\}}\Big(\E\big[|w-\tilde\tau(x_2')|\big]-\E\big[\big|\sqrt N\,\Gamma_{\mathcal N}(x_1,x_2,x_2')\big|\big]\Big)dx_2',\\
\tilde\tau(x_2')&:=\sqrt N\big(\tau_{\mathcal N}(x_1,x_2')-\E[\tau_{\mathcal N}(x_1,x_2')]\big)-\sqrt N\,\E\big[\Gamma_{\mathcal N}(x_1,x_2,x_2')\big],
\end{align*}
a deterministic, $\lambda(\mathcal B_2)$-Lipschitz map (each $w\mapsto\E[|w-\tilde\tau(x_2')|]$ is $1$-Lipschitz), so that $s\mapsto\Lambda(\sqrt{\kappa}\,s)$ is $\lambda(\mathcal B_2)O(\eta^{-1/2})$-Lipschitz.

\textit{(b) The linear covariance.} By the law of total covariance for the compound Poisson sums (conditioning on the counts, as in Section~\ref{osec:aux-poisson}), for $x=(x_1,x_2)$ and with $a(W;x)=\eta^{-1}\xi(W;x)\mathcal K(x;X)$ the uncentered summand of \eqref{oeq:poisson-field},
\begin{align*}
\Cov\Big(\sqrt N\big(\tau_{\mathcal N}(x)-\E[\tau_{\mathcal N}(x)]\big),\,U_{\mathcal N}\Big)
&=\frac{\E\big[a(W;x)\,\ind\{(X_1,X_2)\in\mathcal B(M)\}\big]}{\sqrt{1-\pext}}\\
&=\frac{\E[a(W;x)]}{\sqrt{1-\pext}}=\frac{B_N(x)}{\sqrt{1-\pext}}=O(h_1^{s_1}+h_2^{s_2}),
\end{align*}
the first equality because $U_{\mathcal N}$ counts only $\mathcal B(M)$ and the Poisson covariance of $\sum_{i\le\mathcal N}a(W_i;x)$ with a count is the mean of $a$ times the indicator, the second because the kernel factor of $a$ is supported in $\mathcal W(x)\subset\mathcal B(M)$, and the third by the definition \eqref{oeq:theta} of $B_N$; the covariance is of the type $\E[a\,\ind_{\mathcal B}]$, the smoothing bias. This is where adding the level $\tau(x)$ deterministically, rather than Poissonizing it, pays: a Poissonized field carrying the level $\tau(x)$ in its summands has covariance of order $\tau(x)/\sqrt{1-\pext}=O(1)$ with the count, and the decorrelation then has to be closed by the symmetry of the normal law and the oddness of $d|u|/du$. After standardization,
\begin{equation}\label{oeq:gammastar}
\rho^U(x):=\Cov\big(S_{\mathcal N}(x),U_{\mathcal N}\big)=\frac{B_N(x)}{\sqrt{1-\pext}\,\sqrt{\kappa_{\mathcal N}(x)}}=O\big((h_1^{s_1}+h_2^{s_2})\,\eta^{1/2}\big)
\end{equation}
uniformly in $x\in \mathcal B_1\times \mathcal B_2$, consistently with \eqref{oeq:farzero}--\eqref{oeq:uloc}.

\textit{(c) Gaussian transfer.} Let $(Z_1,Z_2)$ be jointly normal, standardized, with correlation $\rho^U(x_1,x_2)$. Both $S_{\mathcal N}(x_1,x_2)$ and $U_{\mathcal N}$ are standardized sums of the same $N$ i.i.d.\ Poisson$(1)$ blocks, jointly in the sense of \eqref{oeq:blocksjoint}, each block carrying both its $\theta_N$-integral and its count of points in $\mathcal B(M)$, with block third moments $O(\eta^{-1/2})$ and $O(1)$ respectively by \eqref{oeq:thirdmom}. Theorem~1 of \citet{sweeting1977speeds}, the multidimensional smooth-function central limit theorem with rates from which \citet{gine2003bm} obtain their Fact~6.1, then gives, for the function $(s,t)\mapsto\Lambda(\sqrt\kappa\,s)\,t$, whose modulus is controlled by the Lipschitz constant $\lambda(\mathcal B_2)O(\eta^{-1/2})$ in $s$ and linearity in $t$,
\begin{equation}\label{oeq:sweeting}
\Big|\Cov\big(\Lambda(\sqrt\kappa\,S_{\mathcal N}),U_{\mathcal N}\big)-\Cov\big(\Lambda(\sqrt\kappa\,Z_1),Z_2\big)\Big|=O\big(\eta^{-1/2}\big)\cdot O\Big(\frac{\eta^{-1/2}}{\sqrt N}\Big)=o(1),
\end{equation}
the final rate being $O((N\eta^2)^{-1/2})$.

\textit{(d) The Gaussian covariance via Stein's identity.} For jointly normal $(Z_1,Z_2)$ with correlation $\rho^U$ and a Lipschitz $F$, Stein's identity gives $\Cov(F(Z_1),Z_2)=\rho^U\,\E[F'(Z_1)]$; with $F(s)=\Lambda(\sqrt\kappa\,s)$, $|F'(s)|\le\sqrt\kappa\,\lambda(\mathcal B_2)$ almost everywhere, because each $w\mapsto\E[|w-\tilde\tau(x_2')|]$ has derivative $\E[\operatorname{sign}(w-\tilde\tau(x_2'))]$, bounded by one in absolute value. Hence, with \eqref{oeq:gammastar},
\[
\big|\Cov\big(\Lambda(\sqrt\kappa\,Z_1),Z_2\big)\big|\le|\rho^U|\,\sqrt{\kappa_{\mathcal N}(x)}\,\lambda(\mathcal B_2)=\frac{|B_N(x)|\,\lambda(\mathcal B_2)}{\sqrt{1-\pext}}=O(h_1^{s_1}+h_2^{s_2})=o(1)
\]
uniformly over $x\in \mathcal B_1\times \mathcal B_2$, with no use of the symmetry of the normal law, of the sign structure of $d|u|/du$ or of the null hypothesis. Combining (a)--(d) and integrating over the bounded domain completes the proof.
\end{proof}

\textit{Extension to general $\Psi$.} Lemma~\ref{olem:clt} uses the projection field only through \eqref{oeq:projvar}, \eqref{oeq:delta3}, the measurability of $\projfield^{\circ}_{\mathcal N}(x)$ with respect to the window $\mathcal W(x)$ and Lemma~\ref{olem:decorr}; with the two-slot projection field of \eqref{oeq:pf-Delta-proj} it holds for the class and for every configuration set without change, the Lyapunov step using only the moment bound of Lemma~\ref{olem:pf-projection}(d). In Lemma~\ref{olem:decorr}, part (a) uses the Lipschitz property of $w\mapsto\E[\Psi(w-\tilde\tau)]$, part (c) only that Lipschitz constant, and part (d) only $|F'|\le\sqrt\kappa\,\lambda(\mathcal B_2)$, that is, $|\Psi'|\le1$ almost everywhere, which is Assumption~\ref{ass:psi}(a); the bound $\rho^U=O((h_1^{s_1}+h_2^{s_2})\eta^{1/2})$ of part (b) concerns the fields alone. Hence the covariance of $\projfield_{\mathcal N}(x)$ with $U_{\mathcal N}$ is $o(1)$ uniformly in $x$, and $\Cov(\Qproj,U_{\mathcal N})\to0$, for every $\Psi$ in the class, with or without the null; neither the symmetry of $|\cdot|$ nor that of the normal law is used.

\subsection{De-Poissonization and the assembly of the theorem}\label{osec:aux-depoisson}

\textit{Lemma~\ref{olem:depoisson} (de-Poissonization), stated in Section~\ref{osec:aux-moved}.} If $(\mathcal T_{\mathcal N},U_{\mathcal N})$ is independent of $\Uext$ for each $N$ and $(\mathcal T_{\mathcal N},U_{\mathcal N})\dto(\sigma Z_1,\sqrt{1-\pext}Z_2)$ with $Z_1,Z_2$ independent standard normals, then $\Pr(\mathcal T_{\mathcal N}\le t\mid\mathcal N=N)\to\Pr(\sigma Z_1\le t)$ for every $t$.

Here, as in Section~\ref{osec:aux-clt}, $N_{1,\mathcal N}:=\mathcal N(\mathcal B)$ and $N_{2,\mathcal N}:=\mathcal N(\mathcal B^c)$ are independent Poisson variables with means $N(1-\pext)$ and $N\pext$, $\pext\in(0,1)$, $\mathcal N=N_{1,\mathcal N}+N_{2,\mathcal N}$, $U_{\mathcal N}=\{N_{1,\mathcal N}-N(1-\pext)\}/\sqrt N$ and $\Uext=\{N_{2,\mathcal N}-N\pext\}/\sqrt N$. The lemma is Lemma~2.4 of \citet{gine2003bm}, there attributed to \citet{beirlant1995asymptotic}, and is used without proof. The limiting second coordinate has variance $1-\pext$, not $1$: it is the interior count that affects the limit, and the exterior count is required only to be independent of the pair; the version in \citet{gine2003bm} also allows a nonzero limiting correlation, which is not needed here. Three consequences are shown, for $\Dh_2$ under Assumptions~\ref{ass:sampling}--\ref{ass:reg} and $H_0^2$:
(i) with $\bar D_{2N}(\mathcal B_1,\mathcal B_2)$ the fixed-sample analogue of \eqref{oeq:DP},
\begin{equation}\label{oeq:fixedCLT}
\frac{\bar D_{2N}(\mathcal B_1,\mathcal B_2)-\E\big[D^P_{\mathcal N}(\mathcal B_1,\mathcal B_2)\big]}{\sigma_{\mathcal N}(\mathcal B_1,\mathcal B_2)}\dto N(0,1);
\end{equation}
(ii) the same holds with $\dom$ in place of $\mathcal B_1\times \mathcal B_2$, that is, with $D^{\mathrm{lin}}_2$ of Lemma~\ref{olem:pf-linearization}(c), its Poissonized mean and its Poissonized standard deviation; and
(iii) the plug-in variance estimator
\begin{equation}\label{oeq:sigmahat}
\sigh^2_2:=\int_{\Xone}\int_{\Xtwo}\int_{\Xtwo}\int_{\Xtwo}\int_{[-1,1]^{d_2}}\int_{[-1,1]^{d_1}}
g\big(\widehat\rho^*_2\big)\,\widehat\kappa^*_2(x_1,x_2,x_2',x_2'')\;du_1\,du_2\,dx_2''\,dx_2'\,dx_2\,dx_1 ,
\end{equation}
where $\widehat\rho^*_2$ and $\widehat\kappa^*_2$ substitute $\rhoh$ for $\varrho$ in \eqref{oeq:rho2star} and \eqref{oeq:omega}, satisfies $\sigh^2_2-\sigma^2_{2,\dom}=o_p(1)$.

\begin{proof}[Proof of the consequences]
\textit{(i).} Lemmas~\ref{olem:clt} and \ref{olem:decorr} deliver exactly the two hypotheses of Lemma~\ref{olem:depoisson} with $\mathcal T_{\mathcal N}=\mathcal S_{\mathcal N}$ and $\sigma=1$: the split of the sample count in \eqref{oeq:counts} realizes $N_{1,\mathcal N}=\mathcal N(\mathcal B(M))$ and $N_{2,\mathcal N}=\mathcal N(\mathcal B(M)^c)$ as independent Poisson variables with the stated means, Lemma~\ref{olem:clt}(ii) is hypothesis (i), and Lemma~\ref{olem:clt}(i) is hypothesis (ii). The asymptotic independence of the statistic from the interior count, together with the exact independence from the exterior count, makes conditioning on the realized sample size cost no variance in the limit; and conditioning on $\{\mathcal N=N\}$ turns the Poissonized statistic into its fixed-sample analogue, the Poisson points being, given the count, $N$ i.i.d.\ draws. This gives \eqref{oeq:fixedCLT}.

\textit{(ii).} Let $\mathcal B_2\uparrow\Xtwo$ and $\mathcal B_1\uparrow\Xone$ along a sequence of sets satisfying \eqref{oeq:M1}--\eqref{oeq:M3} with $\epsilon_k\downarrow0$. By Sections~\ref{osec:aux-poisson}--\ref{osec:aux-center}, the contribution to the mean and the variance from $(\Xone\times\Xtwo^2)\setminus(\mathcal B_1\times \mathcal B_2^2)$ can be made arbitrarily small uniformly in $N$: for the mean, the proof of \eqref{oeq:center-i} applied with the complement region in place of $\mathcal B_1\times \mathcal B_2^2$ gives that the mean minus its centering is $o(1)$, as its bound is uniform over configurations; for the variance, from Lemma~\ref{olem:variance} applied on the complement, whose limiting integrand is bounded by $(1-2/\pi)\cdot2\sup\varrho$. The standard diagonal argument then extends \eqref{oeq:fixedCLT} to $\dom$. Each member of the sequence carries its own box $\mathcal B(M_k)$ and hence its own exterior mass $\pext^{(k)}\in(0,1)$; since the conclusion of Lemma~\ref{olem:depoisson} is $N(0,1)$ whatever the value of $\pext$, the limit along the diagonal is unaffected and no condition on the rate at which $\pext^{(k)}\downarrow0$ is needed.

\textit{(iii).} By Lemma~\ref{olem:variance}, Lemma~\ref{olem:consis}(c), the continuity and boundedness of $g$ and $\inf\varrho>0$, the map from $\varrho$ to the integrand of \eqref{oeq:sigma2} is uniformly continuous on $\{\varrho\ge\inf\varrho/2\}$, an event of probability tending to one, and the domain of integration is bounded; hence $\sigh^2_2-\sigma^2_{2,\dom}=o_p(1)$.
\end{proof}

The assembly of Theorem~\ref{thm:normality}, given in Section~\ref{osec:proofs}, consists of three steps: Lemma~\ref{olem:pf-linearization} replaces $\Dh_2$ by $D^{\mathrm{lin}}_2$, whose law is that of $D^P_2$ given $\mathcal N=N$; consequence (ii) gives the fixed-sample limit with the Poissonized centering and scale; and Lemma~\ref{olem:centering} together with consequence (iii) replace the centering by $\widehat m^{\,\kappa}_2$ (no side conditions) or, under \eqref{oeq:extra}, by $\widehat m_2$, and the scale by $\sigh_2$, each at a cost $o_p(1)$; Slutsky's theorem then gives $(\Dh_2-\widehat m^{\,\kappa}_2)/\sigh_2\dto N(0,1)$ under $H_0^2$.{} The statistic $\Sh_2$ of the main text, \eqref{eq:S}, is centered and scaled instead by $\widetilde m_2$ and $\widetilde\sigma_{2,N}$ of \eqref{eq:center-feasible} and \eqref{eq:sigmaN}; Lemmas~\ref{prop:centering-corr} and \ref{prop:sigmaN} give $\widetilde m_2-m^{\kappa}_{2,N}=o_p(1)$ and $\widetilde\sigma^2_{2,N}\pto\sigma^2_2$, so the same conclusion holds for $\Sh_2$.

\textit{Extension to general $\Psi$.} Lemma~\ref{olem:depoisson} and consequence (i) do not involve $\Psi$. Part (ii) uses the uniform bounds of the mean and variance on the complement, which hold for the class by the extension paragraphs of Sections~\ref{osec:aux-center} and \ref{osec:aux-var} (the limiting variance integrand being bounded by $\Var(\Psi(Z))\cdot2\sup\varrho\le2\sup\varrho$). Part (iii) uses only the continuity and boundedness of $g$, which hold for $g_\Psi$ by \eqref{oeq:gpsi}; the plug-in for a general configuration set substitutes $\rhoh$ into the general form, equation~\eqref{eq:sigma-gen} of the main text.

\subsection{Further remarks}\label{osec:aux-remarks}

\begin{remark}\label{orem:pivotal}
\textit{Special cases and a bounded-weight variant.} (i) If $\varrho(\cdot,\cdot)\equiv\bar\varrho$ is constant on $\dom$, then $\rho^*_2=\tfrac12\bar\rho_2(u_2)\bar\rho_1(u_1)$, $m_2=\eta^{-1/2}\sqrt{2/\pi}\,\tfrac{\sqrt{2\bar\varrho}}{2}\lambda(\Xtwo)^2\lambda(\Xone)$ and
\[
\sigma^2_2=2\bar\varrho\,\lambda(\Xtwo)^3\lambda(\Xone)\int_{[-1,1]^{d_2}}\int_{[-1,1]^{d_1}}g\big(\tfrac12\bar\rho_2(u_2)\bar\rho_1(u_1)\big)\,du_1\,du_2 :
\]
even under homoskedasticity the unweighted statistic is not pivotal, but both constants involve only the scalar level $\bar\varrho$; the main text does not use this simplification and estimates both constants in pre-limit form from the kernel weights and corrected residuals, \eqref{eq:center-feasible} and \eqref{eq:sigmaN} (details in Section~\ref{osec:centering-details}).(ii) If one instead retains the bounded weight $(\rhoh(x_1,x_2)+\rhoh(x_1,x_2'))^{-1/2}$ in \eqref{oeq:D2}, the same arguments apply with every $\sqrt{\varrho+\varrho'}$ factor canceled: the centering becomes the pivotal $m_2=\eta^{-1/2}\sqrt{2/\pi}\,\tfrac12\lambda(\Xtwo)^2\lambda(\Xone)$ and the variance is \eqref{oeq:sigma2} without the square-root scale factors. The choice between the two is one of finite-sample stability (no estimated ratio at all) versus partial pivotality. The uniform consistency of the weight follows from Lemma~\ref{olem:consis}(c) and $\inf\varrho>0$, in contrast with the inverse-standard-error weight discussed in Section~\ref{sec:stat-stats} of the main text, but the estimated weight affects the statistic and its effect is not analyzed here: the bounded-weight variant is not covered by the results of the paper.
\end{remark}

\begin{remark}\label{orem:structure}
\textit{Summary of the structural features.} Five elements of this section are worth collecting; the first four are specific to the joint smoothing over $(x_1,x_2)$, the fifth to the two-point form of the statistic. (a) The conjunction structure: dependence between fields requires simultaneous $X_2$ and $X_1$ window overlap, so every covariance computation carries the gate $\ind(\|x_1-x_1'\|_\infty\le h_1)$, and the $X_1$ pair $(x_1,x_1')$ collapses to a single free $X_1$ by the change of variables $x_1'=x_1+u_1h_1$. (b) The factorized correlation: the limiting correlation \eqref{oeq:rho2star} carries the product $\bar\rho_2(u_2)\bar\rho_1(u_1)$, the $X_1$ factor interpolating continuously between full correlation ($u_1=0$) and exact independence ($\|u_1\|_\infty\ge1$). (c) The anisotropy condition: the limit-functional centering requires $h_2^{d_2}=o(h_1^{d_1})$ in addition to the pooled-dimension restriction $h_2^2+h_1^2=o(\eta^{1/2})$, because the $X_2$ band's measure cannot repay the $X_1$ part of the bias scale; the smoothed-target centering $\widehat m^{\,\kappa}_2$ removes both; the centering $\widetilde m_2$ of the main text removes the anisotropy condition, and is consistent for the same target under Assumption~\ref{ass:reg}(h), which contains the pooled-dimension restriction (Lemma~\ref{prop:centering-corr}). (d) The local/far count decomposition: the covariance between the projection field and the interior count is controlled by splitting that count at the kernel box, yielding the exact orthogonality \eqref{oeq:farzero}, the self-contained bound \eqref{oeq:gammastar}, and a band-slice covariance bound that is free of the anisotropy condition. (e) No perturbation device: because $\Dh_2$ differences the field at two $X_2$ points and carries no weight, the limiting correlation is bounded away from one uniformly on the configurations that survive the reduction, \eqref{oeq:rhobar}, while the configurations at which it would degenerate are removed in Part~1 of the proof of Lemma~\ref{olem:variance} by measure against the crude bound \eqref{oeq:csb}; the Berry--Esseen pre-factor $(1-\rho^{*2})^{-3/2}$ of \eqref{oeq:BEcov} is therefore a fixed constant, and the regularization that the corresponding one-sided, single-point arguments require is not needed.
\end{remark}

\subsection{The mirror statistic}\label{osec:aux-mirror}

Every result of this section applies verbatim to
\[
\Dh_1=\tfrac12\int_{\Xtwo}\int_{\Xone}\int_{\Xone}\sqrt N\,\big|\tauh(x_1,x_2)-\tauh(x_1',x_2)\big|\,dx_1\,dx_1'\,dx_2
\]
upon exchanging the roles of the two blocks throughout, that is, upon the substitution
\[
\big(x_2,\ \Xtwo,\ d_2,\ h_2,\ K_2,\ \bar\rho_2,\ \tilde z_2,\ u_2,\ \mathcal B_2\big)\;\longleftrightarrow\;\big(x_1,\ \Xone,\ d_1,\ h_1,\ K_1,\ \bar\rho_1,\ \tilde z_1,\ u_1,\ \mathcal B_1\big),
\]
which leaves $\eta=h_1^{d_1}h_2^{d_2}$, the influence function $\theta_N(W;x)$, the pointwise scaled variance $\varrho(x_1,x_2)$ and the function $g$ unchanged. Three points make the transfer immediate rather than a matter of re-deriving the argument.

First, the estimator \eqref{oeq:tauhat} is symmetric in the two blocks: it is a single kernel regression on $(x_1,x_2)$ with the product kernel $K_1(\cdot)K_2(\cdot)$, and no step of Sections~\ref{osec:aux-lin}--\ref{osec:aux-center} distinguishes the two arguments. The linearization, the Poissonization and the window geometry are stated in terms of the joint window $\mathcal W(x_1,x_2)$ of Lebesgue measure $\eta$, which is invariant under the exchange.

Second, the asymmetry affects the argument only through the differencing direction, and it does so in exactly one place: the band on which the two differenced estimators share observations. For $\Dh_2$ that band is $\{\|x_2-x_2'\|_\infty\le h_2\}$, of measure $O(h_2^{d_2})$; for $\Dh_1$ it is $\{\|x_1-x_1'\|_\infty\le h_1\}$, of measure $O(h_1^{d_1})$. Every bound in Sections~\ref{osec:aux-var}--\ref{osec:aux-clt} that invokes the band measure therefore carries $h_1^{d_1}$ in place of $h_2^{d_2}$, and the degenerate remainder of the projection vanishes at rate $O(h_1^{d_1})$ rather than $O(h_2^{d_2})$. Both are $o(1)$ under Assumption~\ref{ass:reg}(g), so the conclusions are unaffected.

Third, the exchange reverses the direction of the auxiliary bandwidth condition \eqref{oeq:extra}(ii) required for the limit-functional centering: $\Dh_2$ requires $h_2^{d_2}=o(h_1^{d_1})$ and $\Dh_1$ requires $h_1^{d_1}=o(h_2^{d_2})$. The two are mutually exclusive, which is why the smoothed-target centering $\widehat m^{\,\kappa}_j$ is the default for both statistics, as discussed in Section~\ref{sec:asym-center} of the main text. No such conflict arises for $\widehat m^{\,\kappa}_j$, whose consistency (Lemma~\ref{olem:centering}(iv)) requires no relation between the two bandwidths beyond Assumption~\ref{ass:reg}(g).

Consequently, under $H_0^1$ and Assumptions~\ref{ass:sampling}--\ref{ass:reg}, $(\Dh_1-\widehat m^{\,\kappa}_1)/\sigh_1\dto N(0,1)${} and, by Lemmas~\ref{prop:centering-corr} and \ref{prop:sigmaN}, $\Sh_1\dto N(0,1)$, with $\sigma^2_1$ the mirror image of \eqref{oeq:sigma2} obtained by the substitution displayed above, Theorem~\ref{thm:variance}(b) of the main text. For a general integrand $\Psi$ and a configuration set $\mathcal C_1\subseteq\Xone^2\times\Xtwo$ the same substitution applies to the extension paragraphs of each subsection.

\providecommand{\projfield}{\Delta}
\providecommand{\pext}{p_{\mathcal B}}\providecommand{\Uext}{U'_{\mathcal N}}\providecommand{\Qproj}{Q_{\mathcal N}}
\providecommand{\movedsec}[3]{\subsubsection{#3}\label{#1}}
\subsection{Routine computations for the proofs in Section~\ref{osec:proofs}}\label{osec:aux-moved}
This subsection documents the routine computations behind the proofs of Theorems~\ref{thm:variance} and \ref{thm:normality} and Lemma~\ref{olem:variance-gen} in Section~\ref{osec:proofs}, in the notation of that section: $j$ is fixed, $\mathcal C:=\mathcal C_j$, $\kappa:=\kappa_{j,N}$, $\Theta:=\Theta_j$, $c_1,c_2$ are the two evaluation points of a configuration $c$, $\mathcal W(x)$ is the kernel window centered at $x$, of Lebesgue measure $\eta$, and $S_{\mathcal N}(c)$ the standardized Poissonized field. The arguments are those of Sections~\ref{osec:aux-poisson}--\ref{osec:aux-clt}, written for a general integrand $\Psi$ of the class of Assumption~\ref{ass:psi} and a general configuration set; the only properties
of $\Psi$ used are the Lipschitz bound, positive homogeneity and $|\Psi(u)|\le|u|$, through the representation $\Psi(u)=a_\Psi|u|+b_\Psi u$ of \eqref{eq:stat-psi-rep}.

\movedsec{osec:moved-BE}{S.2.A}{Berry--Esseen comparisons for general $\Psi$}
Let $(Q_i,Q_i')$, $i\le N$, be i.i.d.\ standardized bivariate blocks with correlation $\rho$, $|\rho|\le1-\delta$, $S:=N^{-1/2}\sum_iQ_i$, $S':=N^{-1/2}\sum_iQ_i'$, and let $(Z_1,Z_2)$ be standard bivariate normal with correlation $\rho$. Fact~6.1 of \citet{gine2003bm}, obtained from Theorem~1 of \citet{sweeting1977speeds}, gives universal constants $C_1,C_2$ with
\begin{equation}\label{oeq:BE-abs}
\big|\E|S|-\E|Z_1|\big|\le C_1\frac{\E|Q|^3}{\sqrt N},\qquad
\big|\E|SS'|-\E|Z_1Z_2|\big|\le\frac{C_2}{(1-\rho^2)^{3/2}}\frac{\E|Q|^3+\E|Q'|^3}{\sqrt N},
\end{equation}
together with the one-sided analog $|\E[S|S'|]-\E[Z_1|Z_2|]|$ bounded by the same right-hand side, which those authors need for their max-type functionals. Since $\Cov(|S|,|S'|)-g_{\mathrm{abs}}(\rho)=\{\E|SS'|-\E|Z_1Z_2|\}-\E|S'|\{\E|S|-\E|Z_1|\}-\E|Z_1|\{\E|S'|-\E|Z_2|\}$ and since $\E|S'|\le1$, $\E|Z_1|\le1$, we obtain the covariance form with constant $C_2(1-\rho^2)^{-3/2}+2C_1$. For $\Psi(u)=a_\Psi|u|+b_\Psi u$,
\[
\Cov(\Psi(S),\Psi(S'))=a_\Psi^2\Cov(|S|,|S'|)+a_\Psi b_\Psi\{\Cov(|S|,S')+\Cov(S,|S'|)\}+b_\Psi^2\rho ,
\]
and $\Cov(S,|S'|)=\E[S|S'|]$ is compared with $\E[Z_1|Z_2|]=0$ by the one-sided inequality, so that
\begin{equation}\label{oeq:BE-gen}
\big|\Cov(\Psi(S),\Psi(S'))-g_\Psi(\rho)\big|\le C_\delta\frac{\E|Q|^3+\E|Q'|^3}{\sqrt N},\qquad
g_\Psi(\rho)=a_\Psi^2g_{\mathrm{abs}}(\rho)+b_\Psi^2\rho ,
\end{equation}
with $C_\delta$ depending on $\delta$ only; likewise $|\E\Psi(S)-m_\Psi|\le C_1\E|Q|^3/\sqrt N$ since $\E S=0$ exactly. These are the two inequalities \eqref{oeq:pf-BE} of Section~\ref{osec:proofs}.

\movedsec{osec:moved-events}{S.2.B}{The four single-overlap events}
Let $n_{cc'}\in\{0,\dots,4\}$ count the overlapping cross pairs $(c_k,c'_l)$, $k,l\in\{1,2\}$, of two configurations. Dependence requires $n_{cc'}\ge1$, and every overlap pins the shared coordinate of $c'$ in the block $j'\neq j$ to within $h_{j'}$ of that of $c$ (both evaluation points of a configuration share their $j'$-coordinate), a set of measure $O(h_{j'}^{d_{j'}})$. For $\iota\in\{1,\dots,4\}$ indexing the cross pairs, the separation indicator $\ind^{\mathrm{sep}}_\iota$ requires every pair of $j$-coordinates other than the $\iota$th cross pair, among the four cross pairs and the two within-configuration pairs $(c_1,c_2)$, $(c_1',c_2')$, to be separated by more than $2h_j$. Three classes are discarded, each bounded by the Cauchy--Schwarz envelope $\eta^{-1}(4\sup\varrho+o(1))$ against its measure: (a) $n_{cc'}\ge2$, (b) $n_{cc'}=1$ with a within-configuration separation at most $2h_j$, (c) $n_{cc'}=1$ with an inactive cross pair in the annulus $(h_j,2h_j]$. Each carries two constraints on distinct $j$-pairs, and whether the two constrained pairs share a coordinate or not, exactly two $j$-coordinates are pinned, so the $j$-measure is $O(h_j^{2d_j})$ and the total measure $O(h_j^{2d_j}h_{j'}^{d_{j'}})=O(\eta h_j^{d_j})$; against the envelope each class contributes $O(h_j^{d_j})=o(1)$. The exact dependence indicator and $\sum_\iota\ind_\iota$ differ only on $\{n_{cc'}\ge2\}$, so the indicator may be replaced by the sum, which yields the four-event form used in Section~\ref{osec:proofs}, with $\mathcal E_{kl}:=\{\ind_\iota\ind^{\mathrm{sep}}_\iota=1\}$ for the pair $\iota=(k,l)$. Using the margin $2h_j$ rather than $h_j$ makes the vanishing statements below exact: a center displaced by at most $h_j$ from $c_k$ is still more than $h_j$ from every point more than $2h_j$ away from $c_k$.

\movedsec{osec:moved-corr}{S.2.C}{Change of variables and the limiting correlation}
On $\mathcal E_{kl}$ substitute $c'_l=c_k+(u_1h_1,u_2h_2)$ in the coordinates of the shared window, $(u_1,u_2)\in[-1,1]^{d_1}\times[-1,1]^{d_2}$, $dc'_l=\eta\,du_1\,du_2$. Let $q(z;x,x'):=\E[\xi(W;x)\xi(W;x')\mid X=z]$, which is bounded and twice continuously differentiable in $z$ and Lipschitz in $(x,x')$ by Assumptions~\ref{ass:reg}(b),(e) and \ref{ass:overlap}. Conditioning on $(X_1,X_2)$,
\begin{align*}
\eta\,\E\big[\Theta(W;c)\Theta(W;c')\big]={}&\sum_{n,n'\in\{1,2\}}(-1)^{n+n'}\frac1\eta\iint f(z)\,q(z;c_n,c'_{n'})\\
&\times K_1\Big(\frac{c_{n,1}-z_1}{h_1}\Big)K_1\Big(\frac{c'_{n',1}-z_1}{h_1}\Big)K_2\Big(\frac{c_{n,2}-z_2}{h_2}\Big)K_2\Big(\frac{c'_{n',2}-z_2}{h_2}\Big)dz ,
\end{align*}
where $c_{n,1}$ and $c_{n,2}$ are the $X_1$- and $X_2$-coordinates of the $n$th point $c_n$ of $c$, and the kernel product vanishes identically whenever the two centers differ by more than $h_1$ in some $X_1$ coordinate or by more than $h_2$ in some $X_2$ coordinate. On $\mathcal E_{kl}$ the three terms with $(n,n')\neq(k,l)$ vanish: for $(n,n')=(k,l')$, $l'\neq l$, and $(k',l)$, $k'\neq k$, because the corresponding $j$-separations exceed $2h_j$; for $(k',l')$ because $\|c_{k'}-c'_{l'}\|\ge\|c_{k'}-c_k\|-\|c_k-c'_{l'}\|>2h_j-h_j=h_j$ in the $j$th block. In the surviving term substitute $z=c_k-(h_1\tilde z_1,h_2\tilde z_2)$, so that the kernel product becomes $K_1(\tilde z_1)K_1(\tilde z_1+u_1)K_2(\tilde z_2)K_2(\tilde z_2+u_2)$, supported in $[-\tfrac12,\tfrac12]^{d}$ whatever $(u_1,u_2)$; the uniform kernel-convolution expansion (Lemma~\ref{olem:gine}, a first-order Taylor expansion of $f(z)q(z;\cdot)$ about $z=c_k$ with the Lipschitz property of $q$ in its evaluation points) gives
\begin{align*}
\eta\,\E\big[\Theta(W;c)\Theta(W;c')\big]={}&(-1)^{k+l}f(c_k)\,\E\big[\xi(W;c_k)^2\mid X=c_k\big]\\
&\times\int K_1(\tilde z_1)K_1(\tilde z_1+u_1)d\tilde z_1\int K_2(\tilde z_2)K_2(\tilde z_2+u_2)d\tilde z_2+O(h_1+h_2)
\end{align*}
uniformly on $\mathcal E_{kl}$ and in $(u_1,u_2)$. Decomposing the conditional second moment over $\{T=1\}$ and $\{T=0\}$, $\E[\xi(W;x)^2\mid X=x]=\sum_t\Pr(T=t\mid X=x)\varsigma_t^2(x)/P_t(x)^2$, and $f(x)\Pr(T=t\mid X=x)/P_t(x)^2=1/P_t(x)$ by the definition of $P_t$ after \eqref{eq:Pt}, so that
\begin{equation}\label{oeq:rho-id}
f(x)\,\E\big[\xi(W;x)^2\mid X=x\big]\int K_1^2\int K_2^2=\varrho(x),
\end{equation}
and the numerator of $\rho_N(c,c')$ is $(-1)^{k+l}\varrho(c_k)\bar\rho_1(u_1)\bar\rho_2(u_2)+O(h_1+h_2)$. For the denominators, $\eta\kappa(c)$ expands into two squared terms and a cross term; the cross term vanishes identically off the band, and each squared term is the convolution of $f q$ against the even kernel $K_1^2K_2^2$ centered at $c_1$, respectively $c_2$, so the first-order Taylor term drops and $\eta\kappa(c)=\varrho(c_1)+\varrho(c_2)+O(h_1^2+h_2^2)$; for $c'$, whose argument $c'_l$ has been displaced by $(u_1h_1,u_2h_2)$, the continuous differentiability of $\varrho$ returns the arguments to $(c_1',c_2')$ at cost $O(h_1+h_2)$. Since $|\bar\rho_1(u_1)\bar\rho_2(u_2)|\le1$ by Cauchy--Schwarz,
\begin{equation}\label{oeq:rhobound}
\sup|\rho^*_{kl}|\le\frac{\sup\varrho}{\sup\varrho+\inf\varrho}<1 ,
\end{equation}
so $|\rho_N|\le1-\delta$ on the surviving pairs for $N$ large and the pre-factor in \eqref{oeq:BE-gen} is a fixed constant. In the Gaussian comparison the integrand $\eta\sqrt{\kappa_{\mathcal N}(c)\kappa_{\mathcal N}(c')}\{\Cov(\Psi(S_{\mathcal N}),\Psi(S'_{\mathcal N}))-g_\Psi(\rho_N)\}$ is $O(1)\cdot O((N\eta)^{-1/2})$ by \eqref{oeq:BE-gen} and the Rosenthal bound $\E|\mathcal Q_{\mathcal N}|^3=O(\eta^{-1/2})$ for the blocks \eqref{oeq:thirdmom}, integrated over a domain of measure at most $2^{d}\lambda(\mathcal C)\lambda(\Xtwo)$ (for $j=2$), hence $o(1)$ by $N\eta\to\infty$; the limit \eqref{eq:sigma-gen} then follows by bounded convergence, the integrand being bounded by $(2\sup\varrho+o(1))$, the boundary indicators of $\mathcal E_{kl}$ converging almost everywhere, and $g_\Psi$ being continuous.

For $\Psi=\Psiabs$ on the symmetric set $\mathcal C_2^{\mathrm{sym}}$ and $j=2$: in the event $(1,1)$ the free coordinates are $(x_1,x_2,x_2',x_2''')$ with $x_2''=x_2+u_2h_2$, $x_1'=x_1+u_1h_1$, and the limit is
\[
\tfrac14\int g_{\mathrm{abs}}(\rho^*_2)\sqrt{[\varrho(x_1,x_2)+\varrho(x_1,x_2')][\varrho(x_1,x_2)+\varrho(x_1,x_2''')]},
\]
the $\tfrac14$ being the square of the factor $\tfrac12$ in $\Dh_2$, the mirror image of \eqref{eq:D2}; the events $(1,2)$, $(2,1)$, $(2,2)$ carry $g_{\mathrm{abs}}(-\rho^*)$, $g_{\mathrm{abs}}(-\rho^*)$, $g_{\mathrm{abs}}(+\rho^*)$ with the shared point's $\varrho$ in the numerator, and renaming $x_2''\mapsto x_2'''$, interchanging $x_2\leftrightarrow x_2'$, or both, reproduces the same integral, so that by the evenness of $g_{\mathrm{abs}}$ the sum is \eqref{oeq:pf-sigma2}. On an ordered set the four events have different limiting domains, e.g.\ for $j=1$ and $(k,l)=(1,1)$ the domain $\{x_1'\le x_1,\ x_1'''\le x_1\}$ and for $(1,2)$ the domain $\{x_1'\le x_1,\ x_1''\ge x_1\}$, and the signs matter when $b_\Psi\neq0$.

\movedsec{osec:moved-lin}{S.2.D}{The linear part}
Let $w_{\mathcal C}$ be the signed slot multiplicity, so that $\int_{\mathcal C}\Theta(W;c)\,dc=\int_{\dom}w_{\mathcal C}(x)\theta_N(W;x)\,dx$; the substitution $x=X+(h_1\tilde z_1,h_2\tilde z_2)$ and the uniform continuity of $\xi(W;\cdot)$ reduce $\int w_{\mathcal C}\theta_N$, up to its mean, to $\xi(W;X)$ times the kernel average $\int w_{\mathcal C}(X+(h_1\tilde z_1,h_2\tilde z_2))K_1(\tilde z_1)K_2(\tilde z_2)\,d\tilde z$. As $|w_{\mathcal C}|\le\lambda(\mathcal X_j)$, this average converges to $w_{\mathcal C}(X)$ in $L_2$ by the continuity of translation in $L_2$, and $\E[\xi(W;X)^2\mid X]$ is bounded; no continuity of $w_{\mathcal C}$ is needed, which matters because on an ordered set $w_{\mathcal C}$ jumps at the boundary of $\dom$. Hence $\int w_{\mathcal C}\theta_N\to w_{\mathcal C}(X)\xi(W;X)-\E[w_{\mathcal C}(X)\xi(W;X)]$ in $L_2$, and $\E[\xi(W;X)\mid X]=0$, so
\begin{equation}\label{oeq:sigma-lin}
\lim_{N\to\infty}\iint_{\mathcal C^2}\E\big[\Theta(W;c)\Theta(W;c')\big]\,dc\,dc'=\E\big[w_{\mathcal C}(X)^2\xi(W;X)^2\big]=\int_{\dom}w_{\mathcal C}(x)^2\sum_{t}\frac{\varsigma_t^2(x)}{P_t(x)}\,dx ,
\end{equation}
which is $\sigma^2_{j,\mathrm{lin}}(\mathcal C)$,
by the same decomposition as in \eqref{oeq:rho-id}. On a symmetric set $w_{\mathcal C}\equiv0$ and the linear part vanishes identically.

\movedsec{osec:moved-clt}{S.2.E}{Grid, Lyapunov and decorrelation steps of the central limit theorem}
Partition $\mathbb R^{d}$ into the half-open cells $C_{\mathbf i}:=\prod_{l\le d_1}(i_lh_1,(i_l+1)h_1]\times\prod_{l\le d_2}(i_{d_1+l}h_2,(i_{d_1+l}+1)h_2]$, $\mathbf i\in\mathbb Z^d$, of volume $\eta$, and index the array by the cells meeting $\mathcal B$, of cardinality $O(\eta^{-1})$. Put $R_{\mathbf i}:=C_{\mathbf i}\cap\dom$, $\alpha_{\mathbf i}:=\sigma^{-1}\int_{R_{\mathbf i}}\projfield_{\mathcal N}$, with $\sigma:=\sigma_{j,\Psi}(\mathcal C)$ of Lemma~\ref{olem:pf-clt}, and $u_{\mathbf i}:=\{\mathcal N(C_{\mathbf i}\cap\mathcal B)-N\mu(C_{\mathbf i}\cap\mathcal B)\}/\sqrt N$, so that $\Qproj=\sum\alpha_{\mathbf i}$ and $U_{\mathcal N}=\sum u_{\mathbf i}$ exactly; indexing by the cells meeting $\mathcal B$ rather than $\dom$ makes the second identity exact. \emph{One-dependence.} $\projfield_{\mathcal N}(x)$ is a measurable function of $\tau_{\mathcal N}(x)$, which is a functional of the increments over $\mathcal W(x)$, whose half-widths $h_1/2,h_2/2$ are below the cell sides; so for $x\in C_{\mathbf i}$, $\mathcal W(x)$ lies in the union of $C_{\mathbf i}$ and its $3^d-1$ neighbors, and cells at lattice distance at least two have disjoint neighborhoods. \emph{Lyapunov.} By Jensen over $R_{\mathbf i}$ and Lemma~\ref{olem:pf-projection}(d), $\E|\alpha_{\mathbf i}|^3\le\sigma^{-3}\lambda(R_{\mathbf i})^2\int_{R_{\mathbf i}}\E|\projfield_{\mathcal N}|^3=O(\eta^2\cdot\eta\cdot\eta^{-3/2})=O(\eta^{3/2})$, whence $\sum_{\mathbf i}\E|\alpha_{\mathbf i}|^3=O(\eta^{1/2})$; the central third absolute moment of a Poisson variable with mean $n_0$ is $O(n_0+n_0^{3/2})$ and $N\mu(C_{\mathbf i}\cap\mathcal B)=O(N\eta)$, so $\sum_{\mathbf i}\E|u_{\mathbf i}|^3=O(\eta^{1/2})+O(N^{-1/2})$. \emph{Variances.} $\Var(\Qproj)\to1$ by Lemma~\ref{olem:pf-projection}(d) and $\Var(U_{\mathcal N})=1-\pext$ exactly. \emph{Decorrelation.} Fix $x\in\dom$ and split $U_{\mathcal N}=U^{\mathrm{loc}}_x+U^{\mathrm{far}}_x$ at the window $\mathcal W(x)\subset\mathcal B$. The far count is independent of $\tau_{\mathcal N}(x)$, hence of $\projfield_{\mathcal N}(x)$, so $\Cov(\projfield_{\mathcal N}(x),U^{\mathrm{far}}_x)=0$ exactly, and $\|U^{\mathrm{loc}}_x\|_2^2=\mu(\mathcal W(x))\le\sup f\cdot\eta$. The band slice of $\projfield_{\mathcal N}(x)$ has $L_2$-norm $O(h_j^{d_j}\eta^{-1/2})$ (Lemma~\ref{olem:pf-projection}(d)), so its covariance with $U_{\mathcal N}$ is $O(h_j^{d_j})$. The off-band slice is $F(\sqrt{\kappa_{\mathcal N}(x)}S_{\mathcal N}(x))$ with $F(v):=\int_{\{c\ \mathrm{off\ band}\}}\{\E\Psi(\pm v\mp H(c))-\E\Psi(\pm\sqrt{\kappa_{\mathcal N}(x)}S_{\mathcal N}(x)\mp H(c))\}\{\delta_x(c_1)+\delta_x(c_2)\}dc$, $H(c)$ the independent other-window term, a deterministic function with Lipschitz constant at most $\lambda(\mathcal C_x)$, where $\mathcal C_x:=\{c\in\mathcal C:c_1=x\text{ or }c_2=x\}$ is the fibre of configurations through $x$, because $\Psi$ is $1$-Lipschitz, so $v\mapsto F(\sqrt{\kappa_{\mathcal N}(x)}v)$ has Lipschitz constant $O(\eta^{-1/2})$. Both $S_{\mathcal N}(x)$ and $U_{\mathcal N}$ are standardized sums of the same $N$ compound-Poisson blocks, with block third moments $O(\eta^{-1/2})$ and $O(1)$; Theorem~1 of \citet{sweeting1977speeds}, applied to $(s,t)\mapsto F(\sqrt\kappa s)t$, gives
\[
\big|\Cov(F(\sqrt\kappa S_{\mathcal N}(x)),U_{\mathcal N})-\Cov(F(\sqrt\kappa Z_1),Z_2)\big|=O(\eta^{-1/2})\cdot O\big(\eta^{-1/2}N^{-1/2}\big)=o(1),
\]
$(Z_1,Z_2)$ standard bivariate normal with correlation $\rho^U(x):=\Cov(S_{\mathcal N}(x),U_{\mathcal N})$. By the law of total covariance for the Poisson process, and since the Poissonized field \eqref{oeq:pf-poisson-field} Poissonizes the uncentered summand $a(W;x):=\eta^{-1}\xi(W;x)K_1K_2$, whose kernel factor is supported in $\mathcal W(x)\subset\mathcal B$, $N\Cov(\tau_{\mathcal N}(x),\mathcal N(\mathcal B))=\E[a(W;x)\ind\{X\in\mathcal B\}]=\E[a(W;x)]=O(h_1^{s_1}+h_2^{s_2})$, the kernel-weighted mean of $\xi(W;x)$ being the smoothing bias; hence $\rho^U(x)=O((h_1^{s_1}+h_2^{s_2})\eta^{1/2})$ uniformly. (The centered Poissonization $\tau(x)+B_N(x)+N^{-1}\sum_{i\le\mathcal N}\theta_N(W_i;x)$ would differ from $\tau_{\mathcal N}(x)$ by $(\mathcal N/N-1)B_N(x)$, which involves the exterior count; the far count would then not be exactly independent of $\tau_{\mathcal N}(x)$, and (b) below would fail. With the uncentered summands, both hold exactly, at the cost of the additive term $B_{j,N}(c)B_{j,N}(c')=O((h_1^{s_1}+h_2^{s_2})^2)$ in the Poissonized second moments, Lemma~\ref{olem:poisson}.) Stein's identity for Lipschitz $F$ gives $\Cov(F(\sqrt\kappa Z_1),Z_2)=\rho^U\E[\sqrt\kappa F'(\sqrt\kappa Z_1)]$ with $|F'|\le\lambda(\mathcal C_x)$, so this covariance is $O((h_1^{s_1}+h_2^{s_2})\eta^{1/2})\cdot O(\eta^{-1/2})=o(1)$ uniformly in $x$, using only $|\Psi'|\le1$. Integrating over $\dom$ gives $\Cov(\Qproj,U_{\mathcal N})\to0$. (If instead{} the Poissonized field carries the level $\tau(x)$ in its summands so that $\rho^U=\E[\zeta(W;x)]/\sqrt\kappa=O(\eta^{1/2})$, the same step is closed by the symmetry of the normal law and the oddness of $d|u|/du$; adding the level deterministically makes that argument unnecessary and covers every $\Psi$ of the class.) \emph{Conclusion.} For fixed $(a,b)$ the array $\{a\alpha_{\mathbf i}+bu_{\mathbf i}\}$ is one-dependent with variance converging to $a^2+b^2(1-\pext)$ and vanishing Lyapunov ratio, so the central limit theorem for $k$-dependent random fields of \citet{shergin1990central}{} and the Cram\'er--Wold device give Lemma~\ref{olem:pf-clt}(a); (b) holds because $D^P_j$, the Poissonized field being local, and $U_{\mathcal N}$ are measurable with respect to the increments over $\mathcal B$ and $\Uext$ with respect to those over $\mathcal B^c$.

\movedsec{osec:moved-lemmas}{S.2.F}{Statements of the Poissonization, centering, variance and de-Poissonization lemmas}
The following four lemmas are referred to by number from Section~\ref{osec:proofs}; their proofs are given in the preceding subsections of this section (Poissonization and moment identities in Section~\ref{osec:aux-poisson}, centering in Section~\ref{osec:aux-center}, variance in Sections~\ref{osec:moved-BE}--\ref{osec:moved-lin} above together with Section~\ref{osec:aux-var}, de-Poissonization in Section~\ref{osec:aux-depoisson}). Assumptions~\ref{ass:sampling}--\ref{ass:overlap}, \ref{ass:psi} and \ref{ass:reg} of the main text are in force, $\mathcal C_j$ is a configuration set of direction $j$, $\Gamma_{\mathcal N}$ and $D^P_j$ are the Poissonized field and statistic of equation~\eqref{oeq:pf-poisson-field}, $\kappa_{j,N}$ is the variance defined in Section~\ref{sec:asym-center}, $m^\kappa_{j,N}$ is \eqref{eq:center}, $\ah^{\,\kappa}_j:=m_\Psi\int_{\mathcal C_j}\widehat\kappa_j^{1/2}$ is the plug-in centering of Section~\ref{osec:aux-center}, formed from the estimated influence function (the centering $\widetilde m_j$ of \eqref{eq:center-feasible}, formed from the kernel weights and corrected residuals, is treated in Section~\ref{osec:from-main}), and $\mathcal W(x)$ is the kernel window at $x$.

\begin{lemma}[Poissonization]\label{olem:poisson}
For $c,c'\in\mathcal C_j$,
\[
N\Cov(\Gamma_{\mathcal N}(c),\Gamma_{\mathcal N}(c'))=\E[\Theta_j(W;c)\Theta_j(W;c')]+B_{j,N}(c)B_{j,N}(c')\quad\text{exactly},
\]
with $\sup|B_{j,N}|^2=O((h_1^{s_1}+h_2^{s_2})^2)$, $\eta\kappa_{j,N}(c)\le4\sup\varrho+o(1)$ uniformly on $\mathcal C_j$, and $\eta\kappa_{j,N}(c)=\varrho(c_1)+\varrho(c_2)+O(h_1^2+h_2^2)$ uniformly off the band. $\tau_{\mathcal N}(x)$ is a deterministic level plus a functional of the Poisson increments over $\mathcal W(x)$, so fields with disjoint windows are independent, and the standardized fields $S_{\mathcal N}(c):=\sqrt N\{\Gamma_{\mathcal N}(c)-\E\Gamma_{\mathcal N}(c)\}/\sqrt{\kappa_{\mathcal N}(c)}$, $\kappa_{\mathcal N}:=\kappa_{j,N}+B_{j,N}^2$, are, jointly over finitely many configurations, standardized sums of $N$ i.i.d.\ compound-Poisson blocks with third absolute moments $O(\eta^{-1/2})$.
\end{lemma}
\noindent\hyperref[oprf:poisson]{\emph{Proof}} (Section~\ref{osec:aux-poisson}).

\begin{lemma}[centering]\label{olem:centering}
If $\Gamma_j\equiv0$ on $\mathcal C_j$, then
\[
\E[D^P_j]-m^\kappa_{j,N}(\mathcal C_j)=O\big(\sqrt N(h_1^{s_1}+h_2^{s_2})\big)+O\big((N\eta^2)^{-1/2}\big)=o(1).
\]
With or without the null, $\ah^{\,\kappa}_j(\mathcal C_j)-m^\kappa_{j,N}(\mathcal C_j)=o_p(1)$, with no condition on the relative sizes of $h_1$ and $h_2$ beyond Assumption~\ref{ass:reg}(g),(h). The plug-in of the limit form $m_2$ of \eqref{oeq:a2}{} may replace $\ah^{\,\kappa}_2$ if in addition $h_1^2+h_2^2=o(\eta^{1/2})$ and $h_2^{d_2}=o(h_1^{d_1})$, and the reverse anisotropy condition is required for $\Dh_1$.
\end{lemma}
\noindent\hyperref[oprf:centering]{\emph{Proof}} (Section~\ref{osec:aux-center}).

\begin{lemma}[Variance, general form]\label{olem:variance-gen}
With $$\rho_N(c,c'):=\E[\Theta_j(W;c)\Theta_j(W;c')]/\sqrt{\kappa_{j,N}(c)\kappa_{j,N}(c')}$$ and $\Gamma_j\equiv0$ on $\mathcal C_j$,
\[
\Var(D^P_j)=\int_{\mathcal C_j}\int_{\mathcal C_j}g_\Psi\big(\rho_N(c,c')\big)\sqrt{\kappa_{j,N}(c)\kappa_{j,N}(c')}\,dc\,dc'+o(1),
\]
and the double integral converges, its limit $\sigma^2_{j,\Psi}(\mathcal C_j)$ of \eqref{eq:sigma-gen} being $a_\Psi^2\sigma^2_{j,\mathrm{abs}}(\mathcal C_j)+b_\Psi^2\sigma^2_{j,\mathrm{lin}}(\mathcal C_j)$; the same holds for the Poissonized noise field $\Gamma_{\mathcal N}-\Gamma_j$ without the null.
\end{lemma}
\noindent\hyperref[oprf:variance-gen]{\emph{Proof}} (Section~\ref{osec:aux-var}, the proof of the explicit form Lemma~\ref{olem:variance}, with the general-$\Psi$ computations of Sections~\ref{osec:moved-BE}--\ref{osec:moved-lin}).

\begin{lemma}[De-Poissonization]\label{olem:depoisson}
Let $U_{\mathcal N}$ and $\Uext$ be the standardized interior and exterior counts of Lemma~\ref{olem:pf-clt}. If $(\mathcal T_{\mathcal N},U_{\mathcal N})$ is independent of $\Uext$ for each $N$ and $(\mathcal T_{\mathcal N},U_{\mathcal N})\dto(\sigma Z_1,\sqrt{1-\pext}Z_2)$ with $Z_1,Z_2$ independent standard normals, then $\Pr(\mathcal T_{\mathcal N}\le t\mid\mathcal N=N)\to\Pr(\sigma Z_1\le t)$ for every $t$ \citep[Lemma~2.4]{gine2003bm}.
\end{lemma}
\noindent No proof is given: the lemma is Lemma~2.4 of \citet{gine2003bm}, there attributed to \citet{beirlant1995asymptotic}, and is used as stated in Section~\ref{osec:aux-depoisson}.

\movedsec{osec:moved-extension}{S.2.G}{Extension from the interior sets to $\dom$}
The Poissonized statements are first proved on $\mathcal C\cap(\mathcal B_1\times\mathcal B_2)^{(2)}$, the configurations in $\mathcal C$ whose two evaluation points both lie in $\mathcal B_1\times\mathcal B_2$, for compact product sets $\mathcal B_1\times\mathcal B_2$ whose windows lie in a box $\mathcal B$ with $\mu(\mathcal B^c)=:\pext\in(0,1)$ and $\mu(\dom)-\mu(\mathcal B_1\times\mathcal B_2)<\epsilon$, which exist by compactness of $\dom$ and the lower bound on $f$. On the complement, the mean of the statistic minus its centering is $o(1)$ uniformly in $\epsilon$, by the Berry--Esseen bound \eqref{oeq:BE1} and the drift bound of the proof of \eqref{oeq:center-i}, whose constants depend only on the kernels and the moment bounds of Assumption~\ref{ass:reg}. Its contribution to the variance is $O(\epsilon)+o(1)$: by the band-valid envelopes $\eta\kappa\le4\sup\varrho+o(1)$ and $|g_\Psi|\le1$, as for each $c$ the pairs with overlapping windows have measure $O(\eta)$, while pairs with disjoint windows contribute $o(1)$ as in Step~2(i) of the proof of Lemma~\ref{prop:sigmaN}, and the conclusion of Lemma~\ref{olem:depoisson} is $N(0,1)$ whatever the value of $\pext$; the standard diagonal argument in $\epsilon\downarrow0$ extends the fixed-sample limit to $\dom$.
\section{Estimated nuisance parameters and uniformity}\label{osec:nui}
This section supplements Section~\ref{sec:asym-nuisance} and Section~\ref{osec:proofs-nui}. Section~\ref{osec:nuisance-ext} gives the linearization of the Nadaraya--Watson field with a drift, the auxiliary lemmas used in the proofs of Lemma~\ref{lem:drift} and Theorems~\ref{thm:nuisance-param} and \ref{thm:nuisance-split}, and the extensions to the max form and to the magnitudes; Section~\ref{osec:uniform} shows the uniformity of the oracle results of Section~\ref{osec:aux} along sequences of laws, on which Theorem~\ref{thm:nuisance-split} rests.
\subsection{Estimated nuisance parameters: linearization and extensions}\label{osec:nuisance-ext}
This subsection supplements Section~\ref{sec:asym-nuisance} and Section~\ref{osec:proofs-nui}, whose notation it uses: $j$ is fixed, $\mathcal C:=\mathcal C_j$, $\mathcal K(x;X):=K_1((x_1-X_1)/h_1)K_2((x_2-X_2)/h_2)$, $\mathcal K_i(x):=\mathcal K(x;X_i)$, $h^s:=h_1^{s_1}+h_2^{s_2}$, $h^{2s}:=h_1^{2s_1}+h_2^{2s_2}$, $\|\cdot\|_1,\|\cdot\|_2$ are Lebesgue norms on $\dom$, $\mathcal U$ is a neighborhood of $\dom$ on which $f\ge\underline p$, and $\theta_N$, $\Theta_j$ are formed with the score $\xi(W;x):=\{\zeta-\tau(x)\}/f(x)$. It gives the linearization of the Nadaraya--Watson field with a drift used in the proof of Theorem~\ref{thm:nuisance-split}, three auxiliary lemmas used in the proofs of Lemma~\ref{lem:drift} and Theorems~\ref{thm:nuisance-param} and \ref{thm:nuisance-split}, and the statements and proofs of the extensions of Theorems~\ref{thm:nuisance-param} and \ref{thm:nuisance-split} to the max form under the monotone null and to the magnitudes.

\begin{lemma}[Linearization with a drift]\label{olem:nui-lin}
Let $\{(\zeta_i,X_i)\}_{i\le N}$ be i.i.d.\ from a law $P_N$ under which $X$ has the density $f$ of Assumption~\ref{ass:reg}(b),(e), $\E_N[\zeta\mid X]=\tau+\bar\pi_N$ with $\tau$ as in Assumption~\ref{ass:reg}(e), and $\sup_{x\in\mathcal U}\E_N[\{\zeta-\tau(X)\}^2\mid X=x]\le C_\xi$. Put $B^\tau_N(x):=\E[\{\tau(X)-\tau(x)\}\mathcal K(x;X)]/\{\eta f(x)\}$ and $\delta_N(x):=\E_N[\bar\pi_N(X)\mathcal K(x;X)]/\{\eta f(x)\}$. Then
\[
\tauh(x)-\tau(x)=N^{-1}\sum_i\theta_N(W_i;x)+B^\tau_N(x)+\delta_N(x)+R_N(x)
\]
on $\dom$, with $\theta_N$ formed under $P_N$ and $\sup_{\dom}|B^\tau_N|=O(h^s)$; moreover $\|\delta_N\|_{L_p(\dom)}\le C\|\bar\pi_N\|_{L_p(\mathcal U)}$ for $p=1,2,\infty$, and
\begin{equation}\label{oeq:nui-R}
\sqrt N\int_{\dom}|R_N|=O_p\Big(\frac{1}{\sqrt N\eta}+\frac{h^s}{\sqrt\eta}+\sqrt Nh^{2s}+\Big(\frac{1}{\sqrt\eta}+\sqrt Nh^s\Big)\|\delta_N\|_2\Big),
\end{equation}
the $O_p$ depending on $P_N$ only through $C_\xi$. If $N\eta^{1/2}\|\delta_N\|_2^2=O(1)$ and $\Gamma_j\equiv0$ on $\mathcal C$, then $\Dh_j^\Psi(\mathcal C)-D^{\mathrm{lin}}_j(\delta_N)=o_p(1)$, with $D^{\mathrm{lin}}_j(\cdot)$ of Lemma~\ref{lem:drift} formed under $P_N$ and $B_{j,N}(c):=B^\tau_N(c_1)-B^\tau_N(c_2)$.
\end{lemma}

\begin{proof}
The law of $X$ is that of Assumption~\ref{ass:reg}, so $\sup_{\dom}|\widehat f-f|=O_p(\epsilon_N)$ by Lemma~\ref{olem:pf-linearization}(a) and $\mathcal E_N':=\{\inf_{\dom}\widehat f\ge\inf f/2\}$ has probability tending to one. On $\mathcal E_N'$, $\tauh-\tau=A/\widehat f$ with $A(x):=(N\eta)^{-1}\sum_i\{\zeta_i-\tau(x)\}\mathcal K_i(x)$; by definition $\E A=f(B^\tau_N+\delta_N)$ and $A-\E A=fN^{-1}\sum_i\theta_N(W_i;\cdot)$, so the identity holds with $R_N:=-A(\widehat f-f)/(f\widehat f)$, and
\[
|R_N|\le C\{|N^{-1}\textstyle\sum_i\theta_N(W_i;\cdot)|+|B^\tau_N|+|\delta_N|\}|\widehat f-f|.
\]
Now $\E\|N^{-1}\sum_i\theta_N(W_i;\cdot)\|_2^2\le N^{-1}\eta^{-2}\int\E[\xi(W;x)^2\mathcal K(x;X)^2]dx\le CC_\xi(N\eta)^{-1}$ and $\E\|\widehat f-f\|_2^2\le C\{(N\eta)^{-1}+h^{2s}\}$ by the kernel orders and the smoothness of $f$; the Cauchy--Schwarz inequality gives \eqref{oeq:nui-R}, the bias part contributing $\sqrt N\sup|B^\tau_N|\,\E\|\widehat f-f\|_1=O(h^s\eta^{-1/2}+\sqrt Nh^{2s})$. The bias bound is the standard expansion of $\int(\tau f)(z)\mathcal K(x;z)dz/\eta-\tau(x)\int f(z)\mathcal K(x;z)dz/\eta$, and the bounds on $\delta_N$ follow from $|\delta_N|\le(\sup f/\inf f)\,|\bar\pi_N|*|\eta^{-1}\mathcal K|$ and Young's inequality in $L_1$, $L_2$ and $L_\infty$, the windows lying in $\mathcal U$ for $N$ large. Under $N\eta^{1/2}\|\delta_N\|_2^2=O(1)$ the right side of \eqref{oeq:nui-R} is $o_p(1)$: $(N\eta^2)^{-1/2}\to0$, $h^s\eta^{-1/2}=\sqrt Nh^s(N\eta)^{-1/2}\to0$, $\sqrt Nh^{2s}\to0$ as $Nh^{2s}\to0$, $\eta^{-1/2}\|\delta_N\|_2=O((N\eta^{3/2})^{-1/2})$ and $\sqrt Nh^s\|\delta_N\|_2=O(h^s\eta^{-1/4})$. Finally $\Gammah_j(c)=B_{j,N}(c)+\delta^\Gamma_N(c)+N^{-1}\sum_i\Theta_j(W_i;c)+R_N(c_1)-R_N(c_2)$ under the null, and the Lipschitz property gives $|\Dh_j^\Psi(\mathcal C)-D^{\mathrm{lin}}_j(\delta_N)|\le2\lambda(\mathcal X_j)\sqrt N\int_{\dom}|R_N|$.
\end{proof}

\begin{lemma}[De-Poissonization of a perturbation]\label{olem:nui-depois}\hfill\break
Under the conditions of Lemma~\ref{lem:drift}, in the notation of its proof in Section~\ref{osec:from-main}, if $\Var(\Delta^P)\to0$ then $D^{\mathrm{lin}}_j(\delta_N)-D^{\mathrm{lin}}_j(0)-\E\Delta^P=o_p(1)$.
\end{lemma}

\begin{proof}
If $a_\Psi=0$, then $\Psi$ is linear, so $\Delta^P=b_\Psi\sqrt N\int_{\mathcal C}\delta^\Gamma_N$ is deterministic and the claim is immediate; otherwise $\sigma^2_{j,\Psi}(\mathcal C)\ge a_\Psi^2\sigma^2_{j,\mathrm{abs}}(\mathcal C)>0$ by Theorem~\ref{thm:variance}. Let $\sigma_N^2:=\Var(D^P(0))\to\sigma^2_{j,\Psi}(\mathcal C)>0$ and $\mathcal Z^0_{\mathcal N}:=\{D^P(0)-\E D^P(0)\}/\sigma_N$. By Lemmas~\ref{olem:pf-projection}(a),(c) and \ref{olem:pf-clt}, $(\mathcal Z^0_{\mathcal N},U_{\mathcal N})\dto(Z_1,\sqrt{1-p_{\mathcal B}}Z_2)$ with $Z_1,Z_2$ independent standard normals, and since $\Var(\Delta^P)\to0$ the same holds, for every fixed $\lambda\in\mathbb R$, for $(\mathcal Z^0_{\mathcal N}+\lambda\{\Delta^P-\E\Delta^P\}/\sigma_N,\,U_{\mathcal N})$. This pair is a functional of the increments over the box $\mathcal B$ of Lemma~\ref{olem:pf-clt}, since the Poissonized noise field $\Gamma^0_{\mathcal N}$ is local, its uncentered summands vanish outside the windows and its deterministic part does not involve the count, and $\delta^\Gamma_N$ is deterministic; it is therefore independent of the exterior count $U'_{\mathcal N}$. (The centered Poissonization $B_{j,N}+N^{-1}\sum_{i\le\mathcal N}\Theta_j(W_i;\cdot)$ would differ from $\Gamma^0_{\mathcal N}$ by $(\mathcal N/N-1)\bar A$, $\bar A$ the mean of the uncentered summand, which depends on the total count; this is why the uncentered summands are Poissonized.) Lemma~\ref{olem:depoisson} yields, under the fixed-sample law,
\[
\mathcal Z^0_N+\lambda R^\delta_N\dto N(0,1)\quad\text{for every }\lambda\in\mathbb R,
\]
where $\mathcal Z^0_N:=\{D^{\mathrm{lin}}_j(0)-\E D^P(0)\}/\sigma_N$ and $R^\delta_N:=\{D^{\mathrm{lin}}_j(\delta_N)-D^{\mathrm{lin}}_j(0)-\E\Delta^P\}/\sigma_N$.
With $\lambda=0$, $\mathcal Z^0_N=O_p(1)$. For $\epsilon>0$ and $\lambda>0$, $\Pr(|R^\delta_N|>\epsilon)\le\Pr(|\mathcal Z^0_N+\lambda R^\delta_N|>\lambda\epsilon/2)+\Pr(|\mathcal Z^0_N|>\lambda\epsilon/2)\to2\Pr(|Z_1|>\lambda\epsilon/2)$, which is arbitrarily small for $\lambda$ large. Hence $R^\delta_N=o_p(1)$, and $\sigma_N\to\sigma_{j,\Psi}(\mathcal C)$ gives the claim.
\end{proof}

\begin{lemma}[centering and scale under a parametric first stage]\label{olem:nui-scale}
Let $\mathcal U'$ be a neighborhood of $\dom$ whose closure lies in $\mathcal U$, and $\mathcal W(x)$ the kernel window at $x$. (a) If $Z_i\ge0$ and $\sup_{x\in\mathcal U}\E[Z^2\mid X=x]<\infty$, then $\sup_{x\in\mathcal U'}(N\eta)^{-1}\sum_iZ_i\ind\{X_i\in\mathcal W(x)\}=O_p(1)$; in particular $\sup_{x\in\mathcal U'}(N\eta)^{-1}\sum_iZ_i|\mathcal K_i(x)|=O_p(1)$. (b) Under the conditions of Theorem~\ref{thm:nuisance-param}, in the notation of its proof in Section~\ref{osec:proofs-nui}, let $\widetilde m_{j*}(\mathcal C)$ and $\widetilde\sigma^2_{j,\Psi,N*}(\mathcal C)$ be the centering \eqref{eq:center-feasible} and the variance \eqref{eq:sigmaN} computed from $\zeta^*_i=\zeta(W_i;\chi_{\vartheta^*})$. Then
\[
\widetilde m_j(\mathcal C)-\widetilde m_{j*}(\mathcal C)=O_p\big((N\eta)^{-1/2}\big),\qquad
\widetilde\sigma^2_{j,\Psi,N}(\mathcal C)-\widetilde\sigma^2_{j,\Psi,N*}(\mathcal C)=O_p\big(N^{-1/4}\big).
\]
Consequently, as $\zeta(W;\chi_{\vartheta^*})$ satisfies the conditions of Proposition~\ref{prop:transfer} by Assumption~\ref{ass:firststage}(a)(iv), Lemmas~\ref{prop:centering-corr} and \ref{prop:sigmaN} give $\widetilde m_j(\mathcal C)-m^\kappa_{j,N}(\mathcal C)=o_p(1)$ and $\widetilde\sigma^2_{j,\Psi,N}(\mathcal C)\pto\sigma^2_{j,\Psi}(\mathcal C)$, with $m^\kappa_{j,N}$ and $\sigma^2_{j,\Psi}$ formed from $\zeta(W;\chi_{\vartheta^*})$; the limit is the variance of the statistic under $\Gamma_j\equiv0$ on $\mathcal C$ and that of the noise-field statistic otherwise.
\end{lemma}

\begin{proof}
\emph{(a)} Partition $\mathbb R^d$ into cells of sides $h_1/2$ in the $X_1$ coordinates and $h_2/2$ in the $X_2$ coordinates. A window, of sides $h_1$ and $h_2$, meets at most $3^d$ cells, and for $N$ large the $O(\eta^{-1})$ cells meeting a window centered in $\mathcal U'$ lie in $\mathcal U$. With $A_{\mathrm{cell}}:=(N\eta)^{-1}\sum_iZ_i\ind\{X_i\in\mathrm{cell}\}$, $\sup_{x\in\mathcal U'}(N\eta)^{-1}\sum_iZ_i\ind\{X_i\in\mathcal W(x)\}\le3^d\max A_{\mathrm{cell}}$, where $\E A_{\mathrm{cell}}\le C$ and $\Var A_{\mathrm{cell}}\le(N\eta)^{-2}N\E[Z^2\ind\{X\in\mathrm{cell}\}]\le C(N\eta)^{-1}$; hence $\E\max_{\mathrm{cell}}(A_{\mathrm{cell}}-\E A_{\mathrm{cell}})^2\le C\eta^{-1}(N\eta)^{-1}=C(N\eta^2)^{-1}\to0$, and $\max A_{\mathrm{cell}}=O_p(1)$. The second claim follows from
\[
|\mathcal K_i(x)|\le\sup|K_1K_2|\,\ind\{X_i\in\mathcal W(x)\}.
\]

\emph{(b)} Work on $\mathcal E_N\cap\{\inf_{\mathcal U'}\widehat f\ge\inf f/2\}$, whose probability is at least $1-2\epsilon$ for $N$ large, and take $N$ so large that $\|\delta_\vartheta\|\le1$ there and that the windows of points of $\dom$ lie in $\mathcal U'$. The weights $\ell_i$, $L_{il}$ and the divisors $q_i$ do not involve the response, and $q_i\ge\tfrac12$ for all $i$ with $X_i\in\mathcal U'$ with probability tending to one (Section~\ref{osec:from-main}). By Taylor's theorem, $|\widehat\zeta_i-\zeta^*_i|\le\|\delta_\vartheta\|(\|\zeta_{\vartheta,i}\|+\|\delta_\vartheta\|M(W_i))\le\|\delta_\vartheta\|\bar M_i$ with $\bar M_i:=\|\zeta_{\vartheta,i}\|+M(W_i)$. Let $\Lambda_N:=\sup_{x\in\mathcal U'}(N\eta)^{-1}\sum_l(1+|\zeta^*_l|+\bar M_l)^2\ind\{X_l\in\mathcal W(x)\}$, which is $O_p(1)$ by (a), the fourth conditional moments of $\zeta^*$ and $M(W)$ being bounded. For $X_i\in\mathcal U'$, $\sum_l|L_{il}|\bar M_l\le\{N\eta\widehat f(X_i)\}^{-1}\sum_l|\mathcal K_l(X_i)|\bar M_l\le C\Lambda_N$ and likewise $\sum_l|L_{il}||\zeta^*_l|\le C\Lambda_N$, so the residuals $\widehat e_i$ and $\widehat e_{i*}$ formed from $\widehat\zeta_i$ and $\zeta^*_i$ satisfy
\[
|\widehat e_i-\widehat e_{i*}|\le\|\delta_\vartheta\|(\bar M_i+C\Lambda_N),\qquad|\widehat e_{i*}|\le|\zeta^*_i|+C\Lambda_N .
\]
With $\ell^\Gamma_i(c):=\ell_i(c_1)-\ell_i(c_2)$, $N \ell^\Gamma_i(c)^2\le CN^{-1}\eta^{-2}\ind\{X_i\in\mathcal W(c_1)\cup\mathcal W(c_2)\}$. Let $\mathbf z(c):=(N^{1/2}\ell^\Gamma_i(c)\widehat e_iq_i^{-1/2})_{i\le N}$ and $\mathbf z_*(c)$ the same vector with $\widehat e_{i*}$, so that $\widetilde\kappa_j(c)=\|\mathbf z(c)\|^2$, $\widetilde\kappa_j(c,c')=\mathbf z(c)'\mathbf z(c')$, and likewise for the starred quantities. By (a), uniformly in $c\in\mathcal C$,
\begin{gather*}
\|\mathbf z(c)-\mathbf z_*(c)\|^2\le\frac{C\|\delta_\vartheta\|^2}{\eta}\sup_{x\in\mathcal U'}\frac1{N\eta}\sum_i(\bar M_i+C\Lambda_N)^2\ind\{X_i\in\mathcal W(x)\}=O_p\Big(\frac1{N\eta}\Big),\\
\|\mathbf z_*(c)\|^2+\|\mathbf z(c)\|^2=O_p\Big(\frac1\eta\Big).
\end{gather*}
\emph{centering.} As $|\|\mathbf z\|-\|\mathbf z_*\||\le\|\mathbf z-\mathbf z_*\|$, $\sup_c|\widetilde\kappa_j(c)^{1/2}-\widetilde\kappa_{j*}(c)^{1/2}|=O_p((N\eta)^{-1/2})$, and integration over $\mathcal C$ gives the first claim. \emph{Variance.} By the Cauchy--Schwarz inequality, $|\mathbf z(c)'\mathbf z(c')-\mathbf z_*(c)'\mathbf z_*(c')|\le\|\mathbf z(c)-\mathbf z_*(c)\|\|\mathbf z(c')\|+\|\mathbf z_*(c)\|\|\mathbf z(c')-\mathbf z_*(c')\|$, so $\sup_{c,c'}\|\widetilde\Sigma(c,c')-\widetilde\Sigma_*(c,c')\|=O_p(N^{-1/2}\eta^{-1})$ for the matrices $\widetilde\Sigma$ of the proof of Lemma~\ref{prop:sigmaN}, whose traces are $O_p(\eta^{-1})$. If $c'\notin\mathcal D(c)$ both off-diagonal entries vanish and both integrands are zero; on $\{c'\in\mathcal D(c)\}$, of measure at most $C\lambda(\mathcal C)\eta$ by (F3) of Section~\ref{osec:from-main}, the bound \eqref{oeq:H-lip} gives $|H(\widetilde\Sigma)-H(\widetilde\Sigma_*)|\le\sqrt2\,O_p(\eta^{-1/2})\,O_p(N^{-1/4}\eta^{-1/2})$ uniformly. Integrating gives $|\widetilde\sigma^2_{j,\Psi,N}-\widetilde\sigma^2_{j,\Psi,N*}|=O_p(N^{-1/4})$. The consequences follow from Lemmas~\ref{prop:centering-corr} and \ref{prop:sigmaN} for $\zeta^*$, whose proofs (Section~\ref{osec:from-main}) supply $\widetilde m_{j*}-m^X_{j,N}=o_p(1)$, $m^X_{j,N}-m^\kappa_{j,N}=o_p(1)$ and $\widetilde\sigma^2_{j,\Psi,N*}-\bar\sigma^2_N=o_p(1)$ with $\bar\sigma^2_N\to\sigma^2_{j,\Psi}(\mathcal C)$.
\end{proof}

\begin{lemma}[Deterministic first stage]\label{olem:nui-seq}
Let the conditions of Theorem~\ref{thm:nuisance-split} hold, let $\chi_N$ be a deterministic sequence of nuisance values satisfying Assumption~\ref{ass:firststage}(b)(ii) and deterministic versions of (b)(iii),(iv), $\zeta_N:=\zeta(W;\chi_N)$, $\bar\pi_N:=\E[\zeta_N-\zeta\mid X]$, and $P_N$ the law of $(\zeta_N,X)$. Then $P_N$ satisfies hypotheses (i)--(iv) of Lemma~\ref{olem:uniform} with $\varsigma_\zeta^2:=\Var(\zeta\mid X)$, and $\sup_{\mathcal U}|\bar\pi_N|=o(\eta^{1/2})$.
\end{lemma}

\begin{proof}
The law of $X$ does not depend on $N$. By Proposition~\ref{prop:id}, $\E[\zeta\mid X]=\tau$, so $\E[\zeta_N\mid X]=\tau+\bar\pi_N$, where $\tau$ and $f$ have the smoothness of Assumption~\ref{ass:reg}(e) by (b)(v) and Proposition~\ref{prop:transfer}, and $\sup_{\mathcal U}|\bar\pi_N|=o(\eta^{1/2})$ by (b)(iv). By Minkowski's inequality for conditional standard deviations, with $\varsigma_{\zeta,N}(x)$ the conditional standard deviation of $\zeta_N$ given $X=x$, $\sup_{\mathcal U}|\varsigma_{\zeta,N}-\varsigma_\zeta|\le\sup_{x\in\mathcal U}\E[(\zeta_N-\zeta)^2\mid X=x]^{1/2}\to0$ by (b)(iii), with $\varsigma_\zeta^2$ continuous and bounded away from zero on $\mathcal U$ by Assumption~\ref{ass:firststage}; replacing $\mathcal U$, if necessary, by a neighborhood of $\dom$ whose closure lies in $\mathcal U$, which keeps $f\ge\underline p$ and the suprema in (b)(iii),(iv), $\varsigma_\zeta^2$ is continuous on $\bar{\mathcal U}$; as $\varsigma_\zeta^2$ is bounded, $\sup_{\mathcal U}|\varsigma_{\zeta,N}^2-\varsigma_\zeta^2|\to0$. Finally $|\zeta_N|\le2C+(|Y|+C)/\underline p$ by (b)(ii), so the bounded sixth conditional moments of $Y$ bound those of $\zeta_N$ uniformly in $N$.
\end{proof}

Let
\[
\omega_1(x):=\int_{\Xone}\mathrm{sgn}\{\tau(x_1,x_2)-\tau(x_1',x_2)\}dx_1',
\]
$\omega_2$ its mirror image, and $\xi^*(W;x):=\{\zeta(W;\chi_{\vartheta^*})-\tau(x)\}/f(x)$.

\begin{proposition}[Extensions]\label{oprop:nui-ext}
(a) Let the conditions of Theorem~\ref{thm:nuisance-param} hold, let $\sqrt N(\widehat\vartheta-\vartheta^*)=N^{-1/2}\sum_i\mathrm{IF}_\vartheta(W_i)+o_p(1)$ with $\E\mathrm{IF}_\vartheta(W)=0$ and $\E\|\mathrm{IF}_\vartheta(W)\|^3<\infty$, and let the conditions of Proposition~\ref{prop:gini-clt} hold for $\zeta(W;\chi_{\vartheta^*})$. Then $\sqrt N(\Dh_j/\sqrt N-D_j)\dto N(0,\Omega_j^\vartheta)$ with $\Omega_j^\vartheta:=\Var\{\omega_j(X)\xi^*(W;X)+\mathrm{IF}_\vartheta(W)'\int_{\dom}\omega_j\tau_\vartheta\}$, provided $\Omega_j^\vartheta>0$. If both working models are correct, $\tau_\vartheta\equiv0$ and $\Omega_j^\vartheta=\Omega_j$. If, in addition, $\widehat{\mathrm{IF}}_i$ are estimates with $N^{-1}\sum_i\|\widehat{\mathrm{IF}}_i-\mathrm{IF}_\vartheta(W_i)\|^2=o_p(1)$, then $\widehat\Omega^\vartheta_{j,N}:=N^{-1}\sum_i\{\widehat u_i\widehat e_iq_i^{-1/2}+\widehat g'\widehat{\mathrm{IF}}_i\}^2\pto\Omega_j^\vartheta$, where $\widehat u_i:=N\int\widehat\omega_j\ell_i$, so that $\widehat\Omega_{j,N}=N^{-1}\sum_i\widehat u_i^{\,2}\widehat e_i^{\,2}/q_i$, $\widehat g:=N^{-1}\sum_i\widehat u_i\,\partial_\vartheta\zeta(W_i;\chi_{\widehat\vartheta})$, and the residuals $\widehat e_i$ are formed from $\widehat\zeta_i$. (b) Under the conditions of Theorem~\ref{thm:nuisance-split}, the conservativeness and consistency conclusions of Proposition~\ref{prop:monotone}(c) hold if in addition $(\sqrt N\int_{\dom}w_{\Cord{j}}\delta_N)_+=o_p(1)$, and the equality at the constant boundary if $\sqrt N\int_{\dom}w_{\Cord{j}}\delta_N=o_p(1)$. (c) Under the conditions of Theorem~\ref{thm:nuisance-split}, the conclusion of Proposition~\ref{prop:gini-clt} holds, with the same $\Omega_j$, if in addition its conditions hold for $\zeta$ and $\sqrt N\|\bar\pi_N\|_{L_1(\mathcal U)}+\sqrt N\sup_{\mathcal U}\bar\pi_N^2=o_p(1)$.
\end{proposition}

By part (a), with a parametric first stage fitted on the same sample, the plug-in standard error of Section~\ref{sec:app-gini} omits the first-stage term unless both working models are correct, so standard errors for the magnitudes must then use the estimate $\widehat\Omega^\vartheta_{j,N}$ of part (a), which adds the first-stage term to $\widehat\Omega_{j,N}$, in place of $\widehat\Omega_{j,N}$ in \eqref{eq:gini-interval}. Under Theorem~\ref{thm:nuisance-param}, the max form on ordered sets, and with it Proposition~\ref{prop:monotone}(c), is not covered in general: the statistic acquires the term $b_\Psi\{\sqrt N(\widehat\vartheta-\vartheta^*)\}'\int w_{\mathcal C_j}\tau_\vartheta$, which is $O_p(1)$. If $\tau_\vartheta\equiv0$, as when both working models are correct, Step~1 of the proof of Theorem~\ref{thm:nuisance-param} gives $\Dh_j^\Psi(\mathcal C_j)=\Dh^\Psi_{j*}(\mathcal C_j)+o_p(1)$ on every configuration set, and with Lemma~\ref{olem:nui-scale} both are covered. The additional conditions in (b) and (c) are root-$N$ conditions, because the max form on ordered sets and the magnitude functional contain root-$N$-estimable linear functionals of $\tau$. For $\zeta^{\mathrm{DR}}$ of \eqref{eq:psi-dr} the root-$N$ product rate $\|\widehat r-r\|_{2}\|\widehat\mu-\mu\|_{2}=o_p(N^{-1/2})$ of double machine learning suffices for the condition of (b), since $\sqrt N|\int w_{\Cord{j}}\delta_N|\le C\sqrt N\|\widehat r-r\|_{2}\|\widehat\mu-\mu\|_{2}$, and for the $L_1$ part of the condition of (c); the latter further requires $\|\widehat r-r\|_\infty\|\widehat\mu-\mu\|_\infty=o_p(N^{-1/4})$.

\phantomsection\label{oprf:nui-ext}
\begin{proof}
\emph{(a)} Take $\mathcal C=\Csym{j}$ and follow the proof of Proposition~\ref{prop:gini-clt} in Section~\ref{osec:proofs-gini} with $u_N:=\Gammah_j-\Gamma_j$, in the notation of the proof of Theorem~\ref{thm:nuisance-param} in Section~\ref{osec:proofs-nui}. Since $\sup_{\dom}|\widehat\tau_\vartheta|=O_p(1)$, $\sup|u_N|=O_p(\epsilon_N+N^{-1/2})$ and the remainder $R_N$ of Step~1 of the proof of Proposition~\ref{prop:gini-clt} is $O_p(\sqrt N(\epsilon_N+N^{-1/2})^2)=o_p(1)$. In the linear term, $\tfrac12\sqrt N\int\mathrm{sgn}(\Gamma_j)\{\delta_\vartheta'\{(\widehat\tau_\vartheta-\tau_\vartheta)(c_1)-(\widehat\tau_\vartheta-\tau_\vartheta)(c_2)\}+\tauh^{(R)}(c_1)-\tauh^{(R)}(c_2)\}=o_p(1)$ by Step~1 of the proof of Theorem~\ref{thm:nuisance-param}, and $\tfrac12\int\mathrm{sgn}(\Gamma_j)\Gamma_\vartheta=\int_{\dom}\omega_j\tau_\vartheta$ by antisymmetry, so $\mathcal L_N=N^{-1/2}\sum_i\{Z_{N,i}+\mathrm{IF}_\vartheta(W_i)'\int\omega_j\tau_\vartheta\}+o_p(1)$, with $Z_{N,i}$ formed from $\zeta^*$. The summands are i.i.d., centered, with bounded third moments, and $Z_{N,i}\to\omega_j(X_i)\xi^*(W_i;X_i)$ in $L_2$; Lyapunov's theorem gives the limit $N(0,\Omega_j^\vartheta)$.

\emph{Consistency of $\widehat\Omega^\vartheta_{j,N}$.} Put $g:=\int_{\dom}\omega_j\tau_\vartheta$, $\widetilde e_i:=\widehat e_iq_i^{-1/2}$, $\widehat\phi_i:=\widehat u_i\widetilde e_i+\widehat g'\widehat{\mathrm{IF}}_i$,
\[
\phi_i:=\omega_j(X_i)\xi^*(W_i;X_i)+g'\mathrm{IF}_\vartheta(W_i),
\]
with $\omega_j:=0$ outside $\dom$, and $\|a\|_{N,p}:=(N^{-1}\sum_i|a_i|^p)^{1/p}$. As $\E[\zeta(W;\chi_{\vartheta^*})\mid X]=\tau$ and $\E\,\mathrm{IF}_\vartheta(W)=0$, $\E\phi_i=0$ and $\E\phi_i^2=\Omega^\vartheta_j$, so $\|\phi\|_{N,2}^2\pto\Omega^\vartheta_j$ by the law of large numbers; since $|\widehat\Omega^{\vartheta\,1/2}_{j,N}-\|\phi\|_{N,2}|\le\|\widehat\phi-\phi\|_{N,2}$, it suffices that $\|\widehat\phi-\phi\|_{N,2}=o_p(1)$. Work on the event $\mathcal E_N$ of the proof of Theorem~\ref{thm:nuisance-param}, intersected with $\{\inf_{\mathcal U'}\widehat f\ge\inf f/2\}$ for the set $\mathcal U'$ of Lemma~\ref{olem:nui-scale}, on which $|\widehat u_i|\le C$, as $|\widehat\omega_j|\le\lambda(\mathcal X_j)$ and $|\ell_i|\le C(N\eta)^{-1}$ on windows of measure $\eta$. (1) \emph{Weights.} As in the proof of the consistency of $\widehat\Omega_{j,N}$ in Section~\ref{osec:from-main}, which uses the margin condition, $\|\widehat u-\omega_j(X)/f(X)\|_{N,1}=o_p(1)$, and so $\|\widehat u-\omega_j(X)/f(X)\|_{N,4}=o_p(1)$, the difference being bounded. (2) \emph{Residuals.} By the proof of Lemma~\ref{olem:nui-scale}(b), $\|\widehat e-\widehat e_*\|_{N,2}=O_p(N^{-1/2})$, where $\widehat e_{*i}=\zeta^*_i-\tau(X_i)-(\tauh_*-\tau)(X_i)$ and $\sup_{\mathcal U'}|\tauh_*-\tau|=O_p(\epsilon_N)$ as on $\dom$, the observations with $\widehat u_i\neq0$ lying in $\mathcal U'$ for $N$ large, and $q_i^{-1/2}=1+O_p((N\eta)^{-1})$ uniformly over $X_i\in\mathcal U'$; hence $\|\widetilde e-f(X)\xi^*\|_{N,2}=o_p(1)$ over the observations with $\widehat u_i\neq0$, and $\|\widehat u\widetilde e-\omega_j(X)\xi^*\|_{N,2}\le C\|\widetilde e-f(X)\xi^*\|_{N,2}+\|\widehat u-\omega_j(X)/f(X)\|_{N,4}\|f(X)\xi^*\|_{N,4}=o_p(1)$, the fourth conditional moments of $\zeta^*$ being bounded. (3) \emph{First-stage coefficient.} By Taylor's theorem on $\mathbb B_0$, $\|\partial_\vartheta\zeta(W_i;\chi_{\widehat\vartheta})-\zeta_{\vartheta,i}\|\le\|\delta_\vartheta\|M(W_i)$, so $\widehat g=N^{-1}\sum_i\widehat u_i\zeta_{\vartheta,i}+O_p(N^{-1/2})$; by (1) and the Cauchy--Schwarz inequality this is $N^{-1}\sum_i\omega_j(X_i)\zeta_{\vartheta,i}/f(X_i)+o_p(1)$, which converges in probability to $\E[\omega_j(X)\tau_\vartheta(X)/f(X)]=g$. (4) \emph{First-stage term.} $\|\widehat g'\widehat{\mathrm{IF}}-g'\mathrm{IF}_\vartheta\|_{N,2}\le\|\widehat g-g\|\,\|\widehat{\mathrm{IF}}\|_{N,2}+\|g\|\,\|\widehat{\mathrm{IF}}-\mathrm{IF}_\vartheta\|_{N,2}=o_p(1)$, as $\|\widehat{\mathrm{IF}}\|_{N,2}=O_p(1)$. By (2) and (4), $\|\widehat\phi-\phi\|_{N,2}=o_p(1)$. For a maximum likelihood or least squares first stage with estimating function $s(W;\vartheta)$, so that $\mathrm{IF}_\vartheta=-H^{-1}s(W;\vartheta^*)$ with $H:=\E[\partial_\vartheta s(W;\vartheta^*)]$ nonsingular, the estimates $\widehat{\mathrm{IF}}_i:=-\widehat H^{-1}s(W_i;\widehat\vartheta)$, $\widehat H:=N^{-1}\sum_i\partial_\vartheta s(W_i;\widehat\vartheta)$, satisfy the condition on $\widehat{\mathrm{IF}}_i$ whenever $\partial_\vartheta s$ is continuous in $\vartheta$ on a ball around $\vartheta^*$ with an envelope of finite second moment there: then $\widehat H\pto H$, and $N^{-1}\sum_i\|s(W_i;\widehat\vartheta)-s(W_i;\vartheta^*)\|^2=O_p(N^{-1})$.

\emph{(b), (c)} The reduction of Step~2 of the proof of Theorem~\ref{thm:nuisance-split} applies to each part, with the additional condition included among the $o_p(1)$ quantities that vanish almost surely along a further subsequence, so it suffices to argue along the deterministic sequences $P_N$ of Step~1 of that proof. (b) In the proof of Proposition~\ref{prop:monotone}(c) (Section~\ref{osec:proofs}) the null affects the proof only through the noise-field statistic $\sqrt N\int_{\Cord{j}}(\Gammah_j-\Gamma_j)_+$. The decomposition in the proof of Lemma~\ref{olem:nui-lin} does not use the null, so this statistic is $\sqrt N\int_{\Cord{j}}\Psimax(\Gamma^{\mathrm{lin}}_j-\Gamma_j+\delta^\Gamma_N)+o_p(1)$; Lemma~\ref{lem:drift} for the noise field, with $b_{\Psimax}=\tfrac12$, makes it $\sqrt N\int_{\Cord{j}}\Psimax(\Gamma^{\mathrm{lin}}_j-\Gamma_j)+\tfrac12\sqrt N\int w_{\Cord{j}}\delta_N+o_p(1)$, to which Lemma~\ref{olem:uniform}(b),(c) applies; the conservativeness argument needs only an upper bound on this statistic, for which the one-sided condition suffices, the equality at the constant boundary needs the two-sided one, consistency uses neither, and the remaining steps are unchanged. (c) Follow the proof of Proposition~\ref{prop:gini-clt} with $u_N:=\Gammah_j-\Gamma_j$. By Lemma~\ref{olem:uniform}(c), $\sup|u_N|=O_p(\epsilon_N)+2\sup|\delta_N|$, so the remainder there is $O_p(\sqrt N\epsilon_N^2+\sqrt N\sup\bar\pi_N^2)=o_p(1)$. In the linear term the bias contributes $O(\sqrt Nh^s)$, the drift at most $\sqrt N\int_{\mathcal C}|\delta^\Gamma_N|\le C\sqrt N\|\delta_N\|_1\le C\sqrt N\|\bar\pi_N\|_{L_1(\mathcal U)}=o(1)$ by Lemma~\ref{olem:nui-lin}, and $R_N$ of Lemma~\ref{olem:nui-lin} at most $C\sqrt N\int|R_N|=o_p(1)$, so $\mathcal L_N=N^{-1/2}\sum_iZ_{N,i}+o_p(1)$ with $Z_{N,i}$ formed under $P_N$. Let $Z^0_N$ be the same functional of $\zeta$. Then $|Z_N-Z^0_N|\le C|\zeta_N-\zeta|+C\E|\zeta_N-\zeta|$, so $\Var(Z_N)-\Var(Z^0_N)\to0$ by Assumption~\ref{ass:firststage}(b)(iii), and $\Var(Z^0_N)\to\Omega_j$ by the proof of Proposition~\ref{prop:gini-clt}; the third moments being bounded, Lyapunov's theorem for triangular arrays completes the proof.
\end{proof}
\subsection{Uniformity of the oracle results}\label{osec:uniform}
Theorem~\ref{thm:nuisance-split} of the main text conditions on the first-stage sample, after which the second stage is a kernel regression of a response whose law changes with $N$. This subsection shows that the oracle results of Sections~\ref{osec:aux-poisson}--\ref{osec:aux-moved} hold along such sequences of laws, and that so do the estimated centering and variance of the main text. It is written for the Nadaraya--Watson regression $\tauh$ of a scalar response $\zeta$ on $X=(X_1,X_2)$, with the score $\xi(W;x):=\{\zeta-\tau(x)\}/f(x)$ in place of \eqref{oeq:theta}, that is, with $(\bar\mu_t,P_t)$ replaced by $(\tau,f)$, and with $\theta_N$, $\Theta_j$, $\kappa_{j,N}$, $m^\kappa_{j,N}(\mathcal C_j)$ formed from it as in Section~\ref{osec:aux-setting}. Let $\mathcal U$ be a neighborhood of $\dom$ with compact closure on which $f\ge\underline p$, $\mathcal K(x;X):=K_1((x_1-X_1)/h_1)K_2((x_2-X_2)/h_2)$ and $B^\tau_N(x):=\E[\{\tau(X)-\tau(x)\}\mathcal K(x;X)]/\{\eta f(x)\}$.

\begin{lemma}[Uniformity along sequences of laws]\label{olem:uniform}
Fix $j$, a configuration set $\mathcal C_j$, $\Psi$ satisfying Assumption~\ref{ass:psi} with $a_\Psi>0$, and kernels and bandwidths satisfying Assumption~\ref{ass:reg}(d),(g),(h). Let $P_N$ be a sequence of laws of $(\zeta,X)$ such that: (i) the law of $X$ does not depend on $N$ and satisfies Assumptions~\ref{ass:overlap} and \ref{ass:reg}(a),(b); (ii) $\E_N[\zeta\mid X=x]=\tau(x)+\bar\pi_N(x)$, where $\tau$ does not depend on $N$, $\tau$ and $f$ satisfy the smoothness of Assumption~\ref{ass:reg}(e), and $\sup_{\mathcal U}|\bar\pi_N|\to0$; (iii) $\varsigma_{\zeta,N}^2(x):=\Var_N(\zeta\mid X=x)$ satisfies $\sup_{\mathcal U}|\varsigma_{\zeta,N}^2-\varsigma_\zeta^2|\to0$ for a $\varsigma_\zeta^2$ continuous on $\bar{\mathcal U}$ with $\inf_{\mathcal U}\varsigma_\zeta^2>0$; (iv) $\sup_N\sup_{x\in\mathcal U}\E_N[|\zeta|^6\mid X=x]<\infty$. Let $B_{j,N}(c):=B^\tau_N(c_1)-B^\tau_N(c_2)$, let $D^P_j$ and $D^{\mathrm{lin}}_j(0)$ be the Poissonized and fixed-sample statistics \eqref{eq:Dgen} built from the noise field $B_{j,N}+N^{-1}\sum_i\Theta_j(W_i;\cdot)$, Poissonized locally as in \eqref{oeq:poisson-field}: the uncentered summand $A_j(W;c):=\eta^{-1}\{\xi(W;c_1)\mathcal K(c_1;X)-\xi(W;c_2)\mathcal K(c_2;X)\}$, which vanishes unless $X$ lies in one of the two windows, has mean $\bar A_{j,N}:=B_{j,N}+\delta^\Gamma_{j,N}$ under $P_N$, with $\delta^\Gamma_{j,N}(c):=\delta_N(c_1)-\delta_N(c_2)$ for $\delta_N$ of (c) below, and the Poissonized noise field is $B_{j,N}-\bar A_{j,N}+N^{-1}\sum_{i\le\mathcal N}A_j(W_i;\cdot)$, let $\varrho_\zeta:=\varsigma_\zeta^2\int K_1^2\int K_2^2/f$, and let $\sigma^2:=\sigma^2_{j,\Psi}(\mathcal C_j)$ be computed from $\varrho_\zeta$. Then, under $P_N$:
(a) Lemmas~\ref{olem:poisson}, \ref{olem:variance-gen}, \ref{olem:pf-projection}, \ref{olem:pf-clt} and \ref{olem:decorr} and the first display of Lemma~\ref{olem:centering} hold, with the rates $O(h_1+h_2)$ and $O(h_1^2+h_2^2)$ of the second-moment expansions replaced by $o(1)$ and the limits computed from $\varrho_\zeta$;
(b) $\{D^{\mathrm{lin}}_j(0)-m^\kappa_{j,N}(\mathcal C_j)\}/\sigma\dto N(0,1)$;
(c) $\sup_{\dom}|\tauh-\tau-\delta_N|=O_p(\epsilon_N)$, where $\delta_N(x):=\E_N[\bar\pi_N(X)\mathcal K(x;X)]/\{\eta f(x)\}$; and if moreover $\sup_{\mathcal U}|\bar\pi_N|=o(\eta^{1/2})$, the centering $\widetilde m_j(\mathcal C_j)$ of \eqref{eq:center-feasible} and the finite-sample variance $\widetilde\sigma^2_{j,\Psi,N}(\mathcal C_j)$ of \eqref{eq:sigmaN}, computed from the Nadaraya--Watson weights and the corrected residuals of the $\zeta_i$, satisfy $\widetilde m_j(\mathcal C_j)-m^\kappa_{j,N}(\mathcal C_j)=o_p(1)$ and $\widetilde\sigma^2_{j,\Psi,N}(\mathcal C_j)\pto\sigma^2$.
\end{lemma}

\noindent In Lemma~\ref{olem:uniform}, uniformity is established by checking each step of the oracle proofs and of the proofs of Lemmas~\ref{prop:centering-corr} and~\ref{prop:sigmaN} along $P_N$; the check is shown in parts (i)--(vii) of the proof. The hypotheses impose no smoothness on $\varsigma_{\zeta,N}^2$, only uniform closeness to a continuous $\varsigma_\zeta^2$, and allow a vanishing drift $\bar\pi_N$, which is carried by the centered score and not by the deterministic part $B_{j,N}$ of the field. The noise field coincides with the linearized field under $\Gamma_j\equiv0$ and is the field $\Gamma^{\mathrm{lin}}_j-\Gamma_j$ of the proof of Proposition~\ref{prop:monotone}(c) in general; no null hypothesis is used below.

\phantomsection\label{oprf:uniform}
\begin{proof}
The proof is an inspection of the oracle proofs and of the proofs of Lemmas~\ref{prop:centering-corr} and \ref{prop:sigmaN}: for each step we show which properties of the law it uses and check that they hold along $P_N$ with constants that do not depend on $N$. Throughout, $\tau^\pi_N:=\tau+\bar\pi_N$ and $C$ denotes such a constant.

\emph{(i) Structural steps.} The exact moment identities, window geometry and block representation of Lemma~\ref{olem:poisson}(a)--(c), the decomposition and degeneracy of Lemma~\ref{olem:pf-projection}(a),(b), the classification of configurations and its measure counts (Sections~\ref{osec:aux-proj} and \ref{osec:moved-events}), the one-dependence of the cell array and the exact independence from the exterior count (Sections~\ref{osec:aux-clt} and \ref{osec:moved-clt}), and Lemma~\ref{olem:depoisson} use only i.i.d.\ sampling, the support of the kernels and the law of $X$, which is fixed: the window masses $\mu(\mathcal W(x))\le\sup f\,\eta$ and the exterior mass $p_{\mathcal B}$ do not depend on $N$. The exact independence holds because the Poissonized noise field is local: its summands $A_j(W_i;\cdot)$ vanish outside the windows and its deterministic part $B_{j,N}-\bar A_{j,N}=-\delta^\Gamma_{j,N}$ does not involve the count, so it is a functional of the Poisson increments over the windows, exactly as \eqref{oeq:poisson-field} under the fixed law. Its exact second moments are $N\Cov=\E_N[\Theta_j\Theta_j']+\bar A_{j,N}\bar A_{j,N}'$, and $\sup|\bar A_{j,N}|\le C(\sup|B^\tau_N|+\sup|\delta_N|)\to0$ by hypothesis (ii), so the additive term is negligible exactly as $B_{j,N}B_{j,N}'$ is under the fixed law, both against $\kappa_{j,N}\asymp\eta^{-1}$ in the correlations and in the centering, where $0\le\{\kappa_{j,N}+\bar A_{j,N}^2\}^{1/2}-\kappa_{j,N}^{1/2}\le|\bar A_{j,N}|$. Lemma~\ref{olem:depoisson} is a statement about arbitrary sequences of random vectors and needs no fixed law.

\emph{(ii) Second-moment expansions.} Under $P_N$ the conditional product moment \eqref{oeq:qdef} is $q_N(z;x,x')=\{\varsigma_{\zeta,N}^2(z)+(\tau^\pi_N(z)-\tau(x))(\tau^\pi_N(z)-\tau(x'))\}/\{f(x)f(x')\}$, so $f(z)q_N(z;x,x')=\Xi^0(z;x,x')+R^q_N(z;x,x')$, where $\Xi^0:=f(z)\{\varsigma_\zeta^2(z)+(\tau(z)-\tau(x))(\tau(z)-\tau(x'))\}/\{f(x)f(x')\}$ does not depend on $N$, is uniformly continuous in $z$ on $\mathcal U$ uniformly in $(x,x')$ and Lipschitz in $(x,x')$ uniformly in $z$, and $\sup|R^q_N|\le C\{\sup_{\mathcal U}|\varsigma_{\zeta,N}^2-\varsigma_\zeta^2|+\sup_{\mathcal U}|\bar\pi_N|\}\to0$. The proof of Lemma~\ref{olem:gine} shows that, for an integrand $\Xi$ uniformly continuous with modulus $\omega_\Xi$, the right side of \eqref{oeq:gine1} is at most $\omega_\Xi(c_H(h_1+h_2))\bar H\lambda(\mathcal C_H)$, and the perturbation $R^q_N$ adds at most $\sup|R^q_N|\bar H\lambda(\mathcal C_H)$. Hence \eqref{oeq:kapunif}, \eqref{oeq:kapupper}, \eqref{oeq:kapadd}, \eqref{oeq:kapsqrt}, \eqref{oeq:num}, \eqref{oeq:kap}, \eqref{oeq:kap2} and \eqref{oeq:rhostar} hold with $o(1)$, uniform in $N$, in place of their rates, and with $\varrho$ replaced by $\varrho_\zeta$: the moment identity \eqref{oeq:rho2id} becomes $f(x)\E_N[\xi(W;x)^2\mid X=x]\int K_1^2\int K_2^2=\{\varsigma_{\zeta,N}^2(x)+\bar\pi_N(x)^2\}\int K_1^2\int K_2^2/f(x)=\varrho_\zeta(x)+o(1)$. These rates are consumed only as $o(1)$: in Parts~1 and 3 of the proof of Lemma~\ref{olem:variance} through bounded convergence, and in the off-band lower bound $\eta\kappa_{j,N}\ge2\inf\varrho_\zeta-o(1)$. Two steps consume a rate of order $\bar h^2$ against the diverging factor $\eta^{-1/2}$: the limit-form centering \eqref{oeq:biasderiv}--\eqref{oeq:extra}, which the lemma does not cover, and the comparison of $m^X_{j,N}$ with $m^\kappa_{j,N}$ for the estimate $\widetilde m_j$ (step (3) of part (vii) below), which Assumption~\ref{ass:reg}(h) controls. The lemma covers the centering $m^\kappa_{j,N}$ and its estimate $\widetilde m_j$ only.

\emph{(iii) Bias and drift.} The deterministic part $B_{j,N}$ depends on $\tau$ and $f$ only, so \eqref{oeq:driftrate} holds verbatim. The kernel-weighted mean of the score, which affects Lemma~\ref{olem:decorr} through \eqref{oeq:gammastar}, is $\E_N[\eta^{-1}\xi(W;x)\mathcal K(x;X)]=B^\tau_N(x)+\delta_N(x)$ with $\sup|\delta_N|\le C\sup_{\mathcal U}|\bar\pi_N|\to0$; for the local Poissonization this is $N\Cov(\tau_{\mathcal N}(x),\mathcal N(\mathcal B))$, so part (b) of that proof gives $\rho^U=O(\sup|B^\tau_N+\delta_N|\,\eta^{1/2})$ and part (d) bounds the covariance by $C\sup|B^\tau_N+\delta_N|/\sqrt{1-p_{\mathcal B}}=o(1)$, uniformly in $x$.

\emph{(iv) Moments.} The envelopes are $|\xi(W;x)|\le C(1+|\zeta|)$ on windows centered in $\dom$ and the conditional moments of $\zeta$ of orders up to six, bounded uniformly in $N$ by hypothesis (iv). They affect the Rosenthal bound \eqref{oeq:thirdmom}, the Berry--Esseen comparisons of Fact~\ref{ofact:BE} and \eqref{oeq:BE-gen}, whose constants are universal, Theorem~1 of \citet{sweeting1977speeds} in Lemma~\ref{olem:decorr}, and the truncation and Bernstein steps of Lemma~\ref{olem:kappahat}, whose constants $\bar C_0,\dots,\bar C_8$ depend on the moment bounds and the kernels only. Theorem~1 of \citet{sweeting1977speeds} concerns standardized sums of i.i.d.\ vectors with mean zero and identity covariance, with a constant depending only on the dimension, and bounds the error through their third absolute moment. In Lemma~\ref{olem:decorr} the block pair $Q:=(Q_1,Q_2)$, the standardized $\theta_N$-integral and count of one block, has covariance $\Sigma$ with unit diagonal and off-diagonal $\rho^U$, whose eigenvalues are $1\pm\rho^U$. The theorem is applied to $\Sigma^{-1/2}Q$, which has identity covariance, and to the composition of the function of \eqref{oeq:sweeting} with $\Sigma^{1/2}$, whose Lipschitz constant in the first argument and linear growth in the second are those of the untransformed function multiplied by at most $\|\Sigma^{1/2}\|\le\sqrt2$. Since $(a^2+b^2)^{3/2}\le\sqrt2(|a|^3+|b|^3)$,
\[
\E\|\Sigma^{-1/2}Q\|^3\le(1-|\rho^U|)^{-3/2}\,\E\|Q\|^3\le\sqrt2\,(1-|\rho^U|)^{-3/2}\big(\E|Q_1|^3+\E|Q_2|^3\big),
\]
so the constant in \eqref{oeq:sweeting} depends on the law only through the dimension, the block third moments, bounded uniformly in $N$ by \eqref{oeq:thirdmom} and hypothesis (iv), and $1-|\rho^U|$, which tends to one uniformly by (iii). This is the reduction used for the bivariate comparison \eqref{oeq:BEcovPsi}.

\emph{(v) Nondegeneracy.} For $N$ large, $\inf\varrho_N\ge\inf\varrho_\zeta/2$ and $\sup\varrho_N\le2\sup\varrho_\zeta$ for
\[
\varrho_N:=f\E_N[\xi(W;\cdot)^2\mid X=\cdot]\int K_1^2\int K_2^2,
\]
so the correlation bounds \eqref{oeq:rhobar} and \eqref{oeq:rhobound} hold with constants formed from $\varrho_\zeta$, and the Berry--Esseen prefactor $(1-\rho^2)^{-3/2}$ is a fixed constant; $\sigma^2>0$ by Theorem~\ref{thm:variance}, $\varrho_\zeta$ being a fixed continuous function bounded away from zero.

\emph{(vi) Limits and assembly.} The variance limit of Lemma~\ref{olem:variance-gen} (Part~3 of the proof of Lemma~\ref{olem:variance} and Section~\ref{osec:moved-corr}) is obtained by bounded convergence of an integrand that converges, by (ii) and (v), to the integrand formed with $\varrho_\zeta$; hence $\Var(D^P_j)\to\sigma^2$. The Lyapunov and variance steps of Lemma~\ref{olem:pf-clt} use (iv) and this limit, the extension to $\dom$ of Section~\ref{osec:moved-extension} uses the band-valid envelope \eqref{oeq:kapupper} and $|g_\Psi|\le1$ against the measure of the complement, uniformly in $N$, and the first display of Lemma~\ref{olem:centering} uses (iii), (iv) and \eqref{oeq:BEmeanPsi}. This proves (a), and (b) follows from (a) as Theorem~\ref{thm:normality} follows from the oracle lemmas.

\emph{(vii) Feasible estimators.} On $\{\inf_{\dom}\widehat f\ge\inf f/2\}$, whose probability tends to one by Lemma~\ref{olem:consis}, $\tauh-\tau=A/\widehat f$ with $A(x):=(N\eta)^{-1}\sum_i\{\zeta_i-\tau(x)\}\mathcal K_i(x)$ and $\E A=f(B^\tau_N+\delta_N)$. The functions $w\mapsto\{\zeta-\tau(x)\}\mathcal K(x;X)$, $x\in\dom$, have envelope $C(1+|\zeta|)$ and are Lipschitz in $x$ with constant $C(1+|\zeta|)(h_1^{-1}+h_2^{-1})$, so Steps~1--3 of the proof of Lemma~\ref{olem:kappahat} (truncation, the Bernstein inequality on a polynomial net, Lipschitz increments), applied to $A$ in place of the second-moment field, give $\sup|A-\E A|=O_p(\{\log N/(N\eta)\}^{1/2})$ with constants depending on hypothesis (iv) only: at $t_N:=c_A\{\log N/(N\eta)\}^{1/2}$ the summands have variance $O(\eta^{-1})$ and, after truncation at $N^{1/4}$, range $O(N^{1/4}\eta^{-1})$, so the Bernstein exponent is at least $cc_A^2\log N$ once $N^{1/4}t_N\to0$, which holds by Assumption~\ref{ass:reg}(g); with $\sup|\widehat f-f|=O_p(\epsilon_N)$ this gives $\sup|\tauh-\tau-\delta_N|=O_p(\epsilon_N)$. For the centering and the variance, the proofs of Lemmas~\ref{prop:centering-corr} and \ref{prop:sigmaN} in Section~\ref{osec:from-main} are followed along $P_N$ with $e_i:=\zeta_i-\tau^\pi_N(X_i)$ and $\varsigma_{\zeta,N}^2$ in place of $\varsigma_\zeta^2$, so that $\E_N[e\mid X]=0$ and $\E_N[e^2\mid X]=\varsigma_{\zeta,N}^2(X)$. (1) The centered second-moment process $N^{-1}\sum_i\varpi_i(e_i^2-\varsigma_{\zeta,N}^2(X_i))$, with the weights $\varpi_i$ of the proof of Lemma~\ref{prop:centering-corr} in Section~\ref{osec:from-main}, and its polarized versions are bounded by Steps~1--3 of the proof of Lemma~\ref{olem:kappahat}, whose constants depend on hypothesis (iv) and the kernels only, and Lemma~\ref{olem:nui-scale}(a) bounds the kernel averages uniformly in $N$, its proof using second conditional moments only. (2) The smoothing error of the residuals is $\Delta_i=\sum_lL_{il}\zeta_l-\tau^\pi_N(X_i)=(\tauh-\tau-\delta_N)(X_i)+(\delta_N-\bar\pi_N)(X_i)$, so, with $\mathcal U'$ as in Lemma~\ref{olem:nui-scale}, $\max_{X_i\in\mathcal U'}|\Delta_i|=O_p(\epsilon_N)+2C\sup_{\mathcal U}|\bar\pi_N|=o_p(\eta^{1/2})$ under the additional hypothesis, which is all that the third term in the first claim of Lemma~\ref{prop:centering-corr} requires. The middle term $\sum_{l\neq i}L_{il}^2\{\varsigma_{\zeta,N}^2(X_l)-\varsigma_{\zeta,N}^2(X_i)\}$ of $\E_N[\widehat e_i^{\,2}\mid X]$, which under a fixed law is $O_p(\bar h(N\eta)^{-1})$ by the Lipschitz property of $\varsigma_\zeta^2$, is $O_p((N\eta)^{-1})$ along $P_N$, as $\varsigma_{\zeta,N}^2$ is bounded and need not be Lipschitz; it changes $\widetilde m_j$ by $O_p((N\eta^{3/2})^{-1})=o_p(1)$. (3) In the second claim of that lemma and in Step~2(ii) of the proof of Lemma~\ref{prop:sigmaN}, $\mathcal K^\tau(X;c)$ has $\tau^\pi_N(X)-\tau(c_k)$ in place of $\tau(X)-\tau(c_k)$, which is bounded by $C\bar h+\sup_{\mathcal U}|\bar\pi_N|$ on the windows; the bounds there hold with $\bar h$ replaced by $\bar h+\sup_{\mathcal U}|\bar\pi_N|$ and still vanish, as $\sup_{\mathcal U}|\bar\pi_N|^2\eta^{-1/2}=o(\eta^{1/2})$ and $\bar h^2\eta^{-1/2}\to0$ by Assumption~\ref{ass:reg}(h). (4) The off-band lower bound $\eta\kappa^X_{j,N}\ge c_0$ uses $\inf \varsigma_{\zeta,N}^2\ge\inf \varsigma_\zeta^2/2$ for $N$ large. (5) In Step~2(i) of the proof of Lemma~\ref{prop:sigmaN}, the uncentered summands have mean $\bar A_{j,N}=B_{j,N}+\delta^\Gamma_{j,N}$ under $P_N$, so pairs with disjoint windows have $\E_N[\Theta_j(W;c)\Theta_j(W;c')]=-\bar A_{j,N}(c)\bar A_{j,N}(c')$ in place of $-B_{j,N}(c)B_{j,N}(c')$; as $\sup\bar A_{j,N}^2\le C(\sup|B^\tau_N|+\sup|\delta_N|)^2\to0$ by hypothesis (ii), these pairs still contribute $o(1)$. Hence $\widetilde m_j-m^\kappa_{j,N}=o_p(1)$ and $\widetilde\sigma^2_{j,\Psi,N}-\iint g_\Psi(\rho_N)\{\kappa_{j,N}(c)\kappa_{j,N}(c')\}^{1/2}=o_p(1)$, and the double integral converges to $\sigma^2$ by (a).
\end{proof}

Two steps of the oracle argument are not covered by this inspection and are replaced rather than inspected. The uniform rates of Lemma~\ref{olem:consis} are stated for a fixed law; for the quantities used here they are rederived in (vii) from the steps of Lemma~\ref{olem:kappahat}, whose constants are explicit. The linearization of the denominator, Lemma~\ref{olem:three}, is the ratio expansion of the proof of Lemma~\ref{olem:pf-linearization} in Section~\ref{osec:from-main}, whose remainder is a product of two uniform errors; for the Nadaraya--Watson regression it is the direct decomposition $\tauh-\tau=A/f-A(\widehat f-f)/(f\widehat f)$ of Lemma~\ref{olem:nui-lin}, which uses second moments only.
\section{Implementation results: proofs and details}\label{osec:impl-proofs}

This section supports Sections~\ref{sec:asym} and \ref{sec:impl} of the main text on the procedure as implemented. Section~\ref{osec:from-main} states, with their proofs where these are not given below, the results that the main text uses in words or whose statements or proofs are given here for reasons of length. Section~\ref{osec:sharp} proves the sharp uniform rate behind Lemma~\ref{olem:pf-linearization}, with which Assumption~\ref{ass:reg}(g) needs no logarithmic strengthening. Section~\ref{osec:centering-details} completes the proof of Lemma~\ref{prop:centering-corr} (the own-weight-corrected centering) and gives a second-order correction of the centering. Section~\ref{osec:grid-formulas} gives the grid formulas used in the computations for the centering, the finite-sample variance of Lemma~\ref{prop:sigmaN}, the scale of the max form (Proposition~\ref{oprop:maxvar}) and the magnitude variance of Proposition~\ref{prop:gini-clt}. Section~\ref{osec:tuning} proves Lemma~\ref{olem:tuning} on data-dependent tuning, Section~\ref{osec:gini-if} gives the influence functions of the density-weighted magnitude and of the index and the coverage of the interval \eqref{eq:gini-interval}, and Section~\ref{osec:jitter} proves Proposition~\ref{oprop:jitter} on integer-valued modifiers. The proof of Proposition~\ref{prop:gini-clt} is in Section~\ref{osec:proofs-gini}.

\subsection{Supporting results stated in the main text}\label{osec:from-main}
This subsection gives the proofs of Lemmas~\ref{olem:pf-linearization}, \ref{prop:centering-corr}, \ref{prop:sigmaN} and \ref{lem:drift} and of Proposition~\ref{prop:transfer} of the main text, the second-order expansion of the estimated centering, the consistency of the variance estimate of Proposition~\ref{prop:gini-clt}(c), and three results that Section~\ref{sec:asym} states in words: the scale of the max form, data-dependent tuning, and integer-valued modifiers. The max form under the monotone null is Proposition~\ref{prop:monotone}(c) of the main text, proved in Section~\ref{osec:proofs}. The notation is that of the main text, with $\widetilde e_i^{\,2}:=\widehat e_i^{\,2}/q_i$ for $q_i$ of \eqref{eq:center-feasible}, $\bar h:=h_1\vee h_2$, $\bar K(z):=K_1(z_1)K_2(z_2)$, so that $\bar K(0)=K_1(0)K_2(0)$ and $\|\bar K\|^2:=\int K_1^2\int K_2^2$.

\emph{The linearized field.} Only Lemma~\ref{olem:pf-linearization} depends on the form of the estimator. The later steps use the linearized field through three properties that the estimators \eqref{eq:tauhat} and \eqref{eq:tauhat-psi} share: $\Theta_j(W;c)$ is centered and, up to its mean, vanishes unless $X\in\mathcal W(c_1)\cup\mathcal W(c_2)$; $|\xi(W;x)|\le C(1+|\zeta|)$ on the windows, with bounded conditional sixth moments; and $f(x)\E[\xi(W;x)^2\mid X=x]\|\bar K\|^2$ is the scaled variance, $\varrho$ or $\varrho_\zeta$.

\begin{lemma}[Second-order expansion of the estimated centering]\label{olem:centering-bias}
Under the conditions of Lemma~\ref{prop:centering-corr}, with $\epsilon^m_N:=m^X_{j,N}(\mathcal C_j)/(N\eta)$,
\[
\widetilde m_j(\mathcal C_j)-m^X_{j,N}(\mathcal C_j)=-\tfrac12C_q\,\phi_N\,\epsilon^m_N\{1+o_p(1)\}+O_p\big(\bar h\epsilon^m_N+\eta^{-1/2}h_1^{2s_1}+\eta^{-1/2}h_2^{2s_2}\big)-J_N+F_N ,
\]
where $\phi_N$ is a weighted mean of $1/f$ over the windows, $\E[F_N\mid X]=0$, $F_N=O_p((N\eta)^{-1/2})$, $0\le J_N=O_p(\epsilon^m_N)$ does not depend on the divisor, and $C_q=2\bar K(0)-\|\bar K\|^2$, $\bar K(0)-\|\bar K\|^2$, $-\|\bar K\|^2$ and $0$ for the uncorrected, HC2, leave-one-out and $q_i$-corrected residuals. For the product fourth-order kernel of Section~\ref{sec:impl}, $\bar K(0)=7.91$ and $\|\bar K\|^2=6.25$, so $\frac12C_q=4.79$, $0.83$, $-3.13$ and $0$.
\end{lemma}

Throughout, $j$ is fixed, $j'\neq j$ and $\mathcal C:=\mathcal C_j$. \emph{Three facts.} For the band $\mathfrak B:=\{c\in\mathcal C:\|c_1-c_2\|_\infty\le h_j\}$, $\delta\in L_2(\dom)$ and $\delta^\Gamma(c):=\delta(c_1)-\delta(c_2)$: (F1) $\lambda(\mathfrak B)\le Ch_j^{d_j}$; (F2) $\|\delta^\Gamma\|^2_{L_2(\mathcal C)}\le4\lambda(\mathcal X_j)\|\delta\|_2^2$, $\|\delta^\Gamma\|^2_{L_2(\mathfrak B)}\le4(2h_j)^{d_j}\|\delta\|_2^2$ and $\int_{\mathcal C}\delta^\Gamma=\int_{\dom}w_{\mathcal C}\delta$; (F3) the set $\mathcal D(c)$ of $c'\in\mathcal C$ such that some window $\mathcal W(c'_k)$ meets some $\mathcal W(c_l)$ has $\lambda(\mathcal D(c))\le C\eta$, as each overlap pins the $j'$-coordinate of $c'$ and one $j$-coordinate to cubes of sides $2h_{j'}$ and $2h_j$.

\phantomsection\label{oprf:pf-linearization}
\begin{proof}[Proof of Lemma~\ref{olem:pf-linearization} and of Proposition~\ref{prop:transfer}]
\emph{(a)} The uniform rate $\{\log N/(N\eta)\}^{1/2}+h_1^{s_1}+h_2^{s_2}$ for kernel averages of a variable with bounded conditional sixth moments holds whenever $N\eta^3\to\infty$ (Lemma~\ref{olem:sharp}); it applies to $\widehat P_t$, to the numerators $\widehat A_t:=(N\eta)^{-1}\sum_iY_i\ind(T_i=t)\mathcal K_i$ and, for Proposition~\ref{prop:transfer}, to $\widehat f$ and $\widehat A:=(N\eta)^{-1}\sum_i\zeta_i\mathcal K_i$; the claim for $\rhoh$ is Lemma~\ref{olem:consis}. Next, $\sqrt N\epsilon_N^2\le3\{\log N/(N\eta^2)^{1/2}+\sqrt Nh_1^{2s_1}+\sqrt Nh_2^{2s_2}\}\to0$ and $\epsilon_N/\eta^{1/2}\le\{\log N/(N\eta^2)\}^{1/2}+(Nh_1^{2s_1}+Nh_2^{2s_2})^{1/2}(N\eta)^{-1/2}\to0$ by Assumption~\ref{ass:reg}(g).
\emph{(b)} For a ratio, $\widehat A/\widehat P=\widehat A/P-A(\widehat P-P)/P^2+R^\tau_N$ with $R^\tau_N:=-(\widehat A-A)(\widehat P-P)/P^2+\widehat A(\widehat P-P)^2/(P^2\widehat P)$ and $A:=\bar\mu_tP_t$, as expanding $\widehat A/\widehat P-\widehat A/P$ shows. On the event $\{\inf_{\dom}\widehat P_t\ge\inf P_t/2,\ \sup_{\dom}|\widehat A_t|\le2\sup|A_t|\}$, whose probability tends to one by (a), $\sup|R^\tau_N|\le C\sup(|\widehat A_t-A_t|+|\widehat P_t-P_t|)^2=O_p(\epsilon_N^2)$. The linear part $\sum_t(-1)^{1-t}\{\widehat A_t/P_t-\bar\mu_t(\widehat P_t-P_t)/P_t\}$ equals $\tau+B_N+N^{-1}\sum_i\theta_N(W_i;\cdot)$ (Section~\ref{osec:aux-setting}); differencing at $c_1,c_2$ gives (b), with $R_{j,N}(c)$ the difference of the $R^\tau_N$ terms. For the Nadaraya--Watson estimator, $\tauh-\tau=\{\widehat A-\tau\widehat f\}/\widehat f$, whose linear part $\{\widehat A-\tau\widehat f\}/f$ is $B_N+N^{-1}\sum_i\theta_N(W_i;\cdot)$ with the score $\xi(W;x)=\{\zeta-\tau(x)\}/f(x)$, and the remainder $-\{\widehat A-\tau\widehat f\}(\widehat f-f)/(f\widehat f)$ is $O_p(\epsilon_N^2)$ uniformly. \emph{(c)} By the Lipschitz property, $|\Dh_j^\Psi(\mathcal C_j)-D_j^{\mathrm{lin}}|\le\sqrt N\lambda(\mathcal C_j)\sup_c|R_{j,N}(c)|=o_p(1)$. No null hypothesis is used. For Proposition~\ref{prop:transfer} it remains to check the three properties listed above: $|\xi|\le C(1+|\zeta|)$ on the windows as $f\ge\underline p$ there, $\zeta$ inherits bounded sixth moments from $Y$ through the bounded inverse propensities, and $f(x)\E[\xi(W;x)^2\mid X=x]\|\bar K\|^2=\varrho_\zeta(x)$ by definition; the auxiliary results of Section~\ref{osec:aux} use $\tauh$ only through these.
\end{proof}

\phantomsection\label{oprf:centering-bias}
\begin{proof}[Proof of Lemma~\ref{prop:centering-corr} and Lemma~\ref{olem:centering-bias}]
Write $\varsigma_{\zeta,i}^2:=\varsigma_\zeta^2(X_i)$, $e_i:=\zeta_i-\tau(X_i)$,
\[
\widehat\varpi_i(c):=N^2\{\ell_i(c_1)-\ell_i(c_2)\}^2=\eta^{-2}\{\mathcal K_i(c_1)/\widehat f(c_1)-\mathcal K_i(c_2)/\widehat f(c_2)\}^2,
\]
so that $\widetilde\kappa_j=N^{-1}\sum_i\widehat\varpi_i\widetilde e_i^{\,2}$ and $\kappa^X_{j,N}=N^{-1}\sum_i\widehat\varpi_i\varsigma_{\zeta,i}^2$, and let $\varpi_i(c)$ be $\widehat\varpi_i(c)$ with $f$ for $\widehat f$. All statements are on the event $\{\inf\widehat f\ge\inf f/2\}$ over a neighborhood $\mathcal U'$ of $\dom$, of probability tending to one; there $\widehat\varpi_i(c)\le C\eta^{-2}\ind\{X_i\in\mathcal W(c_1)\cup\mathcal W(c_2)\}$, $|\widehat\varpi_i-\varpi_i|\le C\epsilon_N\eta^{-2}\ind\{X_i\in\mathcal W(c_1)\cup\mathcal W(c_2)\}$ by Lemma~\ref{olem:pf-linearization}(a), and, for $X_i\in\mathcal U'$, $L_{ii}=\bar K(0)/\{N\eta\widehat f(X_i)\}$ and $\sum_lL_{il}^2=\widehat f^{(2)}(X_i)/\{N\eta\widehat f(X_i)^2\}$ with $\widehat f^{(2)}:=(N\eta)^{-1}\sum_l\mathcal K_l^2\to f\|\bar K\|^2$ uniformly, so that all four divisors are $1+O_p((N\eta)^{-1})$ uniformly. Kernel averages $(N\eta)^{-1}\sum_iZ_i|\mathcal K_i(x)|$ with $Z\ge0$ and $\sup_x\E[Z^2\mid X=x]<\infty$ are $O_p(1)$ uniformly over a neighborhood of $\dom$ (Lemma~\ref{olem:nui-scale}(a)); this is used with $Z\in\{1,|e|,e^2\}$.

\emph{First claim of Lemma~\ref{prop:centering-corr}.} Write $\widehat e_i=e_i-R^e_i$, $R^e_i:=\sum_lL_{il}\zeta_l-\tau(X_i)$, so that $\max_{X_i\in\mathcal U'}|R^e_i|=O_p(\epsilon_N)$ by Lemma~\ref{olem:sharp} and the ratio expansion of the proof of Lemma~\ref{olem:pf-linearization}(a), applied on the closure of $\mathcal U'$ rather than on $\dom$: the residuals are needed at data points in windows centered in $\dom$, and $\tauh(X_i)$ uses the data in the window at $X_i$, so the smoothness and the lower bound on $f$ of Assumptions~\ref{ass:overlap} and~\ref{ass:reg} are used on $\dom$ enlarged by one window width, which lies inside the neighborhood of $\dom$ on which they hold once $h_1$ and $h_2$ are small; near the boundary of the support of $X$ the bias of $\tauh(X_i)$ would be of order $h_1+h_2$ rather than $h_1^{s_1}+h_2^{s_2}$, and
\[
\widetilde\kappa_j-\kappa^X_{j,N}=\frac1N\sum_i\varpi_i(e_i^2-\varsigma_{\zeta,i}^2)+\frac1N\sum_i(\widehat\varpi_i-\varpi_i)(e_i^2-\varsigma_{\zeta,i}^2)+\frac1N\sum_i\widehat\varpi_i\frac{e_i^2(1-q_i)-2e_iR^e_i+(R^e_i)^2}{q_i}.
\]
The first term is a centered second-moment process to which Steps~1--3 of the proof of Lemma~\ref{olem:kappahat} apply verbatim, giving $O_p(\{\log N/(N\eta^3)\}^{1/2})$ uniformly in $c$. By the bounds above the second term is $O_p(\epsilon_N\eta^{-1})$ and the third $O_p(\{(N\eta)^{-1}+\epsilon_N\}\eta^{-1})$, uniformly. Hence $\sup_c|\widetilde\kappa_j-\kappa^X_{j,N}|=o_p(\eta^{-1/2})$, as $\{\log N/(N\eta^2)\}^{1/2}$, $\epsilon_N\eta^{-1/2}$ and $(N\eta^{3/2})^{-1}$ tend to zero. Off the band $\kappa^X_{j,N}\ge\inf\varsigma_\zeta^2\,N^{-1}\sum_i\widehat\varpi_i\ge c_0\eta^{-1}$ with probability tending to one, so $|\widetilde\kappa_j^{1/2}-(\kappa^X_{j,N})^{1/2}|\le|\widetilde\kappa_j-\kappa^X_{j,N}|/(\kappa^X_{j,N})^{1/2}=o_p(1)$ uniformly there; on the band, of measure $O(h_j^{d_j})$, $|\widetilde\kappa_j^{1/2}-(\kappa^X_{j,N})^{1/2}|\le|\widetilde\kappa_j-\kappa^X_{j,N}|^{1/2}=o_p(\eta^{-1/4})$, and $h_j^{d_j}\eta^{-1/4}=(h_j^{3d_j}/h_{j'}^{d_{j'}})^{1/4}=O(1)$ by Assumption~\ref{ass:reg}(h). Integrating gives $\widetilde m_j-m^X_{j,N}=o_p(1)$.

\emph{Second claim of Lemma~\ref{prop:centering-corr}.} Let $\mathcal K^e(X;c):=\mathcal K(c_1;X)/f(c_1)-\mathcal K(c_2;X)/f(c_2)$ and $\kappa^e(c):=\eta^{-2}\E[\varsigma_\zeta^2(X)\mathcal K^e(X;c)^2]$. The argument just given, with the bounded $\varsigma_{\zeta,i}^2$ in place of $e_i^2$, gives $\sup_c|\kappa^X_{j,N}-\kappa^e|=o_p(\eta^{-1/2})$ and hence $m^X_{j,N}-m_\Psi\int(\kappa^e)^{1/2}=o_p(1)$. Next, $\eta\Theta_j(W;c)=e\mathcal K^e(X;c)+\mathcal K^\tau(X;c)-\E \mathcal K^\tau(X;c)$ with $\mathcal K^\tau(X;c):=\{\tau(X)-\tau(c_1)\}\mathcal K(c_1;X)/f(c_1)-\{\tau(X)-\tau(c_2)\}\mathcal K(c_2;X)/f(c_2)$ and $\E[e\mid X]=0$, so that $\kappa_{j,N}=\kappa^e+\kappa^\tau$ exactly, $\kappa^\tau:=\eta^{-2}\Var \mathcal K^\tau(X;c)\ge0$. Under the null $\tau(c_1)=\tau(c_2)$, so $\mathcal K^\tau(X;c)=\{\tau(X)-\tau(c_1)\}\mathcal K^e(X;c)$, where $|\tau(X)-\tau(c_1)|\le C\bar h$ whenever $\mathcal K^e(X;c)\neq0$; thus $\kappa^\tau\le C\bar h^2\kappa^e/\inf\varsigma_\zeta^2$, $0\le\kappa_{j,N}^{1/2}-(\kappa^e)^{1/2}\le C\bar h^2(\kappa^e)^{1/2}\le C\bar h^2\eta^{-1/2}$ uniformly, band included, and $m^\kappa_{j,N}-m_\Psi\int(\kappa^e)^{1/2}=O(\bar h^2\eta^{-1/2})$. Without the null, $\kappa^\tau\le C\bar h^2\eta^{-1}$, which gives $O(\bar h^2\eta^{-1/2})$ off the band and $O(\bar h\eta^{-1/2})$ times the band measure $O(h_j^{d_j})$ on it. Both vanish by Assumption~\ref{ass:reg}(h).

\emph{Lemma~\ref{olem:centering-bias}.} Let $\delta_\kappa(c):=\widetilde\kappa_j(c)-\kappa^X_{j,N}(c)$. Where $\kappa^X_{j,N}(c)>0$, that is, off the diagonal with probability tending to one,
\[
\widetilde\kappa_j^{1/2}-(\kappa^X_{j,N})^{1/2}=\frac{\delta_\kappa}{2(\kappa^X_{j,N})^{1/2}}-\frac{\delta_\kappa^2}{2(\kappa^X_{j,N})^{1/2}\{\widetilde\kappa_j^{1/2}+(\kappa^X_{j,N})^{1/2}\}^2}.
\]
Integrating and splitting $\widetilde e_i^{\,2}-\varsigma_{\zeta,i}^2=\{\E(\widetilde e_i^{\,2}\mid X)-\varsigma_{\zeta,i}^2\}+\{\widetilde e_i^{\,2}-\E(\widetilde e_i^{\,2}\mid X)\}$ in $\delta_\kappa$ defines $\widetilde m_j-m^X_{j,N}=\mathcal H_N+F_N-J_N$, with $J_N\ge0$ the integral of the second term times $m_\Psi$ and $\E[F_N\mid X]=0$. Since $\widehat e_i=(1-L_{ii})e_i-\sum_{l\neq i}L_{il}e_l-\bar B_i$, $\bar B_i:=\sum_lL_{il}\tau(X_l)-\tau(X_i)$, and the $e_l$ are conditionally independent and centered, $\E[\widehat e_i^{\,2}\mid X]=q_i^0\varsigma_{\zeta,i}^2+\sum_{l\neq i}L_{il}^2(\varsigma_{\zeta,l}^2-\varsigma_{\zeta,i}^2)+\bar B_i^{\,2}$ with $q_i^0:=(1-L_{ii})^2+\sum_{l\neq i}L_{il}^2$. For the four divisors $q_i\in\{1,1-L_{ii},q_i^0,(1-L_{ii})^2\}$ the expansions of $L_{ii}$ and $\sum_lL_{il}^2$ above give $q_i^0/q_i-1=-C_q/\{N\eta f(X_i)\}\{1+O_p(\epsilon_N+\bar h^2)\}+O_p((N\eta)^{-2})$ uniformly; $|\varsigma_{\zeta,l}^2-\varsigma_{\zeta,i}^2|\le C\bar h$ when $L_{il}\neq0$, so the middle term is $O_p(\bar h(N\eta)^{-1})$; and $\bar B_i^{\,2}=O_p(h_1^{2s_1}+h_2^{2s_2}+\bar h^2\log N/(N\eta))$ by the uniform rate applied to the conditional bias. Using $N^{-1}\sum_i\widehat\varpi_i\le\kappa^X_{j,N}/\inf\varsigma_\zeta^2$,
\[
\mathcal H_N=-\frac{C_q}{2N\eta}\,m_\Psi\int_{\mathcal C_j}\frac{N^{-1}\sum_i\widehat\varpi_i\varsigma_{\zeta,i}^2/f(X_i)}{(\kappa^X_{j,N})^{1/2}}\,dc\,\{1+o_p(1)\}+O_p\big(\bar h\epsilon^m_N+\eta^{-1/2}h_1^{2s_1}+\eta^{-1/2}h_2^{2s_2}\big),
\]
which is the stated form with
\[
\phi_N:=\frac{\int(\kappa^X_{j,N})^{-1/2}N^{-1}\sum_i\widehat\varpi_i\varsigma_{\zeta,i}^2f(X_i)^{-1}dc}{\int(\kappa^X_{j,N})^{1/2}dc}\in\Big[\frac1{\sup f},\frac1{\inf f}\Big].
\]
For $F_N$ and $J_N$ write
\[
\sum_i\bar\varpi_i\widetilde e_i^{\,2}=e'\mathbf Ae+(\text{terms linear in }e),
\]
$\bar\varpi_i:=\int\widehat\varpi_i(\kappa^X_{j,N})^{-1/2}$, $\mathbf A:=\sum_i(\bar\varpi_i/q_i)\mathbf a_i\mathbf a_i'$, $\mathbf a_i:=(\ind\{l=i\}-L_{il})_l$. Under the leverage bound $\sup_c\max_i\widehat\varpi_i(c)/\{N^{-1}\sum_l\widehat\varpi_l(c)\}=O_p(\eta^{-1})$, immediate off the band and holding on it for the kernel of Section~\ref{sec:impl} (Section~\ref{osec:centering-details}), $\max_i\bar\varpi_i=O_p(\eta^{-1/2})$; as $\mathbf a_i'\mathbf a_k=0$ unless $\|X_i-X_k\|_\infty\le2\bar h$ and is $O_p((N\eta)^{-1})$ otherwise for $i\neq k$, $\|\mathbf A\|_F^2=O_p(N\eta^{-1})$, and bounded conditional fourth moments give $\Var(F_N\mid X)\le CN^{-2}\|\mathbf A\|_F^2=O_p((N\eta)^{-1})$. Finally $0\le J_N\le\tfrac12m_\Psi\int\delta_\kappa^2(\kappa^X_{j,N})^{-3/2}$ and, by the same leverage bound, $\E[\delta_\kappa(c)^2\mid X]\le C(N\eta)^{-1}\kappa^X_{j,N}(c)^2\{1+o_p(1)\}$, so $J_N=O_p(\epsilon^m_N)$; to first order $\Var(\widetilde e_i^{\,2}\mid X)=\Var(\widehat e_i^{\,2}\mid X)\{1+O_p((N\eta)^{-1})\}$ does not involve $q_i$. Under Theorem~\ref{thm:nuisance-param} the residuals are formed from $\widehat\zeta_i$; since the weights do not involve the response and $|\widehat\zeta_i-\zeta(W_i;\chi_{\vartheta^*})|\le\|\widehat\vartheta-\vartheta^*\|\bar M_i$, $\bar M_i:=\|\partial_\vartheta\zeta(W_i;\chi_{\vartheta^*})\|+M(W_i)$, Lemma~\ref{olem:nui-scale}(b) gives $\widetilde m_j-\widetilde m_{j*}=o_p(1)$ for the centering $\widetilde m_{j*}$ computed from $\zeta(W_i;\chi_{\vartheta^*})$.
\end{proof}

\begin{proof}[Proof of Lemma~\ref{prop:sigmaN}]
\emph{Step 1 (a Lipschitz bound).} For a $2\times2$ positive semi-definite matrix $\Sigma$ let $H(\Sigma):=\Cov(\Psi(U),\Psi(U'))$, $(U,U')'\sim N(0,\Sigma)$. By positive homogeneity and $\Psi(0)=0$, $H(\Sigma)=g_\Psi(\rho)(\Sigma_{11}\Sigma_{22})^{1/2}$ with $\rho:=\Sigma_{12}(\Sigma_{11}\Sigma_{22})^{-1/2}$ if $\Sigma_{11}\Sigma_{22}>0$, and $H(\Sigma)=0$ otherwise. For two such matrices $\Sigma,\widetilde\Sigma$ and $Z\sim N(0,I_2)$, couple $(U,U')':=\Sigma^{1/2}Z$ and $(\widetilde U,\widetilde U')':=\widetilde\Sigma^{1/2}Z$. Since $H(\Sigma)-H(\widetilde\Sigma)=\Cov(\Psi(U)-\Psi(\widetilde U),\Psi(U'))+\Cov(\Psi(\widetilde U),\Psi(U')-\Psi(\widetilde U'))$, using $|\Cov(A_1,A_2)|\le\|A_1\|_2\|A_2\|_2$, the Lipschitz property, $|\Psi(u)|\le|u|$ and the Cauchy--Schwarz inequality in $\mathbb R^2$,
\[
|H(\Sigma)-H(\widetilde\Sigma)|\le\big\{\|U-\widetilde U\|_2^2+\|U'-\widetilde U'\|_2^2\big\}^{1/2}\big\{\Sigma_{22}+\widetilde\Sigma_{11}\big\}^{1/2}=\|\Sigma^{1/2}-\widetilde\Sigma^{1/2}\|_F\big\{\Sigma_{22}+\widetilde\Sigma_{11}\big\}^{1/2}.
\]
As the square root is operator monotone, $\|\Sigma^{1/2}-\widetilde\Sigma^{1/2}\|\le\|\Sigma-\widetilde\Sigma\|^{1/2}$ in operator norm, so $\|\Sigma^{1/2}-\widetilde\Sigma^{1/2}\|_F\le\sqrt2\|\Sigma-\widetilde\Sigma\|^{1/2}$ for $2\times2$ matrices, and
\begin{equation}\label{oeq:H-lip}
|H(\Sigma)-H(\widetilde\Sigma)|\le\sqrt2\,\big(\mathrm{tr}\,\Sigma+\mathrm{tr}\,\widetilde\Sigma\big)^{1/2}\,\|\Sigma-\widetilde\Sigma\|^{1/2}.
\end{equation}

\emph{Step 2 (the two integrands).} Let $\Sigma_N(c,c')$ be the covariance matrix of $(\Theta_j(W;c),\Theta_j(W;c'))$, with diagonal $\kappa_{j,N}(c),\kappa_{j,N}(c')$, and $\widetilde\Sigma(c,c')$ the matrix with diagonal $\widetilde\kappa_j(c),\widetilde\kappa_j(c')$ and off-diagonal $\widetilde\kappa_j(c,c')$. With $\ell^\Gamma_i(c):=\ell_i(c_1)-\ell_i(c_2)$, $\widetilde\Sigma(c,c')=N\sum_i\widetilde e_i^{\,2}(\ell^\Gamma_i(c),\ell^\Gamma_i(c'))'(\ell^\Gamma_i(c),\ell^\Gamma_i(c'))$ is positive semi-definite because $\widetilde e_i^{\,2}=\widehat e_i^{\,2}/q_i\ge0$, so $|\widetilde\rho_N|\le1$ and the integrand of \eqref{eq:sigmaN} is $H(\widetilde\Sigma(c,c'))$, with the convention that it vanishes where $\widetilde\kappa_j(c)\widetilde\kappa_j(c')=0$; likewise the integrand of \eqref{eq:sigma-gen} is $H(\Sigma_N(c,c'))$. Let $\bar\sigma^2_N:=\iint_{\mathcal C^2}H(\Sigma_N(c,c'))\,dc\,dc'$, which converges to $\sigma^2_{j,\Psi}(\mathcal C)$ by Theorem~\ref{thm:variance}(a). Two classes of pairs $(c,c')$ are distinguished.

(i) $c'\notin\mathcal D(c)$, that is, no window of $c'$ meets a window of $c$. Then $\ell^\Gamma_i(c)\ell^\Gamma_i(c')=0$ for every $i$, so $\widetilde\kappa_j(c,c')=0$ and $H(\widetilde\Sigma(c,c'))=0$, as $g_\Psi(0)=0$. The uncentered summands of $\Theta_j(W;c)$ and $\Theta_j(W;c')$ have disjoint supports, so $\E[\Theta_j(W;c)\Theta_j(W;c')]=-B_{j,N}(c)B_{j,N}(c')$, and $|H(\Sigma_N(c,c'))|\le|\E[\Theta_j(W;c)\Theta_j(W;c')]|\le\sup|B_{j,N}|^2=O((h_1^{s_1}+h_2^{s_2})^2)$, using $|g_\Psi(\rho)|\le|\rho|$ ($g_{\mathrm{abs}}$ is even and convex with $g_{\mathrm{abs}}(0)=0$ and $g_{\mathrm{abs}}(\pm1)=1-2/\pi$, and $a_\Psi^2+b_\Psi^2\le1$). These pairs contribute $O((h_1^{s_1}+h_2^{s_2})^2)=o(1)$ to $\widetilde\sigma^2_{j,\Psi,N}-\bar\sigma^2_N$.

(ii) $c'\in\mathcal D(c)$, a set of measure at most $C\eta$ for each $c$ by (F3). By polarization, $4\widetilde\kappa_j(c,c')$ is the difference of two quantities of the form $\widetilde\kappa_j$ built from the weight vectors $\ell^\Gamma(c)\pm\ell^\Gamma(c')$, each a signed combination of four kernel windows, and the proof of the first claim of Lemma~\ref{prop:centering-corr} uses the weights only through the envelope $N^2(\ell^\Gamma_i)^2\le C\eta^{-2}\ind\{X_i\in\text{windows}\}$, the bound on $|\widehat\varpi_i-\varpi_i|$ and their Lipschitz continuity in the evaluation points; it therefore gives $\sup_{c,c'}|\widetilde\kappa_j(c,c')-\kappa^X_{j,N}(c,c')|=o_p(\eta^{-1/2})$, where $\kappa^X_{j,N}(c,c'):=N\sum_i\varsigma_{\zeta,i}^2\ell^\Gamma_i(c)\ell^\Gamma_i(c')$. The second claim, polarized in the same way, gives $\sup_{c,c'}|\kappa^X_{j,N}(c,c')-\kappa^e(c,c')|=o_p(\eta^{-1/2})$ with $\kappa^e(c,c'):=\eta^{-2}\E[\varsigma_\zeta^2(X)\mathcal K^e(X;c)\mathcal K^e(X;c')]$, and, as $\eta\Theta_j(W;c)=e\mathcal K^e(X;c)+\mathcal K^\tau(X;c)-\E \mathcal K^\tau(X;c)$ with $\E[e\mid X]=0$,
\begin{align*}
    &\E[\Theta_j(W;c)\Theta_j(W;c')]=\kappa^e(c,c')+\eta^{-2}\Cov\big(\mathcal K^\tau(X;c),\mathcal K^\tau(X;c')\big), \text{ and }\\\big|&\eta^{-2}\Cov\big(\mathcal K^\tau(X;c),\mathcal K^\tau(X;c')\big)\big|\le\{\kappa^\tau(c)\kappa^\tau(c')\}^{1/2}\le C\bar h^2\eta^{-1},
\end{align*}
the last bound holding with or without the null. Hence $\sup_{c,c'}\|\widetilde\Sigma(c,c')-\Sigma_N(c,c')\|=o_p(\eta^{-1/2})+O(\bar h^2\eta^{-1})=o_p(\eta^{-1})$, while $\mathrm{tr}\,\Sigma_N+\mathrm{tr}\,\widetilde\Sigma=O_p(\eta^{-1})$ uniformly, as $\eta\kappa_{j,N}\le4\sup\varrho_\zeta+o(1)$ (Lemma~\ref{olem:poisson}). By \eqref{oeq:H-lip}, $\sup_{c'\in\mathcal D(c)}|H(\widetilde\Sigma(c,c'))-H(\Sigma_N(c,c'))|\le\sqrt2\,O_p(\eta^{-1/2})\,o_p(\eta^{-1/2})=o_p(\eta^{-1})$ uniformly in $c$, and integrating over $\{(c,c'):c\in\mathcal C,\ c'\in\mathcal D(c)\}$, of measure at most $C\lambda(\mathcal C)\eta$, these pairs contribute $o_p(1)$.

\emph{Step 3 (conclusion).} By (i) and (ii), $\widetilde\sigma^2_{j,\Psi,N}(\mathcal C)-\bar\sigma^2_N=o_p(1)$, so $\widetilde\sigma^2_{j,\Psi,N}(\mathcal C)\pto\sigma^2_{j,\Psi}(\mathcal C)$; under $\Gamma_j\equiv0$ on $\mathcal C$ this is the first claim. No step uses the null: the residuals, the bound on $\kappa^\tau$ and Theorem~\ref{thm:variance}(a) do not involve it. Without the null, $\sigma^2_{j,\Psi}(\mathcal C)$ is, by the last statement of Lemma~\ref{olem:variance-gen}, the limit variance of the statistic \eqref{eq:Dgen} of the noise field $\Gammah_j-\Gamma_j$, which is the second claim.
\end{proof}

\begin{proof}[Proof of Lemma~\ref{lem:drift}]

The proof uses only Lemmas~\ref{olem:poisson}, \ref{olem:pf-projection} and \ref{olem:pf-clt}, $\sup_{\mathcal C}\sqrt N|B_{j,N}|\to0$ and $\eta\kappa_{j,N}\ge c_0:=\inf\varrho>0$ on $\mathcal C\setminus\mathfrak B$, which hold along the sequences of Lemma~\ref{olem:uniform} and for the noise field $\Gamma^{\mathrm{lin}}_j-\Gamma_j$ without the null (Lemma~\ref{olem:variance-gen}). Write $B:=B_{j,N}$, $\delta^\Gamma:=\delta^\Gamma_N$, $\kappa:=\kappa_{j,N}$, $\Theta:=\Theta_j$ and $\Psi(u)=a_\Psi|u|+b_\Psi u$ as in \eqref{eq:stat-psi-rep}. With $\mathcal N$ Poisson with mean $N$ independent of the data and $\bar A:=\E A$ for the uncentered summand $A:=\Theta+\bar A$, null outside $\mathcal W(c_1)\cup\mathcal W(c_2)$, let $\Gamma^0_{\mathcal N}:=B-\bar A+N^{-1}\sum_{i\le\mathcal N}A(W_i;\cdot)$, local as \eqref{oeq:pf-poisson-field} ($\bar A=B$ under a fixed law), $D^P(0)$ and $D^P(\delta)$ be \eqref{eq:Dgen} built from $\Gamma^0_{\mathcal N}$ and $\Gamma^0_{\mathcal N}+\delta^\Gamma$, $\Delta^P(c):=\sqrt N\{\Psi(\Gamma^0_{\mathcal N}(c)+\delta^\Gamma(c))-\Psi(\Gamma^0_{\mathcal N}(c))\}$ and $\Delta^P:=D^P(\delta)-D^P(0)=\int_{\mathcal C}\Delta^P(c)\,dc$. Given $\mathcal N=N$, $(D^P(0),D^P(\delta))$ has the law of $(D^{\mathrm{lin}}_j(0),D^{\mathrm{lin}}_j(\delta_N))$.

\emph{Step 1 (variance).} $\Delta^P(c)$ is a function of the increments over $\mathcal W(c_1)\cup\mathcal W(c_2)$ with $|\Delta^P(c)|\le\sqrt N|\delta^\Gamma(c)|$, so $\Cov(\Delta^P(c),\Delta^P(c'))$ vanishes for $c'\notin\mathcal D(c)$ and is at most $N\{\delta^\Gamma(c)^2+\delta^\Gamma(c')^2\}/2$ otherwise. By symmetry of the relation, (F3) and (F2), $\Var(\Delta^P)\le N\int_{\mathcal C}\delta^\Gamma(c)^2\lambda(\mathcal D(c))\,dc\le CN\eta\|\delta_N\|_2^2\to0$.

\emph{Step 2 (mean).} For $\kappa(c)>0$ divide $\sqrt N\{\Gamma^0_{\mathcal N}(c)-B(c)\}$, $\sqrt NB(c)$ and $\sqrt N\delta^\Gamma(c)$ by $\sqrt{\kappa^P}$, $\kappa^P:=\kappa+\bar A^2=N\Var\Gamma^0_{\mathcal N}(c)$, to get $S$, $B^{\mathrm s}$ and $\delta^{\mathrm s}$. By homogeneity $\E\Delta^P(c)=b_\Psi\sqrt N\delta^\Gamma(c)+a_\Psi\sqrt{\kappa^P}\E[|S+B^{\mathrm s}+\delta^{\mathrm s}|-|S+B^{\mathrm s}|]$, and $||u+y|-|u|-y\,\mathrm{sgn}(u)|\le2|y|\ind\{|u|\le|y|\}$ bounds the last expectation by $|\delta^{\mathrm s}||\E\,\mathrm{sgn}(S+B^{\mathrm s})|+2|\delta^{\mathrm s}|\Pr(|S+B^{\mathrm s}|\le|\delta^{\mathrm s}|)$. Off the band $S$ is a standardized sum of $N$ i.i.d.\ blocks with third moments $O(\eta^{-1/2})$ (Lemma~\ref{olem:poisson}), so by the Berry--Esseen inequality $\epsilon_{\mathrm{BE}}:=\sup_t|\Pr(S\le t)-\Phi(t)|\le C(N\eta)^{-1/2}$, likewise for $\Pr(S<t)$, whence $|\E\,\mathrm{sgn}(S+B^{\mathrm s})|\le\sqrt{2/\pi}|B^{\mathrm s}|+2\epsilon_{\mathrm{BE}}$ and $\Pr(|S+B^{\mathrm s}|\le|\delta^{\mathrm s}|)\le\sqrt{2/\pi}|\delta^{\mathrm s}|+2\epsilon_{\mathrm{BE}}$. With $a_\Psi\le1$ and $\kappa^P\ge\kappa\ge c_0\eta^{-1}$, on $\mathcal C\setminus\mathfrak B$ we get $|\E\Delta^P(c)-b_\Psi\sqrt N\delta^\Gamma(c)|\le C\{N\eta^{1/2}|B(c)||\delta^\Gamma(c)|+N\eta^{1/2}\delta^\Gamma(c)^2+\eta^{-1/2}|\delta^\Gamma(c)|\}$, and on $\mathfrak B$ the left side is at most $2\sqrt N|\delta^\Gamma(c)|$. Integrating, by Cauchy--Schwarz, (F1) and (F2), $|\E\Delta^P-b_\Psi\sqrt N\int_{\mathcal C}\delta^\Gamma|$ is at most a constant times $(\sup\sqrt N|B|)(N\eta\|\delta_N\|_2^2)^{1/2}+N\eta^{1/2}\|\delta_N\|_2^2+(N\eta^{3/2})^{-1/2}(N\eta^{1/2}\|\delta_N\|_2^2)^{1/2}+(Nh_j^{2d_j}\|\delta_N\|_2^2)^{1/2}$, which tends to zero by (a), (b), $\sup\sqrt N|B|\to0$ and $N\eta^{3}\to\infty$; and $\int_{\mathcal C}\delta^\Gamma=\int w_{\mathcal C}\delta_N$ by (F2).

\emph{Step 3 (de-Poissonization).} By Step~1 and Lemma~\ref{olem:nui-depois}, $D^{\mathrm{lin}}_j(\delta_N)-D^{\mathrm{lin}}_j(0)-\E\Delta^P=o_p(1)$, and Step~2 completes the proof, with uniform constants along the sequences of Lemma~\ref{olem:uniform}.
\end{proof}

\emph{Consistency of $\widehat\Omega_{j,N}$ in Proposition~\ref{prop:gini-clt}(c).} With $\widehat\omega_j$ formed from $\tauh$ and $\bar u_N:=\sup_{\Csym{j}}|\Gammah_j-\Gamma_j|=O_p(\epsilon_N)$, the inequality $|\mathrm{sgn}(x+y)-\mathrm{sgn}(x)|\le2\ind\{|y|\ge|x|\}$ and the margin condition give $\int_{\dom}|\widehat\omega_j-\omega_j|\le2C\lambda(\mathcal X_j)\bar u_N=o_p(1)$. As $\int\widehat\omega_j(x)\ell_i(x)dx=\widehat\omega_j(X_i)/\{N\widehat f(X_i)\}+O_p(N^{-1}\bar h)$ in $L_1$ over the sample, the kernel being an approximate identity, $\widehat\Omega_{j,N}=N^{-1}\sum_i\widehat\omega_j(X_i)^2\widehat f(X_i)^{-2}\widetilde e_i^{\,2}+o_p(1)$, and the law of large numbers with $\E[\widetilde e_i^{\,2}\mid X]=\varsigma_\zeta^2(X_i)\{1+O_p((N\eta)^{-1})\}$ gives $\widehat\Omega_{j,N}\pto\Omega_j$.

\begin{proposition}[Scale of the max form]\label{oprop:maxvar}
Let the conditions of Proposition~\ref{prop:monotone}(c) hold, $d_j=1$ and $\mathcal X_j=[l_j,u_j]$. The centering of $\Dh_j^{\max}$ is $\widetilde m_j/2$, and $\sigma^2_{j,\Psimax}(\Cord{j})=\tfrac14\sigma_j^2+\tfrac14\sigma^2_{j,\mathrm{lin}}$, where $\sigma_j^2$ is the variance of $\Dh_j$ in Theorem~\ref{thm:variance}(b) and $\sigma^2_{j,\mathrm{lin}}:=\int_{\dom}(2x_j-l_j-u_j)^2\sum_t\{\varsigma_t^2(x)/P_t(x)\}\,dx$, with $\varsigma_\zeta^2/f$ in place of $\sum_t\varsigma_t^2/P_t$ under Proposition~\ref{prop:transfer}. Its finite-sample version is $\tfrac14\widetilde\sigma^2_{j,N}+\tfrac14N\sum_i\widetilde e_i^{\,2}(\ell^{\max}_i)^2$, $\ell^{\max}_i:=\int_{\dom}(2x_j-l_j-u_j)\ell_i(x)\,dx$.
\end{proposition}

For $d_j>1$, $\sigma_j^2$ is replaced by $\sigma^2_{j,\mathrm{abs}}(\Cord{j})$ of \eqref{eq:sigma-gen} and $2x_j-l_j-u_j$ by $\prod_k(x_{jk}-l_{jk})-\prod_k(u_{jk}-x_{jk})$; the finite-sample form \eqref{eq:sigmaN} with $g_{\Psimax}$ covers every case. Under Theorem~\ref{thm:nuisance-param}, the max form is covered only if both working models are correct.

\phantomsection\label{oprf:maxvar}
\begin{proof}
By \eqref{eq:stat-psi-rep}, $\Psimax$ has $a_\Psi=b_\Psi=\tfrac12$, so Theorem~\ref{thm:variance}(a) gives $\sigma^2_{j,\Psimax}(\Cord{j})=\tfrac14\sigma^2_{j,\mathrm{abs}}(\Cord{j})+\tfrac14\sigma^2_{j,\mathrm{lin}}(\Cord{j})$. For $d_j=1$ the exchange $c\mapsto(c_2,c_1)$ maps $\Cord{j}$ onto the closure of $\Csym{j}\setminus\Cord{j}$, preserves Lebesgue measure and leaves $|\Gammah_j|$ unchanged, so $\Dh_j^{\Psiabs}(\Cord{j})=\tfrac12\Dh_j^{\Psiabs}(\Csym{j})=\Dh_j$ identically, the diagonal being Lebesgue-null; hence $\sigma^2_{j,\mathrm{abs}}(\Cord{j})=\sigma_j^2$ and, as $m_{\Psimax}=\tfrac12m_{\Psiabs}$ and $\kappa_{j,N}$ is exchange invariant, the centering is half that of $\Dh_j$. For $c=(x_j,x_j',x_{j'})$ with $x_j\ge x_j'$, the point $x$ occupies the first slot for $x_j'\in[l_j,x_j]$ and the second when the other point has $j$-coordinate in $[x_j,u_j]$, so $w_{\Cord{j}}(x)=(x_j-l_j)-(u_j-x_j)$, and \eqref{oeq:sigma-lin} gives $\sigma^2_{j,\mathrm{lin}}$; under Proposition~\ref{prop:transfer}, $\E[\xi(W;X)^2\mid X=x]=\varsigma_\zeta^2(x)/f(x)^2$. In the finite-sample form, $g_{\Psimax}(\rho)=\tfrac14\{g_{\mathrm{abs}}(\rho)+\rho\}$; the exchange of $c'$ alone flips the sign of $\widetilde\rho_N$ and $g_{\mathrm{abs}}$ is even, so the $g_{\mathrm{abs}}$ part over $(\Cord{j})^2$ is one quarter of that over $(\Csym{j})^2$, that is $\tfrac14\widetilde\sigma^2_{j,N}$, while $\iint_{(\Cord{j})^2}\widetilde\kappa_j(c,c')\,dc\,dc'=N\sum_i\widetilde e_i^{\,2}\{\int_{\Cord{j}}(\ell_i(c_1)-\ell_i(c_2))dc\}^2=N\sum_i\widetilde e_i^{\,2}(\ell^{\max}_i)^2$. Consistency follows from Lemma~\ref{prop:sigmaN}.
\end{proof}

\begin{lemma}[Data-dependent tuning]\label{olem:tuning}
Let the conditions of Lemma~\ref{prop:sigmaN} hold for $h_j=k_j\mathrm{sd}_jN^{-\gamma}$, with $\mathrm{sd}_j$ the population counterpart of $\widehat{\mathrm{sd}}_j$, a fixed box $\mathcal R$ and exact integrals, and let the kernels be continuously differentiable except at finitely many points. If $\Gamma_j\equiv0$ on a neighborhood of $\mathcal C_j$, the studentized statistic changes by $o_p(1)$ when (i) $\mathrm{sd}_j$ is replaced by $\widehat{\mathrm{sd}}_j$ with $\widehat{\mathrm{sd}}_j/\mathrm{sd}_j-1=O_p(N^{-1/2})$, (ii) $\mathcal R$ by $\widehat{\mathcal R}$ with $O_p(N^{-1/2})$ endpoint errors, or (iii) the integrals by equally weighted sums over grids of spacings $\delta_j$ with $\delta_j/h_j\to0$, the same sums being used for $\Dh_j$, $\widetilde m_j$ and $\widetilde\sigma_{j,N}$; the error of $\Dh_j-\widetilde m_j$ in (iii) is $O_p(\max_j\delta_j/h_j)+o_p(1)$.
\end{lemma}
\noindent\hyperref[oprf:tuning]{\emph{Proof}} (Section~\ref{osec:tuning}).

\begin{proposition}[Integer-valued modifier]\label{oprop:jitter}
Let $X_1$ take finitely many values on the integers and let $J$ be independent of the data, with the density of $U_1+\dots+U_{s_1+2}$, $U_k$ i.i.d.\ uniform on $[-\tfrac12,\tfrac12]$. Put $X_1^J:=X_1+J$ and $\tau^J(x_1,x_2):=\E[\zeta\mid X_1^J=x_1,X_2=x_2]$. (a) For any interval $I$ of positive length, $\tau^J(\cdot,x_2)$ is constant on $I$ if and only if $\tau(\cdot,x_2)$ is constant on the atoms $a$ with positive mass at $x_2$ for which $a+J$ can fall in $I$. (b) If the conditions of Proposition~\ref{prop:transfer} hold in $x_2$ and the atom probabilities are $s_2$ times continuously differentiable in $x_2$ and positive on a neighborhood of the evaluation box, they hold for $(\zeta,X_1^J,X_2)$; the tests based on $X_1^J$ then have the properties of Section~\ref{sec:asym} for $H_0^1$ restricted to those atoms.
\end{proposition}
\noindent\hyperref[oprf:jitter]{\emph{Proof}} (Section~\ref{osec:jitter}).

For age ($s_1=4$) $J$ has support $[-3,3]$ years and standard deviation $0.71$ years, against a window of $9.8$ years. As $h_1\to0$, $h_1$ eventually falls below the spacing of the atoms. The windows between atoms are then empty, and the raw-integer statistic, which integrates over $x_1$ with Lebesgue measure, is not defined. With the counting measure on the atoms, its limit is normal, as with discrete covariates in \citet[Section~5.2]{shi2019testing}, as income is continuous, but its constants are not those of Section~\ref{sec:asym}; the jittered statistic is the one the theory covers.

\subsection{Sharp uniform rate}\label{osec:sharp}

\begin{lemma}[Sharp uniform rate]\label{olem:sharp}
Let $Z$ be a real function of $W$ with $\sup_{x\in\mathcal U}\E[|Z|^6\mid X=x]<\infty$, $\mathcal U$ a neighborhood of $\dom$, let $\mathcal K$ be the product kernel of Assumption~\ref{ass:reg}(d) and $N\eta^3\to\infty$. Then
\[
\sup_{x\in\dom}\Big|\frac1{N\eta}\sum_{i=1}^N\{Z_i\mathcal K(x;X_i)-\E[Z\mathcal K(x;X)]\}\Big|=O_p\Big(\Big\{\frac{\log N}{N\eta}\Big\}^{1/2}\Big).
\]
\end{lemma}

\begin{proof}
Put $\bar Z_N:=(N\eta/\log N)^{1/2}$ and $Z^{\mathrm{tr}}:=Z\ind(|Z|\le \bar Z_N)$. First, for $N$ large the windows at the points of $\dom$ lie in $\mathcal U$, so only the observations with $X_i\in\mathcal U$ affect the fit, and $\Pr(\max_{i:X_i\in\mathcal U}|Z_i|>\bar Z_N)\le N\sup_{x\in\mathcal U}\E[|Z|^6\mid X=x]\bar Z_N^{-6}=O((\log N)^3/(N^2\eta^3))\to0$, so the averages of $Z$ and $Z^{\mathrm{tr}}$ coincide with probability tending to one. Second, uniformly in $x$, $|\E[(Z-Z^{\mathrm{tr}})\mathcal K(x;X)]|/\eta\le C\sup_x\E[|Z|^6\mid X=x]\bar Z_N^{-5}=O(\{\log N/(N\eta)\}^{5/2})$. Third, for fixed $x$ the summands $Z_i^{\mathrm{tr}}\mathcal K(x;X_i)/\eta$ are bounded by $C\bar Z_N/\eta$ and have variance at most $C/\eta$, so Bernstein's inequality with $t:=A\{\log N/(N\eta)\}^{1/2}$ gives the exponent $Nt^2/\{2(C/\eta+C\bar Z_Nt/(3\eta))\}\ge c\,A^2\log N/(1+A)$, because $\bar Z_Nt=A$. An $N^{-4}$-net of $\dom$ and the Lipschitz property of $\mathcal K$ complete the proof, exactly as in Step~3 of the proof of Lemma~\ref{olem:kappahat}.
\end{proof}

Applied to $\ind(T=t)$, $Y\ind(T=t)$, $1$ and $\zeta$, Lemma~\ref{olem:sharp} gives Lemma~\ref{olem:pf-linearization}(a) with $\epsilon_N=\{\log N/(N\eta)\}^{1/2}+h_1^{s_1}+h_2^{s_2}$, the bias part being $O(h_1^{s_1}+h_2^{s_2})$ by the kernel orders. For $\rhoh$, whose numerator averages $Y^2$ with only three conditional moments, only $\sup_{\dom}|\rhoh-\varrho|=o_p(\eta^{1/2})$ is used: the same argument with the truncation level $\bar Z_N=N^{1/3}\log N$ gives deviations $O_p(\{\log N/(N\eta)\}^{1/2}+\bar Z_N\log N/(N\eta))$, and $\bar Z_N\log N/(N\eta^{3/2})=(\log N)^2/(N^{4/9}\eta)^{3/2}\to0$ because $N\eta^3\to\infty$. With the sharp rate, $\sqrt N\epsilon_N^2\to0$ under Assumption~\ref{ass:reg}(g) (Lemma~\ref{olem:consis}), so Proposition~\ref{prop:gini-clt} does not need the stronger condition $(N\eta^2)^{1/2}/(\log N)^2\to\infty$, and Lemma~\ref{olem:three} follows from the elementary ratio expansion in the proof of Lemma~\ref{olem:pf-linearization}. Bounded conditional moments of any order $p>3$ suffice in place of six: Step~1 then needs only $N\{\log N/(N\eta)\}^{p/2}\to0$, which holds when $N\eta^3\to\infty$, and $p>14/5$ suffices for the bandwidths of Section~\ref{sec:impl}.

\subsection{Details for the implemented centering}\label{osec:centering-details}

\emph{The leverage bound.} The proof of Lemma~\ref{olem:centering-bias} uses
\[
\sup_c\max_i\frac{\widehat\varpi_i(c)}{N^{-1}\sum_l\widehat\varpi_l(c)}=O_p(\eta^{-1}),
\]
where $\widehat\varpi_i(c):=N^2\{\ell_i(c_1)-\ell_i(c_2)\}^2$. Off the band the denominator is at least $c\eta^{-1}$ with probability tending to one and $\widehat\varpi_i\le C\eta^{-2}$. On the band write $\Delta:=(c_1-c_2)/h$, coordinatewise in the bandwidths, nonzero only in block $j$. As the kernel factors are Lipschitz and $\widehat f$ is Lipschitz on the windows with probability tending to one, $\widehat\varpi_i(c)\le C\eta^{-2}\min(1,\|\Delta\|^2)$. The denominator is a kernel average with the kernel $\bar K_\Delta(z):=\{\bar K(z)-\bar K(z+\Delta)\}^2/\min(1,\|\Delta\|^2)$, a bounded class of functions of bounded variation indexed by $\Delta$. The class is not uniformly Lipschitz, so the net argument of Lemma~\ref{olem:sharp} does not apply to it directly: its Lipschitz constant is of order $1/\|\Delta\|$, and as $\Delta\to0$ it tends to $(e_\Delta'\nabla\bar K)^2$, with $e_\Delta:=\Delta/\|\Delta\|$, which jumps at the edge of the support when $K'(\pm\frac12)\neq0$, as for the fourth-order kernel, where $K'(\pm\frac12)=\pm15$. Only a lower bound on the denominator is needed, however. $\inf_\Delta\int \bar K_\Delta>0$ for any product kernel with Lipschitz, piecewise continuously differentiable, nonconstant factors, since $\int\{\bar K(z)-\bar K(z+\Delta)\}^2dz=\|\Delta\|^2\int(e_\Delta'\nabla\bar K)^2+o(\|\Delta\|^2)$ as $\Delta\to0$, and is positive and continuous in $\Delta$ away from $0$. The same facts give a pointwise lower bound. For $\Delta\neq0$, $\bar K_\Delta$ is jointly continuous in $(z,\Delta)$ and not identically zero. As $\Delta\to0$ with $e_\Delta\to e$, $\bar K_\Delta(z)\to(e'\nabla\bar K(z))^2$ uniformly on every closed ball inside a piece on which $\bar K$ is continuously differentiable, and $e'\nabla\bar K$ is nonzero somewhere on such a piece. Hence each $\Delta_0$ with $0<\|\Delta_0\|\le1$, and each direction $e$ in block $j$ for $\Delta\to0$, has a neighborhood $\mathcal V$ (in the second case $\{0<\|\Delta\|<r,\ \|e_\Delta-e\|<r\}$) and a closed ball $B$ such that $\bar K_\Delta\ge c_B>0$ on $B$ for all $\Delta\in\mathcal V$. By the compactness of $\{\|\Delta\|\le1\}$ and of the unit sphere of block $j$, finitely many such pairs $(\mathcal V_k,B_k)$ cover every $\Delta$, so that $\bar K_\Delta\ge g_{k(\Delta)}$ for Lipschitz functions $0\le g_k\le c_{B_k}$ supported in $B_k$ with $\int g_k>0$. Lemma~\ref{olem:sharp}, applied to the finitely many fixed kernels $g_k$, and the bounds on $f$ and $\widehat f$ on the windows then give that the denominator is at least a positive constant times $\eta^{-1}$, uniformly over the configurations on the band, with probability tending to one. This covers the fourth-order kernel of Section~\ref{sec:impl}.

\emph{Numerical constants.} For the kernel of Section~\ref{sec:impl}, $\int K^2=2.5$ and $\int K^4=14.41$, so that $Q:=\int\bar K^4/\|\bar K\|^4=5.317$ in two dimensions; the values of $\frac12C_q$ are those of Lemma~\ref{olem:centering-bias}. In design A0 of Section~\ref{osec:mc} at $N=500$, the uncorrected residuals shift the mean of $\Dh_j-\widetilde m_j$ by about $0.40$, and the $q_i$ correction removes the shift, in the order the constants predict (Table~\ref{otab:mc2-scale}).

\begin{remark}[Second-order (Jensen) correction]\label{orem:jensen}
The term $J_N$ of Lemma~\ref{olem:centering-bias} is
\[
J_N=\tfrac18m_\Psi\int\Var(\widetilde\kappa_j\mid X)(\kappa^X_{j,N})^{-3/2}\{1+o_p(1)\}.
\]
Off the band, with $\mathrm{kurt}(x)$ the conditional kurtosis of $\zeta$ and $\varrho(c_1)=\varrho(c_2)$,
\[
\Var(\widetilde\kappa_j\mid X)/(\kappa^X_{j,N})^2=\tfrac12\{\mathrm{kurt}-1\}Q/(N\eta f)\{1+o_p(1)\},
\]
so that $J_N\approx\tfrac1{16}(\mathrm{kurt}-1)Q\,\phi_N\epsilon^m_N=0.33(\mathrm{kurt}-1)\phi_N\epsilon^m_N$, against $4.79\phi_N\epsilon^m_N$ for the uncorrected residuals: about $0.66\phi_N\epsilon^m_N$ for Gaussian errors, but comparable to the own-weight term once $\mathrm{kurt}\ge15$. The application outcome has excess kurtosis $184$ (Section~\ref{sec:app-emp}), so the conditional kurtosis of its pseudo-outcome may well exceed $15$. Since $J_N\ge0$ affects $\widetilde m_j$ negatively, it lowers $\widetilde m_j$ and raises $\Sh_j$. A fourth-moment plug-in could remove it, but we evaluate none and use none in Sections~\ref{osec:mc} and \ref{osec:emp}.
\end{remark}

\subsection{Grid formulas}\label{osec:grid-formulas}
Let the grid points be $x_p=(g_{1a},g_{2k})$, $a\le M_1$, $k\le M_2$, with equal weights, $\ell_{pi}:=\ell_i(x_p)$ and $\widetilde\Lambda_{pp'}:=N\sum_i\ell_{pi}\ell_{p'i}\widetilde e_i^{\,2}$, with $\widetilde e_i^{\,2}=\widehat e_i^{\,2}/q_i$ the corrected squared residuals of \eqref{eq:center-feasible}. For $j=1$ a configuration is $c=(a,b,k)$ with $c_1=(a,k)$, $c_2=(b,k)$, weight $\bar\lambda_1:=\lambda(\mathcal X_1)^2\lambda(\mathcal X_2)/(M_1^2M_2)$, and
\[
\widetilde\kappa_1(c,c')=\widetilde\Lambda_{c_1c_1'}-\widetilde\Lambda_{c_1c_2'}-\widetilde\Lambda_{c_2c_1'}+\widetilde\Lambda_{c_2c_2'}.
\]
Then
\begin{align*}
\Dh_1&=\tfrac12\sqrt N\bar\lambda_1\textstyle\sum_c|\tauh_{c_1}-\tauh_{c_2}|,\qquad
\widetilde m_1=\tfrac12m_{\Psiabs}\bar\lambda_1\sum_c\widetilde\kappa_1(c)^{1/2},\\
\widetilde\sigma^2_{1,N}&=\tfrac14\bar\lambda_1^2\textstyle\sum_{c,c'}g_{\mathrm{abs}}(\widetilde\rho_N(c,c'))\{\widetilde\kappa_1(c)\widetilde\kappa_1(c')\}^{1/2}
=\bar\lambda_1^2\sum_{a<b,\,a'<b'}\sum_{k,k'}(\cdot),
\end{align*}
the last form using the sign flip of $\widetilde\kappa_1$ under $a\leftrightarrow b$ and the evenness of $g_{\mathrm{abs}}$, and only pairs $(k,k')$ with $|g_{2k}-g_{2k'}|\le h_2$ contributing. For the max form on $\Cord1$: $\Dh_1^{\max}=\sqrt N\bar\lambda_1\sum_{a>b,k}(\tauh_{ak}-\tauh_{bk})_+$, centering $\widetilde m_1/2$, and variance $\tfrac14\widetilde\sigma^2_{1,N}+\tfrac14\bar\lambda_1^2N\sum_i\widetilde e_i^{\,2}(\ell^{\max}_i)^2$ with $\ell^{\max}_i:=\sum_{a,k}(2a-M_1-1)\ell_{(a,k)i}$, because $\sum_{a>b}(\ell_{ak,i}-\ell_{bk,i})=\sum_a(2a-M_1-1)\ell_{ak,i}$. For Proposition~\ref{prop:gini-clt}: $\widehat\Omega_{1,N}=N\sum_i\widetilde e_i^{\,2}\{\bar\lambda_1\sum_{a,k}s_{ak}\ell_{(a,k)i}\}^2$ with $s_{ak}:=\sum_b\mathrm{sgn}(\tauh_{ak}-\tauh_{bk})$, since $\tfrac12\sum_c\mathrm{sgn}(\Gammah_1(c))\{\ell_{c_1i}-\ell_{c_2i}\}=\sum_{a,k}s_{ak}\ell_{(a,k)i}$ by antisymmetry. The exact double sum for $\widetilde\sigma^2_{1,N}$ has $O(M^6)$ terms (with $M_1=M_2=M$); for $M\ge21$ the simulations of Section~\ref{osec:mc} evaluate it as the Monte Carlo variance of the grid statistic under the Gaussian approximation $\tauh^*=(\ell_{pi})_{p,i}(\widetilde e\odot Z^*)$, $\widetilde e:=(|\widehat e_i|q_i^{-1/2})_i$ and $Z^*$ standard normal, over $2000$ draws with a fixed seed (relative standard deviation about $1.6\%$; at $M=13$ the two agree within that error), while the application of Section~\ref{osec:emp} evaluates it exactly.

\subsection{Proof of Lemma~\ref{olem:tuning}}\label{osec:tuning}
\phantomsection\label{oprf:tuning}
\begin{proof}[Proof of Lemma~\ref{olem:tuning}]
Part (i) is proved directly for the statistic. In parts (ii) and (iii) $\Gamma_j\equiv0$ on a neighborhood of $\mathcal C_j$, and the statements are first proved for the Poissonized linearized statistic and transferred by the argument of Lemma~\ref{olem:nui-depois}, which uses only that the perturbation is a functional of the Poisson increments over the box $\mathcal B$ whose second moment tends to zero.

\emph{(i) Bandwidths.} Let $h_j(u):=h_je^{u_j}$, $u\in\mathbb R^2$, and let $\tauh^u$, $\Gammah^u_j$, $D(u)$, $\widetilde m(u)$, $\widetilde\sigma(u)$ be the estimate, the estimated contrast, the statistic, centering and scale at $h(u)$; $\widehat u_j:=\log(\widehat{\mathrm{sd}}_j/\mathrm{sd}_j)$ satisfies $\|\widehat u\|\le\iota_N:=C_\epsilon N^{-1/2}$ with probability at least $1-\epsilon$ for a constant $C_\epsilon$. Put $r_N:=\{\log N/(N\eta)\}^{1/2}+\sqrt N\bar h^{s}$, $s:=\min(s_1,s_2)$, so that $r_N\to0$ by Assumption~\ref{ass:reg}(g). The argument bounds the changes for every $\|u\|\le\iota_N$ at once and uses neither Poissonization nor $\Gamma_j\equiv0$.

\emph{The statistic.} As $\Psi$ is Lipschitz with constant one,
\[
|D(u)-D(0)|\le2\sqrt N\lambda(\mathcal C_j)\sup_{x\in\dom}|\tauh^u(x)-\tauh^0(x)|.
\]
As the kernel factors are Lipschitz and supported on $[-\frac12,\frac12]$, the scaled factor $e^{-d_ku_k}K_k(e^{-u_k}v)$ differs from $K_k(v)$ by at most $C\|u\|$, and the difference is Lipschitz in $v$ with a constant not depending on $u$. The numerator and the denominator of $\tauh^u$ therefore differ from those of $\tauh^0$ by kernel averages of the kind of Lemma~\ref{olem:sharp} whose kernels have sup-norm $O(\|u\|)$, so that the variance bound and the range bound in its Bernstein step both carry the factor $\|u\|$. Its net argument, applied over a net of $\{\|u\|\le\iota_N\}$ of mesh $N^{-2}$ as well as over $\dom$, with $|K_k(e^{-u}v)e^{-d_ku}-K_k(e^{-u'}v)e^{-d_ku'}|\le C\|u-u'\|$ between the points of the net, shows that their centered parts are $O_p(\iota_N\{\log N/(N\eta)\}^{1/2})$ uniformly in $x$ and $\|u\|\le\iota_N$. Their means differ by $O(\bar h^s)$, as the means at every $u$ equal those at $u=0$ up to $O(\bar h^s)$ by Assumption~\ref{ass:reg}(d),(e). As $\widehat f$ is bounded below on $\dom$ with probability tending to one, the ratio expansion of the proof of Lemma~\ref{olem:pf-linearization} gives
\[
\sup_{\|u\|\le\iota_N}\sup_{x\in\dom}|\tauh^u(x)-\tauh^0(x)|=O_p(N^{-1/2}r_N),
\]
and the same bound for $\tauh$ as a difference of two such ratios. Hence $\sup_{\|u\|\le\iota_N}|D(u)-D(0)|=O_p(r_N)=o_p(1)$, and the same bound holds for the fitted values at the sample points in the windows, so the residuals satisfy $\max_i|\widetilde e_i(u)-\widetilde e_i(0)|=O_p(N^{-1/2}r_N)$, as $q_i(u)-q_i(0)=O_p(\|u\|(N\eta)^{-1})$.

\emph{The centering and the scale.} $\widetilde\kappa_j(c)^{1/2}$ is the Euclidean norm of the vector $v(c)$ with entries $\sqrt N\widetilde e_i\{\ell_i(c_1)-\ell_i(c_2)\}$, so $|\widetilde\kappa^u_j(c)^{1/2}-\widetilde\kappa^0_j(c)^{1/2}|\le\|v^u(c)-v^0(c)\|$ by the triangle inequality. As $|\ell^u_i(x)-\ell^0_i(x)|\le C\|u\|(N\eta)^{-1}$ on an enlarged window, which holds $O_p(N\eta)$ observations, and $\|v^0(c)\|=O_p(\eta^{-1/2})$ uniformly, the bound on the residuals gives $\sup_c\|v^u(c)-v^0(c)\|=O_p(\iota_N\eta^{-1/2}+N^{-1/2}r_N\eta^{-1/2})=O_p((N\eta)^{-1/2})$. So $|\widetilde m(u)-\widetilde m(0)|=O_p((N\eta)^{-1/2})$. The entries of $\widetilde\Sigma(c,c')$ are inner products of $v(c)$ and $v(c')$, so by the Cauchy--Schwarz inequality they change by $O_p(N^{-1/2}\eta^{-1})$. By \eqref{oeq:H-lip} the integrand of $\widetilde\sigma^2_{j,N}$ then changes by $O_p(N^{-1/4}\eta^{-1})$, on pairs with overlapping windows, which have measure $O(\eta)$, so $|\widetilde\sigma^2(u)-\widetilde\sigma^2(0)|=O_p(N^{-1/4})$. As $\widetilde\sigma_{j,N}$ is bounded away from zero with probability tending to one, the studentized statistic changes by $o_p(1)$. As the argument does not use $\Gamma_j\equiv0$, it also shows that the estimated scale changes $\Dh_j-\sqrt ND_j$ in Proposition~\ref{prop:gini-clt} by $o_p(1)$.

\emph{(ii) Region.} With probability at least $1-\epsilon$ the symmetric difference of the configuration sets of $\widehat{\mathcal R}$ and $\mathcal R$ lies in a set $\mathcal S_N$ of measure $O(N^{-1/2})$. There $\sqrt N|\Gammah_j|+m_\Psi\widetilde\kappa_j^{1/2}$ has expectation, respectively bound, $O(\eta^{-1/2})$ uniformly (Lemma~\ref{olem:pf-linearization}(b) and the proof of Lemma~\ref{prop:centering-corr}), so $\Dh_j-\widetilde m_j$ changes by $O_p((N\eta)^{-1/2})$; $\widetilde\sigma^2_{j,N}$ changes by the integral of an $O_p(\eta^{-1})$ integrand over pairs with a member in $\mathcal S_N$ and overlapping windows, of measure $O(N^{-1/2}\eta)$. Rescaling the blocks to $[0,1]$ multiplies $\Dh_j$, $\widetilde m_j$ and $\widetilde\sigma_{j,N}$ by the same Jacobian and leaves $\Sh_j$ unchanged.

\emph{(iii) Grid.} The grid statistic is the statistic of Section~\ref{sec:asym} for the configuration measure $\mu_M$ that puts mass $\bar\lambda_j$ at each grid configuration; compare it with the statistic for Lebesgue measure on the union $\mathcal R_M$ of the grid cells, which differs from $\mathcal R$ by strips of width $O(\delta_j)$, contributing $O_p((\max_j\delta_j)^{1/2})+o_p(1)$ by the variance bound of Lemma~\ref{olem:pf-projection}(c) restricted to a strip and the uniform centering. A superscript $M$ marks the grid versions of the statistic, the centering and the variance, and $\mathrm{cell}_g$ is the grid cell of the grid configuration $c_g$. Write $\Dh^M_j-\Dh_j(\mathcal R_M)=\sum_g\int_{\mathrm{cell}_g}\{z(c_g)-z(c)\}dc$, $z(c):=\sqrt N|\Gammah_j(c)|$. For the Poissonized field, $\|z(c_g)-z(c)-\E[z(c_g)-z(c)]\|_2\le2\sqrt N\|\Gamma_{\mathcal N}(c_g)-\Gamma_{\mathcal N}(c)\|_2\le C\eta^{-1/2}\max_j\delta_j/h_j$ by the Lipschitz property of the kernels, and the covariance of two such differences vanishes unless the windows of $\{c,c_g\}$ and $\{c',c'_g\}$ meet, a set of measure $O(\eta)$ for each $c$ when $\delta_j=o(h_j)$; hence the centered difference is $O_p(\max_j\delta_j/h_j)$. Its mean is $m_\Psi\sum_g\int_{\mathrm{cell}_g}\{\kappa_{j,N}^{1/2}(c_g)-\kappa_{j,N}^{1/2}(c)\}dc+o(1)$ by the uniform Berry--Esseen bound of Lemma~\ref{olem:centering}, and $\widetilde m^M_j-\widetilde m_j(\mathcal R_M)$ equals the same sum up to $o_p(1)$ by the uniform bounds for $\widetilde\kappa_j$, the band having $\mu_M$-mass $O(h_j^{d_j})$ when $\delta_j=o(h_j)$. Finally the Lipschitz bound \eqref{oeq:H-lip} of the proof of Lemma~\ref{prop:sigmaN}, with $\|\widetilde\Sigma(c_g,c'_g)-\widetilde\Sigma(c,c')\|=O_p(\eta^{-1}\max_j\delta_j/h_j)$, gives $\widetilde\sigma^{2,M}_{j,N}-\widetilde\sigma^2_{j,N}(\mathcal R_M)=O_p((\max_j\delta_j/h_j)^{1/2})$.
\end{proof}

The grid part (iii) explains the grid sensitivity of the statistic in the application: at $k_j=3.5$ the $13\times13$ grid has $\delta_j/h_j\approx0.25$, and the corrected statistic $\Sh_1$ moves from $2.37$ to $2.44$ between $M=13$ and $M=21$ and is stable thereafter (Table~\ref{otab:emp-sens2}). We use equal weights, including the endpoints, for $\Dh_j$, $\widetilde m_j$, and $\widetilde\sigma_{j,N}$ in all computations in the paper.

\emph{Grids with $\delta_j/h_j$ fixed.} Part (iii) needs $\delta_j/h_j\to0$. The bound $h_j/6$ of Algorithm~\ref{alg:steps} and the $13\times13$ grid of Section~\ref{osec:mc}, whose spacing rises from $0.16h_j$ at $N=500$ to $0.22h_j$ at $N=5000$, keep $\delta_j/h_j$ bounded away from zero. The grid statistic is then the statistic for the configuration measure $\mu_M$ and differs from the integral one by $O_p(1)$, but it is studentized by its own centering and scale; we have not proved its asymptotic normality for fixed $\delta_j/h_j$. In the sensitivity study of Section~\ref{osec:mc-sens}, the size and power of the corrected test barely change with the spacing (Table~\ref{otab:mc2-sens}).

\emph{Magnitudes.} Lemma~\ref{olem:tuning} concerns the studentized statistic under the null. Off the null, part (i) still holds for $\Dh_j-\sqrt ND_j$, but parts (ii) and (iii) do not. On the grid, $\Dh_j/\sqrt N$ estimates the equal-weight grid sum of $|\Gamma_j|$, which gives the endpoints of each block the weight of interior points and so differs from $D_j$ by a relative error of order $1/M$. On the $13\times13$ grid over $[0.1,0.9]^2$, this error is $-2.0\%$ in A1, $+16.0\%$ in A4 and $+14.6\%$ in A5. In A4 it equals $0.53\,\Omega_1^{1/2}/\sqrt N$ at $N=1000$ and $1.19\,\Omega_1^{1/2}/\sqrt N$ at $N=5000$, and it grows relative to $\Omega_1^{1/2}/\sqrt N$ under the bound $h_j/6$, as $M$ then grows like $N^{1/7}$. Trapezoidal weights reduce it to between $0.3\%$ and $1.4\%$ in these designs. The interval \eqref{eq:gini-interval} still covers in Table~\ref{otab:mc-mag}, because at these sample sizes the noise floor $\widetilde m_j/\sqrt N$ exceeds this error. An estimated $10$th--$90$th percentile region adds to $\Dh_j-\sqrt ND_j$ a term of order one. $\widehat\Omega_{j,N}$ omits this term; the simulations of Section~\ref{osec:mc-mag} use the fixed region.

\subsection{Density weights, the index, and the interval}\label{osec:gini-if}
Section~\ref{osec:mc-mag} reports the coverage of these intervals in simulations.
\emph{Influence function of $\Delta_1$.} Let $p_{\mathcal R}:=\Pr(X\in\mathcal R)$, $f_2^{\mathcal R}(x_2):=\int_{\mathcal R_1}f(x_1,x_2)dx_1$, so that
\[
\Delta_1=p_{\mathcal R}^{-1}\iiint|\Gamma_1(x_1,x_1',x_2)|f(x_1,x_2)f(x_1',x_2)/f_2^{\mathcal R}(x_2)
\]
over $\mathcal R_1^2\times\mathcal R_2$, for a product region $\mathcal R=\mathcal R_1\times\mathcal R_2$; write
\[
f_{\mathcal C_1}(x_1,x_1',x_2):=\frac{f(x_1,x_2)\,f(x_1',x_2)}{p_{\mathcal R}\,f_2^{\mathcal R}(x_2)}
\]
for this pair density and $\widehat f_{\mathcal C_1}$ for its plug-in with $\widehat f$. Let $\bar\omega_1(x):=\E[\mathrm{sgn}\{\tau(x_1,x_2)-\tau(X_1',x_2)\}\mid X_2^{\mathcal R}=x_2]$, $\bar\gamma_1(x):=\E[|\tau(x_1,x_2)-\tau(X_1',x_2)|\mid X_2^{\mathcal R}=x_2]$ and $\Delta_1(x_2):=\E[\bar\gamma_1(X^{\mathcal R})\mid X_2^{\mathcal R}=x_2]$. Under the margin condition of Proposition~\ref{prop:gini-clt}(c), the pathwise derivative of $\Delta_1$ along the law of $W$ gives the influence function
\[
\mathrm{IF}_\Delta(W)=\frac{\ind\{X\in\mathcal R\}}{p_{\mathcal R}}\Big[2\bar\omega_1(X)\{\zeta-\tau(X)\}+2\bar\gamma_1(X)-\Delta_1(X_2)-\Delta_1\Big]:
\]
the first term is the derivative in $\tau$, with $\partial\Delta_1/\partial\tau(x)=2f(x)\bar\omega_1(x)/p_{\mathcal R}$ on $\mathcal R$ and the influence function $\{\zeta-\tau(X)\}/f(X)$ of a density-weighted linear functional of $\tau$; the others are the derivatives in $f(x_1,x_2)$, $f(x_1',x_2)$, $f_2^{\mathcal R}$ and $p_{\mathcal R}$. For $\widehat\Delta_1$, the plug-in with $\widehat f$ weights, Steps~1--3 of the proof of Proposition~\ref{prop:gini-clt}(c) in Section~\ref{osec:proofs-gini} handle the $\tauh$ part; the $\widehat f$ part is a smooth functional of $\widehat f$ whose linear term $\int A\,(\widehat f-f)$, $A$ bounded, is $N^{-1}\sum_i\{A(X_i)-\E A(X)\}+O_p(N^{-1/2}\bar h+h^s)$ and whose quadratic remainder is $O_p(\|\widehat f-f\|_{L_2}^2)=O_p((N\eta)^{-1}+h^{2s})=o_p(N^{-1/2})$; the cross term $\int|\Gammah_1-\Gamma_1||\widehat f_{\mathcal C_1}-f_{\mathcal C_1}|=O_p(\epsilon_N(N\eta)^{-1/2})=o_p(N^{-1/2})$. Hence $\sqrt N(\widehat\Delta_1-\Delta_1)\dto N(0,\Var\mathrm{IF}_\Delta)$, and the sample variance of the plug-in $\widehat{\mathrm{IF}}_\Delta(W_i)$, with $\tauh$ and its grid values as a function of $X_1$ given $X_2$, is consistent by the argument for $\widehat\Omega_{j,N}$. Without the margin condition, part (a) of Proposition~\ref{prop:gini-clt} for $\Delta_1$ follows as in the proof in Section~\ref{osec:proofs-gini}, with weights $\widehat f_{\mathcal C_1}$ and the linear $\widehat f$ term, which is $O_p(N^{-1/2})$. 

\emph{The index.} With $\mu_{\mathcal R}:=\E\tau(X^{\mathcal R})$ and $\tau$ of one sign on $\mathcal R$, $\E|\tau(X^{\mathcal R})|=|\mu_{\mathcal R}|$, estimated by $\widehat\mu_{\mathcal R}:=\sum_i\ind\{X_i\in\mathcal R\}\widehat\zeta_i/\sum_i\ind\{X_i\in\mathcal R\}$ with influence function $\mathrm{IF}_\mu=\ind\{X\in\mathcal R\}(\zeta-\mu_{\mathcal R})/p_{\mathcal R}$, and $\mathrm{IF}_G=\{\mathrm{IF}_\Delta-2G_1\mathrm{sgn}(\mu_{\mathcal R})\mathrm{IF}_\mu\}/(2|\mu_{\mathcal R}|)$. If the fitted CATE function changes sign on $\mathcal R$, $\E|\tau(X^{\mathcal R})|$ is itself a noisy functional of the type of part (a), and only the bracket should be reported; this is the case in the application (Section~\ref{osec:emp-mag}). Under Theorem~\ref{thm:nuisance-param} each influence function acquires the term $\mathrm{IF}_\vartheta(W)'\partial_\vartheta(\cdot)$ of Proposition~\ref{oprop:nui-ext}(a) unless both working models are correct.

\emph{Coverage of the interval \eqref{eq:gini-interval}.} Since $\widetilde D_j=\Dh_j/\sqrt N-\widetilde m_j/\sqrt N$, the interval fails to cover $D_j$ only if $\sqrt N(\widetilde D_j-D_j)>z\,\widehat{\mathrm{se}}_j$ or $\sqrt N(\Dh_j/\sqrt N-D_j)<-z\,\widehat{\mathrm{se}}_j$, $z:=z_{1-\alpha/2}$. Under $H_0^j$, $D_j=0\le\Dh_j/\sqrt N$, and $\Pr(\sqrt N\widetilde D_j>z\,\widehat{\mathrm{se}}_j)\le\Pr(\Sh_j>z)\to\alpha/2$ as $\widehat{\mathrm{se}}_j\ge\widetilde\sigma_{j,N}$. Under (c), $\limsup_N\Pr(\sqrt N(\Dh_j/\sqrt N-D_j)<-z\,\widehat{\mathrm{se}}_j)\le\alpha/2$ as $\widehat{\mathrm{se}}_j\ge\widehat\Omega_{j,N}^{1/2}\pto\Omega_j^{1/2}$, and $\sqrt N(\widetilde D_j-D_j)=O_p(1)-\widetilde m_j\to-\infty$. Under (d), $\sqrt N(\Dh_j/\sqrt N-D_j)=m^\kappa_{j,N}(F_j)+O_p(1)\to+\infty$ and $\sqrt N(\widetilde D_j-D_j)=-m^\kappa_{j,N}(\Csym{j}\setminus F_j)+O_p(1)\to-\infty$ when $\lambda(\Csym{j}\setminus F_j)>0$ (otherwise $H_0^j$ holds). In all three cases, the asymptotic coverage is at least $1-\alpha$. Without any of the three structures, Proposition~\ref{prop:gini-clt}(a) still gives consistency of both ends, but no coverage statement is claimed. All statements are pointwise in the law, not uniform near the null.

\subsection{Proof of Proposition~\ref{oprop:jitter}}\label{osec:jitter}
\phantomsection\label{oprf:jitter}
\begin{proof}[Proof of Proposition~\ref{oprop:jitter}]
Let $n:=s_1+2$ and $\varphi$ the density of $J$, the centered cardinal B-spline of order $n$: $\varphi\in C^{n-2}=C^{s_1}$, $\mathrm{supp}\,\varphi=[-n/2,n/2]$, $\sum_{a\in\mathbb Z}\varphi(u-a)=1$, and the integer translates of $\varphi$ whose supports meet an open interval $I$ are linearly independent on $I$ \citep[local linear independence of B-splines; e.g.][Ch.~IX]{deboor2001practical}. Let $p(a,x_2):=\Pr(X_1=a\mid X_2=x_2)f_2(x_2)$, with $f_2$ the density of $X_2$. As $J$ is independent of $(\zeta,X)$, $(X_1^J,X_2)$ has the density $f^J(x_1,x_2)=\sum_ap(a,x_2)\varphi(x_1-a)$ and
\[
\tau^J(x_1,x_2)=\frac{\sum_ap(a,x_2)\varphi(x_1-a)\tau(a,x_2)}{f^J(x_1,x_2)},
\]
because $\E[\zeta\mid X_1=a,X_2,J]=\tau(a,X_2)$. (a) $\tau^J(\cdot,x_2)\equiv c$ on $I$ if and only if $\sum_ap(a,x_2)\{\tau(a,x_2)-c\}\varphi(x_1-a)=0$ on $I$, if and only if $p(a,x_2)\{\tau(a,x_2)-c\}=0$ for every $a$ with $(a-n/2,a+n/2)\cap I\neq\emptyset$, by local linear independence. (b) $f^J$, $\tau^J$ and $\Var(\zeta\mid X^J)=\sum_ap^J_a\{\varsigma_\zeta^2(a,\cdot)+\tau(a,\cdot)^2\}-(\tau^J)^2$, $p^J_a:=p(a,x_2)\varphi(x_1-a)/f^J$, are $s_1$ times continuously differentiable in $x_1$, through $\varphi$, and $s_2$ times in $x_2$, through $p(a,\cdot)$, $\tau(a,\cdot)$, $\varsigma_\zeta^2(a,\cdot)$; by the partition of unity $f^J\ge\min\{p(a,x_2):|a-x_1|\le n/2\}$, bounded below on a neighborhood of the box under the stated positivity; conditional moments of $\zeta$ given $X^J$ are mixtures of those given $X$. So the conditions of Proposition~\ref{prop:transfer} hold for $(\zeta,X_1^J,X_2)$, and under $H_0^1$ on the atoms, $\tau^J$ is constant in $x_1$ by (a), so Theorem~\ref{thm:normality} applies; consistency follows from (a) and Proposition~\ref{prop:size}(b). Under Theorem~\ref{thm:nuisance-param} the first stage uses the unjittered $X_1$ and is unaffected; $\tau^J_\vartheta(x):=\E[\partial_\vartheta\zeta(W;\chi_{\vartheta^*})\mid X^J=x]$ is continuous by the same mixture representation.
\end{proof}

\begin{remark}[Uniform jitter and the raw lattice]\label{orem:jitter-uniform}
A jitter uniform on $[-1/2,1/2]$ also preserves the null exactly, since $X_1$ is recovered by rounding, but it gives a step-function density in $x_1$, so that Assumption~\ref{ass:reg}(b),(e) fails; the B-spline jitter of order $s_1+2$ is the smallest that delivers $s_1$ continuous derivatives. As $h_1\to0$ with the lattice fixed, the raw integer design falls under a coarse-discrete limit, as in \citet[Section~5]{shi2019testing}, which is normal when the other block is continuous but has other constants, so the continuous theory is not the asymptotics of the raw-age procedure; a shrinking lattice, under which $h_1$ divided by the atom spacing diverges, would restore it but does not describe the application. Since a single jitter draw adds randomness unrelated to the data, the application reports the average of the studentized statistics over independent draws (Section~\ref{osec:emp-jitter}).
\end{remark}
\section{Monte Carlo design and detailed results}\label{osec:mc}

In this section, we provide the full design and the complete results of the simulation study summarized in Section~\ref{sec:mc} and Figure~\ref{fig:mc} of the main text. The joint $L_1$ test of the paper is computed in the \textit{corrected} calibration of Lemmas~\ref{prop:centering-corr} and \ref{prop:sigmaN} and, for comparison, in the \textit{uncorrected} calibration (in-sample residuals and the limit-form variance), and is compared with the tests of the same directional nulls discussed in Section~\ref{sec:intro}.

The results come from two sets of runs. The \textit{main runs}, behind Figure~\ref{fig:mc}, Table~\ref{otab:mc-main} and Figure~\ref{ofig:unadj}, are computed at $N=1000$ with $1{,}000$ replications per design in both panels. As a benchmark, we compute the PCM and WGCM with the replication code of \citet{lundborg2024projected} and its generalized additive model (GAM) learners, and the kernel significance test of Lavergne, Maistre and Patilea (LMP) with the kernels and bandwidths of \citet[Section~4]{lavergne2015significance}. All other results of this section, namely the other sample sizes, the detailed tables at $N=1000$ with the variants of the joint test, the size diagnostics, the additional designs, the sensitivity study, the first-stage designs and the coverage study, come from the \textit{supplementary runs}, in which every method, the benchmarks included, is computed as in the main runs (Section~\ref{osec:mc-bench}), so that the results for a design and sample size common to both sets of runs differ only by Monte Carlo error.

In Section~\ref{osec:mc-design}, we specify the Monte Carlo designs, Section~\ref{osec:mc-paper} describes the joint and marginal $L_1$ tests and Section~\ref{osec:mc-bench} the benchmarks, Section~\ref{osec:mc-results} reports the rejection frequencies (rates) in the base designs, Section~\ref{osec:mc-diag} the size diagnostics of the studentized statistic, Section~\ref{osec:mc-new} the additional designs (an estimated propensity with misspecified outcome models, correlated modifier blocks and heavy tails), Section~\ref{osec:mc-sens} the sensitivity of the tests to the grid and the bandwidth, Section~\ref{osec:mc-fs} the designs with an estimated first stage, Section~\ref{osec:mc-blocks} covariate blocks with several covariates, and Section~\ref{osec:mc-mag} the coverage of the magnitude intervals.

\subsection{Designs}\label{osec:mc-design}
We consider four primary classes of designs described as follows.
\paragraph{1. Base designs.} Covariates are $X_1,X_2\stackrel{\mathrm{iid}}{\sim}U[0,1]$; treatment is randomized with known propensity $\Pr(T=1\mid X_1,X_2)=r_1=1/2$; and
\begin{equation}\label{oeq:mc-dgp}
Y=\mu_0(X_1,X_2)+T\,\tau(X_1,X_2)+\epsilon,
\qquad \mu_0(x_1,x_2)=1+\tfrac12x_1+\tfrac12x_2,
\qquad \epsilon\sim N(0,1),
\end{equation}
with $\epsilon$ independent of $(X_1,X_2,T)$. We consider six CATE functions:
\begin{align*}
\text{A0:}\quad &\tau=1
 &&\text{both nulls true;}\\
\text{A1:}\quad &\tau=1+0.8\sin(2\pi x_1)
 &&H_0^{1}\ \text{false},\ H_0^{2}\ \text{true;}\\
\text{A2:}\quad &\tau=1+0.8\sin(2\pi x_2)
 &&H_0^{1}\ \text{true},\ H_0^{2}\ \text{false;}\\
\text{A3:}\quad &\tau=1+0.8\{\sin(2\pi x_1)+\sin(2\pi x_2)\}
 &&\text{both nulls false;}\\
\text{A4:}\quad &\tau=1+\beta\,(x_1-\tfrac12)(x_2-\tfrac12),\ \beta=13.0016
 &&\text{both nulls false, marginal nulls true;}\\
\text{A5:}\quad &\tau=1+\beta\,\{(x_1-\tfrac12)^2-\tfrac1{12}\},\ \beta=13.0016
 &&H_0^{1}\ \text{false},\ H_0^{2}\ \text{true}.
\end{align*}
Designs A1 and A2 are the informative ones for attribution: in each, exactly one null holds, so they test simultaneously whether a procedure has power in the direction where the effect varies and whether it keeps its level in the direction where it does not. Design A4 is the \textit{bilinear interaction} function \eqref{eq:interaction-example} of the main text. Because $\E[X_j-1/2]=0$ and $X_1\perp X_2$, both marginalized CATE functions $\tau^{\mathrm{marg}}_1(x_1):=\E[\tau(x_1,X_2)\mid X_1=x_1]$ and $\tau^{\mathrm{marg}}_2$ are identically one, so both marginal nulls hold, yet $D_1=D_2>0$. The constant $\beta$ is calibrated so that the integrated pairwise absolute difference parameter $D_1$ in A4 coincides with its value in A1, namely $0.178$ on the evaluation region $[0.1,0.9]^2$. Since $D_1$ is linear in $\beta$, the calibrated value is the ratio of the target to the value at $\beta=1$. The marginal counterparts of the parameters $D_1$ and $D_2$ in A4 are zero. Every method uses the IPW pseudo-outcome, which with $r_1=1/2$ known is $$\zeta_i=\{\tfrac{T_i}{r_1}-\tfrac{(1-T_i)}{(1-r_1)}\}Y_i=2(2T_i-1)Y_i,$$ so that $\E[\zeta\mid X_1,X_2]=\tau(X_1,X_2)$ and there is no first stage.

\paragraph{2. Designs with estimated propensity (Designs E1, E2).} A confounder $V\sim N(0,1)$, independent of $(X_1,X_2)$, affects treatment and outcome: $$\Pr(T=1\mid X,V)=\mathrm{logit}^{-1}(-1+X_1+X_2+0.75V), \quad \mu_0(x,v)=1+\tfrac12x_1+\tfrac12x_2+v, \quad \epsilon\sim N(0,4)$$  with the CATE functions of the base designs (resulting in designs E1-A$k$; the main runs use A0, A1, A3, A4 and A5). The error standard deviation $2$ makes the conditional variance of the doubly robust (DR) pseudo-outcome, about $4\E[1/r_1+1/(1-r_1)]\approx20$, comparable to that of the IPW pseudo-outcome in the base designs, so that power is comparable. Every method uses the DR pseudo-outcome \eqref{eq:psi-dr} of the main text with a logistic propensity on $(1,X_1,X_2,V)$, fitted by maximum likelihood without trimming, and linear outcome regressions on $(1,X_1,X_2,V)$ within each arm, all fitted on the full sample. This corresponds to the setting of Theorem~\ref{thm:nuisance-param} with a correct propensity model. In design E1, the propensity model and the control-arm outcome model are correctly specified, since $\mu_0$ is linear in $(x_1,x_2,v)$; the treated-arm model, a linear model for $\mu_0+\tau$, is correct only in E1-A0 and is misspecified in the other designs, where $\tau$ is not linear in $(x_1,x_2)$. 

In an augmented class of designs denoted as E2 (E2-A0, E2-A1), $\mu_0$ is augmented by $1/2\cdot(V^2-1)+\sin(\pi x_1)$, so that both outcome models are misspecified in E2-A0 as well, while the propensity model remains correct. The fitted propensities range from $0.009$ to $0.988$ over the supplementary replications of E1 and E2.

\paragraph{3. Designs with correlated modifier blocks (Designs C0, C1, C4).} Let $X_j=\Phi(Z_j)$ with $(Z_1,Z_2)$ standard bivariate normal with correlation $0.5$ (Spearman correlation of $X$ around $0.48$), so that the marginals remain $U[0,1]$ and the bandwidths and grid are unchanged while the density of $X$ is not a product. $r_1=1/2$ is known and the pseudo-outcome is as in the base designs. Design C0 has $\tau=1$; C1 has $\tau=2x_2$, so that $H_0^1$ is true while the marginal null for $X_1$ is false, and a rejection of $H_0^1$ is a false attribution to $X_1$; C4 has the CATE function of A4, for which the marginal nulls fail because the correlation breaks the cancellation.

\paragraph{4. Designs with heavy-tailed outcomes (T3, LN).} In another augmented class of design, we let $\epsilon=t_3/\sqrt3$, which has unit variance but no third moment with all else as in the base designs A0 and A1 (resulting in designs T3-A0, T3-A1), or with the standardized lognormal $\epsilon=\{e^{Z}-e^{1/2}\}/\{e(e-1)\}^{1/2}$, $Z\sim N(0,1)$, with skewness $6.2$ and kurtosis $114$ (resulting in designs  LN-A0, LN-A1, LN-A4).

\paragraph{Sample sizes and replications.} The \textit{main runs} use the base designs A0, A1, A3, A4 and A5 (Panel~A of Table~\ref{otab:mc-main}) and design E1 with the same CATE functions (Panel~B) at $N=1000$, with $1{,}000$ replications per design and panel; the Monte Carlo standard error of a rejection frequency is at most $0.016$, and about $0.007$ near $0.05$. Each design and replication has its own seed, and all methods are applied to the same data set.

In the \textit{supplementary runs}, the base designs A0--A4 are run at $N\in\{500,1000,2000\}$ with $500$, $500$ and $300$ replications, and A0, A1 and A4 at $N=5000$ with $500$ replications (the PCM and WGCM in A0 and A1 only). We run the additional designs at $N=1000$ with $500$ replications. Since the cost of every method grows quickly with $N$, the supplementary runs use $500$ replications rather than the $1{,}000$ of the main runs, and the smaller numbers, $300$ at $N=2000$ here and at most $300$ at $N=5000$ in Section~\ref{osec:mc-fs}, serve only to confirm at a larger $N$ the patterns that the runs with $500$ or more replications establish. The Monte Carlo standard error of a rejection frequency of $0.05$ is $0.010$ and at most $0.022$ for any frequency with $500$ replications. All tests are one-sided at the nominal level $5\%$ unless stated otherwise; the Holm procedure is applied to the two $p$-values of every method; the familywise error rate (FWER) is the frequency of rejecting at least one true null; and the exact attribution rate (``both correct'') is the frequency with which the pair of Holm decisions coincides with the true pair. Each replication of each design and sample size has its own seed, so each cell is reproducible in isolation and all methods apply to the same data sets.

\subsection{The joint and marginal $L_1$ tests}\label{osec:mc-paper}

\paragraph{Estimator and tuning.} Let $\tauh$ denote the Nadaraya--Watson regression of $\zeta_i$ on $(X_{1i},X_{2i})$ with the product of the fourth-order kernel
\begin{equation}\label{oeq:mc-kernel}
K(u)=\tfrac{45}{16}\big(1-\tfrac{28}{3}u^2\big)\big(1-4u^2\big)\ind\{|u|\le\tfrac12\},
\end{equation}
whose first three moments vanish as Assumption~\ref{ass:reg} of the main text requires. Bandwidths are $h_j=k_j\widehat{\mathrm{sd}}_jN^{-1/7}$ with $k_j=3.5$ and $\widehat{\mathrm{sd}}_j$ the sample standard deviation of $X_j$, so that $\eta\asymp N^{-2/7}$ and $N\eta^{3}\asymp N^{1/7}\to\infty$ (Remark~\ref{orem:mc-bw}). The CATE function is evaluated on an $M\times M$ uniform grid spanning $[0.1,0.9]^2$, $M=13$ unless stated; $\Dh_j$, the centering and the variance use the same equal weights, and the inner double integrals of $\Dh_j$ are computed exactly by the sorting identity \eqref{eq:gini-sort}. The grid spacing $1/15$ is $0.16h_j$ at $N=500$, $0.18h_j$ at $N=1000$, $0.20h_j$ at $N=2000$ and $0.22h_j$ at $N=5000$, so the $13\times13$ grid meets the spacing of at most $h_j/6$ recommended in Section~\ref{sec:impl} only at $N=500$ and is slightly coarser than recommended at the larger sample sizes. In the grid study of Section~\ref{osec:mc-sens} (Table~\ref{otab:mc2-sens}), at $N=1000$, the grids with $21$ and $31$ points per axis (spacing $0.11h_j$ and $0.07h_j$, within the recommendation) give practically the same corrected test as the $13\times13$ grid.

\paragraph{Windows at the edge of the evaluation region.} The kernel is supported on $[-1/2,1/2]$, so the window at a grid point $x_j$ is $[x_j-h_j/2,x_j+h_j/2]$, and $X_j$ is uniform on $[0,1]$ in every design (in the C designs through the Gaussian copula). With $k_j=3.5$ the half-window $h_j/2$ is $0.208$, $0.188$, $0.171$ and $0.150$ at $N=500$, $1000$, $2000$ and $5000$, against a margin of $0.1$ between the evaluation region and the edge of the support. The windows at the outermost grid points therefore reach up to $0.108$, $0.088$, $0.071$ and $0.050$ beyond $[0,1]$: at $N\le2000$ the windows of $4$ of the $13$ grid values per axis (the two outermost on each side) leave the support, so that $88$ of the $169$ grid points ($52\%$) have a window that does, and at $N=5000$ the windows of $2$ of $13$ grid values per axis ($48$ of $169$ grid points, $28\%$). At most $26\%$ of a window lies outside the support ($23\%$ at $N=1000$, $17\%$ at $N=5000$). In the bandwidth study at $N=1000$ the corresponding shares of the $169$ grid points are $28\%$ at $k_j=3$, $71\%$ at $k_j=4.5$ and $85\%$ at $k_j=6$, with up to $19\%$, $29\%$ and $35\%$ of a window outside the support. Assumption~\ref{ass:overlap}, which requires the density to be bounded away from zero on a neighborhood of the evaluation region, holds in the limit, because the windows shrink into such a neighborhood, but not at these bandwidths. At the affected grid points, the kernel is truncated, loses its vanishing moments, and the smoothing bias of $\tauh$ is of order $h_j$ rather than $h_j^{4}$. Fewer observations lie in these windows, so $\tauh$ is noisier there; the corrected centering and variance account for this because they are computed from the realized weights, whereas the boundary bias is not. The simulations therefore describe the joint test with this boundary effect, outside the theory's conditions at the edge grid points. This is a limitation of the design, and we did not simulate an evaluation region farther from the edge of the support.

\paragraph{Uncorrected calibration.} 
The comparison calibration uses in-sample residuals for centering
and the limit-form variance for studentization.
The centering estimator is $\widehat m^{\,\kappa}_j$ in
\eqref{eq:center-feasible}, based on
$\widehat\kappa_j(c)
=N\sum_i\{\ell_i(c_1)-\ell_i(c_2)\}^2\widehat e_i^{\,2}$,
where $\widehat e_i=\zeta_i-\tauh(X_{1i},X_{2i})$
are the in-sample residuals and $\ell_i$ are the
Nadaraya--Watson weights.
For studentization, we use the plug-in estimator $\sigh_j$
of the limit variance in \eqref{oeq:pf-sigma2}.
The variance function $\varrho$ is estimated on the grid by
$\widehat\varrho(x)
=\{\tauh^{(\zeta^2)}(x)-\tauh(x)^2\}
(\int K^2)^2/\widehat f(x)$,
where $\tauh^{(\zeta^2)}$ is the kernel regression of
$\zeta_i^2$ and $\widehat f$ is the product-kernel
density estimator.
For numerical stability, the estimated conditional variance
is floored at $10^{-6}$ times its maximum over the grid,
and $\widehat f$ is floored at $0.001$.
The double integral of $g_{\mathrm{abs}}$ over the kernel
offsets is tabulated once at $1001$ values of its argument.

\paragraph{Corrected calibration.} 

The test of Section~\ref{sec:impl} uses the corrected
centering estimator $\widetilde m_j$ of
Lemma~\ref{prop:centering-corr}, based on the
own-weight-corrected squared residuals
$\widetilde e_i^{\,2}=\widehat e_i^{\,2}/q_i$, where
$q_i=(1-L_{ii})^2+\sum_{l\neq i}L_{il}^2
=1-2L_{ii}+\sum_lL_{il}^2$ and $L_{il}=\ell_l(X_i)$.
For studentization, we use the finite-sample variance
estimator $\widetilde\sigma^2_{j,N}$ of
Lemma~\ref{prop:sigmaN}, computed from the same
weights and corrected residuals using the exact double
sum in Section~\ref{osec:grid-formulas}.
The resulting test statistic is
$\Sh_j=(\Dh_j-\widetilde m_j)/\widetilde\sigma_{j,N}$,
with one-sided $p$-value $p_j=1-\Phi(\Sh_j)$.
The code floors $q_i$ at $0.25$, and the HC2 variant below trims $L_{ii}$ at $0.5$.
To isolate the effects of the centering and variance
corrections, Section~\ref{osec:mc-results} also reports
three variants. The first replaces $q_i$ with the HC2
factor $1-L_{ii}$, giving
$\widetilde e_i^{\,2}=\widehat e_i^{\,2}/(1-L_{ii})$.
The second combines corrected centering with the limit
variance (``centering only''). The third combines the
finite-sample variance with uncorrected centering and
residuals (``variance only'').

\paragraph{Denominator guard.} A replication is discarded by all joint tests if the Nadaraya--Watson denominator is not positive at some grid point or sample point. This happened in $7$ of the $6{,}500$ replications of the base designs at $N\le2000$, all at $N=500$ (one in A0 and three each in A3 and A4; Table~\ref{otab:mc2-500}), never at $N\ge1000$ in the base designs, once in the grid and bandwidth study, at $k_j=3$ (Table~\ref{otab:mc2-sens}), and in $10$ to $16$ of the $500$ replications of each correlated design (Table~\ref{otab:mc2-newsize}). The benchmarks use all replications.

\paragraph{The marginal $L_1$ test.} The same functional is applied to the one-dimensional kernel estimator of $\zeta_i$ on $X_1$ alone,
\begin{equation}\label{oeq:mc-marg-est}
\tauh^{\,\mathrm{marg}}_1(x_1)
:=\frac{\sum_i\zeta_i K\big((x_1-X_{1i})/h_1\big)}
        {\sum_i K\big((x_1-X_{1i})/h_1\big)} ,\qquad
\Dh^{\mathrm{marg}}_1
=\tfrac12\sqrt N\iint\big|\tauh^{\,\mathrm{marg}}_1(x_1)
-\tauh^{\,\mathrm{marg}}_1(x_1')\big|\,dx_1\,dx_1',
\end{equation}
with the mirror definition in direction~2, the same kernel, bandwidth, grid and sorting identity, the centering built from the one-dimensional weights and residuals, and the variance specialized to a single block, in which the second kernel autocorrelation is identically one.
It tests the marginal null that $\tau^{\mathrm{marg}}_1$
is constant and \textit{has no power against A4 by construction}.
It differs from the joint test only in the covariates
used for kernel smoothing, with an effective window
count of $Nh_1$ rather than $Nh_1h_2$.

\begin{remark}[Bandwidth rate, kernel order and the denominator guard]\label{orem:mc-bw}
(i) \textit{The rate.} Assumption~\ref{ass:reg}(g) requires $N\eta^{3}\to\infty$ and $Nh_j^{2s_j}\to0$, so $h_j\asymp N^{-\gamma}$ needs $1/(2s)<\gamma<1/\{3(d_1+d_2)\}$ (Section~\ref{sec:impl}). With $d_1=d_2=1$ and $s=4$, $\gamma=1/7$ is admissible; the rule $h\asymp N^{-2/7}$ of \citet{chang2015nonparametric}, calibrated to a single one-dimensional smooth, is not, since it gives $N\eta^3\asymp N^{-5/7}$. The slow growth of $N\eta^{3}\asymp N^{1/7}$ is one reason why the normal approximation is reached only gradually (Section~\ref{osec:mc-diag}).\\ (ii) \textit{The kernel order.} With $\gamma=1/7$, $2s_j\gamma>1$ forces $s_j\ge4$: second-order kernels are inadmissible even in the lowest-dimensional case, as \citet[Remark~3.4]{shi2019testing} note for their test; $d_1+d_2=3$ needs a sixth-order kernel.\\ (iii) \textit{The denominator guard.} Bias-reducing kernels take negative values, so the denominator $\sum_iK_1K_2$ can approach or cross zero in sparse regions. In the base designs, at $k_j=2$ roughly $60\%$ of the replications contained at least one such grid point, at $k_j=2.5$ about $12\%$, and at $k_j\ge3$ under $1\%$. Because the kernels are supported on $[-1/2,1/2]$, a bandwidth $h_j$ yields a window of width $h_j$, so admissible values of $k_j$ are larger than the $[-1,1]$ convention suggests. We recommend that practitioners check the \textit{minimum effective window count} over the grid and increase $k_j$ until it is comfortably positive; Section~\ref{sec:impl} recommends $k_j\in[3,5]$.
\end{remark}

\subsection{Benchmarks}\label{osec:mc-bench}

Each benchmark is applied with $(Y,X,Z)=(\zeta,X_1,X_2)$ for direction~1 and $(\zeta,X_2,X_1)$ for direction~2, so that it tests $\E[\zeta\mid X_1,X_2]=\E[\zeta\mid X_2]$, which is $H_0^{1}$. The symbols of this subsection ($Y$, $X$, $Z$, the kernels $K$, $L$, $\psi$, the bandwidths $g$, $h$ and the basis dimension $k$) follow the cited papers and are local to it; $n$ is the size of the sample on which a test is computed, $N$ is the half-sample size.

\paragraph{Benchmarks of main runs.} The joint and marginal $L_1$ tests are computed as in Section~\ref{osec:mc-paper}. The PCM, single split and derandomized over six splits, the WGCM with estimated weight and the generalized covariance measure (GCM) of \citet{shah2020hardness} are computed with the functions \texttt{pcm\_test}, \texttt{wGCM\_est} and \texttt{gcm\_test} of the replication code of \citet{lundborg2024projected}, with its generalized additive model learners (\texttt{mgcv}): \texttt{gam\_reg\_method} for the regressions, \texttt{gam\_gtilde} for the auxiliary fit of the PCM and \texttt{gam\_vhat\_reg\_method} for its variance regression. We use their code unchanged, except for the number of basis functions used in the nonparametric regressions and the scaling of the WGCM statistic. Their choice, $k=\lfloor(n-1)/d\rfloor$, is designed for seven covariates and yields $k$ close to $n$ when $d\le2$. At $n=500$, a single thin-plate fit with $k=499$ takes approximately nine minutes.
We therefore cap $k$ at $40$, while retaining generalized
cross-validation to select the smoothing parameter.  These learners are additive in $X$ and $Z$, so the PCM's fit of $\E[\zeta\mid X_1,X_2]$ cannot represent an interaction between the blocks, which is the only heterogeneity in A4. The LMP test uses the statistic $\tilde I_n$ and the variance estimator $\widehat\omega_n$ of \citet{lavergne2015significance} with the settings of their Section~4: covariates scaled by their sample standard deviations, the standard normal density for the kernels $L$ and $K$ on $Z$ and for $\psi$ on $X$, and bandwidths $g=n^{-1/(p+4)}$ and $h=2n^{-2.1/(p+4)}$, their $n^{-1/6}$ and $n^{-2.1/6}$ for a two-dimensional conditioning variable, here with $p=1$, which satisfies their condition $nh^{p/2}g^{2s}\to0$ with a second-order kernel $L$; the $p$-value is $1-\Phi(\mathcal T_{\mathrm{LMP}})$, with $\mathcal T_{\mathrm{LMP}}:=nh^{1/2}\tilde I_n/\widehat\omega_n$.
\paragraph{Benchmarks and variants of supplementary runs.} The benchmarks are computed as in the main runs, with the same code, learners and settings. Tables~\ref{otab:mc2-500}--\ref{otab:mc2-5000} also report four rows that do not appear in the main paper. Three are calibrations of the joint test that show the contribution of each correction (Section~\ref{osec:mc-paper}). ``Joint, corr.\ (HC2)'' replaces the factor $q_i$ in the corrected residuals with $1-L_{ii}$, the familiar HC2 leverage adjustment, to check that the results do not hinge on the form of the own-weight correction. ``Joint, centering only'' combines the corrected centering with the limit variance \eqref{oeq:pf-sigma2}, and ``Joint, variance only'' combines the uncorrected centering with the finite-sample variance, so that each isolates one of the two corrections. The fourth, ``PCM, single split'', is the PCM computed on one random split of the sample into halves, whereas the derandomized PCM of the main paper averages the statistic over six splits. It shows the effect of derandomization: the single split has size closer to nominal, but its decision depends on the split, and averaging reduces that dependence at the price of conservative size. Three features matter for the comparison. (a) The PCM, the WGCM and the GCM use the learners of the replication code of \citet{lundborg2024projected}; with other learners, their power may differ in either direction. The additive fit of $\E[\zeta\mid X_1, X_2]$ in the PCM cannot represent the interaction in A4, whereas the WGCM needs no such fit, because its weight follows the sign of the covariance of $X$ and $Y$ given $Z$. The GCM is reported only in the first-stage designs (Section~\ref{osec:mc-fs}). (b) In the designs with an estimated propensity, every method receives the same DR pseudo-outcome; the benchmarks' level results are for an observed response, and Theorem~\ref{thm:nuisance-param} covers only the joint test. (c) The sample splits of the PCM and the WGCM are random, so their decisions for a given data set depend on the split, as in the published procedures.

\begin{remark}[Scope of the comparison]\label{orem:mc-scope}
The designs are two-dimensional and, apart from E1 and E2, have a known propensity, so the comparison says little about higher-dimensional blocks, where the rate of the joint smoother depends on both blocks, and those of the benchmarks depend on the conditioning block only or attain the minimax rate \citep{lundborg2024projected}. Section~\ref{osec:mc-blocks} extends the comparison to blocks of up to two covariates each.
\end{remark}

\subsection{Rejection rates}\label{osec:mc-results}

\begin{figure}[h]
\centering
\includegraphics[width=0.97\textwidth]{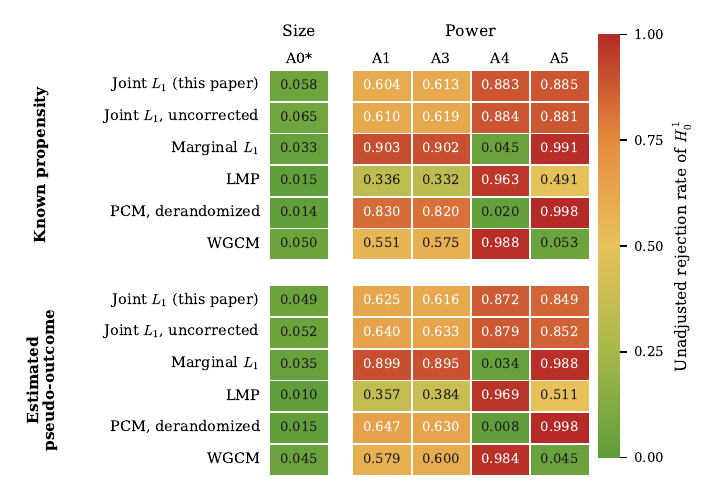}
\caption{\label{ofig:unadj}Unadjusted rejection rate of $H_0^1$ at the $5\%$ level, the frequency of $p_1<0.05$ without multiplicity adjustment, in the main runs at $N=1000$, for $1000$ replications. A0, where $H_0^1$ is true (marked $^*$): size; A1, A3, A4 and A5, where it is false: power. Top: known propensity; bottom: estimated pseudo-outcome. Methods, designs and replications as in Table~\ref{otab:mc-main}.}
\end{figure}

Table~\ref{otab:mc-main} and Figure~\ref{ofig:unadj} report the main runs. From the supplementary runs, Tables~\ref{otab:mc2-500}--\ref{otab:mc2-5000} report, for each sample size, the Holm-adjusted and the raw rejection frequencies of every method in both directions, and Table~\ref{otab:mc2-fwer} reports the FWER and the exact attribution rate across sample sizes.

\begin{table}[htbp]
\caption{\label{otab:mc-main}Size, direction selectivity and attribution, $N=1000$ for $1000$ replications (the numbers behind Figure~\ref{fig:mc} of the main text).}
\begin{center}
\small\setlength{\tabcolsep}{2pt}
\begin{tabular*}{\textwidth}{@{\extracolsep{\fill}}lccccccccccc@{}}
\hline\hline
 & A0 & \multicolumn{2}{c}{A1} & \multicolumn{2}{c}{A3} & \multicolumn{2}{c}{A4} & \multicolumn{2}{c}{A5} & \multicolumn{2}{c}{Both correct} \\
Method & FWER & $H_0^{1}$ & $H_0^{2}$ & $H_0^{1}$ & $H_0^{2}$ & $H_0^{1}$ & $H_0^{2}$ & $H_0^{1}$ & $H_0^{2}$ & A1 & A4 \\
\hline
\multicolumn{12}{@{}l}{\textit{Panel A: known propensity}} \\
Joint $L_1$, corrected & 0.053 & 0.513 & 0.060 & 0.575 & 0.553 & 0.853 & 0.867 & 0.842 & 0.066 & 0.454 & 0.843 \\
Joint $L_1$, uncorrected & 0.051 & 0.494 & 0.061 & 0.570 & 0.550 & 0.850 & 0.861 & 0.827 & 0.065 & 0.434 & 0.841 \\
Marginal $L_1$ & 0.039 & 0.852 & 0.038 & 0.894 & 0.890 & 0.029 & 0.022 & 0.985 & 0.047 & 0.816 & 0.002 \\
LMP, asymptotic & 0.008 & 0.242 & 0.012 & 0.267 & 0.283 & 0.950 & 0.942 & 0.373 & 0.018 & 0.235 & 0.929 \\
PCM, derandomized & 0.010 & 0.699 & 0.012 & 0.782 & 0.788 & 0.013 & 0.007 & 0.995 & 0.015 & 0.687 & 0.000 \\
WGCM & 0.049 & 0.465 & 0.037 & 0.522 & 0.534 & 0.988 & 0.987 & 0.026 & 0.028 & 0.439 & 0.975 \\
\hline
\multicolumn{12}{@{}l}{\textit{Panel B: estimated pseudo-outcome}} \\
Joint $L_1$, corrected & 0.048 & 0.513 & 0.054 & 0.554 & 0.536 & 0.846 & 0.850 & 0.780 & 0.056 & 0.461 & 0.831 \\
Joint $L_1$, uncorrected & 0.048 & 0.519 & 0.057 & 0.561 & 0.546 & 0.845 & 0.848 & 0.778 & 0.057 & 0.464 & 0.831 \\
Marginal $L_1$ & 0.044 & 0.852 & 0.028 & 0.889 & 0.890 & 0.017 & 0.018 & 0.979 & 0.034 & 0.826 & 0.003 \\
LMP, asymptotic & 0.010 & 0.261 & 0.006 & 0.312 & 0.288 & 0.962 & 0.962 & 0.360 & 0.018 & 0.256 & 0.945 \\
PCM, derandomized & 0.009 & 0.509 & 0.005 & 0.570 & 0.558 & 0.004 & 0.004 & 0.992 & 0.003 & 0.505 & 0.000 \\
WGCM & 0.047 & 0.496 & 0.048 & 0.550 & 0.539 & 0.984 & 0.986 & 0.019 & 0.022 & 0.456 & 0.974 \\
\hline\hline
\end{tabular*}
\end{center}
\footnotesize
\renewcommand{\baselineskip}{11pt}
\textbf{Note:} Holm-adjusted rejection frequencies at the $5\%$ level. FWER: frequency of rejecting at least one true null; Both correct: frequency with which both decisions are correct. $H_0^2$ is true in A1 and A5, both nulls are false in A3 and A4, and both marginal nulls are true in A4. 
\end{table}

\paragraph{Rejection rates from the main runs.} Table~\ref{otab:mc-main} gives the numbers behind Figure~\ref{fig:mc}, which the main text discusses, and Figure~\ref{ofig:unadj} shows each test of $H_0^1$ on its own, without the Holm adjustment. Three points stand out. First, our proposed integrated absolute pairwise difference test (Joint $L_1$ in Table~\ref{otab:mc-main}) has size close to nominal in both panels, and the two calibrations give nearly the same rates at this sample size. Second, it is the only test of $H_0^1$ with substantial power against every alternative in both panels. Each benchmark has a blind spot: the derandomized PCM and the marginal test in the interaction A4, the first because its additive learners cannot represent the interaction and the second because its estimand is zero there; the WGCM in the symmetric design A5; and the LMP test in A1 and A3. Third, a benchmark has higher power where the alternative suits it: the derandomized PCM and the marginal test in the additive designs, the WGCM and the LMP test in A4. The proposed test rejects more often than the WGCM in A1 and A3 in both panels, and with the estimated pseudo-outcome its rates are close to those of the derandomized PCM there. The WGCM keeps its level. Figure~\ref{ofig:unadj} also shows that the conservativeness of the derandomized PCM and the LMP test is present before the multiplicity step, so it does not come from the Holm adjustment.

\begin{table}[p]
\caption{\label{otab:mc2-500}Rejection frequencies, $N=500$ (supplementary runs).}
\begin{center}
\footnotesize
\setlength{\tabcolsep}{3pt}
\begin{tabular*}{\textwidth}{@{\extracolsep{\fill}}lcccccccccc@{}}
\hline\hline
 & \multicolumn{2}{c}{A0} & \multicolumn{2}{c}{A1} & \multicolumn{2}{c}{A2} & \multicolumn{2}{c}{A3} & \multicolumn{2}{c}{A4} \\
Method & $H_0^{1}$ & $H_0^{2}$ & $H_0^{1}$ & $H_0^{2}$ & $H_0^{1}$ & $H_0^{2}$ & $H_0^{1}$ & $H_0^{2}$ & $H_0^{1}$ & $H_0^{2}$ \\
\hline
\multicolumn{11}{@{}l}{Panel A: Holm-adjusted rejection frequencies}\\
Joint, uncorrected & 0.066 & 0.072 & 0.340 & 0.110 & 0.112 & 0.322 & 0.298 & 0.340 & 0.594 & 0.584 \\
Joint $L_1$, corrected & 0.054 & 0.056 & 0.288 & 0.088 & 0.082 & 0.276 & 0.252 & 0.292 & 0.527 & 0.537 \\
Joint, corr.\ (HC2) & 0.062 & 0.064 & 0.320 & 0.102 & 0.096 & 0.298 & 0.270 & 0.306 & 0.569 & 0.567 \\
Joint, centering only & 0.036 & 0.032 & 0.196 & 0.048 & 0.046 & 0.210 & 0.185 & 0.229 & 0.404 & 0.414 \\
Joint, variance only & 0.134 & 0.134 & 0.486 & 0.176 & 0.176 & 0.452 & 0.455 & 0.487 & 0.718 & 0.716 \\
Marginal $L_1$ & 0.030 & 0.024 & 0.548 & 0.042 & 0.024 & 0.482 & 0.550 & 0.574 & 0.020 & 0.012 \\
LMP, asymptotic & 0.008 & 0.006 & 0.098 & 0.004 & 0.012 & 0.102 & 0.122 & 0.076 & 0.600 & 0.598 \\
PCM, single split & 0.036 & 0.020 & 0.340 & 0.040 & 0.030 & 0.300 & 0.350 & 0.368 & 0.038 & 0.046 \\
PCM, derandomized & 0.010 & 0.008 & 0.274 & 0.008 & 0.002 & 0.290 & 0.328 & 0.324 & 0.008 & 0.008 \\
WGCM & 0.022 & 0.038 & 0.260 & 0.024 & 0.032 & 0.238 & 0.260 & 0.246 & 0.820 & 0.846 \\
\hline
\multicolumn{11}{@{}l}{Panel B: raw rejection frequencies}\\
Joint, uncorrected & 0.102 & 0.112 & 0.446 & 0.122 & 0.130 & 0.426 & 0.392 & 0.419 & 0.668 & 0.650 \\
Joint $L_1$, corrected & 0.080 & 0.076 & 0.374 & 0.100 & 0.090 & 0.364 & 0.314 & 0.356 & 0.606 & 0.602 \\
Joint, corr.\ (HC2) & 0.090 & 0.084 & 0.408 & 0.110 & 0.102 & 0.386 & 0.334 & 0.370 & 0.632 & 0.624 \\
Joint, centering only & 0.052 & 0.052 & 0.288 & 0.062 & 0.056 & 0.282 & 0.239 & 0.278 & 0.491 & 0.501 \\
Joint, variance only & 0.168 & 0.160 & 0.562 & 0.186 & 0.190 & 0.556 & 0.493 & 0.539 & 0.779 & 0.755 \\
Marginal $L_1$ & 0.044 & 0.050 & 0.644 & 0.046 & 0.038 & 0.578 & 0.614 & 0.622 & 0.052 & 0.038 \\
LMP, asymptotic & 0.014 & 0.012 & 0.146 & 0.012 & 0.016 & 0.158 & 0.162 & 0.156 & 0.670 & 0.668 \\
PCM, single split & 0.056 & 0.048 & 0.448 & 0.046 & 0.046 & 0.414 & 0.452 & 0.454 & 0.064 & 0.064 \\
PCM, derandomized & 0.020 & 0.014 & 0.450 & 0.014 & 0.008 & 0.428 & 0.414 & 0.420 & 0.022 & 0.018 \\
WGCM & 0.050 & 0.056 & 0.334 & 0.044 & 0.060 & 0.308 & 0.338 & 0.300 & 0.842 & 0.856 \\
\hline\hline
\end{tabular*}
\end{center}
\footnotesize
\renewcommand{\baselineskip}{11pt}
\textbf{Note:} $N=500$, 500 replications (499 in A0 and 497 in A3 and A4 for the joint tests); all methods on the same data sets.
\end{table}
\begin{table}[p]
\caption{\label{otab:mc2-1000}Rejection frequencies, $N=1000$ (supplementary runs).}
\begin{center}
\footnotesize
\setlength{\tabcolsep}{3pt}
\begin{tabular*}{\textwidth}{@{\extracolsep{\fill}}lcccccccccc@{}}
\hline\hline
 & \multicolumn{2}{c}{A0} & \multicolumn{2}{c}{A1} & \multicolumn{2}{c}{A2} & \multicolumn{2}{c}{A3} & \multicolumn{2}{c}{A4} \\
Method & $H_0^{1}$ & $H_0^{2}$ & $H_0^{1}$ & $H_0^{2}$ & $H_0^{1}$ & $H_0^{2}$ & $H_0^{1}$ & $H_0^{2}$ & $H_0^{1}$ & $H_0^{2}$ \\
\hline
\multicolumn{11}{@{}l}{Panel A: Holm-adjusted rejection frequencies}\\
Joint, uncorrected & 0.038 & 0.042 & 0.516 & 0.080 & 0.054 & 0.504 & 0.576 & 0.530 & 0.878 & 0.872 \\
Joint $L_1$, corrected & 0.040 & 0.044 & 0.526 & 0.076 & 0.054 & 0.508 & 0.574 & 0.526 & 0.880 & 0.876 \\
Joint, corr.\ (HC2) & 0.044 & 0.048 & 0.544 & 0.082 & 0.054 & 0.528 & 0.590 & 0.558 & 0.886 & 0.884 \\
Joint, centering only & 0.020 & 0.026 & 0.428 & 0.038 & 0.026 & 0.396 & 0.468 & 0.442 & 0.794 & 0.788 \\
Joint, variance only & 0.076 & 0.090 & 0.646 & 0.120 & 0.102 & 0.628 & 0.674 & 0.654 & 0.936 & 0.934 \\
Marginal $L_1$ & 0.016 & 0.028 & 0.876 & 0.034 & 0.040 & 0.846 & 0.860 & 0.878 & 0.018 & 0.036 \\
LMP, asymptotic & 0.002 & 0.002 & 0.274 & 0.004 & 0.010 & 0.248 & 0.270 & 0.268 & 0.964 & 0.962 \\
PCM, single split & 0.022 & 0.030 & 0.696 & 0.042 & 0.066 & 0.658 & 0.740 & 0.702 & 0.052 & 0.042 \\
PCM, derandomized & 0.006 & 0.000 & 0.714 & 0.012 & 0.020 & 0.702 & 0.790 & 0.776 & 0.014 & 0.008 \\
WGCM & 0.014 & 0.014 & 0.482 & 0.054 & 0.044 & 0.484 & 0.504 & 0.512 & 0.984 & 0.988 \\
\hline
\multicolumn{11}{@{}l}{Panel B: raw rejection frequencies}\\
Joint, uncorrected & 0.064 & 0.066 & 0.622 & 0.080 & 0.056 & 0.610 & 0.634 & 0.578 & 0.910 & 0.908 \\
Joint $L_1$, corrected & 0.064 & 0.066 & 0.612 & 0.076 & 0.054 & 0.606 & 0.624 & 0.574 & 0.910 & 0.906 \\
Joint, corr.\ (HC2) & 0.068 & 0.074 & 0.634 & 0.082 & 0.058 & 0.622 & 0.642 & 0.590 & 0.916 & 0.916 \\
Joint, centering only & 0.034 & 0.038 & 0.530 & 0.044 & 0.030 & 0.506 & 0.520 & 0.510 & 0.854 & 0.842 \\
Joint, variance only & 0.108 & 0.140 & 0.722 & 0.120 & 0.106 & 0.710 & 0.698 & 0.696 & 0.942 & 0.956 \\
Marginal $L_1$ & 0.032 & 0.046 & 0.924 & 0.034 & 0.044 & 0.886 & 0.874 & 0.894 & 0.040 & 0.058 \\
LMP, asymptotic & 0.004 & 0.008 & 0.354 & 0.008 & 0.016 & 0.344 & 0.318 & 0.344 & 0.972 & 0.964 \\
PCM, single split & 0.034 & 0.048 & 0.772 & 0.052 & 0.074 & 0.764 & 0.770 & 0.750 & 0.074 & 0.074 \\
PCM, derandomized & 0.010 & 0.018 & 0.842 & 0.016 & 0.022 & 0.836 & 0.826 & 0.826 & 0.022 & 0.022 \\
WGCM & 0.048 & 0.022 & 0.586 & 0.066 & 0.056 & 0.584 & 0.556 & 0.568 & 0.986 & 0.988 \\
\hline\hline
\end{tabular*}
\end{center}
\footnotesize
\renewcommand{\baselineskip}{11pt}
\textbf{Note:} $N=1000$, 500 replications; all methods on the same data sets.
\end{table}
\begin{table}[h]
\caption{\label{otab:mc2-2000}Rejection frequencies, $N=2000$ (supplementary runs).}
\begin{center}
\footnotesize
\setlength{\tabcolsep}{3pt}
\begin{tabular*}{\textwidth}{@{\extracolsep{\fill}}lcccccccccc@{}}
\hline\hline
 & \multicolumn{2}{c}{A0} & \multicolumn{2}{c}{A1} & \multicolumn{2}{c}{A2} & \multicolumn{2}{c}{A3} & \multicolumn{2}{c}{A4} \\
Method & $H_0^{1}$ & $H_0^{2}$ & $H_0^{1}$ & $H_0^{2}$ & $H_0^{1}$ & $H_0^{2}$ & $H_0^{1}$ & $H_0^{2}$ & $H_0^{1}$ & $H_0^{2}$ \\
\hline
\multicolumn{11}{@{}l}{Panel A: Holm-adjusted rejection frequencies}\\
Joint, uncorrected & 0.040 & 0.033 & 0.853 & 0.040 & 0.040 & 0.857 & 0.867 & 0.880 & 0.997 & 0.997 \\
Joint $L_1$, corrected & 0.050 & 0.040 & 0.870 & 0.057 & 0.050 & 0.877 & 0.890 & 0.897 & 0.997 & 1.000 \\
Joint, corr.\ (HC2) & 0.053 & 0.040 & 0.873 & 0.060 & 0.053 & 0.883 & 0.893 & 0.903 & 0.997 & 1.000 \\
Joint, centering only & 0.017 & 0.020 & 0.800 & 0.020 & 0.023 & 0.807 & 0.837 & 0.833 & 0.993 & 0.993 \\
Joint, variance only & 0.077 & 0.057 & 0.920 & 0.080 & 0.080 & 0.913 & 0.930 & 0.927 & 0.997 & 1.000 \\
Marginal $L_1$ & 0.037 & 0.020 & 0.997 & 0.043 & 0.050 & 1.000 & 1.000 & 0.997 & 0.010 & 0.017 \\
LMP, asymptotic & 0.010 & 0.000 & 0.657 & 0.013 & 0.017 & 0.607 & 0.627 & 0.647 & 1.000 & 1.000 \\
PCM, single split & 0.030 & 0.013 & 0.957 & 0.053 & 0.047 & 0.953 & 0.957 & 0.980 & 0.040 & 0.027 \\
PCM, derandomized & 0.010 & 0.007 & 0.990 & 0.017 & 0.017 & 0.987 & 0.993 & 0.990 & 0.007 & 0.013 \\
WGCM & 0.020 & 0.013 & 0.800 & 0.050 & 0.047 & 0.807 & 0.813 & 0.840 & 1.000 & 1.000 \\
\hline
\multicolumn{11}{@{}l}{Panel B: raw rejection frequencies}\\
Joint, uncorrected & 0.063 & 0.053 & 0.917 & 0.040 & 0.040 & 0.910 & 0.890 & 0.893 & 0.997 & 1.000 \\
Joint $L_1$, corrected & 0.073 & 0.060 & 0.923 & 0.057 & 0.050 & 0.920 & 0.900 & 0.900 & 0.997 & 1.000 \\
Joint, corr.\ (HC2) & 0.077 & 0.060 & 0.933 & 0.060 & 0.053 & 0.930 & 0.903 & 0.910 & 0.997 & 1.000 \\
Joint, centering only & 0.037 & 0.033 & 0.870 & 0.020 & 0.023 & 0.880 & 0.857 & 0.870 & 0.997 & 1.000 \\
Joint, variance only & 0.117 & 0.087 & 0.950 & 0.080 & 0.083 & 0.943 & 0.937 & 0.930 & 0.997 & 1.000 \\
Marginal $L_1$ & 0.060 & 0.053 & 0.997 & 0.043 & 0.050 & 1.000 & 1.000 & 0.997 & 0.047 & 0.043 \\
LMP, asymptotic & 0.020 & 0.003 & 0.753 & 0.017 & 0.017 & 0.720 & 0.673 & 0.690 & 1.000 & 1.000 \\
PCM, single split & 0.063 & 0.030 & 0.973 & 0.053 & 0.053 & 0.977 & 0.957 & 0.980 & 0.070 & 0.047 \\
PCM, derandomized & 0.027 & 0.007 & 0.997 & 0.017 & 0.017 & 0.990 & 0.993 & 0.990 & 0.013 & 0.020 \\
WGCM & 0.053 & 0.047 & 0.833 & 0.050 & 0.050 & 0.857 & 0.820 & 0.847 & 1.000 & 1.000 \\
\hline\hline
\end{tabular*}
\end{center}
\footnotesize
\renewcommand{\baselineskip}{11pt}
\textbf{Note:} $N=2000$, 300 replications; all methods on the same data sets.
\end{table}
\begin{table}
\caption{\label{otab:mc2-5000}Rejection frequencies, $N=5000$ (supplementary runs).}
\begin{center}
\footnotesize
\setlength{\tabcolsep}{3pt}
\begin{tabular*}{\textwidth}{@{\extracolsep{\fill}}lcccccc@{}}
\hline\hline
 & \multicolumn{2}{c}{A0} & \multicolumn{2}{c}{A1} & \multicolumn{2}{c}{A4} \\
Method & $H_0^{1}$ & $H_0^{2}$ & $H_0^{1}$ & $H_0^{2}$ & $H_0^{1}$ & $H_0^{2}$ \\
\hline
\multicolumn{7}{@{}l}{Panel A: Holm-adjusted rejection frequencies}\\
Joint, uncorrected & 0.014 & 0.016 & 1.000 & 0.042 & 1.000 & 1.000 \\
Joint $L_1$, corrected & 0.024 & 0.034 & 1.000 & 0.054 & 1.000 & 1.000 \\
Joint, corr.\ (HC2) & 0.024 & 0.036 & 1.000 & 0.056 & 1.000 & 1.000 \\
Joint, centering only & 0.010 & 0.012 & 1.000 & 0.038 & 1.000 & 1.000 \\
Joint, variance only & 0.036 & 0.046 & 1.000 & 0.070 & 1.000 & 1.000 \\
Marginal $L_1$ & 0.016 & 0.026 & 1.000 & 0.042 & 0.022 & 0.016 \\
PCM, single split & 0.022 & 0.032 & 1.000 & 0.040 & -- & -- \\
PCM, derandomized & 0.002 & 0.008 & 1.000 & 0.008 & -- & -- \\
WGCM & 0.024 & 0.016 & 0.996 & 0.050 & -- & -- \\
\hline
\multicolumn{7}{@{}l}{Panel B: raw rejection frequencies}\\
Joint, uncorrected & 0.028 & 0.040 & 1.000 & 0.042 & 1.000 & 1.000 \\
Joint $L_1$, corrected & 0.040 & 0.052 & 1.000 & 0.054 & 1.000 & 1.000 \\
Joint, corr.\ (HC2) & 0.042 & 0.056 & 1.000 & 0.056 & 1.000 & 1.000 \\
Joint, centering only & 0.022 & 0.032 & 1.000 & 0.038 & 1.000 & 1.000 \\
Joint, variance only & 0.056 & 0.068 & 1.000 & 0.070 & 1.000 & 1.000 \\
Marginal $L_1$ & 0.032 & 0.048 & 1.000 & 0.042 & 0.034 & 0.042 \\
PCM, single split & 0.050 & 0.056 & 1.000 & 0.040 & -- & -- \\
PCM, derandomized & 0.006 & 0.012 & 1.000 & 0.008 & -- & -- \\
WGCM & 0.044 & 0.032 & 0.996 & 0.050 & -- & -- \\
\hline\hline
\end{tabular*}
\end{center}
\footnotesize
\renewcommand{\baselineskip}{11pt}
\textbf{Note:} $500$ replications; PCM and WGCM in A0 and A1 only; LMP not run. Otherwise as in Table~\ref{otab:mc2-500}.
\end{table}
\begin{table}
\caption{\label{otab:mc2-fwer}Familywise error and attribution across sample sizes (supplementary runs).}
\begin{center}
\footnotesize
\setlength{\tabcolsep}{3pt}
\begin{tabular*}{\textwidth}{@{\extracolsep{\fill}}lcccccccccccc@{}}
\hline\hline
 & \multicolumn{4}{c}{FWER, A0} & \multicolumn{4}{c}{Holm rejection of true $H_0^{2}$, A1} & \multicolumn{4}{c}{Both correct, A1} \\
Method & 500 & 1000 & 2000 & 5000 & 500 & 1000 & 2000 & 5000 & 500 & 1000 & 2000 & 5000 \\
\hline
Joint, uncorrected & 0.088 & 0.050 & 0.043 & 0.018 & 0.110 & 0.080 & 0.040 & 0.042 & 0.236 & 0.442 & 0.813 & 0.958 \\
Joint, corrected & 0.068 & 0.054 & 0.053 & 0.036 & 0.088 & 0.076 & 0.057 & 0.054 & 0.212 & 0.456 & 0.813 & 0.946 \\
Joint, corr. (HC2) & 0.078 & 0.060 & 0.057 & 0.038 & 0.102 & 0.082 & 0.060 & 0.056 & 0.228 & 0.468 & 0.813 & 0.944 \\
Marginal $L_1$ & 0.052 & 0.042 & 0.050 & 0.040 & 0.042 & 0.034 & 0.043 & 0.042 & 0.516 & 0.842 & 0.953 & 0.958 \\
LMP, asymptotic & 0.012 & 0.004 & 0.010 & -- & 0.004 & 0.004 & 0.013 & -- & 0.098 & 0.272 & 0.647 & -- \\
PCM, derand. & 0.018 & 0.006 & 0.017 & 0.010 & 0.008 & 0.012 & 0.017 & 0.008 & 0.272 & 0.704 & 0.973 & 0.992 \\
WGCM & 0.054 & 0.026 & 0.033 & 0.038 & 0.024 & 0.054 & 0.050 & 0.050 & 0.242 & 0.448 & 0.757 & 0.946 \\
\hline\hline
\end{tabular*}
\end{center}
\footnotesize
\renewcommand{\baselineskip}{11pt}
\textbf{Note:} Holm-adjusted, $5\%$ level. Replications: $500$, $500$, $300$ and $500$ at $N=500$, $1000$, $2000$ and $5000$; LMP not run at $N=5000$.
\end{table}

\paragraph{Rejection rates from the supplementary runs.}
\quad   

\textit{I.Size across sample sizes.} The corrected test is liberal at $N=500$, mildly liberal in A1 at $N=1000$ and close to nominal at larger $N$ (Table~\ref{otab:mc2-fwer}). The uncorrected calibration is more liberal at $N=500$, comes as close to $5\%$ as the corrected test at $N=1000$ and $2000$ only because two errors offset there, and is conservative at $N=5000$ (Section~\ref{osec:mc-diag}). The two corrections work together: either one alone moves the size away from nominal, the centering downwards and the variance upwards (Table~\ref{otab:mc2-1000}).\\

\textit{II. Power and attribution.} Power of the proposed test rises quickly with $N$: the test detects the false $H_0^1$ in A1 in about half of the replications at $N=1000$ and in nearly all at $N=5000$, and the interaction A4 already at $N=1000$ (Tables~\ref{otab:mc2-500}--\ref{otab:mc2-5000}). From $N=1000$, the corrections cost no power, because they move $\Sh_j$ in opposite directions. Exact attribution in A1, the frequency with which the Holm procedure rejects $H_0^1$ and not $H_0^2$ in the same replication, improves accordingly (Table~\ref{otab:mc2-fwer}) in all tests, and at $N\le1000$ most attribution errors are failures to detect the true source rather than false attributions to the other block.\\

\textit{III. Rejection rates of the benchmarks.} The benchmarks behave as in the main runs. Our test has power against both additive and interaction alternatives with size near nominal, which no benchmark does. The derandomized PCM and the marginal test have no power against A4 at any $N$ at which they were run, the WGCM fails in A5 (Table~\ref{otab:mc-main}), and the LMP test is undersized and has low power against additive alternatives (Tables~\ref{otab:mc2-500}--\ref{otab:mc2-5000} and~\ref{otab:mc2-fwer}). In the additive designs, the derandomized PCM and the marginal test have higher power at $N\ge1000$, at the price of a blind spot in A4 and, for the PCM, of conservative size; our proposed test has rates that exceed those of the WGCM in every cell at $N\ge1000$ and in all but one at $N=500$, and by $N=5000$ every conditional test detects A1. The power of the marginal test equals its size in A4, which relies on $X_1\perp X_2$ and the symmetry of the CATE function, and Section~\ref{osec:mc-new} shows the outcome without it.

\subsection{Size diagnostics of the studentized statistic}\label{osec:mc-diag}

Table~\ref{otab:mc2-size} reports the Monte Carlo mean, standard deviation, and skewness of $\Sh_j$ in every direction in which the null is true, with raw rejection frequencies at $1\%$, $5\%$, and $10\%$, and Table~\ref{otab:mc2-scale} decomposes the distortion of the uncorrected calibration into centering and scale. Table~\ref{otab:mc2-newsize} reports the same diagnostics for the additional designs in Section~\ref{osec:mc-new}.

\begin{table}
\caption{\label{otab:mc2-size}Null distribution of the studentized statistic: uncorrected and corrected calibration, with the skewness of the corrected statistic.}
\begin{center}
\footnotesize
\setlength{\tabcolsep}{3pt}
\begin{tabular*}{\textwidth}{@{\extracolsep{\fill}}lcccccccccc@{}}
\hline\hline
 & & \multicolumn{3}{c}{Uncorrected calibration} & \multicolumn{6}{c}{Corrected, $N(0,1)$} \\
$N$ & Null & Mean & SD & $5\%$ & Mean & SD & Skew. & $1\%$ & $5\%$ & $10\%$ \\
\hline
500 & A0, $H_0^{1}$ & 0.526 & 0.887 & 0.102 & 0.125 & 1.041 & 0.46 & 0.030 & 0.080 & 0.130 \\
500 & A0, $H_0^{2}$ & 0.504 & 0.903 & 0.112 & 0.099 & 1.060 & 0.41 & 0.030 & 0.076 & 0.128 \\
500 & A1, $H_0^{2}$ & 0.532 & 0.905 & 0.122 & 0.133 & 1.062 & 0.18 & 0.026 & 0.100 & 0.148 \\
500 & A2, $H_0^{1}$ & 0.547 & 0.917 & 0.130 & 0.150 & 1.079 & 0.25 & 0.032 & 0.090 & 0.150 \\
\hline
1000 & A0, $H_0^{1}$ & 0.321 & 0.863 & 0.064 & 0.052 & 1.024 & 0.22 & 0.016 & 0.064 & 0.098 \\
1000 & A0, $H_0^{2}$ & 0.331 & 0.904 & 0.066 & 0.065 & 1.073 & 0.22 & 0.026 & 0.066 & 0.130 \\
1000 & A1, $H_0^{2}$ & 0.327 & 0.871 & 0.080 & 0.059 & 1.034 & 0.22 & 0.016 & 0.076 & 0.116 \\
1000 & A2, $H_0^{1}$ & 0.300 & 0.844 & 0.056 & 0.027 & 1.001 & 0.11 & 0.010 & 0.054 & 0.102 \\
\hline
2000 & A0, $H_0^{1}$ & 0.197 & 0.878 & 0.063 & 0.015 & 1.041 & 0.22 & 0.020 & 0.073 & 0.140 \\
2000 & A0, $H_0^{2}$ & 0.218 & 0.842 & 0.053 & 0.040 & 0.999 & 0.42 & 0.020 & 0.060 & 0.110 \\
2000 & A1, $H_0^{2}$ & 0.227 & 0.814 & 0.040 & 0.051 & 0.963 & 0.00 & 0.010 & 0.057 & 0.097 \\
2000 & A2, $H_0^{1}$ & 0.169 & 0.868 & 0.040 & -0.018 & 1.028 & 0.03 & 0.010 & 0.050 & 0.097 \\
\hline
5000 & A0, $H_0^{1}$ & 0.030 & 0.848 & 0.028 & -0.092 & 0.996 & 0.08 & 0.004 & 0.040 & 0.084 \\
5000 & A0, $H_0^{2}$ & 0.084 & 0.847 & 0.040 & -0.028 & 0.995 & 0.13 & 0.008 & 0.052 & 0.098 \\
5000 & A1, $H_0^{2}$ & 0.134 & 0.849 & 0.042 & 0.030 & 0.998 & 0.27 & 0.018 & 0.054 & 0.116 \\
\hline\hline
\end{tabular*}
\end{center}
\footnotesize
\renewcommand{\baselineskip}{11pt}
\textbf{Note:} Studentized statistic $\widehat S_j$ in the directions where the null is true; N(0,1) is the target. Percentages: raw (unadjusted) rejection frequencies at the $1\%$, $5\%$ and $10\%$ levels. Skew.: skewness of the corrected $\widehat S_j$.
\end{table}
\begin{table}
\caption{\label{otab:mc2-scale}centering and scale of the joint $L_1$ statistic under true nulls.}
\begin{center}
\small
\setlength{\tabcolsep}{3pt}
\begin{tabular*}{\textwidth}{@{\extracolsep{\fill}}lcccccc@{}}
\hline\hline
 & & \multicolumn{2}{c}{Mean $\widehat D_j-\widehat m_j$} & \multicolumn{2}{c}{SD$_{\rm MC}/\bar{\widehat\sigma}_j$} & Own weight \\
$N$ & Null & Uncorrected &Corrected & \eqref{oeq:pf-sigma2} & Finite-$N$ & mean $L_{ii}$ \\
\hline
500 & A0, $H_0^{1}$ & 0.500 & 0.094 & 0.838 & 1.034 & 0.099 \\
500 & A0, $H_0^{2}$ & 0.479 & 0.073 & 0.855 & 1.054 & 0.099 \\
500 & A1, $H_0^{2}$ & 0.506 & 0.100 & 0.853 & 1.058 & 0.099 \\
\hline
1000 & A0, $H_0^{1}$ & 0.303 & 0.037 & 0.835 & 1.022 & 0.061 \\
1000 & A0, $H_0^{2}$ & 0.313 & 0.047 & 0.878 & 1.071 & 0.061 \\
1000 & A1, $H_0^{2}$ & 0.310 & 0.044 & 0.847 & 1.032 & 0.061 \\
\hline
2000 & A0, $H_0^{1}$ & 0.185 & 0.009 & 0.853 & 1.040 & 0.037 \\
2000 & A0, $H_0^{2}$ & 0.205 & 0.030 & 0.821 & 0.997 & 0.037 \\
2000 & A1, $H_0^{2}$ & 0.214 & 0.038 & 0.785 & 0.965 & 0.037 \\
\hline
5000 & A0, $H_0^{1}$ & 0.027 & -0.075 & 0.836 & 0.996 & 0.019 \\
5000 & A0, $H_0^{2}$ & 0.079 & -0.024 & 0.837 & 0.995 & 0.019 \\
5000 & A1, $H_0^{2}$ & 0.126 & 0.023 & 0.837 & 0.998 & 0.019 \\
\hline\hline
\end{tabular*}
\end{center}
\footnotesize
\renewcommand{\baselineskip}{11pt}
\textbf{Note:} Uncorrected: centering with in-sample residuals; corrected: own-weight-corrected residuals ($\widehat m_j$ is the centering $\widehat m^{\,\kappa}_j$ of the uncorrected calibration in the uncorrected column and $\widetilde m_j$ of \eqref{eq:center-feasible} in the corrected column). SD$_{\rm MC}$: Monte Carlo standard deviation of $\widehat D_j-\widehat m_j$; $\bar{\widehat\sigma}_j$: mean of the plug-in of the limit variance \eqref{oeq:pf-sigma2} or of the finite-$N$ variance. A ratio of one means the variance estimate is right on average.
\end{table}

\paragraph{The test based on the uncorrected calibration.} Under a true null, the studentized statistic based on the uncorrected calibration is shifted upwards and too concentrated (Table~\ref{otab:mc2-size}). The shift comes from the centering: the in-sample residual is shrunk towards zero by the weight of the own observation, so that $\widehat m_j$ is too small, and the shift fades as that weight falls with $N$ (Table~\ref{otab:mc2-scale}). The concentration comes from the scale, with no trend in $N$: the limit variance \eqref{oeq:pf-sigma2} drops the band configurations and the kernel-expansion terms that the finite-sample variance keeps (Lemma~\ref{prop:sigmaN} and the discussion after it). The two errors act in opposite directions and offset near $\alpha=0.05$ only at $N=1000$ and $2000$: at $N=500$ the shift dominates, and the test is liberal; at $N=5000$ the concentration dominates, and it is conservative; and they do not offset at other levels. Its near-nominal size at intermediate $N$ is therefore a coincidence, not a property of the calibration.

\paragraph{The test based on the corrected calibration.} The own-weight correction removes most of the location error and the finite-sample variance removes the scale error on average (Table~\ref{otab:mc2-scale}); a small excess of the Monte Carlo spread over $\widetilde\sigma_{j,N}$ at small $N$ is gone by $N=2000$. The statistic stays right-skewed, as a sum of absolute Gaussian terms over six to nine effective cells should be, so rejections in the upper tail exceed their nominal rates at small $N$, above all at $N=500$ (Table~\ref{otab:mc2-size}). From $N=1000$ on, its first two moments are close to those of $N(0,1)$, and the corrected test with normal critical values is adequate for $ N \ ge 2000$.

\begin{table}
\caption{\label{otab:mc2-newsize}Null distribution of the studentized statistic in the additional designs, $N=1000$.}
\begin{center}
\small
\setlength{\tabcolsep}{2.5pt}
\begin{tabular*}{\textwidth}{@{\extracolsep{\fill}}llccccccccc@{}}
\hline\hline
 & & \multicolumn{3}{c}{Uncorrected calibration} & \multicolumn{5}{c}{Corrected, $N(0,1)$} & \\
Design & Null & Mean & SD & $5\%$ & Mean & SD & $1\%$ & $5\%$ & $10\%$ & $R$ \\
\hline
E2-A0 & $H_0^{1}$ & 0.409 & 0.757 & 0.056 & 0.178 & 0.860 & 0.012 & 0.050 & 0.098 & 500 \\
E2-A0 & $H_0^{2}$ & 0.420 & 0.762 & 0.058 & 0.192 & 0.865 & 0.006 & 0.054 & 0.112 & 500 \\
E2-A1 & $H_0^{2}$ & 0.391 & 0.735 & 0.050 & 0.159 & 0.836 & 0.006 & 0.042 & 0.104 & 500 \\
\hline
C0 & $H_0^{1}$ & 0.323 & 1.386 & 0.072 & 0.028 & 1.002 & 0.017 & 0.050 & 0.101 & 484 \\
C0 & $H_0^{2}$ & 0.307 & 1.382 & 0.079 & 0.007 & 1.008 & 0.010 & 0.058 & 0.118 & 484 \\
\hline
C1 & $H_0^{1}$ & 0.347 & 1.098 & 0.097 & 0.035 & 1.068 & 0.025 & 0.082 & 0.119 & 486 \\
\hline
T3-A0 & $H_0^{1}$ & 0.290 & 0.833 & 0.040 & 0.017 & 0.984 & 0.018 & 0.040 & 0.096 & 500 \\
T3-A0 & $H_0^{2}$ & 0.309 & 0.856 & 0.056 & 0.039 & 1.011 & 0.012 & 0.054 & 0.112 & 500 \\
T3-A1 & $H_0^{2}$ & 0.358 & 0.831 & 0.060 & 0.100 & 0.982 & 0.008 & 0.058 & 0.128 & 500 \\
\hline
LN-A0 & $H_0^{1}$ & 0.376 & 0.854 & 0.078 & 0.124 & 1.002 & 0.024 & 0.072 & 0.118 & 500 \\
LN-A0 & $H_0^{2}$ & 0.389 & 0.838 & 0.082 & 0.139 & 0.984 & 0.022 & 0.078 & 0.124 & 500 \\
LN-A1 & $H_0^{2}$ & 0.365 & 0.793 & 0.062 & 0.111 & 0.931 & 0.016 & 0.058 & 0.106 & 500 \\
\hline\hline
\end{tabular*}
\end{center}
\footnotesize
\renewcommand{\baselineskip}{11pt}
\textbf{Note:} Studentized statistic $\widehat S_j$ in the directions where the null is true, designs of Section~\ref{osec:mc-design}; $N(0,1)$ is the target. Percentages: raw (unadjusted) rejection frequencies at the $1\%$, $5\%$ and $10\%$ levels. $R$: replications used by the joint tests (C0 and C1: $16$ and $14$ of $500$ discarded, Section~\ref{osec:mc-paper}). Monte Carlo standard error of a $5\%$ rate: $0.010$.
\end{table}

In the additional designs (Table~\ref{otab:mc2-newsize}), the corrected statistic is close to $N(0,1)$ under correlated blocks and $t_3$ errors, and its rejection rates are close to nominal except for a mild excess in C1 (Section~\ref{osec:mc-new}). Under the estimated propensity with misspecified outcome models (E2), it shifts upward and becomes too concentrated, and the two errors offset at the $5\%$ level. Under lognormal errors, its spread is close to one, but its location shifts upward, making it mildly liberal in A0 and leaving it close to nominal in A1. The large spread of the uncorrected statistic in C0 comes from a few extreme replications, which do not affect the corrected statistic.

\subsection{Additional designs}\label{osec:mc-new}

Table~\ref{otab:mc2-new} summarizes the additional designs at $N=1000$, and Tables~\ref{otab:mc2-new-a} and \ref{otab:mc2-new-b} give every method and cell.

\begin{table}
\caption{\label{otab:mc2-new}Correlated blocks and heavy tails, $N=1000$.}
\begin{center}
\setlength{\tabcolsep}{3pt}
\begin{tabular*}{\textwidth}{@{\extracolsep{\fill}}lcccccc@{}}
\hline\hline
 & \multicolumn{3}{c}{Correlated blocks (C)} & \multicolumn{3}{c}{Lognormal errors (LN)} \\
 & C0 & \multicolumn{2}{c}{C1} & A0 & \multicolumn{2}{c}{A1} \\
Method & FWER & $H_0^{1}$ & $H_0^{2}$ & FWER & $H_0^{1}$ & $H_0^{2}$ \\
\hline
Joint $L_1$, corrected & 0.050 & 0.070 & 0.257 & 0.070 & 0.540 & 0.054 \\
\hline
Marginal $L_1$ & 0.050 & 0.128 & 0.646 & 0.058 & 0.856 & 0.024 \\
LMP, asymptotic & 0.004 & 0.004 & 0.112 & 0.014 & 0.252 & 0.014 \\
PCM, derandomized & 0.020 & 0.018 & 0.694 & 0.004 & 0.650 & 0.006 \\
WGCM & 0.070 & 0.044 & 0.558 & 0.042 & 0.508 & 0.044 \\
\hline\hline
\end{tabular*}
\end{center}
\footnotesize
\renewcommand{\baselineskip}{11pt}
\textbf{Note:} $N=1000$; Holm-adjusted rejection frequencies at the $5\%$ level; $500$ replications. C: $(X_1,X_2)$ from a Gaussian copula with correlation $0.5$, $r_1=1/2$ known; C0: $\tau=1$; C1: $\tau=2x_2$, so $H_0^{1}$ is true while the marginal null for $X_1$ is false, and a rejection of $H_0^{1}$ is a false attribution to $X_1$; the joint test discards the $16$ (C0) and $14$ (C1) replications in which a Nadaraya--Watson denominator at a corner grid point is not positive. LN: base design with standardized lognormal errors. Otherwise as in Table~\ref{otab:mc-main}; the designs E2, C4, T3 and LN-A4 are in Tables~\ref{otab:mc2-new-a} and~\ref{otab:mc2-new-b}.
\end{table}
\begin{table}[h]
\caption{\label{otab:mc2-new-a}Rejection frequencies in the additional designs, $N=1000$.}
\begin{center}
\scriptsize
\setlength{\tabcolsep}{3pt}
\begin{tabular*}{\textwidth}{@{\extracolsep{\fill}}lcccccccc@{}}
\hline\hline
 & \multicolumn{2}{c}{E2-A0} & \multicolumn{2}{c}{E2-A1} & \multicolumn{2}{c}{C0} & \multicolumn{2}{c}{C1} \\
Method & $H_0^{1}$ & $H_0^{2}$ & $H_0^{1}$ & $H_0^{2}$ & $H_0^{1}$ & $H_0^{2}$ & $H_0^{1}$ & $H_0^{2}$ \\
\hline
Joint, uncorrected & 0.036 & 0.038 & 0.364 & 0.044 & 0.048 & 0.048 & 0.084 & 0.288 \\
Joint, corrected & 0.032 & 0.034 & 0.352 & 0.040 & 0.041 & 0.039 & 0.070 & 0.257 \\
Joint, corr. (HC2) & 0.036 & 0.036 & 0.380 & 0.048 & 0.039 & 0.039 & 0.084 & 0.270 \\
Joint, centering only & 0.016 & 0.010 & 0.258 & 0.016 & 0.027 & 0.019 & 0.049 & 0.183 \\
Joint, variance only & 0.066 & 0.068 & 0.478 & 0.082 & 0.093 & 0.095 & 0.142 & 0.387 \\
Marginal $L_1$ & 0.014 & 0.014 & 0.706 & 0.036 & 0.026 & 0.028 & 0.128 & 0.646 \\
LMP, asympt. & 0.006 & 0.010 & 0.154 & 0.012 & 0.004 & 0.000 & 0.004 & 0.112 \\
PCM, single & 0.026 & 0.022 & 0.350 & 0.014 & 0.026 & 0.040 & 0.040 & 0.642 \\
PCM, derand. & 0.002 & 0.006 & 0.320 & 0.000 & 0.010 & 0.014 & 0.018 & 0.694 \\
WGCM & 0.036 & 0.034 & 0.390 & 0.032 & 0.034 & 0.042 & 0.044 & 0.558 \\
\hline\hline
\end{tabular*}
\end{center}
\footnotesize
\renewcommand{\baselineskip}{11pt}
\textbf{Note:} $N=1000$, Holm-adjusted rejection frequencies at $5\%$; $500$ replications. Truth: A0, C0 both nulls true; A1 $H_0^{2}$ true; C1 $H_0^{1}$ true (marginal null for $X_1$ false); A4, C4 both false. E2: as E1, with $\mu_0$ augmented by $(V^2-1)/2+\sin(\pi x_1)$, so both linear outcome models are misspecified while the propensity model is correct. T3: $\epsilon=t_3/\sqrt3$. LN: $\epsilon=(e^{Z}-e^{1/2})/\{e(e-1)\}^{1/2}$.
\end{table}
\begin{table} [h]
\caption{\label{otab:mc2-new-b}Rejection frequencies in the additional designs, $N=1000$ (continued).}
\begin{center}
\scriptsize
\setlength{\tabcolsep}{3pt}
\begin{tabular*}{\textwidth}{@{\extracolsep{\fill}}lcccccccccccc@{}}
\hline\hline
 & \multicolumn{2}{c}{C4} & \multicolumn{2}{c}{T3-A0} & \multicolumn{2}{c}{T3-A1} & \multicolumn{2}{c}{LN-A0} & \multicolumn{2}{c}{LN-A1} & \multicolumn{2}{c}{LN-A4} \\
Method & $H_0^{1}$ & $H_0^{2}$ & $H_0^{1}$ & $H_0^{2}$ & $H_0^{1}$ & $H_0^{2}$ & $H_0^{1}$ & $H_0^{2}$ & $H_0^{1}$ & $H_0^{2}$ & $H_0^{1}$ & $H_0^{2}$ \\
\hline
Joint, uncorrected & 0.816 & 0.812 & 0.024 & 0.032 & 0.534 & 0.060 & 0.050 & 0.048 & 0.534 & 0.058 & 0.870 & 0.868 \\
Joint, corrected & 0.796 & 0.788 & 0.024 & 0.036 & 0.536 & 0.058 & 0.054 & 0.050 & 0.540 & 0.054 & 0.872 & 0.868 \\
Joint, corr. (HC2) & 0.796 & 0.792 & 0.026 & 0.036 & 0.556 & 0.070 & 0.060 & 0.052 & 0.574 & 0.062 & 0.878 & 0.874 \\
Joint, centering only & 0.724 & 0.724 & 0.018 & 0.020 & 0.428 & 0.020 & 0.030 & 0.024 & 0.436 & 0.034 & 0.808 & 0.802 \\
Joint, variance only & 0.876 & 0.871 & 0.058 & 0.066 & 0.642 & 0.126 & 0.090 & 0.088 & 0.680 & 0.106 & 0.932 & 0.924 \\
Marginal $L_1$ & 0.250 & 0.290 & 0.014 & 0.026 & 0.886 & 0.044 & 0.030 & 0.028 & 0.856 & 0.024 & 0.010 & 0.038 \\
LMP, asympt. & 0.540 & 0.556 & 0.006 & 0.010 & 0.244 & 0.004 & 0.008 & 0.006 & 0.252 & 0.014 & 0.954 & 0.960 \\
PCM, single & 0.284 & 0.314 & 0.018 & 0.018 & 0.656 & 0.058 & 0.038 & 0.028 & 0.602 & 0.062 & 0.038 & 0.046 \\
PCM, derand. & 0.238 & 0.278 & 0.004 & 0.010 & 0.730 & 0.016 & 0.002 & 0.002 & 0.650 & 0.006 & 0.002 & 0.010 \\
WGCM & 0.932 & 0.930 & 0.020 & 0.036 & 0.536 & 0.024 & 0.022 & 0.022 & 0.508 & 0.044 & 0.980 & 0.978 \\
\hline\hline
\end{tabular*}
\end{center}
\footnotesize
\renewcommand{\baselineskip}{11pt}
\textbf{Note:} As Table~\ref{otab:mc2-new-a}; in C4 the joint test discards $10$ replications.
\end{table}

\paragraph{Estimated propensity.} Design E1, a same-sample logistic first stage with the DR pseudo-outcome, is the design of Panel~B of Figure~\ref{fig:mc}, whose results from the main runs are in Table~\ref{otab:mc-main}. When both outcome models are misspecified, and only the propensity model is correct (E2), the corrected test keeps its level, as the double robustness of Theorem~\ref{thm:nuisance-param} predicts, and every method loses power because the pseudo-outcome is noisier (Table~\ref{otab:mc2-new-a}). The independent-sample first stage of Theorem~\ref{thm:nuisance-split} is simulated in the designs F2 of Section~\ref{osec:mc-fs}

\paragraph{Correlated blocks.} Design C1 separates conditional from marginal testing. Because $X_1$ is correlated with $X_2$, the marginal test falsely attributes to $X_1$ heterogeneity that depends only on $x_2$. The conditional tests do not: the benchmarks keep their level, and the corrected joint test is mildly above it, about two standard errors above $5\%$ and closer than the uncorrected calibration (Tables~\ref{otab:mc2-new} and~\ref{otab:mc2-new-a}). In C4, the correlation breaks the exact cancellation that makes A4 a knife-edge for marginal tests, so the marginal test and the derandomized PCM, which have no power against A4, gain some, but remain well below our test (Table~\ref{otab:mc2-new-b}). In C1, in the direction where the null is false, the CATE function is linear in $x_2$ and fully captured by a projection, which is where the projection-based tests are strongest.

\paragraph{Heavy tails.} With $t_3$ errors, which have no third moment and so lie outside the theory's requirement of bounded conditional moments of order above five (Remark~\ref{orem:moments}), the corrected test maintains its level in T3-A0 and T3-A1 and maintains its power (Table~\ref{otab:mc2-new-b}). With lognormal errors, it is mildly above its level in LN-A0, at its level in LN-A1, and keeps its power against A1 and A4 (Tables~\ref{otab:mc2-new} and~\ref{otab:mc2-new-b}). The mild excess thus arises under strongly skewed errors rather than under the symmetric $t_3$ errors. With outcomes such as the one of Section~\ref{osec:emp}, which has finite moments but extreme kurtosis, the size of the joint test should be read with this caveat (Remark~\ref{orem:heavytails}).

\subsection{Sensitivity to the grid and the bandwidth}\label{osec:mc-sens}

Table~\ref{otab:mc2-sens} varies the grid size $M\in\{13,21,31\}$ at $k_j=3.5$ and the bandwidth constant $k_j\in\{3,4.5,6\}$ at $M=13$, in A0 and A1 at $N=1000$ on the data sets of Table~\ref{otab:mc2-1000}.

\begin{table}[h]
\caption{\label{otab:mc2-sens}Sensitivity of the joint $L_1$ test to the grid and the bandwidth.}
\begin{center}
\small
\setlength{\tabcolsep}{3pt}
\begin{tabular*}{\textwidth}{@{\extracolsep{\fill}}lccccccccc@{}}
\hline\hline
 & & \multicolumn{3}{c}{A0} & \multicolumn{3}{c}{A1} & &  \\
$(M,k_j)$ & Calibration & Mean $\widehat S_1$ & SD $\widehat S_1$ & FWER & $H_0^{1}$ & $H_0^{2}$ & Both & Disc. & $\delta/h$ \\
\hline
(13, 3.5) & Corrected & 0.052 & 1.024 & 0.054 & 0.526 & 0.076 & 0.456 & 0 & 0.18 \\
 & Uncorrected & 0.321 & 0.863 & 0.050 & 0.516 & 0.080 & 0.442 & 0 &  \\
\hline
(21, 3.5) & Corrected & 0.051 & 1.028 & 0.046 & 0.532 & 0.076 & 0.460 & 0 & 0.11 \\
 & Uncorrected & 0.320 & 0.887 & 0.048 & 0.532 & 0.078 & 0.458 & 0 &  \\
\hline
(31, 3.5) & Corrected & 0.056 & 1.029 & 0.048 & 0.522 & 0.074 & 0.452 & 0 & 0.07 \\
 & Uncorrected & 0.324 & 0.899 & 0.050 & 0.532 & 0.078 & 0.458 & 0 &  \\
\hline
(13, 3) & Corrected & 0.063 & 1.051 & 0.060 & 0.426 & 0.078 & 0.352 & 1 & 0.21 \\
 & Uncorrected & 0.506 & 0.922 & 0.088 & 0.516 & 0.118 & 0.402 & 1 &  \\
\hline
(13, 4.5) & Corrected & 0.039 & 0.984 & 0.050 & 0.666 & 0.062 & 0.606 & 0 & 0.14 \\
 & Uncorrected & 0.157 & 0.784 & 0.032 & 0.588 & 0.042 & 0.548 & 0 &  \\
\hline
(13, 6) & Corrected & 0.028 & 0.964 & 0.056 & 0.820 & 0.066 & 0.754 & 0 & 0.10 \\
 & Uncorrected & 0.073 & 0.716 & 0.016 & 0.688 & 0.020 & 0.668 & 0 &  \\
\hline\hline
\end{tabular*}
\end{center}
\footnotesize
\renewcommand{\baselineskip}{11pt}
\textbf{Note:} $N=1000$, $500$ replications, same data sets as Table~\ref{otab:mc2-1000}. $M$: grid points per axis; $k_j$: bandwidth constant in $h_j=k_j\,\widehat{\mathrm{sd}}_jN^{-1/7}$; $\delta/h$: grid spacing over bandwidth. Disc.: replications discarded because a Nadaraya--Watson denominator was not positive (A0 and A1 together). For $M\ge21$ the finite-$N$ variance is evaluated by $2000$ Gaussian draws instead of the exact $O(M^6)$ double sum.
\end{table}

The corrected test is insensitive to the grid (Table~\ref{otab:mc2-sens}): its size and power barely change between $M=13$ and $31$, so the coarser $13\times13$ grid of the main study (Section~\ref{osec:mc-paper}) costs little. The rejection rates of the uncorrected calibration also change little with $M$, although the spread of its statistic rises, because its limit variance changes with the grid. Over the bandwidths studied, the corrected test stays centered, and its size stays close to nominal, while wider windows substantially raise its power against the smooth alternative A1. The uncorrected calibration is liberal at $k_j=3$ and conservative at $k_j=6$: its two errors cancel only near $k_j=3.5$. One Nadaraya--Watson denominator failed at $k_j=3$ with $N=1000$.

\subsection{Estimated first stage}\label{osec:mc-fs}

This subsection reports the designs with an estimated first stage that Section~\ref{sec:mc} of the main text refers to: F1, F1b, F1c, and F2. They ask whether an estimated first stage moves the level of the tests of $H_0^1$, for our integrated absolute pairwise difference test under the conditions of Section~\ref{sec:asym-nuisance} and for the benchmarks, whose level results are for an observed response. The study has its own designs and seeds (Section~\ref{osec:mc-fs-design}), and we fixed the designs before a pilot, after which we changed only the slope of the power alternative (Section~\ref{osec:mc-fs-dev}).

\subsubsection{Predictions}\label{osec:mc-fs-pred}

Let $\widehat\zeta$ be the DR pseudo-outcome \eqref{eq:psi-dr} with estimated nuisances, $\bar\pi_N(x)=\E[\widehat\zeta-\zeta\mid X=x,\widehat\chi]$ the average drift of Section~\ref{sec:asym-nuisance}, and $\bar\pi_N^\perp(x):=\bar\pi_N(x)-\int_0^1\bar\pi_N(u,x_2)\,du$ its part that varies with $x_1$ at fixed $x_2$; a drift that does not vary with $x_1$ leaves $H_0^1$ intact. By Lemma~\ref{lem:drift} our proposed test keeps its level if $N\eta^{1/2}\|\bar\pi_N^\perp\|^2\to0$, $\|\cdot\|$ the norm of $L_2([0,1]^2)$. A root-$N$ test of conditional mean independence (GCM, WGCM, PCM) treats $\bar\pi_N^\perp$ as a local alternative and needs $\sqrt N\|\bar\pi_N^\perp\|\to0$ in the directions it detects. Two consequences are tested.

(a) \textit{Same-sample parametric first stage} (Theorem~\ref{thm:nuisance-param}). With one working model correct, $\|\bar\pi_N^\perp\|=O_p(N^{-1/2})$, so $N\eta^{1/2}\|\bar\pi_N^\perp\|^2=O_p(\eta^{1/2})\to0$ for our proposed test, whereas for a root-$N$ test $\sqrt N\|\bar\pi_N^\perp\|=O_p(1)$ is an estimation effect that does not shrink with $N$.

(b) \textit{First stage of controlled rate, independent of the test sample} (Theorem~\ref{thm:nuisance-split}). With first-stage errors of order $N^{-\gamma_\chi}$ in both nuisances, the DR drift is of order $N^{-2\gamma_\chi}$. The root-$N$ tests need $2\gamma_\chi>1/2$; our proposed test, with $\eta\propto N^{-2/7}$, needs $2\gamma_\chi>1/2-1/4\cdot2/7=3/7\approx0.43$; the LMP test, a degenerate $U$-statistic smoothing over $X_2$ only with bandwidth of order $N^{-2.1/5}$, needs by the same heuristic $2\gamma_\chi>1/2-1/4\cdot2.1/5\approx0.40$. The window $3/7<2\gamma_\chi<1/2$, where our proposed test is protected and the root-$N$ tests are not, is narrow: $\eta^{-1/4}=1.63$, $1.71$ and $1.83$ at $N=1000$, $2000$ and $5000$.

\subsubsection{Designs and methods}\label{osec:mc-fs-design}\quad\\

\textit{Common data.} $X_1,X_2\sim U[0,1]$ and $V\sim N(0,1)$ independent, $T=\ind\{U_T<r_1(X,V)\}$ with $U_T\sim U[0,1]$, $Y=\mu_0(X,V)+T\tau(X)+\epsilon$, $\epsilon\sim N(0,1)$, and under the null $\tau=1+0.5x_2$, so that $H_0^1$ holds and $H_0^2$ does not; only $H_0^1$ is tested. Every method receives the same pseudo-outcome $\widehat\zeta_i=\widehat\mu_1-\widehat\mu_0+T_i(Y_i-\widehat\mu_1)/\widehat r_1-(1-T_i)(Y_i-\widehat\mu_0)/(1-\widehat r_1)$, evaluated at $(X_i,V_i)$. The data of replication $r$ at sample size $N$ are drawn from a seed that depends on $(N,r)$ only and are shared by all designs (common random numbers); the random splits of the PCM and the WGCM use separate streams.\\

\textit{Designs.} With $\Lambda_a:=a(x_1-\frac12)+0.5(x_2-\frac12)+0.5v$ and $\mu^{\mathrm L}:=1+0.5x_1+0.5x_2+v$, the propensity $r_1$ and the outcome regression $\mu_0$ (with $\mu_1=\mu_0+\tau$) are
\begin{align*}
\text{F1:}\quad &\mathrm{logit}\,r_1=\Lambda_2,\ \ \mu_0=\mu^{\mathrm L}+1.5\{\sin(2\pi x_1)+\cos(2\pi x_1)\}
 &&\text{outcome models misspecified;}\\
\text{F1b:}\quad &\mathrm{logit}\,r_1=\Lambda_2+1+1.5\cos(2\pi x_1),\ \ \mu_0=\mu^{\mathrm L}
 &&\text{propensity misspecified;}\\
\text{F1c:}\quad &\mathrm{logit}\,r_1=\Lambda_2,\ \ \mu_0=\mu^{\mathrm L}
 &&\text{both correct (control);}\\
\text{F2:}\quad &\mathrm{logit}\,r_1=\Lambda_1,\ \ \mu_0=\mu^{\mathrm L}
 &&\text{independent first stage.}
\end{align*}

\textit{First stage in F1, F1b and F1c.} A logit of $T$ on $(1,X_1,X_2,V)$ fitted by maximum likelihood and ordinary least squares of $Y$ on $(1,X_1,X_2,V)$ within each arm, all on the full test sample, the setting of Theorem~\ref{thm:nuisance-param}. The linear outcome models miss the trigonometric term of F1, and the linear logit absorbs the intercept shift of F1b but cannot represent $1.5\cos(2\pi x_1)$. The fitted propensities are not trimmed; they lie in $[0.03,0.98]$ in F1 and F1c and in $[0.11,0.97]$ in F1b.\\

\textit{First stage in F2.} The true nuisances are perturbed as $\widehat r_1=\min\{\max\{r_1+N^{-\gamma_\chi}(2x_1-1)(1+Z_r),0.05\},0.95\}$ and $\widehat\mu_t=\mu_t+N^{-\gamma_\chi}(1+x_1)(1+Z_\mu)$, $t=0,1$, with $Z_r,Z_\mu\sim N(0,0.5^2)$ drawn once per replication, independently of the test sample, and $\gamma_\chi$ equal to $0.15$, $0.20$, $0.25$, $0.30$ or $0.35$. The perturbation stands in for a first stage computed on an independent sample: before trimming, the drift is $N^{-2\gamma_\chi}(2x_1-1)(1+x_1)(1+Z_r)(1+Z_\mu)\{1/\widehat r_1+1/(1-\widehat r_1)\}$, which varies with $x_1$ and is of order $N^{-2\gamma_\chi}$. The median share of trimmed propensities is $13\%$ at $\gamma_\chi=0.15$ and $N=1000$, $1.4\%$ at $\gamma_\chi=0.25$ and at most $0.3\%$ for $\gamma_\chi\ge0.30$.\\

\textit{Power.} $\tau=1+0.5x_2+1.5(x_1-\frac12)$, in F1 (outcome models still misspecified) and in F2 with $\gamma_\chi=0.25$ (the perturbation of the null design), on the data of the matching null cells.\\

\textit{Methods.} Our test in the corrected calibration of Section~\ref{osec:mc-paper} ($k_j=3.5$, $13\times13$ grid on $[0.1,0.9]^2$, normal critical value) and, for comparison, in the uncorrected calibration; the LMP test with its asymptotic critical value; the PCM with a single split and derandomized over six splits; and the WGCM with estimated weight, all as in the supplementary runs (Section~\ref{osec:mc-bench}); and the GCM, built from the same components: $\Xi_i$$:=\{X_{1i}-\widehat E[X_1\mid X_{2i}]\}\{\widehat\zeta_i-\widehat E[\widehat\zeta\mid X_{2i}]\}$, with $\widehat E[X_1\mid X_2]$ and $\widehat E[\widehat\zeta\mid X_2]$ the regressions on $X_2$ of Section~\ref{osec:mc-bench} fitted on the full sample, $\mathcal T_{\mathrm{GCM}}=N^{1/2}\bar\Xi/\widehat{\mathrm{sd}}(\Xi)$, and the two-sided $p$-value $2\{1-\Phi(|\mathcal T_{\mathrm{GCM}}|)\}$ of \citet{shah2020hardness}. It is the test-set step of the WGCM with weight one, applied to the whole sample.\\

\textit{Drift diagnostics.} For each replication $\bar\pi_N$ is computed numerically from the fitted (F1, F1b, F1c) or perturbed (F2) nuisances, by Gauss--Legendre quadrature with $40\times40$ nodes on $[0,1]^2$ and $40$-point Gauss--Hermite quadrature over $V$, and $\bar\pi_N^\perp$ by subtracting its $x_1$-average. Reported are $\sqrt N\|\bar\pi_N^\perp\|$, $N\eta^{1/2}\|\bar\pi_N^\perp\|^2$ with $\eta=(3.5\cdot12^{-1/2}N^{-1/7})^2$, the population counterpart of the bandwidths of our proposed test, and $\sqrt N\langle x_1-\frac12,\bar\pi_N^\perp\rangle/\mathrm{sd}(X_1)$, the drift in the direction seen by the GCM. Dividing the first two by the standard deviation and the variance of $\widehat\zeta$ puts them on the scale of the noise. In F1 and F1b, the pseudo-true parameters $\vartheta^*$ of the working models were also computed from one sample of $4\times10^6$ draws, and our proposed test, the GCM and the WGCM, were recomputed on the same replications from $\zeta(W;\chi_{\vartheta^*})$, for which $\E[\zeta\mid X]=\tau$ up to the sampling error of $\vartheta^*$; the difference isolates the estimation effect.\\

\textit{Replications.} $500$ at $N=1000$ and $2000$ for every method; at $N=5000$, $300$ for our proposed test, on the first $100$ replications for the GCM, the WGCM and the single-split PCM, and $50$ for the derandomized PCM; the LMP test was not run at $N=5000$. Power: $500$ replications at $N=1000$ and $2000$. Monte Carlo standard errors of a rejection frequency of $0.05$ are $0.010$, $0.013$, $0.022$ and $0.031$ for $500$, $300$, $100$ and $50$ replications.

\subsubsection{Results}\label{osec:mc-fs-results}

Table~\ref{otab:fs-main} summarizes the study, and Table~\ref{otab:fs-drift} provides the drift diagnostics and the moments of the statistics.\\

\textit{Control.} In F1c, where both working models are correct, the drift is negligible and the statistics of our proposed test, the GCM and the WGCM keep their null distributions at every $N$, with the derandomized PCM and the LMP test conservative as in the base designs (Table~\ref{otab:fs-drift}). The distortions below therefore come from the first stage.\\

\textit{Same-sample parametric first stage.} The root-$N$ tests carry an estimation effect that does not shrink with $N$, and its sign depends on which working model is misspecified. With the propensity misspecified (F1b), the GCM and the WGCM over-reject with an inflated spread of their statistics; the single-split PCM does so mildly, and the derandomized PCM's conservativeness protects it (Tables~\ref{otab:fs-main} and~\ref{otab:fs-drift}). The comparison at the pseudo-true parameters on the same data isolates the cause: there the GCM and the WGCM return to their level, rejecting in $0.048$ and $0.046$ and in $0.054$ and $0.060$ of the replications at $N=1000$ and $2000$, and the spread of their statistics returns to about one. With the outcome models misspecified and the propensity correct (F1) the GCM and the derandomized PCM are conservative and the WGCM is at or below its level, while at the pseudo-true parameters the GCM rejects in $0.038$ and $0.050$ of the replications, as is familiar for a correctly specified propensity estimated by maximum likelihood: the estimation effect is minus the projection of the influence function of the statistic on the logit score, which lowers the variance that the studentization, computed as if the response were observed, does not see.

Our proposed test is unaffected in F1b, where it behaves as at the pseudo-true parameters ($0.054$ and $0.050$ at $N=1000$ and $2000$). In F1, it is mildly conservative (Table~\ref{otab:fs-main}), whereas at the pseudo-true parameters on the same data it rejects in $0.042$ and $0.054$ of the replications, so the shortfall comes from estimating the first stage. Lemma~\ref{lem:drift} does not predict this effect, whose sign is opposite to that of its quadratic drift term; it is the same kind as the benchmarks but less than half as large in rejection frequency, and it does not visibly shrink up to $N=5000$. \\

\textit{Independent first stage of controlled rate.} For $2\gamma_\chi=0.4$ the conditions of the root-$N$ tests and of our proposed test fail, the LMP test is at its threshold, and every test over-rejects (Table~\ref{otab:fs-main}). Our test over-rejects less than the GCM and the WGCM in every cell, and its over-rejection grows with $N$ while theirs does not; it over-rejects more than the PCM, which computes its statistic on half of the sample, and about as much as the LMP test. At $2\gamma_\chi=0.6$ our proposed test is mildly above its level at $N=1000$ and close to it at larger $N$, the derandomized PCM and the LMP test stay below their level, the GCM over-rejects at every $N$ and the WGCM at $N\le2000$; the predicted decline of the root-$N$ distortion, $\sqrt N\|\bar\pi_N^\perp\|\propto N^{-0.1}$, is too slow to show by $N=5000$.\\

\textit{Power.} The alternative $\tau=1+0.5x_2+1.5(x_1-\frac12)$ is linear in $x_1$, the direction that the GCM targets and in which the drift itself lies, so the GCM has the highest power in every cell (Table~\ref{otab:fs-main}). In F1, where every test rejects at or below its level under the null, our proposed test rejects more often than the LMP test and less often than the WGCM and the derandomized PCM, and at $N=2000$ it matches the single-split PCM. In F2 with $\gamma_\chi=0.25$ it rejects more often than the PCM and about as often as the WGCM, although these rates include the over-rejection that the drift causes under the null. A first-stage drift and a heterogeneity of the same shape are indistinguishable to any test, and our statistic reacts to both only through a term of order $N\eta^{1/2}\|\cdot\|^2$: its robustness to drift and its response to such alternatives have the same source. In these designs, the robustness yields less over-rejection than the GCM and the WGCM when the first stage converges slowly, and no over-rejection when the parametric propensity is misspecified.

\begin{table}[h]
\caption{\label{otab:fs-main}First-stage estimation: summary of size and power for $H_0^{1}$.}
\begin{center}
\setlength{\tabcolsep}{3pt}
\footnotesize
\begin{tabular*}{\textwidth}{@{\extracolsep{\fill}}llcccccccc@{}}
\hline\hline
Design & $N$ & $\frac{\sqrt N\|\bar\pi_N^\perp\|}{\mathrm{sd}(\widehat\zeta)}$ & $\frac{N\eta^{1/2}\|\bar\pi_N^\perp\|^2}{\mathrm{var}(\widehat\zeta)}$ & Our test & LMP & PCM-DR & PCM-1 & WGCM & GCM \\
\hline
\multicolumn{10}{@{}l}{\emph{Size} (nominal $0.05$)} \\
F1 & 1000 & 1.20 & 0.54 & 0.032 & 0.016 & 0.010 & 0.044 & 0.046 & 0.010 \\
 & 2000 & 1.24 & 0.53 & 0.038 & 0.014 & 0.010 & 0.044 & 0.036 & 0.014 \\
 & 5000 & 1.21 & 0.44 & 0.030 & -- & 0.020 & 0.030 & 0.020 & 0.030 \\
F1b & 1000 & 0.90 & 0.31 & 0.048 & 0.024 & 0.030 & 0.074 & 0.102 & 0.172 \\
 & 2000 & 0.92 & 0.29 & 0.054 & 0.010 & 0.008 & 0.052 & 0.108 & 0.162 \\
 & 5000 & 0.92 & 0.26 & 0.053 & -- & 0.020 & 0.070 & 0.130 & 0.230 \\
F2, $\gamma_\chi=0.20$ & 1000 & 3.22 & 3.91 & 0.362 & 0.354 & 0.244 & 0.284 & 0.448 & 0.604 \\
 & 2000 & 3.52 & 4.22 & 0.336 & 0.342 & 0.254 & 0.300 & 0.508 & 0.656 \\
 & 5000 & 3.85 & 4.44 & 0.400 & -- & 0.240 & 0.320 & 0.550 & 0.720 \\
F2, $\gamma_\chi=0.30$ & 1000 & 0.88 & 0.29 & 0.092 & 0.026 & 0.022 & 0.088 & 0.094 & 0.166 \\
 & 2000 & 0.83 & 0.23 & 0.068 & 0.032 & 0.020 & 0.066 & 0.096 & 0.156 \\
 & 5000 & 0.73 & 0.16 & 0.063 & -- & 0.000 & 0.040 & 0.060 & 0.150 \\
\hline
\multicolumn{10}{@{}l}{\emph{Power}, $\tau=1+0.5x_2+1.5(x_1-\frac12)$} \\
F1 & 1000 &  &  & 0.414 & 0.366 & 0.516 & 0.492 & 0.648 & 0.954 \\
 & 2000 &  &  & 0.748 & 0.706 & 0.808 & 0.746 & 0.904 & 1.000 \\
F2, $\gamma_\chi=0.25$ & 1000 &  &  & 0.910 & 0.910 & 0.818 & 0.782 & 0.922 & 0.990 \\
 & 2000 &  &  & 0.992 & 0.986 & 0.942 & 0.910 & 0.988 & 1.000 \\
\hline\hline
\end{tabular*}
\end{center}
\footnotesize
\renewcommand{\baselineskip}{11pt}
\textbf{Note:} Rejection frequencies of $H_0^{1}$ ($\tau$ free of $x_1$) at the $5\%$ level. ``Our test'' is the integrated absolute pairwise difference test in the corrected calibration. Replications: $500$ at $N=1000$, $2000$; at $N=5000$, $300$ (our proposed test), $100$ (PCM-1, WGCM, GCM), $50$ (PCM-DR); LMP not run at $N=5000$. Monte Carlo standard errors at a rate of $0.05$: $0.010$, $0.013$, $0.022$, $0.031$ for $500$, $300$, $100$, $50$ replications. Columns 3 and 4: medians of the drift diagnostics in Section~\ref{osec:mc-fs-design}. 
\end{table}

\begin{table}[h]
\caption{\label{otab:fs-drift}First-stage drift and the null distribution of the statistics.}
\begin{center}
\setlength{\tabcolsep}{0.5pt}
\fontsize{7}{8.5}\selectfont
\begin{tabular*}{\textwidth}{@{\extracolsep{\fill}}llcccccccccc@{}}
\hline\hline
 & & & & & & & \multicolumn{5}{c}{Mean (sd) of the statistic} \\ \cline{8-12}
Design & $N$ & $\sqrt N\|\bar\pi_N^\perp\|$ & $N\eta^{1/2}\|\bar\pi_N^\perp\|^2$ & $\sqrt N\langle x_1-\frac12,\bar\pi_N^\perp\rangle$ & $\mathrm{var}(\widehat\zeta)$ & $\widehat r_1$ range & Our test & PCM-DR & WGCM & GCM & LMP \\
\hline
F1 & 1000 & 4.32 & 7.02 & [-3.6, 2.9] & 12.9 & [0.03, 0.98] & -0.07 (0.86) & 0.01 (0.69) & -0.08 (0.99) & 0.03 (0.78) & -0.08 (0.71) \\
 & 2000 & 4.47 & 6.81 & [-3.5, 2.8] & 12.9 & [0.03, 0.97] & -0.13 (0.95) & -0.01 (0.71) & -0.14 (0.98) & -0.01 (0.83) & -0.08 (0.72) \\
 & 5000 & 4.38 & 5.74 & [-3.0, 2.7] & 13.0 & [0.03, 0.96] & -0.13 (1.05) & 0.10 (0.74) & -0.06 (0.91) & 0.06 (0.85) & -- \\
F1b & 1000 & 1.97 & 1.47 & [-1.7, 1.8] & 4.8 & [0.11, 0.97] & 0.10 (0.96) & 0.13 (0.78) & 0.22 (1.14) & 0.06 (1.46) & 0.07 (0.66) \\
 & 2000 & 2.02 & 1.39 & [-1.7, 1.8] & 4.8 & [0.12, 0.97] & 0.08 (1.00) & 0.03 (0.68) & 0.22 (1.18) & 0.01 (1.42) & 0.01 (0.62) \\
 & 5000 & 2.03 & 1.24 & [-1.6, 1.8] & 4.8 & [0.15, 0.96] & 0.10 (0.99) & 0.12 (0.81) & 0.18 (1.34) & 0.13 (1.54) & -- \\
F1c & 1000 & 0.12 & 0.01 & [-0.1, 0.2] & 4.7 & [0.03, 0.98] & 0.10 (0.96) & -0.04 (0.59) & -0.01 (1.00) & -0.03 (0.95) & 0.01 (0.64) \\
 & 2000 & 0.09 & 0.00 & [-0.1, 0.1] & 4.7 & [0.03, 0.97] & 0.10 (0.97) & -0.01 (0.58) & 0.02 (1.01) & -0.10 (1.00) & 0.06 (0.71) \\
 & 5000 & 0.05 & 0.00 & [-0.1, 0.1] & 4.7 & [0.03, 0.96] & -0.02 (1.01) & -0.04 (0.51) & -0.15 (1.02) & -0.08 (1.03) & -- \\
F2, $\gamma_\chi=0.15$ & 1000 & 30.55 & 351.57 & [3.5, 79.1] & 34.6 & [0.05, 0.95] & 2.90 (2.66) & 1.60 (1.24) & 2.91 (2.30) & 3.99 (2.46) & 3.31 (3.17) \\
 & 2000 & 32.66 & 363.96 & [4.6, 91.1] & 25.5 & [0.05, 0.95] & 3.63 (3.47) & 1.92 (1.37) & 3.67 (2.61) & 4.80 (2.96) & 4.26 (4.29) \\
 & 5000 & 33.88 & 343.49 & [4.8, 104.9] & 18.4 & [0.05, 0.95] & 5.17 (4.78) & 2.37 (1.54) & 5.19 (3.25) & 6.43 (3.82) & -- \\
F2, $\gamma_\chi=0.20$ & 1000 & 12.56 & 59.43 & [1.7, 35.4] & 15.2 & [0.05, 0.95] & 1.28 (1.49) & 0.89 (0.98) & 1.49 (1.76) & 2.49 (1.77) & 1.34 (1.66) \\
 & 2000 & 11.84 & 47.85 & [2.1, 34.2] & 11.3 & [0.05, 0.95] & 1.29 (1.70) & 0.98 (1.03) & 1.80 (1.82) & 2.75 (2.00) & 1.43 (1.91) \\
 & 5000 & 11.23 & 37.75 & [1.8, 34.7] & 8.5 & [0.05, 0.95] & 1.53 (1.89) & 1.07 (0.97) & 2.22 (2.07) & 3.34 (2.22) & -- \\
F2, $\gamma_\chi=0.25$ & 1000 & 5.06 & 9.65 & [0.8, 14.2] & 8.5 & [0.05, 0.95] & 0.54 (1.05) & 0.42 (0.82) & 0.64 (1.37) & 1.50 (1.36) & 0.46 (0.98) \\
 & 2000 & 4.58 & 7.16 & [0.9, 12.5] & 7.0 & [0.05, 0.95] & 0.41 (1.06) & 0.40 (0.80) & 0.67 (1.34) & 1.46 (1.42) & 0.42 (0.97) \\
 & 5000 & 4.13 & 5.12 & [0.7, 11.3] & 5.9 & [0.05, 0.95] & 0.35 (1.04) & 0.42 (0.73) & 0.74 (1.30) & 1.64 (1.40) & -- \\
F2, $\gamma_\chi=0.30$ & 1000 & 2.18 & 1.79 & [0.4, 5.7] & 6.1 & [0.05, 0.95] & 0.25 (0.97) & 0.16 (0.71) & 0.22 (1.11) & 0.86 (1.14) & 0.13 (0.75) \\
 & 2000 & 1.93 & 1.27 & [0.4, 4.8] & 5.5 & [0.05, 0.95] & 0.14 (0.98) & 0.12 (0.69) & 0.20 (1.07) & 0.72 (1.13) & 0.11 (0.75) \\
 & 5000 & 1.63 & 0.80 & [0.3, 4.0] & 5.1 & [0.05, 0.95] & 0.07 (0.99) & 0.15 (0.61) & 0.23 (0.97) & 0.77 (1.11) & -- \\
F2, $\gamma_\chi=0.35$ & 1000 & 1.00 & 0.38 & [0.2, 2.4] & 5.3 & [0.05, 0.95] & 0.16 (0.98) & 0.04 (0.65) & 0.06 (1.04) & 0.48 (1.04) & 0.04 (0.70) \\
 & 2000 & 0.85 & 0.25 & [0.2, 2.0] & 5.0 & [0.05, 0.95] & 0.07 (0.98) & 0.00 (0.63) & 0.01 (1.02) & 0.31 (1.02) & 0.04 (0.71) \\
 & 5000 & 0.67 & 0.13 & [0.1, 1.6] & 4.7 & [0.05, 0.95] & 0.02 (0.99) & 0.05 (0.52) & 0.14 (0.89) & 0.36 (1.04) & -- \\
\hline\hline
\end{tabular*}
\end{center}
\footnotesize
\renewcommand{\baselineskip}{11pt}
\textbf{Note:} Columns 3, 4 and 6: medians over replications; $\bar\pi_N^\perp$ as in Section~\ref{osec:mc-fs-pred}, $\|\cdot\|$ the $L_2([0,1]^2)$ norm, $\eta=h_1h_2=(3.5\cdot12^{-1/2}N^{-1/7})^2$. Column 5: $10\%$ and $90\%$ quantiles of $\sqrt N\langle x_1-\frac12,\bar\pi_N^\perp\rangle/\mathrm{sd}(X_1)$, the drift in the direction seen by the GCM (before division by the standard deviation of $\widehat\zeta$). Column 7: range of the fitted (F1, F1b, F1c) or perturbed and trimmed (F2) propensity over all replications. Statistics: $\Sh_1$ of our proposed test and the studentized statistics of the benchmarks; under the null without first-stage effect, each is approximately $N(0,1)$, except the derandomized PCM (average of six correlated $N(0,1)$ statistics, standard deviation below one) and LMP (standard deviation below one in finite samples).
\end{table}

\subsubsection{Deviations}\label{osec:mc-fs-dev}\quad\\

(i) \textit{Calibration and pilot.} The strengths of the misspecifications ($1.5$ for the trigonometric term of F1, $1.5$ and intercept $1$ for the cosine term of F1b) and the scale $1$ of the F2 perturbation were chosen once, before the pilot, with an oracle proxy of the GCM statistic, among $\{0,0.75,1.5,2.5\}$, $\{0,1,1.5,2.5\}\times\{-1,0,1\}$ and $\{0.5,1\}$. A pilot of $100$ replications at $N=1000$, on replication numbers disjoint from those of the full run, confirmed the mechanism. After that, we changed only the slope of the power alternative once, from $1.0$ to $1.5$, because the power of our proposed test in F1 was $0.13$. We did not run a second scale: a scale of $0.5$ is equivalent to shifting $\gamma_\chi$ by $\log2/\log N$, from $0.08$ to $0.10$, which the $\gamma_\chi$ grid covers.\\

(ii) \textit{Replication counts.} The replication counts are given in Section~\ref{osec:mc-fs-design}. We did not run the LMP test at $N=5000$ due to the compute budget. We did not compute power at $N=5000$, as it is already high at $N=2000$ in most cells.

\subsection{\texorpdfstring{Covariate blocks with several covariates}{Covariate blocks with several covariates}}\label{osec:mc-blocks}

This subsection reports the size and power of the tests of $H_0^{1}$ when the blocks hold several covariates, $X_1\in[0,1]^{d_1}$ and $X_2\in[0,1]^{d_2}$ with $(d_1,d_2)=(1,2)$ and $(2,2)$, for our proposed test, the LMP test, and the PCM.

\paragraph{\texorpdfstring{Designs.}{Designs.}} All coordinates are independent $U[0,1]$, the propensity $r_1=1/2$ is known, and every method uses the IPW pseudo-outcome of the base designs. With the block indices $c_j:=d_j^{-1/2}\sum_{k=1}^{d_j}(x_{jk}-\frac12)$,
$$Y=1+\tfrac12\big(\tfrac12+c_1\big)+\tfrac12\big(\tfrac12+c_2\big)+T\,\tau(X_1,X_2)+\epsilon,\qquad \epsilon\sim N(0,1),$$
design A0 has $\tau=1$ (size) and design A1 has $\tau=1+0.8\,d_1^{-1/2}\sum_{k=1}^{d_1}\sin(2\pi x_{1k})$ (power). The factors $d_j^{-1/2}$ keep the first two moments of the baseline and of $\tau$ at their values in the base designs, which are the case $d_1=d_2=1$, so that the strength of the signal does not change with the dimension. We run each configuration at $N=2000$ with $1{,}000$ replications per design; each replication has its own seed, and all methods are applied to the same data sets.

\paragraph{\texorpdfstring{Tests.}{Tests.}} Our test is computed as in Algorithm~\ref{alg:steps}, in the corrected calibration of Section~\ref{osec:mc-paper}, with products over all coordinates of the kernels $K(u)=2(1-4u^2)P(4u^2)\ind\{|u|\le\frac12\}$ of order $s$, where $P$ is the polynomial of degree $s/2-1$ fixed by the moment conditions; \eqref{oeq:mc-kernel} is the case $s=4$. The bandwidths are $h_{jk}=k\,\widehat{\mathrm{sd}}_{jk}N^{-\gamma}$. At $(1,2)$, $s=6$, $\gamma=1/10$ and $k=6.5$, the choice of Section~\ref{sec:impl} for $d_1+d_2=3$, which satisfies Assumption~\ref{ass:reg}(g),(h). At $(2,2)$, $s=4$, $\gamma=1/13$, and $k=5.6$; this configuration lies beyond Assumption~\ref{ass:reg}(g),(h) and shows how the test performs outside the theory's conditions. The constants $k$ give bandwidths of about $0.88$ and $0.90$ and a mean own weight $L_{ii}$ of $0.063$ and $0.071$, close to its value $0.061$ in the base designs at $N=1000$ (Table~\ref{otab:mc2-scale}). The grid has $7$ points per coordinate on $[0.1,0.9]$, so $7^3=343$ and $7^4=2401$ grid points, with spacing $0.15h_{jk}$, within the recommended $h/6$, and the finite-$N$ variance is computed from $2000$ Gaussian draws, as for $M\ge21$ in Table~\ref{otab:mc2-sens}. The LMP test uses the settings of Section~\ref{osec:mc-bench} for a conditioning variable of dimension $p=d_2=2$, namely $g=n^{-1/6}$ and $h=2n^{-2.1/6}$, those of \citet[Section~4]{lavergne2015significance}. The PCM is computed on a single split with the generalized additive model learners of Section~\ref{osec:mc-bench}; the single split keeps the cost of the study down, since the derandomized PCM fits six times as many models, and in the base designs its size is closer to nominal than that of the derandomized version. All tests are one-sided, and we test only $H_0^{1}$, so we apply no multiplicity adjustment. The Monte Carlo standard error of a rejection frequency of $0.05$ is $0.007$.

\begin{table}[!htbp]
\caption{\label{otab:mc-blocks}Size and power for $H_0^{1}$ with covariate blocks of dimensions $(d_1,d_2)$, $N=2000$.}
\begin{center}
\small
\setlength{\tabcolsep}{3pt}
\begin{tabular*}{\textwidth}{@{\extracolsep{\fill}}llcccccccc@{}}
\hline\hline
 & & \multicolumn{5}{c}{A0 (size)} & \multicolumn{2}{c}{A1 (power), $5\%$} & \\ \cline{3-7}\cline{8-9}
$(d_1,d_2)$ & Method & $1\%$ & $5\%$ & $10\%$ & Mean & SD & Nominal & Size-adj. & Disc. \\
\hline
$(1,2)$ & Our test & 0.009 & 0.060 & 0.107 & 0.036 & 0.975 & 0.858 & 0.850 & 1 / 4 \\
 & LMP & 0.002 & 0.018 & 0.041 & 0.015 & 0.700 & 0.218 & 0.374 & 0 / 0 \\
 & PCM, single split & 0.012 & 0.046 & 0.092 & $-0.029$ & 1.002 & 0.970 & 0.971 & 0 / 0 \\
 & Our test, eval.-point centering & 0.008 & 0.044 & 0.095 & 0.004 & 0.942 & 0.850 & 0.861 & 1 / 4 \\
\hline
$(2,2)$ & Our test & 0.011 & 0.075 & 0.126 & 0.093 & 1.016 & 0.872 & 0.846 & 2 / 1 \\
 & LMP & 0.001 & 0.016 & 0.057 & $-0.004$ & 0.775 & 0.208 & 0.317 & 0 / 0 \\
 & PCM, single split & 0.008 & 0.049 & 0.106 & 0.059 & 0.999 & 0.937 & 0.938 & 0 / 0 \\
 & Our test, eval.-point centering & 0.010 & 0.058 & 0.115 & 0.056 & 0.984 & 0.868 & 0.857 & 2 / 1 \\
\hline\hline
\end{tabular*}
\end{center}
\footnotesize
\renewcommand{\baselineskip}{11pt}
\textbf{Note:} $1{,}000$ replications per design; all methods on the same data sets. Columns $1\%$--$10\%$: rejection frequencies in A0 at the nominal level; Mean, SD: moments of the studentized statistic in A0. Nominal: rejection frequency in A1 at the $5\%$ critical value of $N(0,1)$; Size-adj.: at the $95\%$ quantile of the statistic in A0. Disc.: replications discarded by our test in A0 / A1 because a Nadaraya--Watson denominator was not positive. Our test: corrected calibration; $(1,2)$: kernel of order $6$, $h_{jk}=6.5\,\widehat{\mathrm{sd}}_{jk}N^{-1/10}$, within Assumption~\ref{ass:reg}(g),(h); $(2,2)$: kernel of order $4$, $h_{jk}=5.6\,\widehat{\mathrm{sd}}_{jk}N^{-1/13}$, beyond Assumption~\ref{ass:reg}(g),(h); $7$ grid points per coordinate on $[0.1,0.9]$. Eval.-point centering: residuals $\zeta_i-\tauh(c)$ at the evaluation point in the centering, same variance. LMP: asymptotic critical value. Monte Carlo standard error at a rate of $0.05$: $0.007$.
\end{table}

\paragraph{\texorpdfstring{Results.}{Results.}} Within the conditions of the theory, at $(1,2)$, our test holds its level at all three nominal levels, and its statistic is close to standard normal (Table~\ref{otab:mc-blocks}). At $(2,2)$, beyond these conditions, the statistic retains a standard-normal spread, but its center shifts up by about a tenth of a standard deviation, which puts the test somewhat above its level at $5\%$ and $10\%$ while it holds the $1\%$ level. The single-split PCM is at its level in both configurations, and the LMP test is conservative, with its statistic's spread well below one, as in the base designs.

Our test detects A1 in most replications in both configurations, and its power hardly changes from $(1,2)$ to $(2,2)$, although at $(2,2)$ the signal is spread over two coordinates. This power is not a by-product of the size at $(2,2)$: at the critical values that give exact size in A0 it is the same in both configurations. The LMP test loses most of its power against A1 in the base designs (Table~\ref{otab:mc2-2000}); size adjustment recovers only part of the gap, so smoothing over the two-dimensional conditioning block costs power beyond its conservative calibration. The PCM has the highest power. Design A1 is additive across the coordinates of $X_1$, a structure that the additive learners of the PCM represent exactly; the same learners cannot represent an interaction between the blocks, against which the PCM has no power in the base designs (design A4, Tables~\ref{otab:mc2-1000} and~\ref{otab:mc2-2000}). The ranking of the three tests in A1 is that of the base designs. Our test is also inexpensive here: it took $4$ to $7$ seconds per replication, about as long as or less than the LMP test ($7$ to $11$) and less than a third of the time of the single-split PCM ($15$ to $33$).

Computing the residuals in the centering at the evaluation point, $\zeta_i-\tauh(c)$ with a leave-one-out rescaling analogous to $q_i$, instead of at the observations, brings the size at $(2,2)$ close to nominal at the same power (rows ``eval.-point centering'' of Table~\ref{otab:mc-blocks}). This centering also removes the term that Assumption~\ref{ass:reg}(h) controls, and a theory for higher-dimensional covariate blocks built on it is left for future research.

\subsection{Coverage of the magnitude intervals}\label{osec:mc-mag}
This subsection supplements Proposition~\ref{prop:gini-clt} and the interval \eqref{eq:gini-interval} in the main text, whose coverage theory appears in Section~\ref{osec:gini-if}; the statistics are computed using the implementation of the joint test used throughout this section.

The interval \eqref{eq:gini-interval} for $D_j$ is designed to cover in every case of Proposition~\ref{prop:gini-clt}: under the null (b), when $\tau$ varies everywhere (c), and when it is flat on part of the domain (d). Its coverage is assessed in the base designs with a known propensity, A0, A1, A4 and A5, and in a design A6 in which $\tau=1+c_6\,b\{(x_1-0.75)/0.25\}$, with $b(u)=\exp\{1-1/(1-u^2)\}$ for $|u|<1$ and $0$ otherwise, a smooth bump that is flat for $x_1\le0.5$, so that a quarter of the configurations in direction $1$ form the flat set $F_1$ of part (d); $c_6=1.597$ gives the same $D_1$ ($0.178$) as A1, A4 and A5. The target is the population $D_j$ on $[0.1,0.9]^2$; the statistic, the centering and $\widetilde\sigma_{j,N}$ are those of Section~\ref{sec:mc-tests}, $\widehat\Omega_{j,N}$ is formed with $\widehat\omega_j$ computed from $\tauh$ on the grid, and $z=1.96$. Table~\ref{otab:mc-mag} reports the coverage of \eqref{eq:gini-interval} and, for comparison, that of the two intervals it combines: $\widetilde D_j\pm z\widetilde\sigma_{j,N}/\sqrt N$, which part (b) justifies under the null, and $\Dh_j/\sqrt N\pm z\widehat\Omega_{j,N}^{1/2}/\sqrt N$, which part (c) justifies when $\tau$ varies everywhere.

\begin{table}[!htbp]
\caption{\label{otab:mc-mag}Coverage of the magnitude intervals, nominal level $95\%$.}
\begin{center}
\small\setlength{\tabcolsep}{3pt}
\begin{tabular*}{\textwidth}{@{\extracolsep{\fill}}lcccccccccc@{}}
\hline\hline
 & & & & \multicolumn{4}{c}{Coverage} & \multicolumn{1}{c}{Width} & \multicolumn{2}{c}{Single-case, $N=5000$} \\
\cline{5-8}\cline{10-11}
Design & $j$ & Target & Case & $500$ & $1000$ & $2000$ & $5000$ & $500$ / $5000$ & (b) & (c) \\
\hline
\multicolumn{11}{@{}l}{\textit{Panel A: $D_j$, interval \eqref{eq:gini-interval}}} \\
A0 & 1 & 0 & (b) & 0.992 & 0.991 & 0.996 & 0.995 & 0.50 / 0.20 & 0.941 & 0.000 \\
A1 & 1 & 0.178 & (c) & 1.000 & 1.000 & 1.000 & 1.000 & 0.51 / 0.21 & 0.000 & 0.458 \\
A1 & 2 & 0 & (b) & 0.994 & 0.995 & 0.997 & 0.998 & 0.50 / 0.20 & 0.952 & 0.000 \\
A4 & 1 & 0.178 & (c) & 1.000 & 1.000 & 1.000 & 1.000 & 0.52 / 0.21 & 0.015 & 0.006 \\
A5 & 1 & 0.178 & (c) & 1.000 & 1.000 & 1.000 & 1.000 & 0.49 / 0.20 & 0.008 & 0.034 \\
A5 & 2 & 0 & (b) & 0.988 & 0.996 & 0.995 & 0.994 & 0.47 / 0.19 & 0.954 & 0.000 \\
A6 & 1 & 0.178 & (d) & 1.000 & 1.000 & 1.000 & 1.000 & 0.58 / 0.24 & 0.002 & 0.281 \\
A6 & 2 & 0 & (b) & 0.989 & 0.995 & 0.997 & 0.995 & 0.56 / 0.22 & 0.952 & 0.000 \\
\hline
\multicolumn{11}{@{}l}{\textit{Panel B: $\Delta_j$}} \\
A0 & 1 & 0 & (b) & 1.000 & 1.000 & 1.000 & 1.000 & 2.18 / 0.84 &  &  \\
A1 & 1 & 0.695 & (c) & 1.000 & 1.000 & 1.000 & 1.000 & 2.21 / 0.87 &  &  \\
A1 & 2 & 0 & (b) & 0.999 & 1.000 & 0.998 & 1.000 & 2.19 / 0.85 &  &  \\
A4 & 1 & 0.695 & (c) & 1.000 & 1.000 & 1.000 & 1.000 & 2.22 / 0.86 &  &  \\
A5 & 1 & 0.695 & (c) & 1.000 & 1.000 & 1.000 & 1.000 & 2.08 / 0.81 &  &  \\
A5 & 2 & 0 & (b) & 0.999 & 1.000 & 0.999 & 1.000 & 2.06 / 0.80 &  &  \\
A6 & 1 & 0.695 & (d) & 1.000 & 1.000 & 1.000 & 1.000 & 2.50 / 0.98 &  &  \\
A6 & 2 & 0 & (b) & 1.000 & 1.000 & 0.999 & 0.999 & 2.44 / 0.94 &  &  \\
\hline
\multicolumn{11}{@{}l}{\textit{Panel C: $G_j$}} \\
A0 & 1 & 0 & (b) & 1.000 & 1.000 & 1.000 & 1.000 & 0.88 / 0.42 &  &  \\
A1 & 1 & 0.348 & (c) & 1.000 & 1.000 & 1.000 & 1.000 & 0.85 / 0.41 &  &  \\
A1 & 2 & 0 & (b) & 0.999 & 1.000 & 0.998 & 1.000 & 0.84 / 0.40 &  &  \\
A4 & 1 & 0.329 & (c) & 1.000 & 1.000 & 1.000 & 1.000 & 0.80 / 0.38 &  &  \\
A5 & 1 & 0.548 & (c) & 0.999 & 1.000 & 1.000 & 0.990 & 0.88 / 0.45 &  &  \\
A5 & 2 & 0 & (b) & 0.999 & 1.000 & 0.999 & 1.000 & 0.87 / 0.45 &  &  \\
A6 & 1 & 0.223 & (d) & 1.000 & 1.000 & 1.000 & 1.000 & 0.74 / 0.32 &  &  \\
A6 & 2 & 0 & (b) & 1.000 & 1.000 & 0.999 & 0.999 & 0.72 / 0.31 &  &  \\
\hline\hline
\end{tabular*}
\end{center}
\footnotesize
\renewcommand{\baselineskip}{11pt}
\textbf{Note:} $1000$ replications (at most five discarded where a kernel denominator vanished); known propensity, IPW pseudo-outcome, $k_j=3.5$, $13\times13$ grid. Target: population value on $[0.1,0.9]^2$. Case: part of Proposition~\ref{prop:gini-clt} that applies. Width: mean width at $N=500$ and $5000$. Panel~A, single-case intervals: (b) $\widetilde D_j\pm1.96\,\widetilde\sigma_{j,N}/\sqrt N$; (c) $\Dh_j/\sqrt N\pm1.96\,\widehat\Omega_{j,N}^{1/2}/\sqrt N$. Panel~B: $[\widetilde\Delta_j-1.96\,\widehat{\mathrm{se}}^\Delta_j,\widehat\Delta_j+1.96\,\widehat{\mathrm{se}}^\Delta_j]$, with $\widehat\Delta_j$ the density-weighted magnitude and $\widetilde\Delta_j$ its noise-corrected version (Section~\ref{osec:mc-mag}), $\widehat{\mathrm{se}}^\Delta_j$ the larger of the null standard deviation of $\widetilde\Delta_j$ and the influence-function standard error of Section~\ref{osec:gini-if}. Panel~C: the interval of Panel~B divided by twice the estimated $\E|\tau(X^{\mathcal R})|$. A4 in direction $2$ and A0 in direction $2$ are as in direction $1$.
\end{table}

The interval covers $D_j$ in every design, direction, and sample size, including the partially flat A6 (Panel~A of Table~\ref{otab:mc-mag}), and it is conservative, as Proposition~\ref{prop:gini-clt} allows. Its width is close to the noise-floor rate $(N\eta)^{-1/2}$, and it typically excludes a zero magnitude for $N=1000$ in A4 and A5, $N=2000$ in A1, and $N=5000$ in A6. Neither single-case interval would do (last two columns of the table). The interval of part (b) covers under the null but not when $D_j>0$, because the centering removes the full noise floor where $\tau$ varies, so that $\widetilde D_j$ understates $D_j$. The interval of part (c) undercovers because the noise floor is still of the order of $D_j$, so that $\Dh_j/\sqrt N$ overstates it; its coverage rises with $N$ only in A1 and A6. In every replication $\widehat\Omega_{j,N}^{1/2}$ exceeded $\widetilde\sigma_{j,N}$, which explains the coverage above $0.95$ under the null. The bracket $[\widetilde D_j,\Dh_j/\sqrt N]$ alone almost always contains $D_j$ when $D_j>0$; the extra margin $z\,\widehat{\mathrm{se}}_j/\sqrt N$ covers $D_j=0$.

The density-weighted magnitudes are computed as in the application: $\widehat\Delta_j$ weights the pairs of grid points by the kernel density estimate of $X$, $\widetilde\Delta_j$ subtracts the correspondingly weighted noise floor, and the interval for $\Delta_j$ is $[\widetilde\Delta_j-z\,\widehat{\mathrm{se}}^\Delta_j,\widehat\Delta_j+z\,\widehat{\mathrm{se}}^\Delta_j]$, with $\widehat{\mathrm{se}}^\Delta_j$ the larger of the null standard deviation of $\widetilde\Delta_j$ and the influence-function standard error of Section~\ref{osec:gini-if}. The interval for $G_j$ divides both ends by twice $\sum_g\widehat\pi_g|\tauh(x_g)|$, a sum over the grid points $x_g$ with normalized kernel-density weights $\widehat\pi_g$, the density-weighted estimate of $\E|\tau(X^{\mathcal R})|$ (Panels~B and~C of Table~\ref{otab:mc-mag}). The intervals for $\Delta_j$ and $G_j$ also cover in nearly every replication (Panels~B and~C of Table~\ref{otab:mc-mag}). The interval for $\Delta_j$ is wide because the influence-function standard error also carries the estimated density. The one visible shortfall for $G_j$ is in A5 at $N=5000$, where every miss has the upper end below $G_1$: where $\tau$ is close to zero on part of the box, $|\tauh|$ has a noise floor of its own, so that the denominator overstates $\E|\tau(X^{\mathcal R})|$ and pulls the interval down. The interval for $G_j$ treats the denominator as known and inherits this bias; in A4 and A5, where $\tau$ changes sign on the box, $G_j$ is moreover not a Gini coefficient.

\section{Additional details on the empirical application}\label{osec:emp}

This section supplements Section~\ref{sec:app-emp}
and Table~\ref{tab:empirical} of the main text.
In Section~\ref{osec:emp-data}, we describe the data and
examine the tails of the outcome distribution.
In Section~\ref{osec:emp-id}, we discuss the exogeneity
of eligibility. In Sections~\ref{osec:emp-first} and
\ref{osec:emp-second}, we report the first-stage estimates
and diagnostics and the second-stage results, respectively.
In Section~\ref{osec:emp-jitter}, we describe the jittering
of age and the jitter-averaged statistic.
In Section~\ref{osec:emp-rob}, we report robustness checks.
Finally, in Section~\ref{osec:emp-mag}, we report the
magnitudes of treatment effect heterogeneity.
All $p$-values are asymptotic and one-sided, with Holm
adjustment across the two directions.

\subsection{Data and sample}\label{osec:emp-data}

The data are the 1991 Survey of Income and Program Participation (SIPP) extract constructed by \citet{chernozhukov2004effects} and distributed with \citet{chernozhukov2018double}: $N=9{,}915$ households with a reference person aged 25 to 64, of which $3{,}682$ are eligible for a 401(k) plan, and $6{,}233$ are not. The treatment $T_i$ is eligibility, the outcome $Y_i$ is net financial assets in US dollars, the modifiers $X_1$ is age in years and $X_2$ is family income in US dollars, and the confounder block $V$ collects education, family size, marital status, two-earner status, a defined-benefit pension indicator, IRA participation, and home ownership. Eligible households hold more net financial assets and earn more than ineligible households, and are more often married, two-earner, covered by a defined-benefit pension, and homeowners (Table~\ref{otab:emp-data}); $70.5\%$ participate in the plan. Age is recorded in integer years, with $40$ distinct values of which $30$ lie inside the evaluation region; five households have nonpositive income.

\begin{table}[H]
\caption{\label{otab:emp-data}Summary statistics of the 401(k) sample.}
\begin{center}
\small\setlength{\tabcolsep}{4pt}
\begin{tabular*}{\textwidth}{@{\extracolsep{\fill}}lcccc@{}}
\hline\hline
 & \multicolumn{2}{c}{All households} & Eligible & Ineligible \\
\cline{2-3}
Variable & Mean & SD & Mean & Mean \\
\hline
Net financial assets (USD) & 18{,}052 & 63{,}519 & 30{,}347 & 10{,}788 \\
Family income (USD) & 37{,}201 & 24{,}773 & 46{,}862 & 31{,}494 \\
Age (years) & 41.1 & 10.3 & 41.5 & 40.8 \\
Education (years) & 13.2 & 2.8 & 13.8 & 12.9 \\
Family size & 2.9 & 1.5 & 2.9 & 2.8 \\
Married & 0.60 & 0.49 & 0.67 & 0.56 \\
Two earners & 0.38 & 0.49 & 0.48 & 0.32 \\
Defined-benefit pension & 0.27 & 0.44 & 0.42 & 0.19 \\
IRA participation & 0.24 & 0.43 & 0.32 & 0.20 \\
Home owner & 0.64 & 0.48 & 0.74 & 0.57 \\
Participates in a 401(k) & 0.26 & 0.44 & 0.70 & 0.00 \\
Households & 9{,}915 &  & 3{,}682 & 6{,}233 \\
\hline
\multicolumn{5}{@{}l}{\textit{Tails of net financial assets}} \\
Skewness & 10.6 &  & 8.7 & 13.2 \\
Excess kurtosis & 183.8 &  & 129.1 & 262.8 \\
Minimum (USD) & $-$502{,}302 &  & $-$502{,}302 & $-$409{,}000 \\
Maximum (USD) & 1{,}536{,}798 &  & 1{,}536{,}798 & 1{,}324{,}445 \\
\hline\hline
\end{tabular*}
\end{center}
\footnotesize
\renewcommand{\baselineskip}{11pt}
\end{table}

The outcome is heavy-tailed (Table~\ref{otab:emp-data}), and the DR pseudo-outcome has standard deviation $110{,}774$ and excess kurtosis $194.7$. Winsorizing the outcome at its 1st and 99th percentiles ($-23{,}453$ and $219{,}922$ USD; $200$ values changed) reduces the skewness to $3.16$ and the excess kurtosis to $11.5$; at the 0.5th and 99.5th percentiles ($-36{,}462$ and $328{,}548$ USD; $100$ values changed), to $4.01$ and $20.8$.

\begin{remark}[Heavy tails]\label{orem:heavytails}
Bounded conditional moments of order six, or of any order above five (Remark~\ref{orem:moments}), affect the asymptotic theory only through the truncation in the uniform consistency of $\widetilde\kappa_j$ and through the Rosenthal and Berry--Esseen bounds; extreme kurtosis is allowed asymptotically but enlarges the constants, notably the Jensen term $J_N$ of Lemma~\ref{olem:centering-bias} (Remark~\ref{orem:jensen}). In the Monte Carlo study, the corrected test stays below its level under $t_3$ errors, which violate the moment condition, and is mildly above it in one design with lognormal errors of kurtosis $114$ (Section~\ref{osec:mc-new}).

Winsorizing the outcome at fixed thresholds changes the estimand
to $\E[\mathrm{wins}(Y(1))-\mathrm{wins}(Y(0))\mid X]$.
Constancy of this estimand neither implies nor is implied by
$H_0^1$. We therefore treat winsorization as a sensitivity
analysis rather than a robustness check of the original
hypothesis, and refit the first stage using the winsorized outcome.
Winsorizing the pseudo-outcome has no such interpretation, since $\E[\mathrm{wins}(\zeta)\mid X]$ need not be constant in $x_1$ under $H_0^1$, and is not done.
\end{remark}

\subsection{The exogeneity of eligibility}\label{osec:emp-id}

Section~\ref{sec:app-emp} of the main text states the three caveats that qualify the estimand: it is an \textit{intention-to-treat effect} of eligibility, unconfoundedness given income and $V$ is contested, and $\tau$ measures effect modification rather than a causal interaction. \citet{poterba1996retirement} and \citet{engen1996illusory} set out the two sides of the exogeneity debate. The main objection of \citet{engen1996illusory} is that eligibility is positively correlated with unobserved tastes for saving, because employers respond to employees' preferences and strong savers seek out employers that offer 401(k) plans; another is that \citet{poterba1995do} omit pension coverage, with eligible families being $24$ to $33$ percentage points more likely to have a defined-benefit pension. Our $V$ includes defined-benefit pension status, which answers the second objection but not the first. \citet{poterba1994plans} first treat eligibility as a natural experiment given income, and \citet{gelber2011how} exploits changes in eligibility after tenure thresholds, finding that eligibility raises 401(k) balances while its effects on other financial assets are imprecise. The heterogeneity attribution tests in this paper are conditional on Assumption~\ref{ass:unconf} given age, income, and $V$; they do not address this debate, and we do not study the sensitivity of their conclusions to a latent confounder.

\subsection{First stage  specification and propensity diagnostics}\label{osec:emp-first}

\textit{Specification.} The CATE is estimated with the DR pseudo-outcome \eqref{eq:psi-dr}. The propensity is a logistic regression of $T$ on cubic polynomials in standardized age and income, their product, and $V$ entered linearly, fitted by maximum likelihood on the full sample. The outcome regressions $\mu_t$ are linear regressions of $Y$ on the same regressors within each arm. The fitted propensities lie in $[0.0350,0.9546]$, with 1st, 5th, 50th, 95th and 99th percentiles $0.065$, $0.098$, $0.354$, $0.733$ and $0.773$. No trimming is applied, and a floor of $0.02$ and a cap of $0.98$ would bind for no household, so that Assumption~\ref{ass:firststage}(a)(i) is not affected by a non-differentiable trim. Sixteen households lie outside $[0.05,0.95]$, fifteen below $0.05$ (one of them eligible) and one above $0.95$ (not eligible). The sample mean of $\widehat\zeta^{\mathrm{DR}}_i$ is $7{,}429$ USD with standard error $1{,}113$ ($1{,}185$ with the first-stage term); the IPW estimate is $5{,}222$ (standard error $3{,}222$), the DR estimate without IRA participation and home ownership in $V$ is $8{,}244$ ($1{,}165$), and the DR mean over the evaluation box, the sample mean of $\widehat\zeta^{\mathrm{DR}}_i$ over the households in the box of Section~\ref{osec:emp-second}, is $7{,}677$ USD averaged over the $50$ jitter draws of age, on which the box depends ($7{,}436$ to $7{,}988$ across draws; $7{,}694$, standard error $920$, in the first draw).\\

\textit{Specification tests and balance.} The \textit{cubic logit} is formally rejected. Against a nesting flexible series logit with $24$ more parameters (cubic regression splines in age and income with four interior knots, three tensor terms, $V$, and $V$ interacted with age and income; $39$ terms), the likelihood-ratio statistic is $86.4$ ($p=6\times10^{-9}$), and the \textit{Hosmer--Lemeshow} statistic over deciles is $37.3$ on $8$ degrees of freedom ($p=1\times10^{-5}$). The \textit{flexible logit} is itself rejected against a richer nesting basis that adds $V\times\mathrm{age}^2$, $V\times\mathrm{income}^2$, all pairwise products of $V$ and $\mathrm{age}^2\times\mathrm{income}^2$ (likelihood ratio $80.6$ on $36$ degrees of freedom, $p=3\times10^{-5}$), but passes the Hosmer--Lemeshow test ($14.1$ on $8$ degrees of freedom, $p=0.079$). Balance after weighting is good under both: the largest absolute inverse-probability-weighted standardized difference over $13$ covariate terms is $0.027$ for the cubic logit and $0.034$ for the flexible one, against $0.64$ before weighting. With $N=9{,}915$ these tests detect small departures.\\

\textit{Choice of the main first stage.} The rule was fixed before the second-stage results were computed: the flexible first stage would be the main one if its logit passed both the likelihood-ratio test against the richer basis and the Hosmer--Lemeshow test at $5\%$. It fails the former, so the cubic first stage remains the main specification and the flexible one, with the same series basis in the propensity and in both outcome regressions, is a robustness specification (Table~\ref{otab:emp-sens}, Panels C and D). Its fitted propensities lie in $[0.021,0.854]$, with $127$ households below $0.05$ (two of them eligible), and its average effects are $8{,}318$ (DR, standard error $1{,}145$) and $8{,}117$ (IPW, $1{,}407$). The second-stage results under the two first stages are practically identical, so the choice \textit{does not} seem to affect any conclusion. The implication should nevertheless be stated plainly: the premise of Theorem~\ref{thm:nuisance-param} that one working model is correctly specified is not literally tenable for either logit, and the linear outcome models are not credible as correct mean models for this outcome. The sample-split route of Theorem~\ref{thm:nuisance-split}, which needs neither (Panel~D of Table~\ref{otab:emp-sens2}), confirms the income result but has too little power to confirm the age result.

\subsection{Second stage}\label{osec:emp-second}

The evaluation region is the rectangle between the 10th and 90th percentiles of each modifier (jittered age $28.1$ to $56.7$, income $12{,}240$ to $69{,}607$ USD), which holds $63.9\%$ of the households, and each modifier is mapped affinely from its region to $[0,1]$ before the second stage. All bandwidths, densities, and Lebesgue measures are in these standardized units, so $\lambda(\Xone)=\lambda(\Xtwo)=1$; we report the CATE function and magnitudes in dollars. The second stage is the Nadaraya--Watson regression of Proposition~\ref{prop:transfer} with products of the fourth-order kernel \eqref{oeq:mc-kernel} and $h_j=k_j\widehat{\mathrm{sd}}_jN^{-1/7}$, $\widehat{\mathrm{sd}}_j$ is the standard deviation of the standardized modifier over the full sample. At $k_j=3.5$ this gives $h=(0.341,0.406)$, windows of $9.8$ years and $23{,}289$ USD with about nine to ten integer ages per window, $\eta=0.138$ and $N\eta^3=26.3$; at $k_j=5$, the upper end of the recommended range, the windows are $13.9$ years and $33{,}270$ USD. We evaluate the CATE function on a $31\times31$ grid, with spacing $h_1/10$ and $h_2/12$, within the spacing of at most $h_j/6$ recommended in Section~\ref{sec:impl} over the whole range $k_j\ge3$ (spacing at most $0.11h_j$ at $k_j=3$). The number of observations in a joint window ranges from $153$ to $1{,}711$ with median $753$. Among the $9{,}035$ households in some grid window, the own weight $L_{ii}$ has mean $0.0116$ and maximum $0.22$, so that $q_i\ge0.77$. The centering $\widetilde m_j$ uses the \textit{corrected} squared residuals $\widetilde e_i^{\,2}=\widehat e_i^{\,2}/q_i$ of Lemma~\ref{prop:centering-corr}, the scale $\widetilde\sigma_{j,N}$ is the finite-sample variance of Lemma~\ref{prop:sigmaN} computed exactly from the realized weights on the grid (Section~\ref{osec:grid-formulas}), $\Dh_j$, $\widetilde m_j$ and $\widetilde\sigma_{j,N}$ use the same equal grid weights, the inner integrals of $\Dh_j$ use the sorting identity \eqref{eq:gini-sort}, and $p_j=1-\Phi(\Sh_j)$, Holm-adjusted across the two directions. The density-weighted magnitudes weight the CATE of $X_j$ given $X_{j'}$ by the product-kernel density estimate on the grid, floored at zero.\\

\textit{Windows at the edge of the evaluation box.} The grid includes the edges of the box, and the window at a grid point extends half a window, $h_j/2$, to each side: $4.9$ years and $11{,}644$ USD at $k_j=3.5$, and from $4.2$ to $7.0$ years and from $9{,}981$ to $16{,}635$ USD over $k_j\in[3,5]$. Age is observed between $25$ and $64$ years (jittered age between $23.2$ and $66.0$ in the first draw; the jitter has support $[-3,3]$), and income between $-2{,}652$ and $242{,}124$ USD. At $k_j=3.5$ the age windows of the two lowest grid values of age reach down to $23.2$ years, $1.8$ years below the youngest household, so that up to $18\%$ of such a window lies below the observed ages; the jittered age has a positive but thin density, since only $136$ of the $9{,}915$ jittered ages fall below $25$. These are $2$ of the $31$ grid values of age and $62$ of the $961$ grid points; at the upper end of age and at both ends of income, the windows stay within the observed range (the lowest income window reaches $596$ USD, a region with few households: the 1st percentile of income is $4{,}080$ USD). With wider windows more grid points are affected: relative to the observed range of age and income, $2$, $3$, $4$ and $5$ of the $31$ age values at $k_j=3$, $4$, $4.5$ and $5$, and $1$ of the $31$ income values for $k_j\ge4.5$; at $k_j=5$ the age window reaches $21.1$ years, below the support of jittered age, and the income window $-4{,}395$ USD, and $181$ of the $961$ grid points ($19\%$) have a window that leaves the observed range, with up to $28\%$ of a window outside it. Assumption~\ref{ass:overlap}, which requires the density of the modifiers to be bounded away from zero on a neighborhood of the evaluation box, holds in the limit, as the windows shrink into such a neighborhood, but not at these bandwidths: the lowest age windows reach beyond the observed ages, into a region where the density of jittered age is close to zero (and, for $k_j\ge4.5$, beyond its support). At the affected grid points, the kernel is truncated or weighted towards one side, loses its vanishing moments, and the smoothing bias of $\tauh$ is of order $h_j$ rather than $h_j^{4}$. Fewer households lie in these windows, so $\tauh$ is noisier there; the corrected centering and variance reflect this through the realized weights, but the boundary bias is not accounted for. This is a limitation of the application: the fitted CATE function for the youngest households in the box (about 28 to 29 years) is subject to boundary bias.

\subsection{Jittered age and the jitter-averaged statistic}\label{osec:emp-jitter}

Age is integer-valued, so the continuous theory does not cover the raw-age statistic (Remark~\ref{orem:jitter-uniform}). Following Proposition~\ref{oprop:jitter}, the second stage uses $X_1^J:=X_1+J$, where $J=U_1+\dots+U_6$ with $U_k$ independent and uniform on $[-1/2,1/2]$ and independent of the data: a centered B-spline density of order six, with support $[-3,3]$ years and standard deviation $0.71$ years (realized standard deviation $0.709$, range $[-2.26,2.35]$ in the first draw). The first stage uses integer-valued age and is unaffected.

A single jitter draw adds randomness that is unrelated to the data: at $k_j=3.5$, over $20$ independent draws, $\Sh_1$ has mean $2.10$, median $2.19$, standard deviation $0.29$ and range $1.54$--$2.57$, and the Holm-adjusted $p$-value for age ranges from $0.005$ to $0.062$, with $18$ of $20$ draws rejecting at $5\%$; the income statistic ranges from $3.04$ to $4.05$ and every draw rejects. The first draw, used for the figure and for the single-draw rows of Table~\ref{otab:emp-sens2}, is on the favorable side ($\Sh_1=2.44$). The reported statistic is therefore the average over $n_J$ independent draws,
\[
\bar S_j:=\frac1{n_J}\sum_{b=1}^{n_J}\Sh_j^{(b)},\qquad \bar p_j:=1-\Phi(\bar S_j),
\]
with Holm's adjustment applied to $(\bar p_1,\bar p_2)$, $n_J=50$ at $k_j=3.5$ for the DR pseudo-outcome and $n_J=20$ otherwise; the draws are the same in all specifications, the first $20$ of the $50$ being those of the single-draw study. $\bar S_j$ approximates the expectation of the randomized statistic over the jitter given the data. Under $H_0^j$, each $\Sh_j^{(b)}$ is asymptotically standard normal by Proposition~\ref{oprop:jitter}, so the average has asymptotic mean zero and, by Minkowski's inequality, standard deviation at most one, with equality only if the draws are perfectly correlated; referring $\bar S_j$ to the standard normal table is then conservative, provided the vector of draw-wise statistics is jointly asymptotically normal, so that the average is asymptotically normal with variance at most one.  Alongside $\bar S_j$, the tables report the standard deviation of $\Sh_j^{(b)}$ across draws and the share of draws whose own Holm-adjusted $p$-value is at most $0.05$; the median-draw statistics are stored in the results file. The magnitudes and their standard errors are averaged over the same draws; an average of standard errors bounds the standard error of the average from above. Across the $50$ draws, the fitted CATE function changes marginally: the standard deviation of $\Dh_1/\sqrt N$ is $146$ USD, against a mean of $5{,}143$.

\subsection{Robustness}\label{osec:emp-rob}

Table~\ref{otab:emp-sens} reports the jitter-averaged statistics over the recommended bandwidth range for both pseudo-outcomes and both first stages; Table~\ref{otab:emp-sens2} reports single-draw checks of integer age, the grid, the outcome, the adjustment set, overlap, and the sample-split route.

\begin{table}
\caption{\label{otab:emp-sens}Robustness of the 401(k) application.}
\begin{center}
\setlength{\tabcolsep}{4pt}
\begin{tabular*}{\textwidth}{@{\extracolsep{\fill}}lccccccc@{}}
\hline\hline
& \multicolumn{3}{c}{$H_0^{1}$ (age)} & \multicolumn{2}{c}{$H_0^{2}$ (income)} & \multicolumn{2}{c}{$\widetilde D_j$ (USD)} \\
\cline{2-4}\cline{5-6}\cline{7-8}
Specification & $\bar S_1$ & $\bar p_1$ & Rej. & $\bar S_2$ & $\bar p_2$ & age & income \\
\hline
\multicolumn{8}{@{}l}{\textit{A. Jitter-averaged, DR, cubic first stage}} \\
$k_j=3$ & 2.10 & 0.018 & 0.90 & 3.45 & 0.001 & 1{,}078 & 1{,}819 \\
$k_j=3.5$ (main) & 2.15 & 0.016 & 0.96 & 3.68 & $<$0.001 & 1{,}045 & 1{,}835 \\
$k_j=4$ & 1.98 & 0.024 & 0.80 & 3.67 & $<$0.001 & 921 & 1{,}749 \\
$k_j=4.5$ & 1.85 & 0.032 & 0.65 & 3.78 & $<$0.001 & 828 & 1{,}729 \\
$k_j=5$ & 1.76 & 0.039 & 0.50 & 4.01 & $<$0.001 & 759 & 1{,}764 \\
\multicolumn{8}{@{}l}{\textit{B. Jitter-averaged, inverse probability weighting, cubic first stage}} \\
$k_j=3$ & 2.34 & 0.010 & 1.00 & 3.73 & $<$0.001 & 1{,}520 & 2{,}475 \\
$k_j=3.5$ & 2.38 & 0.009 & 1.00 & 3.92 & $<$0.001 & 1{,}464 & 2{,}468 \\
$k_j=4$ & 2.21 & 0.014 & 1.00 & 3.91 & $<$0.001 & 1{,}298 & 2{,}349 \\
$k_j=4.5$ & 2.00 & 0.023 & 0.95 & 3.91 & $<$0.001 & 1{,}129 & 2{,}245 \\
$k_j=5$ & 1.82 & 0.034 & 0.60 & 3.99 & $<$0.001 & 987 & 2{,}200 \\
\multicolumn{8}{@{}l}{\textit{C. Jitter-averaged, DR, flexible first stage}} \\
$k_j=3$ & 2.12 & 0.017 & 0.95 & 3.71 & $<$0.001 & 1{,}056 & 1{,}888 \\
$k_j=3.5$ & 2.08 & 0.019 & 0.94 & 3.85 & $<$0.001 & 984 & 1{,}858 \\
$k_j=5$ & 1.70 & 0.045 & 0.50 & 4.02 & $<$0.001 & 716 & 1{,}728 \\
\multicolumn{8}{@{}l}{\textit{D. Jitter-averaged, inverse probability weighting, flexible first stage}} \\
$k_j=3.5$ & 2.46 & 0.007 & 1.00 & 4.24 & $<$0.001 & 1{,}481 & 2{,}604 \\
\hline\hline
\end{tabular*}
\end{center}
\footnotesize
\renewcommand{\baselineskip}{11pt}
\textbf{Note:} Corrected statistics (corrected centering and finite-$N$ variance), averaged over jitter draws (50 at $k_j=3.5$ for the DR rows, 20 otherwise; $31\times31$ grid): $\bar S_j$, the Holm-adjusted $\bar p_j=1-\Phi(\bar S_j)$, the share of draws rejecting $H_0^1$ at $5\%$ after Holm (``Rej.''; income is rejected in every draw) and the average $\widetilde D_j=(\Dh_j-\widetilde m_j)/\sqrt N$. The flexible first stage uses one series basis (cubic regression splines in age and income with four interior knots, three tensor terms, $V$, and $V$ interacted with age and income; 39 terms) for the logistic propensity and both outcome regressions. Its logit nests the cubic one, which it rejects (likelihood ratio 86.4 on 24 degrees of freedom); it is itself rejected against a richer basis (likelihood ratio 80.6 on 36, $p<0.001$) but passes the Hosmer--Lemeshow test ($p=0.079$; cubic logit $p<0.001$).
\end{table}
\begin{table}
\caption{\label{otab:emp-sens2}Further robustness checks of the 401(k) application.}
\begin{center}
\setlength{\tabcolsep}{4pt}
\begin{tabular*}{\textwidth}{@{\extracolsep{\fill}}lcccccc@{}}
\hline\hline
& \multicolumn{2}{c}{$H_0^{1}$ (age)} & \multicolumn{2}{c}{$H_0^{2}$ (income)} & \multicolumn{2}{c}{$\widetilde D_j$ (USD)} \\
\cline{2-3}\cline{4-5}\cline{6-7}
Specification & $\Sh_1$ & $\tilde p_1$ & $\Sh_2$ & $\tilde p_2$ & age & income \\
\hline
\multicolumn{7}{@{}l}{\textit{A. One jitter draw and integer-valued age, DR, cubic first stage}} \\
Jitter draw 1, $k_j=3.5$ & 2.44 & 0.007 & 4.00 & $<$0.001 & 1{,}203 & 2{,}028 \\
Jitter draw 1, $k_j=1.75$ (outside range) & 2.28 & 0.011 & 2.63 & 0.009 & 1{,}417 & 1{,}679 \\
Integer age, $k_j=3.5$ & 2.30 & 0.011 & 3.73 & $<$0.001 & 1{,}119 & 1{,}859 \\
Integer age, $k_j=5$ & 1.89 & 0.030 & 4.05 & $<$0.001 & 812 & 1{,}774 \\
\multicolumn{7}{@{}l}{\textit{B. Grid size}} \\
$k_j=3.5$, $M=13$ & 2.37 & 0.009 & 3.78 & $<$0.001 & 1{,}178 & 1{,}955 \\
$k_j=3.5$, $M=21$ & 2.44 & 0.007 & 3.97 & $<$0.001 & 1{,}207 & 2{,}022 \\
$k_j=3.5$, $M=31$ & 2.44 & 0.007 & 4.00 & $<$0.001 & 1{,}203 & 2{,}028 \\
$k_j=3.5$, $M=41$ & 2.39 & 0.008 & 4.01 & $<$0.001 & 1{,}177 & 2{,}027 \\
$k_j=5$, $M=13$ & 2.24 & 0.012 & 3.77 & $<$0.001 & 976 & 1{,}696 \\
$k_j=5$, $M=21$ & 2.40 & 0.008 & 4.21 & $<$0.001 & 1{,}039 & 1{,}871 \\
$k_j=5$, $M=31$ & 2.40 & 0.008 & 4.37 & $<$0.001 & 1{,}035 & 1{,}928 \\
\multicolumn{7}{@{}l}{\textit{C. Outcome, adjustment set and overlap}} \\
Winsorized 1\%/99\%, $k_j=3.5$ & 4.30 & $<$0.001 & 4.93 & $<$0.001 & 1{,}658 & 1{,}964 \\
Winsorized 1\%/99\%, $k_j=5$ & 4.40 & $<$0.001 & 5.27 & $<$0.001 & 1{,}481 & 1{,}819 \\
Winsorized 0.5\%/99.5\%, $k_j=3.5$ & 3.72 & $<$0.001 & 4.41 & $<$0.001 & 1{,}506 & 1{,}865 \\
Winsorized 0.5\%/99.5\%, $k_j=5$ & 3.89 & $<$0.001 & 4.91 & $<$0.001 & 1{,}371 & 1{,}797 \\
Reduced $V$, $k_j=3.5$ & 2.30 & 0.011 & 4.12 & $<$0.001 & 1{,}216 & 2{,}247 \\
Reduced $V$, $k_j=5$ & 1.91 & 0.028 & 4.73 & $<$0.001 & 885 & 2{,}240 \\
trimmed, $\widehat r_1\in[0.05,0.95]$ & 2.44 & 0.007 & 4.00 & $<$0.001 & 1{,}203 & 2{,}028 \\
Trimmed, $\widehat r_1\in[0.05,0.95]$ & 2.43 & 0.008 & 3.98 & $<$0.001 & 1{,}196 & 2{,}014 \\
\multicolumn{7}{@{}l}{\textit{D. Sample split, Theorem~\ref{thm:nuisance-split}}} \\
Half B, flexible, $k_j=3.5$ & 1.74 & 0.041 & 2.21 & 0.027 & 1{,}364 & 1{,}693 \\
Half B, flexible, $k_j=5$ & 0.74 & 0.229 & 2.13 & 0.033 & 501 & 1{,}395 \\
Half A, flexible, $k_j=3.5$ & 1.10 & 0.136 & 2.57 & 0.010 & 694 & 1{,}772 \\
Half A, flexible, $k_j=5$ & 1.12 & 0.132 & 2.50 & 0.013 & 612 & 1{,}497 \\
Combined, $k_j=3.5$ &  & 0.081 &  & 0.020 &  &  \\
Combined, $k_j=5$ &  & 0.263 &  & 0.025 &  &  \\
Half B, parametric, $k_j=3.5$ & 1.77 & 0.038 & 2.37 & 0.018 & 1{,}421 & 1{,}859 \\
Half A, parametric, $k_j=3.5$ & 0.80 & 0.211 & 2.46 & 0.014 & 517 & 1{,}721 \\
\hline\hline
\end{tabular*}
\end{center}
\footnotesize
\renewcommand{\baselineskip}{11pt}
\textbf{Note:} Corrected statistics for a single jitter draw (the first of those averaged in Table~\ref{otab:emp-sens}), DR pseudo-outcome, cubic first stage, $k_j=3.5$ and a $31\times31$ grid unless stated; ``Jitter draw 1'' at $k_j=3.5$ is that draw alone; the integer-age rows use integer-valued age without jitter, which the theory does not cover (Proposition~\ref{oprop:jitter}); $k_j=1.75$ lies outside the recommended range $[3,5]$. Panel C re-fits the first stage in every row; reduced $V$ drops IRA participation and home ownership; winsorizing the outcome changes the estimand. 16 households (15 below $0.05$, 1 above $0.95$) have fitted propensities outside $[0.05,0.95]$; trimming changes their weights, trimming drops them and re-fits the first stage. Panel D splits the sample at random within treatment arms ($4957$ and $4958$ households), fits the flexible first stage on one half and tests on the other, with the evaluation region of the full sample and bandwidths for the half-sample size; the combined rows use $\min\{1,2\min(p^A,p^B)\}$ in each direction before the Holm step, with $p^A$ and $p^B$ the p-values of the two assignments of the halves to the first stage and to the test; the ``parametric'' rows fit the cubic first stage on the same half, so that their difference from the full-sample rows is the cost of halving the sample.
\end{table} 

\textit{Income.} The effect varies with income at fixed age in every specification: every bandwidth in $[3,5]$, both pseudo-outcomes, both first stages, every grid size and every jitter draw, with the winsorized outcome, the reduced adjustment set, trimming, and in each half of the sample split (Tables~\ref{otab:emp-sens} and~\ref{otab:emp-sens2}).\\

\textit{Age.} The decision does not depend on the bandwidth, although the evidence is moderate: the jitter-averaged test rejects $H_0^1$ at every $k_j\in[3,5]$ with both pseudo-outcomes and both first stages, and the share of individual draws that reject falls as the windows widen (Table~\ref{otab:emp-sens}). The smaller statistics reflect the window width: at $k_j=5$ a window spans about half of the age range of the evaluation box, so variation in age that is local within the box is averaged away, while income magnitude is nearly unchanged; wider windows are therefore not used. The sample-split route does not confirm age, but this is the cost of halving the sample, not of the flexible first stage: the cubic first stage fitted on the same halves gives almost the same statistics, and at a fixed alternative $\Sh_j$ grows roughly like $\sqrt N$. Winsorizing the outcome strengthens the age result markedly, so a few extreme households hold it back rather than drive it, although winsorizing changes the estimand (Remark~\ref{orem:heavytails}). Dropping IRA participation and home ownership from $V$, which may respond to eligibility, and trimming the households with extreme fitted propensities leave the results essentially unchanged (Table~\ref{otab:emp-sens2}).\\

\textit{Grid.} The corrected statistics have converged from $M=21$, and any remaining movement affects no decision (Table~\ref{otab:emp-sens2}). The $13\times13$ grid is coarser than the spacing of at most $h_j/6$ recommended in Section~\ref{sec:impl}, (Section~\ref{osec:tuning}).\\

\textit{Max form.} The max form tests the null $H_0^{j,\ge}$ that $\tau$ is nonincreasing in block $j$, with the centering and variance of Proposition~\ref{oprop:maxvar}. At $k_j=3.5$ its studentized statistics are $2.60$ (age, $p=0.005$) and $4.62$ (income, $p<0.001$); including the first-stage term in the linear part of the variance gives $2.61$ and $4.61$. Under a same-sample parametric first stage, the max form is covered by the theory only if both working models are correct (Proposition~\ref{oprop:nui-ext} and the discussion after it), which is not credible here, so these values are reported as descriptive only.\\

\textit{Summary.} Heterogeneity in income at fixed age is clear and robust. Heterogeneity in age at fixed income is detected at conventional levels over the recommended bandwidths with the jitter-averaged statistic, but it weakens as the windows widen, is not detected in half samples, and its magnitude is not distinguishable from zero (Section~\ref{osec:emp-mag}).

\subsection{Magnitudes}\label{osec:emp-mag}

Table~\ref{otab:emp-mag} reports the Lebesgue-weighted magnitude $D_j$, the density-weighted magnitude $\Delta_j$ and the index $G_j$ of Section~\ref{sec:app-gini}, each as the bracket of Proposition~\ref{prop:gini-clt}, uncorrected (upper end) and noise-corrected (lower end), averaged over the jitter draws.

\begin{table}
\caption{\label{otab:emp-mag}Magnitudes of directional heterogeneity in the 401(k) application.}
\begin{center}
\setlength{\tabcolsep}{4pt}
\begin{tabular*}{\textwidth}{@{\extracolsep{\fill}}lcccccc@{}}
\hline\hline
& \multicolumn{3}{c}{Age ($j=1$)} & \multicolumn{3}{c}{Income ($j=2$)} \\
\cline{2-4}\cline{5-7}
& Uncorrected & Corrected & s.e. & Uncorrected & Corrected & s.e. \\
\hline
\multicolumn{7}{@{}l}{\textit{$k_j=3$, cubic first stage}} \\
$D_j$ (USD) & 5{,}891 & 1{,}078 & 745 & 6{,}707 & 1{,}819 & 788 \\
$\Delta_j$ (USD) & 8{,}528 & 1{,}528 & 1{,}359 & 9{,}739 & 2{,}714 & 1{,}338 \\
$G_j$ & 0.47 & 0.08 & 0.07 & 0.54 & 0.15 & 0.06 \\
\multicolumn{7}{@{}l}{\textit{$k_j=3.5$, cubic first stage}} \\
$D_j$ (USD) & 5{,}143 & 1{,}045 & 690 & 5{,}989 & 1{,}835 & 736 \\
$\Delta_j$ (USD) & 7{,}380 & 1{,}369 & 1{,}332 & 8{,}708 & 2{,}704 & 1{,}315 \\
$G_j$ & 0.43 & 0.08 & 0.07 & 0.51 & 0.16 & 0.07 \\
\multicolumn{7}{@{}l}{\textit{$k_j=5$, cubic first stage}} \\
$D_j$ (USD) & 3{,}571 & 759 & 561 & 4{,}604 & 1{,}764 & 643 \\
$\Delta_j$ (USD) & 5{,}050 & 881 & 1{,}126 & 6{,}838 & 2{,}719 & 1{,}233 \\
$G_j$ & 0.33 & 0.06 & 0.07 & 0.45 & 0.18 & 0.07 \\
\multicolumn{7}{@{}l}{\textit{$k_j=3.5$, flexible first stage}} \\
$D_j$ (USD) & 4{,}989 & 984 & 682 & 5{,}916 & 1{,}858 & 727 \\
$\Delta_j$ (USD) & 7{,}386 & 1{,}512 & 1{,}289 & 8{,}785 & 2{,}919 & 1{,}271 \\
$G_j$ & 0.42 & 0.09 & 0.07 & 0.51 & 0.17 & 0.06 \\
\hline\hline
\end{tabular*}
\end{center}
\footnotesize
\renewcommand{\baselineskip}{11pt}
\textbf{Note:} DR pseudo-outcome, $31\times31$ grid; averages over the jitter draws of Table~\ref{otab:emp-sens}, with standard errors averaged likewise. ``Uncorrected'' is the upper end of the bracket ($\Dh_j/\sqrt N$, $\widehat\Delta_j$, $\widehat G_j$), ``corrected'' the lower end, which subtracts the estimated noise floor (Proposition~\ref{prop:gini-clt}). Standard errors from the plug-in influence functions, including the first-stage term; they assume the margin condition. $G_j=\Delta_j/(2\E|\tau(X^{\mathcal R})|)\in[0,1]$ on the evaluation box, with $\E|\tau(X^{\mathcal R})|$ estimated by 8{,}496 USD at $k_j=3.5$ (DR mean over the box 7{,}677 USD; both averaged over the jitter draws); the fitted CATE function is negative on about 16\% of the density mass of the box, so $G_j$ is not a Gini coefficient of a single-signed CATE function.
\end{table}

\textit{The noise floor.} The noise floor accounts for most of the uncorrected magnitudes (Table~\ref{otab:emp-mag}), which is why the bracket of Proposition~\ref{prop:gini-clt} matters here. The age interval includes zero, although the test rejects $H_0^1$: the interval is valid in each of the regimes of that proposition and is not a test inversion, so it is wide. In words, two households in the same income stratum but of different ages differ in their eligibility effect by about $1{,}400$ USD on average at the lower end of the bracket and at most about $7{,}400$ at the upper end; two households of the same age but different incomes differ by about $2{,}700$ USD at the lower end and at most about $8{,}700$. Only the income magnitude is distinguishable from zero. The standard errors include the first-stage term of Proposition~\ref{oprop:nui-ext}(a), which changes them by less than $1\%$, and assume the margin condition of Proposition~\ref{prop:gini-clt}(c).\\

\textit{The bandwidth signature.} The uncorrected magnitudes fall steeply with the bandwidth while the corrected ones are comparatively stable (Table~\ref{otab:emp-mag}). Most of the decline comes from the noise floor, which is of order $(N\eta)^{-1/2}$ and thus falls as the windows widen. The remainder, a modest decline of $\widetilde D_1$, comes from smoothing that averages away variation in age local to the evaluation box (Section~\ref{osec:emp-rob}); $\widetilde D_2$ shows no such decline.\\

\textit{The index.} $G_j=\Delta_j/\{2\E|\tau(X^{\mathcal R})|\}$ is computed on the evaluation box, with numerator and denominator on the same population, and lies in $[0,1]$ by the triangle inequality; the estimated denominator is $8{,}496$ USD at $k_j=3.5$ (the DR mean over the box is $7{,}677$; both are averages over the $50$ jitter draws). The fitted CATE function is negative on about $16\%$ of the density mass of the box ($\tauh$ ranges from $-31.7$ to $50.4$ thousand USD on the grid), so $G_j$ is not the Gini coefficient of a single-signed CATE function, and $\E|\tau(X^{\mathcal R})|$ is itself estimated with an upward noise bias (Section~\ref{osec:gini-if}); for this reason the main text reports $D_j$ and $\Delta_j$ and not $G_j$, which is given here as a descriptive summary only. The uncorrected index does not exceed $0.67$ in any specification. Dividing instead by the absolute full-sample average effect would not bound the index by one: with the IPW pseudo-outcome at $k_j=3$ (first jitter draw) it would be $1.11$ for income.

\section{Further discussion of the literature}\label{osec:lit}
This section collects literature that supports Sections~\ref{sec:intro},  \ref{sec:app-hom}, and  \ref{sec:asym} of the main text.

\subsection{Relation to the one-point DR test of Shi, Song and Lu}\label{osec:lit-ssl}

\citet{shi2019testing} consider a one-window DR kernel estimator of the density-weighted conditional effect, built from parametric propensity and outcome models fitted on the same sample, with root-$n$ linear representations, assuming either model is correctly specified. They show that a centered and studentized one-point Poissonized $L_1$-type statistic has asymptotic size at most the nominal level, with equality on a nondegenerate contact set, for the null of no conditional qualitative treatment effect, which states that the effect has one sign in the covariates of interest given the others, and that the test is consistent; their scale contains no first-stage term. They use the same fourth-order kernel and the bandwidth rate $h\propto n^{-1/7}$. Theorem~\ref{thm:nuisance-param} is the two-point counterpart of their Theorem~4.3, and adds three things. First, a two-point functional whose two windows are coupled at arbitrary separations: the Poissonization and one-dependent-cell argument of \citet{gine2003bm}, \citet{chang2015nonparametric} and \citet{shi2019testing} applies only after the H\'ajek-type projection of Lemma~\ref{olem:pf-projection}. Second, the directional constancy null $H_0^j$: under $H_0^1$ their contact set is empty, so the rejection probability of their test tends to zero there (their Remark~3.7); their test is a test of a sign null, not of $H_0^1$. Third, Lemma~\ref{lem:drift} characterizes the effect of first-stage drift on an integrated $L_1$ functional under the null.
For the absolute-value functional $a_\Psi|\cdot|$ on symmetric
configuration sets, near-symmetry of the noise field eliminates
the first-order effect. The remaining contribution is bounded
by a multiple of $N\eta^{1/2}\|\delta_N\|_2^2$.
For positive-part functionals on ordered configuration sets,
the drift instead contributes at root-$N$ order through the
linear term $b_\Psi\sqrt N\int w_{\mathcal C_j}\delta_N$. The same-sample parametric case follows by a uniformity argument in the parameter. Theorem~\ref{thm:nuisance-param} adds no new DR or misspecification theory: it follows the same ``either model correct'' structure and root-$N$ first-stage representations, does not cover cross-fitting, and excludes the max form under a same-sample parametric first stage.

\subsection{Second-stage regression of pseudo-outcomes}\label{osec:lit-second}

The relevant rate benchmark for Theorem~\ref{thm:nuisance-split} is the second-stage kernel or series regression of a DR pseudo-outcome on a covariate subvector \citep{fan2022estimation,kennedy2023towards,semenova2021debiased,zimmert2019nonparametric}, in which pointwise inference requires the smoothed first-stage drift to be $o((N\eta)^{-1/2})$. For $\zeta^{\mathrm{DR}}$, the drift condition of Theorem~\ref{thm:nuisance-split} holds if $\|\widehat r-r\|_4\|\widehat\mu-\mu\|_4=o_p(N^{-1/2}\eta^{-1/4})$ and $\|\widehat r-r\|_\infty\|\widehat\mu-\mu\|_\infty=o_p(\eta^{1/2})$ (Remark~\ref{rem:nuisance-rate}). The first part is stronger by a factor of $\eta^{1/4}$ than the pointwise requirement, as the drift affects the integrated statistics through the term of order $N\eta^{1/2}\|\delta_N\|_2^2$ in Lemma~\ref{lem:drift}. This condition and the root-$N$ product rate $\|\widehat r-r\|_2\|\widehat\mu-\mu\|_2=o_p(N^{-1/2})$ of double machine learning \citep{chernozhukov2018double} do not imply each other: the first part of the condition is weaker than the latter by the factor $\eta^{-1/4}$ only when the $L_4$ and $L_2$ norms of the first-stage errors are of the same order, as when the errors are spread over the covariate space, and the sup-norm part has no counterpart in double machine learning. Cross-fitted first stages are covered for series projections of the CATE by \citet{semenova2021debiased} and for the integrated conditional moment test of \citet{lapenta2026machine}. Under cross-fitting, the nuisances for $\widehat\zeta_i$ depend on observations of other folds, and the conditioning argument behind Theorem~\ref{thm:nuisance-split} fails. The extension is left open, and the claim that the same rates suffice is a conjecture.

\subsection{Monotonicity}\label{osec:lit-mono}

The max form tests a monotonicity hypothesis for the CATE of $X_j$ given $X_{j'}$, at every value of $X_{j'}$. Tests of regression monotonicity include \citet{ghosal2000testing}, \citet{hall2000testing}, \citet{chetverikov2019testing} and \citet{lee2018testing}, and \citet{lee2009testing} test stochastic monotonicity; pairwise differences enter \citet{ghosal2000testing} and \citet{lee2009testing} as kernel-weighted signs of pairs of observations in a U-process, not as integrals of differences of a kernel field. For treatment effects, \citet{hsu2021testing} test monotonicity of the conditional average effect at a regression discontinuity cutoff, with uniform size control, and \citet{hsu2026testing} test monotonicity of mean potential outcomes in a continuous treatment with machine-learning nuisances.

\subsection{Multiple testing and finer classifications}\label{osec:lit-multiple}

With two hypotheses, the second step of Holm's procedure already runs at level $\alpha$, so a procedure that exploits the dependence between $\Sh_1$ and $\Sh_2$ can improve only the first-step cutoff, from $\alpha/2$ to at most $\alpha$. The step-up procedure of \citet{hochberg1988sharper}, which rejects both hypotheses when $p_{(2)}\le\alpha$, realizes part of this gain and is valid under independence or positive dependence \citep{sarkar1998some}, but not in general. With more than two blocks, the resampling stepdown methods of \citet{romano2005stepwise} could be used, provided their critical values are monotone across subsets of hypotheses, which must be verified for the statistics at hand; \citet{romano2010hypothesis} survey these methods, and \citet{lee2014multiple} apply them to subgroup treatment effects. An additivity test \citep{sperlich2002nonparametric} combined with two main-effect tests would classify more finely, separating a pure interaction such as \eqref{eq:interaction-example} from main effects, at the cost of three tests and an additive fit. Other nonparametric tests of treatment effect heterogeneity in observational studies \citep{dai2023nonparametric,sant2021nonparametric}, randomization inference on treatment effect variation in experiments \citep{ding2016randomization,ding2019decomposing} and variable selection for qualitative interactions \citep{gunter2011variable} address related but different questions: whether the effect varies at all, how much of its variation the covariates explain, and which covariates change its sign.

\setlength{\bibsep}{0pt plus 0.3ex}
{\small\SupplementReferences

}

\end{document}